%% file: main.tex
\providecommand{\JCPLayout}{reprint}
\documentclass[aip,jcp,floatfix,longbibliography,amsmath,amssymb,\JCPLayout]{revtex4-1}

\input{preamble_jcp}

\input{build/canonical_values}

\input{revision_tracking}

\graphicspath{{figures/output/}}

\newcommand{\kifstar}{k_{\mathrm i\rightarrow\mathrm f}^{*}}
\newcommand{\kfistar}{k_{\mathrm f\rightarrow\mathrm i}^{*}}
\newcommand{\keffstar}{k_{\mathrm{eff}}^{*}}

\begin{document}

\title{\revphasenine{First-fracture kinetics in periodic chains:
from free-energy landscapes and dynamical recrossing to absolute rates}}
\author{Aristotelis P. Sgouros}
\email{asgouros@eie.gr}
\affiliation{Theoretical and Physical Chemistry Institute,
National Hellenic Research Foundation,
Vass. Constantinou 48, GR-11635 Athens, Greece}

\begin{abstract}
\setlength{\parindent}{0pt}
\revphasetwelve{

\noindent The nearly collinear configuration of highly stretched polymer chains
before fracture motivates one-dimensional models. At fixed extension, stretching
one bond redistributes length among the others, coupling rupture to the whole
chain. In this work, I develop a framework for fracture--healing rates in periodic Morse
chains, combining deterministic theories, equilibrium simulations and a
quasi-deterministic approach using equilibrium fluxes and barrier-top
trajectories. Constrained-partition calculations closely reproduce simulated
populations and intact-state tensions. An intact-reservoir potential of mean
force (PMF) supplies the first-fracture surface probability, while recrossing
corrects the attempt flux without a rate-fitted prefactor. Independent rate
tests identify an operational chain-length boundary for accurate two-state
kinetics; failed-return durations decrease relative to population relaxation
as chains grow. Kramers--Grote--Hynes (KGH), nonlinear reactive-flux and
equilibrium rates show excellent agreement in this regime. KGH predicts
rates away from coexistence and a chain-dependent local activation length dominated
by the barrier response. Rates decrease approximately exponentially before
declining more gently for longer chains: barrier growth dominates, additional
fracture sites partly compensate, and transmission varies more weakly. Beyond
coexistence, long chains can fracture practically irreversibly because healing
becomes too rare to observe. Projected committor crossings, where the
configuration-averaged probability of reaching the fractured core first is
one half, lie near site-PMF maxima, although configurations at equal bond
length retain different fracture probabilities. Multiple fractures remain
negligible near coexistence through at least \(N=50\). The framework could supply
state-dependent rates to mesoscopic fracture models of polymer networks.}
\end{abstract}

\maketitle

\input{sections/01_introduction}
\input{sections/02_model_and_theory}
\input{sections/03_computational_methods}
\input{sections/04_results}

\input{sections/07_conclusions}

\section*{Supplementary material}
\revphaseseven{The supplementary material contains supporting derivations,
numerical checks, and additional results. References
\citep{KaplanMeier1958,Ayer1955,Byrd1995,Metropolis1953,RomanoWolf2005,PetersenPedersen2012,Bartsch2005}
are cited therein.}

\begin{acknowledgments}

The development of the framework was realized on the Helios HPC cluster
in the framework of the project ``Center of Excellence for Advanced Organic
Materials, Theranostics in Cancer and Bioelectronics'' (Project ID: 5173962).
The author greatly acknowledges fruitful discussions with Prof. Doros N. Theodorou.
AI-based tools were used to assist with code refinement and language editing.
All algorithms and results were independently developed, verified, and
validated by the author.

\end{acknowledgments}

\section*{Author declarations}
\subsection*{Conflict of interest}
The author has no conflicts to disclose.

\subsection*{Author contributions}
Aristotelis P. Sgouros: Conceptualization; Methodology; Software;
Validation; Formal analysis; Investigation; Visualization;
Writing -- original draft; Writing -- review \& editing.

\section*{Data and code availability}
The data supporting the figures and tables, together with a frozen snapshot
of the code and reproduction scripts, are available on Zenodo at
\url{https://doi.org/10.5281/zenodo.23318388}. The archive reproduces the article and supplementary
material from the final analysis results.

\bibliography{references}

\end{document}


\maketitle
\tableofcontents

\pretocmd{\section}{\clearpage}{}{\PackageError{SI}{Cannot add section page breaks}{}}

\input{si_sections/notation_table}
\input{si_sections/notation_and_estimands}
\input{si_sections/constrained_thermodynamics}
\input{si_sections/conditional_equilibrium_profiles}
\input{si_sections/coexistence_midpoint}
\input{si_sections/site_barriers}
\input{si_sections/langevin_dynamics}
\input{si_sections/deterministic_rates}
\input{si_sections/equilibrium_simulation_rates}
\input{si_sections/barrier_top_simulations}
\input{si_sections/fracture_rate_results}
\input{si_sections/committor_diagnostics}
\input{si_sections/kinetic_crossover}

\bibliographystyle{unsrtnat}
\bibliography{references}

%% file: preamble_jcp.tex
\usepackage[T1]{fontenc}
\usepackage[utf8]{inputenc}
\usepackage{mathptmx}
\usepackage{microtype}
\usepackage{mathtools}
\usepackage{bm}
\usepackage{graphicx}
\graphicspath{{figures/output/}{figures/assets/}}
\usepackage{booktabs}
\usepackage{tabularx}
\usepackage{multirow}
\usepackage[dvipsnames]{xcolor}
\definecolor{linkblue}{RGB}{25,79,128}
\definecolor{issuebg}{RGB}{255,247,230}
\usepackage[colorlinks=true,allcolors=linkblue]{hyperref}
\usepackage[nameinlink,noabbrev]{cleveref}
\usepackage{etoolbox}
\makeatletter
\def\@email#1#2{%
 \endgroup
 \patchcmd{\titleblock@produce}
  {\frontmatter@RRAPformat}
  {\frontmatter@RRAPformat{\produce@RRAP{*#1\href{mailto:#2}{#2}}}\frontmatter@RRAPformat}
  {}{\PackageError{JCP}{Cannot place corresponding-author email}{}}
}
\makeatother
\input{notation}

%% file: notation.tex
\ExplSyntaxOn
\NewDocumentCommand{\ScientificValue}{m}{
  \group_begin:
  \tl_set:Nx \l_tmpa_tl { \fp_to_scientific:n {#1} }
  \seq_set_split:NnV \l_tmpa_seq {e} \l_tmpa_tl
  \fp_eval:n {\seq_item:Nn \l_tmpa_seq {1}}
  \times 10^{\seq_item:Nn \l_tmpa_seq {2}}
  \group_end:
}
\ExplSyntaxOff

\newcommand{\De}{D_{\mathrm e}}
\newcommand{\Tstar}{T^*}
\newcommand{\Fstar}{F^*}
\newcommand{\lhalf}{\lambda_{1/2}}

%% file: build/canonical_values.tex
\newcommand{\CanonicalNtenFixedSurfaceLambda}{1.1073538970587944}

\newcommand{\RegisteredPmfCoexistenceLambda}{1.107354}

\newcommand{\RegisteredCombinedOccupancyRmse}{0.003637}
\newcommand{\RegisteredCombinedVirialRmse}{0.0000607}

\newcommand{\RegisteredCrossoverTauSix}{442.04}
\newcommand{\RegisteredCrossoverMedianSixI}{78}

\newcommand{\RegisteredCrossoverMedianSixF}{118}

\newcommand{\RegisteredCrossoverTauTen}{40291.20}
\newcommand{\RegisteredCrossoverMedianTenI}{91}

\newcommand{\RegisteredCrossoverMedianTenF}{162}

\newcommand{\RegisteredCommittorSdSix}{0.084}

\newcommand{\RegisteredCommittorSdEight}{0.109}

\newcommand{\RegisteredCommittorSdTen}{0.123}

\newcommand{\RegisteredCommittorPmfOffsetSix}{0.0053}
\newcommand{\RegisteredCommittorPmfOffsetEight}{0.0030}
\newcommand{\RegisteredCommittorPmfOffsetTen}{0.0014}

%% file: revision_tracking.tex
\definecolor{revisionphaseone}{RGB}{0,0,0}

\definecolor{revisionphasetwo}{RGB}{0,0,0}

\definecolor{revisionphasethree}{RGB}{0,0,0}

\definecolor{revisionphasefour}{RGB}{0,0,0}

\definecolor{revisionphasefive}{RGB}{0,0,0}

\definecolor{revisionphasesix}{RGB}{0,0,0}

\definecolor{revisionphaseseven}{RGB}{0,0,0}
\DeclareRobustCommand{\revphaseseven}[1]{#1}
\definecolor{revisionphaseeight}{RGB}{0,0,0}

\definecolor{revisionphasenine}{RGB}{0,0,0}
\DeclareRobustCommand{\revphasenine}[1]{#1}
\definecolor{revisionphaseten}{RGB}{0,0,0}

\definecolor{revisionphaseeleven}{RGB}{0,0,0}

\definecolor{revisionphasetwelve}{RGB}{0,0,0}
\DeclareRobustCommand{\revphasetwelve}[1]{#1}
\definecolor{revisionphasethirteen}{RGB}{0,0,0}

\definecolor{revisionphasefourteen}{RGB}{0,0,0}

\definecolor{revisionphasefifteen}{RGB}{0,0,0}

%% file: sections/01_introduction.tex
\section{Introduction}
\label{sec:introduction}

\revphasetwelve{The rupture of a load-bearing polymer strand is a molecular
event with consequences for network damage, material durability and
macroscopic stress \citep{Theodorou2024,Vernerey2018,White2025}.
Understanding polymer fracture therefore requires connecting strand-level
failure to the redistribution of stress through a network.
Experiments link molecular damage to the toughness of multiple-network
elastomers \citep{Slootman2022} and to the different energies required
for cutting and tearing \citep{Zhao2025}.
Experiments on end-linked networks also show that strand reactivity and
network connectivity jointly influence tearing energy \citep{Beech2023}.
Statistical-physics models address this connection at several scales.
Extensible-chain statistical mechanics relates bond-deformation energetics
to elastomer fracture through a critical-energy scission criterion
\citep{Mao2017}, while network-level descriptions relate loop defects and
energy released by surrounding strands to the fracture energy of model
end-linked gels \citep{Barney2022}.
At high strain rates, simulations of polymer gels show how entanglements
and scission of network and solvent chains contribute to fracture
resistance \citep{Sliozberg2014}.
Mesoscopic models link molecular mechanisms to interfacial fracture
\citep{Terzis2002} and polymer-mediated adhesion \citep{Sgouros2025}.
Network models describe reversible chain attachment and detachment
\citep{Kothari2018,Lamont2021}, as well as their competition with irreversible
chain rupture \citep{Lamont2021}; discrete mesoscale models distinguish chain rupture from cohesive
instability under different loading paths \citep{Lamont2025}.
For coarse-grained models that resolve stochastic scission and stress
redistribution \citep{Arora2022,AraujoBrassart2025},
strand-level rupture rates provide the kinetic inputs needed to describe
when chains break \citep{Theodorou2024,Anogiannakis2015}.

For strongly stretched strands, an idealized one-dimensional chain
provides a simple setting for studying how bond energetics, chain length
and dissipation influence rupture \citep{Theodorou2024}. At fixed
extension, stretching one bond redistributes the available length among
the others, coupling local rupture to the mechanical and thermal response
of the remaining chain. Obtaining a rate directly from molecular dynamics
remains difficult because fracture can be rare \citep{Lazzeri2023},
its coordinate is collective, and the observed rate
combines equilibrium access to the transition region with dynamical
recrossing.}

\revphasetwelve{Earlier studies developed quantitative rupture-rate theories
for finite chains using transition-state theory (TST)
\citep{Eyring1935,PuthurSebastian2002}, effective-coordinate escape descriptions
\citep{Oliveira1998}, and multidimensional Kramers--Langer or threshold-escape
approaches \citep{Langer1969,Sain2006,Ghosh2010,Lee2009}, with comparisons to simulation.
Together these studies examined activated rupture, collective motion,
dissipation, boundary and fracture-site effects, and recombination.
\citet{Paturej2013} give a broader account of anharmonic-chain rupture
and comparisons with Kramers--Langer theory.
Earlier coupled-Morse-chain work established an approximately \(N^{-1/2}\)
decrease of an energetic critical strain, with activation barriers increasing
along the corresponding critical-extension sequence \citep{Crist1984}.
These quantitative theories provide the foundation for the present study,
while addressing related systems under different loading and boundary
conditions.}

\revphasetwelve{\citet{Theodorou2024}
introduced the periodic fixed-extension Morse chain and
established the unrestricted single-bond potential of mean force (PMF) and its
adiabatic branches, reversible intact and fractured states,
rate extraction from the tension autocorrelation function (ACF) at \(N=10\),
and a local Zhurkov
representation. The present work builds on this model and equilibrium
description to connect constrained first-fracture thermodynamics with
absolute kinetics over a range of chain lengths and extensions.
A complementary treatment of displacement-controlled scission and healing
was developed by \citet{Ansari2027}, who calculated free-energy barriers
for a freely jointed chain with one extensible Lennard--Jones link and
rigid remaining links, and incorporated these barriers into a two-state
kinetic model.}

\revphasetwelve{
The present study examines a periodic Morse chain held at fixed extension,
in which every bond can stretch, fracture and heal. The length constraint couples
local rupture to the mechanical and thermal response of the remaining chain.
Constrained first-fracture thermodynamics are connected to absolute rates under
prescribed Langevin dynamics, accounting explicitly for dynamical recrossing.
Comparisons with equilibrium simulations test these predictions across chain
lengths and extensions, while committor calculations and independent kinetic
tests assess the reaction coordinate and the validity of a two-state description.}

The fixed-extension constraint raises several connected physical questions.
How does chain length shift coexistence and reshape the free-energy landscape?
Why can a longer, finitely extended chain fracture more slowly at coexistence
despite offering more rupture sites, and how does its rate respond to additional
stretching?
Can a single bond length identify the average transition while distinguishing
individual configurations' fracture probabilities? Finally, under what conditions
does the microscopic fracture--healing motion admit a reliable description by
two constant directional rates? The thermodynamic questions are addressed through
constrained-partition calculations and comparisons with simulated populations and
intact-state tensions. Resolving the partition weight by the number of
fractured bonds also tests when a one-fracture approximation adequately
represents the fractured population.

The central kinetic question is how constrained equilibrium information and
dynamical recrossing can predict absolute first-fracture rates across chain
lengths and extensions. An unrestricted tagged-bond PMF includes
configurations already fractured elsewhere and therefore does not give the
appropriate surface probability for first fracture. Conditioning
the remaining bonds to be intact gives a site PMF that supplies the
equilibrium weight of the fracture surface; combined with the constrained
coordinate's kinetic metric, it determines the attempt flux. Trajectories
launched from the dividing surface (barrier-top simulations) supply the
nonlinear recrossing correction \citep{Bennett1977,Chandler1978}, without a
kinetic prefactor fitted to measured rates. Rates from equilibrium
trajectories test this prediction, while separating the barrier, fracture-site
and dynamical contributions explains the chain-length dependence of the rate.

A state-specific Kramers--Grote--Hynes (KGH) reduction
\citep{Kramers1940,GroteHynes1980} provides a compact prediction from the
site-PMF barrier, curvatures and prescribed dynamics. This study uses its
Markovian, locally harmonic form along the symmetric path and tests it
at coexistence and across sampled extensions. A local Zhurkov-type
representation \citep{Zhurkov1965,Suhir2013} describes the rate's tension
dependence through an activation length, which measures the sensitivity
of the logarithmic rate to force. A further question is how this length depends on chain
size and how much of the response comes from the barrier rather than the
dynamical prefactor. Such descriptions could supply mesoscopic fracture
models with rates that retain the loading-state and chain-length dependence
of the underlying molecular system \citep{Buche2021}.

Committor analysis tests how well tagged-bond length describes the transition
\citep{PetersTrout2006}. At a fixed configuration, the committor is the
probability of reaching the fractured core before the intact core, averaged
over thermal velocities and dynamical noise. Averaging further over
configurations with the same tagged-bond length gives the projected
committor. Repeated trajectory launches estimate both quantities. Comparing
the projected half-probability crossing with the adiabatic and site-PMF
saddles tests which landscape locates the average transition. The spread of
configuration committors at that length tests how well bond length alone
distinguishes individual fracture probabilities.

A complementary question is when a pair of constant directional rates
adequately describes switching. A metastable two-state picture relies on
rapid local motion and loss of memory relative to population relaxation
\citep{Chandler1978}.
\revphasetwelve{For systems with memory, rapid state recrossing can lead to
different mean passage times under different event-sampling conventions
\citep{Zhou2025}.}
The kinetic accuracy of the two-state picture is tested by comparing
independently estimated rates across observation lags, core definitions
and recording intervals. The durations of failed
excursions that return to their original core are compared with the population-relaxation time;
this comparison measures the separation of excursion and population timescales.
All in all, these rate and duration tests assess when microscopic
recrossing is compatible with a useful two-state description.

Section~\ref{sec:model-theory} defines the model, states and constrained
thermodynamics; Sec.~\ref{sec:methods} describes the calculations and rate
estimators. Section~\ref{sec:results} presents equilibrium, kinetic,
committor and crossover results, and Sec.~\ref{sec:conclusions} summarizes
their implications. Derivations and numerical checks are provided in the
Supplementary Information.

%% file: sections/02_model_and_theory.tex
\section{Model, state definitions, and equilibrium theory}
\label{sec:model-theory}

\subsection{Periodic fixed-extension Morse chain}
\label{sec:model}

\begin{figure*}[!tbp]
  \centering
  \includegraphics[width=16cm]{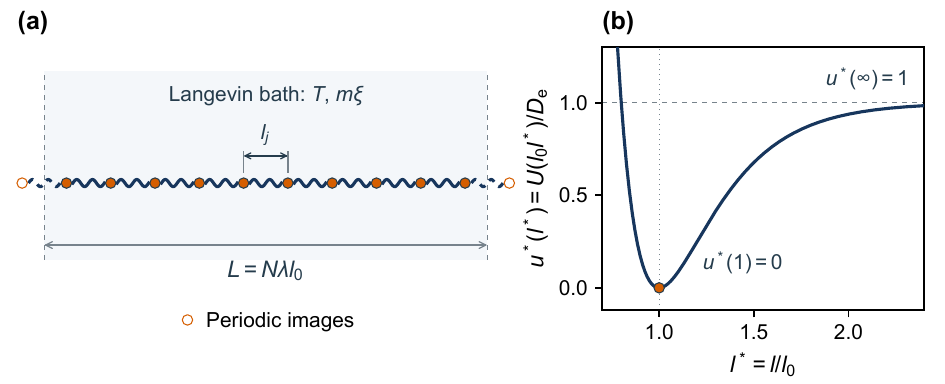}
  \caption{(a) Periodic one-dimensional Morse chain at fixed total extension
  \(L=N\lambda l_0\).  Open circles denote periodic images, and dashed end
  bonds indicate continuation across the cell boundaries.
  (b) Corresponding single-bond potential, with the minimum-zero energy
  reference of \cref{eq:morse-paper}.}
  \label{fig:model-chain}
\end{figure*}

The chain contains \(N\) equal-mass particles and \(N\) nearest-neighbor
bonds in a periodic one-dimensional cell of length \(L\)
(Fig.~\ref{fig:model-chain}a).
With bond lengths
\(l_{j}\), isolated-bond equilibrium length \(l_{0}\), and mean stretch
\(\lambda=L/(Nl_{0})\), the reduced lengths \(l_{j}^{*}=l_{j}/l_{0}\) satisfy
\begin{equation}
  \sum_{j=1}^{N}l_{j}^{*}=N\lambda\equiv L^{*}=\frac{L}{l_0}.
  \label{eq:fixed-extension-paper}
\end{equation}
The particle positions define the configuration \(\bm r=(x_1,\ldots,x_N)\).
Their reduced coordinates are \(x_j^*=x_j/l_0\), with
\(l_j^*=x_j^*-x_{j-1}^*\), using the periodic closure for the boundary bond.
All bonds are equivalent candidate fracture sites by cyclic symmetry.

Each bond is described by a Morse potential \citep{Morse1929} of depth
\(\De\) and inverse width \(\alpha\) (Fig.~\ref{fig:model-chain}b). Its reduced form is
\begin{equation}
  u^{*}(l^{*})=\frac{U(l_{0}l^{*})}{\De}
  =\left[1-e^{-a(l^{*}-1)}\right]^2,
  \qquad a=\alpha l_{0},
  \label{eq:morse-paper}
\end{equation}
The reduced chain energy is \(\sum_j u^{*}(l_{j}^{*})\), with \(u^{*}(1)=0\) and
\(u^{*}(\infty)=1\). An elongated bond loses load-bearing capacity but retains
its interaction, allowing the chain to heal.

Langevin dynamics \citep{Kramers1940} couples the particles to a bath at temperature \(T\)
with friction rate \(\xi\) (SI Sec.~S7.1). The reduced temperature,
tension, time, and rate are defined by
\begin{equation}
  \Tstar=\frac{k_{\mathrm B}T}{\De},\qquad
  \Fstar=\frac{Fl_{0}}{\De},\qquad
  t^*=\xi t,\qquad
  k^*=\frac{k}{\xi}.
  \label{eq:reduced-variables-paper}
\end{equation}
The parameter \(c=\De/(m\xi^2l_{0}^{2})\), with particle mass \(m\), sets the
Morse-force strength in the reduced dynamics. The physical model parameters
follow \citet{Theodorou2024} and are motivated by C--C bonding and
CH$_2$-scale dynamics in polyethylene at 450~K. The baseline parameters
and adopted numerical timestep are listed in
\cref{tab:baseline-parameters}.
Hereafter, stars distinguish reduced quantities from their dimensional counterparts,
whereas the dimensionless groups \(a,c,\lambda\), probabilities and ratios remain
unstarred. SI Sec.~S1, Table~S1, reports the symbols and their local conventions.

\begin{table}[tbp]
  \centering
  \caption{Baseline production parameters. Mass and Morse depth are listed
  in molar units; conversion to particle quantities is specified in SI
  Sec.~S2.1. Numerical settings and checks are detailed in SI Sec.~S7;
  trajectory lengths and sampling are specified in SI Secs.~S9, S10,
  S12 and S13.}
  \label{tab:baseline-parameters}
  \begin{tabular}{llll}
    \toprule
    Quantity & Symbol & Value & Unit \\
    \midrule
    Molar mass & \(N_{\mathrm A}m\) & 14.0 & \(\mathrm{g\,mol^{-1}}\) \\
    Temperature & \(T\) & 450.0 & \(\mathrm K\) \\
    Morse depth & \(N_{\mathrm A}\De\) & 88.62332696 & \(\mathrm{kcal\,mol^{-1}}\) \\
    Inverse width & \(\alpha\) & 2.248 & \(\mathrm{\mathring A^{-1}}\) \\
    Equilibrium length & \(l_{0}\) & 1.54 & \(\mathrm{\mathring A}\) \\
    Shape & \(a=\alpha l_{0}\) & 3.46192 & -- \\
    Friction rate & \(\xi\) & 0.022 & \(\mathrm{fs^{-1}}\) \\
    Reduced temperature & \(\Tstar\) & 0.0100903673 & -- \\
    Dynamic parameter & \(c=\De/(m\xi^2l_{0}^{2})\) & 2.3074106163 & -- \\
    Time step & \(\Delta t\) & 0.25 & \(\mathrm{fs}\) \\
    \bottomrule
  \end{tabular}
\end{table}

\begin{samepage}
\subsection{State classification}
\label{sec:state-classification}

The chain is classified using either its instantaneous configuration or
the basin core most recently visited. Both approaches use the intact
minimum \(l_{\mathrm i}^{\mathrm{ad},*}=\lambda\), fractured minimum
\(l_{\mathrm f}^{\mathrm{ad},*}\), and intervening saddle
\(d_{N}^{*}(\lambda)\equiv l_{\mathrm{ad}}^{\ddagger,*}(N,\lambda)\)
of the adiabatic energy profile.

Following \citet{Theodorou2024}, one bond is held at length \(l^{*}\),
and the remaining length is shared equally among the other \(N-1\) bonds.
The energy relative to the uniformly stretched chain is
\begin{equation}
\begin{aligned}
 E_{\mathrm{ad}}^{(N-1),*}(l^{*};N,\lambda)
 &=u^{*}(l^{*})+(N-1)u^{*}\!\left(\frac{N\lambda-l^{*}}{N-1}\right)\\
 &\quad-Nu^{*}(\lambda).
\end{aligned}
  \label{eq:adiabatic-paper}
\end{equation}

\end{samepage}

\textit{Instantaneous classification.}
The chain is intact when every bond lies at or below the saddle
threshold, and fractured when any bond exceeds it. For
\(\bm l^{*}=(l_{1}^{*},\ldots,l_{N}^{*})\), the state indicators are
\begin{equation}
  h_{\mathrm f}(\bm l^{*};N,\lambda)
  =\mathbf 1\!\left[\max_j l_{j}^{*}>d_{N}^{*}(\lambda)\right],
  \qquad h_{\mathrm i}=1-h_{\mathrm f}.
  \label{eq:state-paper}
\end{equation}
Here \(\mathbf 1[\mathcal A]\) is one when condition \(\mathcal A\)
holds and zero otherwise. This instantaneous classification defines the
equilibrium basin populations.

\textit{Core classification.}
To count transitions between regions inside the basins, intact and fractured
cores are introduced. A core fraction \(0<\eta<1\) sets their
boundaries between each adiabatic minimum and the saddle:
\begin{equation}
 c_{b}^{*}(\eta)=l_{b}^{\mathrm{ad},*}+\eta(d_{N}^{*}-l_{b}^{\mathrm{ad},*}),
 \qquad b\in\{\mathrm i,\mathrm f\}.
 \label{eq:cores-paper-c}
\end{equation}
Larger \(\eta\) extends the cores toward the saddle. Unless otherwise
stated, \(\eta=0.5\), placing each boundary halfway between its minimum
and the saddle.
\begin{equation}
 \mathcal C_b(\eta)=
 \begin{cases}
  \{\bm l^{*}:\max_jl_{j}^{*}\le c_{b}^{*}(\eta)\}, & b=\mathrm i,\\
  \{\bm l^{*}:\max_jl_{j}^{*}\ge c_{b}^{*}(\eta)\}, & b=\mathrm f.
 \end{cases}
 \label{eq:cores-paper-C}
\end{equation}
A trajectory is labelled intact after visiting \(\mathcal C_{\mathrm i}(\eta)\)
and fractured after visiting \(\mathcal C_{\mathrm f}(\eta)\). Between the
cores it retains its last label, so a transition requires reaching the
opposite core. Repeated saddle crossings alone do not count as transitions;
the label remains undefined before the first core visit.

\subsection{Constrained equilibrium and the fracture midpoint}
\label{sec:equilibrium-theory}

Following the fixed-extension bond-coordinate formulation of
\citet{Theodorou2024}, kinetic and translational factors are removed, and
the configurational partition density is expressed in the reduced variables used here:
\begin{equation}
  Z_{N}^{*}(L^{*})=\int_{(0,\infty)^N}
  e^{-\sum_j u^{*}(l_{j}^{*})/\Tstar}
  \delta\!\left(\sum_jl_{j}^{*}-L^{*}\right)\,d\bm l^{*}.
  \label{eq:partition-paper}
\end{equation}
Here \(d\bm l^{*}=\prod_{j=1}^{N}dl_{j}^{*}\). The domain
\((0,\infty)^N\) includes all positive bond lengths, while the Dirac delta
restricts their sum to the imposed chain length \(L^{*}=N\lambda\). Inserting the product of intact-bond
indicators restricts the ensemble to configurations with every bond
at or below the divider:
\begin{equation}
\begin{aligned}
 Z_{N}^{\mathrm i,*}(L^{*})
 &=\int_{(0,\infty)^N}\prod_j\Theta[d_{N}^{*}(\lambda)-l_{j}^{*}]\\
 &\quad\times e^{-\sum_j u^{*}(l_{j}^{*})/\Tstar}
 \delta\!\left(\sum_jl_{j}^{*}-L^{*}\right)\,d\bm l^{*}.
\end{aligned}
  \label{eq:intact-partition-paper}
\end{equation}
The step function \(\Theta(y)\) is one for \(y\ge0\) and zero otherwise,
so the product excludes configurations with any bond above \(d_{N}^{*}\).
The fractured partition is \(Z_{N}^{\mathrm f,*}=Z_{N}^{*}-Z_{N}^{\mathrm i,*}\).
The equilibrium probabilities and their logarithmic ratio (logit) are
\begin{equation}
  P_{\mathrm i}=\frac{Z_{N}^{\mathrm i,*}}{Z_{N}^{*}},
  \qquad P_{\mathrm f}=1-P_{\mathrm i},
  \qquad
  g_N(\lambda)\equiv\ln\frac{P_{\mathrm f}(\lambda)}
                              {P_{\mathrm i}(\lambda)},
  \label{eq:probabilities-paper}
\end{equation}
The thermodynamic midpoint \(\lhalf(N)\), where
\(P_{\mathrm i}=P_{\mathrm f}=1/2\), therefore solves
\begin{equation}
  g_N\!\left(\lhalf(N)\right)=0.
  \label{eq:midpoint-paper}
\end{equation}
\revphasetwelve{These statistical-mechanical quantities are evaluated by
deterministic quadrature; residual grid and truncation errors are assessed in
the SI.}

Writing \(Z_{b}^{*}=Z_{N}^{b,*}\) for \(b\in\{\mathrm i,\mathrm f\}\), the
reduced basin free energy is \(-\Tstar\ln Z_b^*\), up to a constant independent
of length. The conjugate tension measures the change in basin free energy
per unit increase in total chain length, at fixed \(N\) and temperature:
\begin{equation}
  F_{b,\mathrm{conj}}^*
  =-\frac{\Tstar}{N}\frac{d}{d\lambda}
    \ln Z_{b}^{*}\!\left(N\lambda,d_{N}^{*}(\lambda)\right).
  \label{eq:conjugate-force-paper}
\end{equation}
The factor \(1/N\) follows from \(L^*=N\lambda\). Since changing \(\lambda\)
also moves \(d_N^*(\lambda)\), the derivative includes the motion of the
boundary defining the basin.
Relative to the basin-averaged length-weighted virial tension, it includes
a common entropic term from the constrained measure and a moving-boundary
contribution. The common term cancels in the difference between basins.
Differentiating the
logarithmic population ratio gives the state--force identity
\begin{equation}
  g_N'(\lambda)
  =\frac{N}{\Tstar}
   \left(F_{\mathrm i,conj}^*-F_{\mathrm f,conj}^*\right).
  \label{eq:state-force-paper}
\end{equation}
SI Sec.~S3.3 gives the explicit boundary contribution and its numerical
verification.

When configurations with two or more bonds above the divider have
negligible weight, the fractured partition is approximated by \(N\)
equivalent channels, each with one specified bond above the divider and
all others below it:
\begin{equation}
  Z_{N}^{\mathrm f,*}\simeq N Z_{N,\mathrm{site}}^{\mathrm f,*}.
  \label{eq:multiplicity-paper}
\end{equation}
For one labelled site, the one-fracture partition is
\begin{equation}
  Z_{N,\mathrm{site}}^{\mathrm f,*}(L^{*})
  =\int_{d_{N}^{*}(\lambda)}^{L^{*}}
   e^{-u^{*}(l^{*})/\Tstar}
   Z_{N-1}^{\mathrm i,*}(L^{*}-l^{*};d_{N}^{*})\,dl^{*},
  \label{eq:one-fracture-partition-paper}
\end{equation}
Here \(Z_{N-1}^{\mathrm i,*}(L^{*}-l^{*};d_{N}^{*})\) is the partition density of the
remaining \(N-1\) bonds: they share length \(L^{*}-l^{*}\), and each lies below
the full chain's divider \(d_{N}^{*}(\lambda)\). This one-fracture approximation
estimates coexistence by solving
\(N Z_{N,\mathrm{site}}^{\mathrm f,*}=Z_{N}^{\mathrm i,*}\).
SI Sec.~S4.2 derives the reservoir weight and its integral over the long bond.
The factor \(N\) counts the equivalent choices of the fractured bond.
The same cyclic multiplicity converts the rate at one equivalent site into
a chain-level rate and is applied once.

\subsection{Adiabatic, tagged-bond, and first-fracture site profiles}
\label{sec:site-channel-theory}

One bond is selected as the tagged bond; the remaining
\(N-1\) bonds form its reservoir. The equal-sharing reduced energy in
\cref{eq:adiabatic-paper} is the
\(E_{\mathrm{ad}}^{(N-1),*}\) branch: all \(N-1\) reservoir bonds share the
length left by the tagged bond. A complementary one-bond branch keeps
\(N-2\) reservoir bonds at their isolated minimum and lets one bond absorb the
remaining length,
\begin{equation}
\begin{aligned}
 E_{\mathrm{ad}}^{(1),*}(l^{*};N,\lambda)
 &=u^{*}(l^{*})+u^{*}\!\left[N\lambda-l^{*}-(N-2)\right]\\
 &\quad+(N-2)u^{*}(1)-Nu^{*}(\lambda).
\end{aligned}
  \label{eq:one-bond-adiabatic-paper}
\end{equation}
\revphasetwelve{The common reference \(-Nu^{*}(\lambda)\) shifts both branches
equally, leaving the locations of their extrema and barrier differences unchanged.}
\revphasetwelve{Following \citet{Theodorou2024}, the lower envelope
\(E_{\mathrm{env}}^{*}=\min\{E_{\mathrm{ad}}^{(1),*},
E_{\mathrm{ad}}^{(N-1),*}\}\) selects the lower energy of these two reservoir
configurations at each tagged-bond length.}

Fixing bond 1 at \(l^{*}\) leaves an \((N-1)\)-bond reservoir of total length
\(L^{*}-l^{*}\). The unrestricted tagged-bond density and PMF are
\begin{equation}
\begin{gathered}
 p_{N}^{\mathrm{tag},*}(l^{*}\mid L^{*})
 =\frac{e^{-u^{*}(l^{*})/\Tstar}Z_{N-1}^{*}(L^{*}-l^{*})}{Z_{N}^{*}(L^{*})},\\
 G_{N}^{\mathrm{tag},*}(l^{*})=-\Tstar\ln p_{N}^{\mathrm{tag},*}(l^{*}\mid L^{*})+C_{\mathrm{tag}}^{*}.
\end{gathered}
  \label{eq:tagged-pmf-paper}
\end{equation}
The numerator combines the tagged bond's Boltzmann weight with the
partition density of the reservoir; \(Z_{N}^{*}(L^{*})\) normalizes the density.
\revphasetwelve{As \(\Tstar\to0\), this PMF approaches, up to an additive
constant, the minimum chain energy compatible with the fixed tagged-bond
length. The envelope \(E_{\mathrm{env}}^{*}\) approximates that limit using
the two specified reservoir configurations. At finite temperature, the
partition density also accounts for the statistical weight of the compatible
configurations.}
The unrestricted tagged PMF \(G_{N}^{\mathrm{tag},*}\) permits a reservoir bond to fracture
while the tagged bond remains short. It is therefore a valid single-bond
equilibrium PMF, but its short-bond basin is not the intact basin. For this
reason, it is not used as a first-fracture rate barrier.

To isolate first fracture at the tagged site, every competing bond is
restricted below \(d_{N}^{*}\), the adiabatic divider of the full \(N\)-bond chain:
\begin{equation}
  G_{N}^{\mathrm{site},*}(l^{*};L^{*})
  =u^{*}(l^{*})-\Tstar\ln
  Z_{N-1}^{\mathrm i,*}(L^{*}-l^{*};d_{N}^{*})+C_{\mathrm{site}}^{*}.
  \label{eq:site-pmf-paper}
\end{equation}
SI Sec.~S4.2 derives \cref{eq:tagged-pmf-paper,eq:site-pmf-paper} by
integrating over their respective reservoir ensembles.
The intact minimum \(l_{\mathrm i}^{\mathrm{site},*}\) and maximum \(l_{\mathrm s}^{\mathrm{site},*}\) of this
site PMF define the forward one-site dimensionless barrier,
with \(l_{\mathrm s}^{\mathrm{site},*}\) distinct from the reservoir cutoff \(d_{N}^{*}\):
\begin{equation}
  B_N(\lambda)
  =\frac{G_{N}^{\mathrm{site},*}(l_{\mathrm s}^{\mathrm{site},*})-G_{N}^{\mathrm{site},*}(l_{\mathrm i}^{\mathrm{site},*})}{\Tstar}.
  \label{eq:site-barrier-paper}
\end{equation}
The barrier contains no factor \(N\); multiplicity enters only when the
equivalent forward site rates are summed. Both PMFs defined here
are expressed in units of \(D_e\), with arbitrary energy zeros \(C_{\mathrm{tag}}^{*}\)
and \(C_{\mathrm{site}}^{*}\).

\begin{figure*}[!tbp]
  \centering
  \includegraphics[width=16cm]{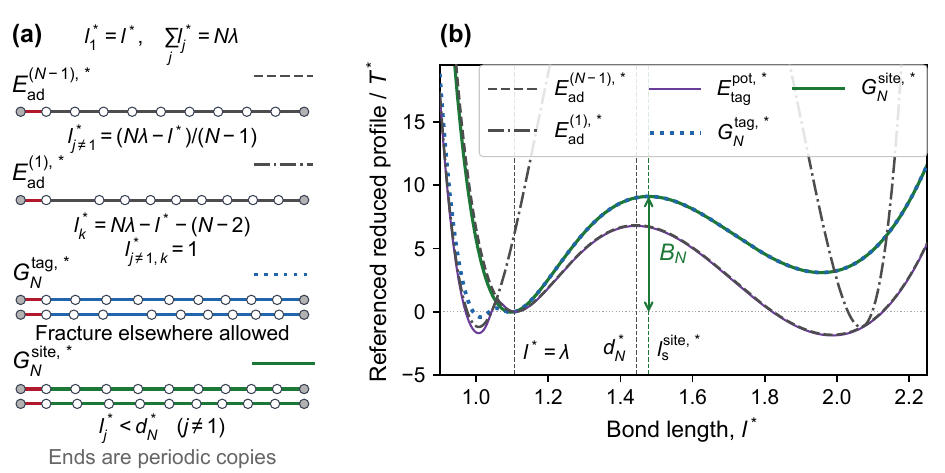}
  \input{figures/pmf_reconciliation/caption.tex}
  \label{fig:pmf-constructions}
\end{figure*}

In \cref{fig:pmf-constructions}, the lower of the two adiabatic branches
closely follows the conditional potential energy: the mean total potential
energy at fixed tagged-bond length, averaged over the unrestricted reservoir
(SI Sec.~S4.3). The conditional potential energy is a Boltzmann-weighted
energy average, whereas the PMFs are free-energy profiles. Restricting the reservoir to
intact bonds gives the single-site barrier used in the rate calculation.

It should be worth mentioning that choosing the adiabatic saddle
\(d_N^*\) provides a straightforward mechanical divider and retains the
state convention of \citet{Theodorou2024}, facilitating direct comparisons.
Another approach would be to align this divider with the site-PMF maximum.
However, since the site PMF depends on the choice of \(d_N^*\)
(Eq.~\ref{eq:site-pmf-paper}), determining a divider satisfying this
condition would require a self-consistent calculation.
Nevertheless, the current PMF is already the equilibrium profile of the
specified restricted ensemble; coincidence of its maximum with the divider
is not required, since the reactive-flux treatment described below accounts
for recrossing of the chosen surface.

%% file: figures/pmf_reconciliation/caption.tex
\caption{\revphaseseven{Energy and free-energy profiles at coexistence for
\(N=10\), \(\lambda\simeq\RegisteredPmfCoexistenceLambda\).
(a) Reservoir configurations underlying the adiabatic branches and the two
PMFs. The red bond is fixed at \(l^*=l/l_0\); filled endpoints are periodic
copies of the same particle. The tagged ensemble permits fracture elsewhere;
the site ensemble keeps all reservoir bonds intact.
(b) Reduced adiabatic energies \(E_{\mathrm{ad}}^{(N-1),*}\) and
\(E_{\mathrm{ad}}^{(1),*}\), conditional mean potential energy
\(E_{\mathrm{tag}}^{\mathrm{pot},*}\) (SI Sec.~S4.3), and PMFs
\(G_N^{\mathrm{tag},*}\) and \(G_N^{\mathrm{site},*}\), all divided by \(T^*\), equivalently in thermal units \(k_{\mathrm B}T\).
The adiabatic branches share the zero of their lower envelope at
\(l^*=\lambda\); the other profiles are independently zeroed there.
Dashed vertical lines mark \(l^*=\lambda\), the adiabatic divider
\(l_{\mathrm{ad}}^{\ddagger,*}=d_N^*\), and the site-PMF saddle \(l_{\mathrm s}^{\mathrm{site},*}\).
The dotted horizontal line marks zero energy; the green arrow shows the
one-site barrier \(B_N\) defined in \cref{eq:site-barrier-paper}.}}

%% file: sections/03_computational_methods.tex
\section{\revphaseseven{Methods}}
\label{sec:methods}

\subsection{Deterministic calculations}
\label{sec:deterministic-method}

For each chain length and extension, the constrained partition densities give
the equilibrium populations and free-energy profiles. With
\(w(l^{*})=\exp[-u^{*}(l^{*})/\Tstar]\), these densities obey the recursion
\begin{equation}
  Z_{1}^{*}(L^{*})=w(L^{*}),\qquad
  Z_{n}^{*}(L^{*})=\int_0^{L^{*}} w(l^{*})Z_{n-1}^{*}(L^{*}-l^{*})\,dl^{*}.
  \label{eq:partition-recursion-paper}
\end{equation}
Each step assigns length \(l^{*}\) to one bond and the remaining length to the
others. The intact partition uses
\(w_{\mathrm i}(l^{*};d_{N}^{*})=w(l^{*})\Theta[d_{N}^{*}(\lambda)-l^{*}]\), with the full-chain
divider fixed throughout the recursion. The integrals are evaluated by direct
summation on a uniform grid, accumulating the terms logarithmically to retain
very small densities. Between grid points, SciPy's piecewise cubic
Hermite interpolator (PCHIP) is applied to the logarithm of the density
\citep{FritschButland1984,Virtanen2020}. Exponentiating this log-PCHIP
interpolant preserves positive finite densities.
The partition ratios give the state populations; solving
\(P_{\mathrm f}=1/2\) gives the coexistence midpoint.

The same partition densities give the tagged and site PMFs in
\cref{fig:pmf-constructions} through the definitions in Sec.~\ref{sec:site-channel-theory}.
Local fourth-degree polynomial fits locate the site-PMF minimum and maximum
and determine their curvatures. The barrier is their free-energy difference;
curvatures are obtained from these fits, not by differentiating the
partition interpolant. Equilibrium populations and these PMFs use direct
logarithmic convolution, whereas the one-fracture midpoints in
\cref{fig:scaled-midpoint-paper} and site barriers in \cref{fig:barrier-paper}
use tilted FFT convolution of the restricted reservoirs (SI Sec.~S3.2).
The one-fracture weight then follows by integrating over the long bond.

Grid refinement and enlarged integration domains test the sensitivity of
these quantities. Independent extended-precision sums check the direct
logarithmic convolution, and direct quadrature checks its off-grid
evaluation at the tested states. SI Sec.~S3.2 distinguishes these checks
from those for the tilted calculations; SI Sec.~S4 gives the profile
constructions.

\subsection{Langevin simulations}
\label{sec:langevin-method}

All simulations use one-dimensional Langevin dynamics at fixed chain
extension. Let \(x_{j}\) be the position of particle \(j\) and
\(l_{j}=x_{j}-x_{j-1}\) the length of its bond to the preceding particle, measured along
the periodic chain. These coordinates and the following equation of motion
are dimensional \citep{Theodorou2024}:
\begin{equation}
  m\ddot x_{j}
  =U'(l_{j+1})-U'(l_{j})
  -m\xi\dot x_{j}+R_j(t),
  \label{eq:langevin-paper}
\end{equation}
where \(m\) is the particle mass, \(U'\) is the derivative of the bond
potential, and \(\xi\) is the friction rate. The thermal force \(R_j(t)\)
has zero mean and correlation
\begin{equation}
  \left\langle R_i(t)R_j(t')\right\rangle
  =2m\xi k_{\mathrm B}T\,\delta_{ij}\delta(t-t'),
\end{equation}
where \(\delta_{ij}\) and \(\delta(t-t')\) are the Kronecker and Dirac
delta functions, respectively.

The equations of motion are integrated in LAMMPS \citep{Thompson2022} using velocity-Verlet
and a Langevin thermostat, with motion confined to the chain direction.
Unless otherwise stated, all simulations use the model parameters and
timestep listed in \cref{tab:baseline-parameters}. Initial ensembles and
sampling procedures are specified with each method below; implementation
details are given in SI Sec.~S7.1.

Simulations span chain lengths from \(N=4\) to \(20\).
The sensitivity of equilibrium populations and ACF rates to the timestep
was assessed at \(N=10\) by comparing the adopted \(0.25\,\mathrm{fs}\) step with
\(0.125\,\mathrm{fs}\). Conservative forces, energies and periodic geometry
were also checked. SI Sec.~S7.2 presents these checks and the remaining
rate uncertainty.

\subsection{Rate framework}

\subsubsection{Rate definitions and normalization}
\label{sec:rate-estimands}
\label{sec:extension-dependent-rates}

A constant-rate two-state description assumes that relaxation within the
basins and recrossing occur on shorter timescales than population relaxation.
At fixed chain length, extension and temperature, two directional rates
describe switching between intact and fractured states:
\(k_{\mathrm i\to\mathrm f}^{*}\) for fracture and
\(k_{\mathrm f\to\mathrm i}^{*}\) for return to the intact state. Each rate is
conditional on the chain occupying its starting state. In a stationary
two-state description, the forward rate is
\begin{equation}
  \kifstar=\frac{J_{\mathrm{if}}^*}{P_{\mathrm i}}.
  \label{eq:static-rate-paper}
\end{equation}
Here \(J_{\mathrm{if}}^*\) is the one-way intact-to-fractured transition
frequency per unit reduced time, measured over the whole equilibrium
ensemble, and \(P_{\mathrm i}\) is the instantaneous intact population. Dividing by
\(P_{\mathrm i}\) expresses this frequency per unit time spent intact.
The reverse rate \(k_{\mathrm f\to\mathrm i}^{*}\) and population
\(P_{\mathrm f}\) are defined analogously.

With core classification, a chain
retains the label of its most recently visited core; the corresponding
populations \(\Pi_{\mathrm i}^{\mathrm{LC}}\) and
\(\Pi_{\mathrm f}^{\mathrm{LC}}\) are the fractions of time carrying each
label. These history-dependent populations need not equal the instantaneous
basin populations. For core-defined transitions \citep{Schutte2011},
\begin{equation}
  k_{\mathrm{i}\to\mathrm f}^{\mathrm{core},*}
  =\frac{J_{\mathrm{i}\to\mathrm f}^{\mathrm{core},*}}
         {\Pi_{\mathrm i}^{\mathrm{LC}}}
  =\frac{N_{\mathrm{i}\to\mathrm f}}
         {\mathcal T_{\mathrm i}^{\mathrm{LC},*}},
  \label{eq:core-rate-normalization-paper}
\end{equation}
where \(N_{\mathrm i\to\mathrm f}\) counts completed label changes,
\(\mathcal T_{\mathrm i}^{\mathrm{LC},*}\) is the total time carrying the intact
label, and \(J_{\mathrm i\to\mathrm f}^{\mathrm{core},*}\) is the number
of these changes divided by the total observation time.

At equilibrium, detailed balance requires
\(P_{\mathrm i}k_{\mathrm i\to\mathrm f}^{*}
=P_{\mathrm f}k_{\mathrm f\to\mathrm i}^{*}\); thus,
the forward and reverse rates coincide when the two populations
are equal. For core-defined rates, the same relation uses the last-core
populations instead. All populations and rates refer to the specified
\((N,\lambda,T^{*})\).

The following sections describe deterministic calculations based on
transition-state theory (TST), harmonic reactive-flux (RF) theory and
Kramers--Grote--Hynes (KGH) theory, followed by simulation estimates from tension
autocorrelations (ACF), core residence times and barrier-top RF trajectories.

\subsection{Deterministic rate calculations}
\subsubsection{Exact tagged-coordinate transition-state flux}
\label{sec:tst-flux}

Transition-state theory (TST) \citep{Eyring1935} estimates the fracture rate by counting
outward crossings of the dividing surface and dividing by the intact
population. The usual surface is \(\max_jl_{j}^{*}=d_{N}^{*}\). To compare alternative
launch surfaces, the surface \(\max_jl_{j}^{*}=s^{*}\) is also considered, keeping the basin
divider \(d_{N}^{*}\) fixed. On each tagged component, one bond has length
\(s^{*}\) and every other bond is shorter than \(s^{*}\).

Fixing the tagged bond and integrating over the remaining lengths gives its
unnormalized surface density \(\rho_{N,s^*}^{*}\). The total outward crossing
frequency is
\begin{equation}
 \begin{gathered}
  \rho_{N,s^*}^{*}(L^{*})=e^{-u^{*}(s^{*})/\Tstar}Z_{N-1}^{\mathrm i,*}(L^{*}-s^{*};s^{*}),\\
  J_{N,s^*}^{+,*}=N\sqrt{\frac{c\Tstar}{\pi}}
    \frac{\rho_{N,s^*}^{*}(L^{*})}{Z_{N}^{*}(L^{*})},
 \end{gathered}
  \label{eq:positive-flux-paper}
\end{equation}
where \(L^{*}=N\lambda\). This expression combines the equilibrium surface
density, the Maxwell average of the positive part of the tagged-bond velocity,
\(\langle v_{l}^{*}\Theta(v_{l}^{*})\rangle=\sqrt{c\Tstar/\pi}\), and the
\(N\) equivalent bonds. Here \(v_{l}^{*}=dl^{*}/dt^*\); the velocity average
includes both signs, with negative velocities contributing zero.
SI Sec.~S8.1 derives this factor from the particle velocities.

Dividing by \(P_{\mathrm i}(d_{N}^{*})=Z_{N}^{\mathrm i,*}(L^{*};d_{N}^{*})/Z_{N}^{*}(L^{*})\) gives
\begin{equation}
  K_{N,s^*\mid d_N^*}^{*}
  \equiv\frac{J_{N,s^*}^{+,*}}{P_{\mathrm i}(d_{N}^{*})}
  =N\sqrt{\frac{c\Tstar}{\pi}}
   \frac{\rho_{N,s^*}^{*}(L^{*})}{Z_{N}^{\mathrm i,*}(L^{*};d_{N}^{*})}.
  \label{eq:exact-tst-paper}
\end{equation}
The reservoir cutoff at launch is \(s^{*}\), while the intact population
uses the fixed divider \(d_{N}^{*}\). At \(s^{*}=d_{N}^{*}\), the TST rate and its reactive-flux correction
are \citep{Chandler1978}
\begin{equation}
 k_{N}^{\mathrm{TST},*}\equiv K_{N,d_N^*\mid d_N^*}^{*},
 \qquad
 k_{N}^{\mathrm{RF},*}=\kappa_N k_{N}^{\mathrm{TST},*}.
 \label{eq:rf-rate-paper}
\end{equation}
Here \(\kappa_N\) is the transmission coefficient accounting for recrossing.
The equilibrium crossing flux is exact for the specified surface.
Without recrossing, every outward crossing succeeds and \(\kappa_N=1\),
recovering TST. The coefficient is evaluated using either the harmonic
approximation below or barrier-top simulations
(Sec.~\ref{sec:barrier-top-method}); SI Sec.~S8.2 gives the reactive-flux
construction.

\subsubsection{Full-coordinate harmonic dynamics}
\label{sec:harmonic-dynamics-paper}

The transmission coefficient is estimated by approximating the potential
near the adiabatic saddle with its quadratic expansion, retaining the
coupled motion of all particles. Within this approximation, the linear
Langevin dynamics determines which side of the dividing surface a
trajectory ultimately reaches. Averaging the resulting signed flux over
thermal initial positions on the surface, Maxwell velocities and
thermostat noise gives \(\kappa_N^{\mathrm{harm}}\) (SI Sec.~S8.3).
Substitution into \cref{eq:rf-rate-paper} yields
\begin{equation}
 k_{N}^{\mathrm{harm\,RF},*}
 =\kappa_N^{\mathrm{harm}}k_{N}^{\mathrm{TST},*},
 \label{eq:harmonic-rf-paper}
\end{equation}
where the equilibrium TST factor retains the full partition calculation.

\subsubsection{Symmetric-path Kramers--Grote--Hynes theory}
\label{sec:kgh-method}

Finally, the Kramers--Grote--Hynes (KGH) approximation
\citep{Kramers1940,GroteHynes1980} describes escape along one coordinate.
The site PMF supplies the free-energy profile, and a symmetric path supplies
the kinetic model: as the tagged bond lengthens, the other \(N-1\) bonds
share the remaining length equally. The associated particle motion gives
an effective mass \(m_{\mathrm{eff}}\), with
\begin{equation}
 \rho_N=\frac{m_{\mathrm{eff}}}{m}=\frac{N(N+1)}{12(N-1)}.
 \label{eq:kgh-path-mass-paper}
\end{equation}
The site-PMF curvatures at its intact minimum and maximum,
\(\mathcal K_{\mathrm i}^{*}=G_{N}^{\mathrm{site},(2),*}(l_{\mathrm i}^{\mathrm{site},*})>0\) and
\(\mathcal K_{\mathrm s}^{*}=G_{N}^{\mathrm{site},(2),*}(l_{\mathrm s}^{\mathrm{site},*})<0\), determine the
well and barrier frequencies. With friction proportional to
instantaneous velocity, the transmission factor is \(\chi_N\):
\begin{equation}
\begin{gathered}
 \omega_{\mathrm i}^*=\sqrt{\frac{c\mathcal K_{\mathrm i}^{*}}{\rho_N}},
 \qquad
 \omega_{\mathrm s}^*=\sqrt{\frac{c|\mathcal K_{\mathrm s}^{*}|}{\rho_N}},\\
 \chi_N=\frac{\sqrt{1/4+(\omega_{\mathrm s}^*)^2}-1/2}{\omega_{\mathrm s}^*}.
\end{gathered}
  \label{eq:kgh-definitions-paper}
\end{equation}
Here \(\omega_\alpha^*=\omega_\alpha/\xi\) for \(\alpha\in\{\mathrm i,\mathrm s\}\).
The attempt frequency \(\omega_{\mathrm i}^*/(2\pi)\), transmission
\(\chi_N\) and barrier factor \(e^{-B_N}\) give the rate at one site;
summing over the \(N\) equivalent sites yields
\begin{equation}
 k_{N}^{\mathrm{KGH},*}=NA_{N}^{\mathrm{KGH},*}(\lambda)e^{-B_N(\lambda)},
 \qquad A_{N}^{\mathrm{KGH},*}=\frac{\omega_{\mathrm i}^*}{2\pi}\chi_N.
 \label{eq:kgh-rate-paper}
\end{equation}
Here \(B_N\) is the site-PMF barrier in \cref{eq:site-barrier-paper}.
Both the attempt frequency and transmission use the one-coordinate
approximation. SI Sec.~S8.5 derives the path mass and transmission.
The PMF barrier and curvatures are evaluated at each \((N,\lambda)\).

\subsection{Simulation-based rate estimators}

\begin{figure*}[!tbp]
  \centering
  \includegraphics[width=16cm]{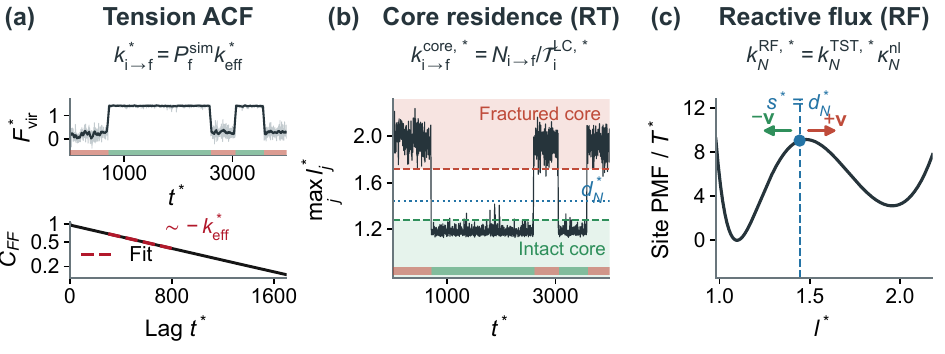}
  \input{figures/rate_workflow_schematic/caption.tex}
  \label{fig:rate-workflows}
\end{figure*}

This section presents simulation-based rate estimates from tension
autocorrelation functions (ACF), core residence times (RT) and barrier-top
reactive flux (RF), as illustrated in \cref{fig:rate-workflows}.

\subsubsection{Tension-autocorrelation estimator}
\label{sec:acf-method}

Switching between intact and fractured states changes the chain tension.
The resulting slow tension fluctuations provide an estimate of the switching
rates from equilibrium trajectories. The instantaneous observable is the
length-weighted virial tension,
\begin{equation}
 F_{\mathrm{vir}}^{*}(t^*)=\frac{1}{N\lambda}\sum_{j=1}^N
 l_{j}^{*}(t^*)f^*\!\left[l_{j}^{*}(t^*)\right].
 \label{eq:virial-observable-paper}
\end{equation}
Here \(f^*(l^*)=u^{(1),*}(l^*)\) is the reduced single-bond force.
The relation between the virial tension and the restricted conjugate force in
\cref{eq:conjugate-force-paper} is given in SI Sec.~S3.
Writing \(\delta F^{*}=F_{\mathrm{vir}}^{*}-\langle F_{\mathrm{vir}}^{*}\rangle\),
the normalized autocorrelation at lag \(t^*\) is \citep{Theodorou2024}
\begin{equation}
\begin{gathered}
 C_{FF}(t^*)=\frac{\langle\delta F^{*}(0)\delta F^{*}(t^*)\rangle}
 {\langle\bigl(\delta F^{*}\bigr)^{2}\rangle}
 \simeq C_0e^{-\keffstar t^*},\\
 \keffstar=\kifstar+\kfistar.
\end{gathered}
  \label{eq:acf-paper}
\end{equation}
When relaxation within each basin is fast compared with switching, the
slow ACF follows the population relaxation of the two-state model.
The tail is fitted with amplitude \(C_0\) and decay rate \(\keffstar\);
\cref{fig:rate-workflows}(a) illustrates the construction.
Equilibrium detailed balance then gives the directional rates \citep{Theodorou2024},
\begin{equation}
 \kifstar=P_{\mathrm f}^{\mathrm{sim}}\keffstar,
 \qquad \kfistar=(1-P_{\mathrm f}^{\mathrm{sim}})\keffstar.
 \label{eq:acf-forward-paper}
\end{equation}
Here \(P_{\mathrm f}^{\mathrm{sim}}\) is the fraction of recorded
configurations classified as fractured by the instantaneous criterion
in Sec.~\ref{sec:state-classification}.
The ACF fit and occupancy are recomputed together in each whole-trajectory
bootstrap sample, retaining their covariance \citep{FieldWelsh2007}. SI Secs.~S9.1--S9.2 describe the
sampling, estimator and sensitivity to the choice of observable and fitting window.

\subsubsection{Last-core residence estimator}
\label{sec:core-residence-method}

Alongside the tension-ACF method, switching rates are estimated by following
visits to the intact and
fractured cores defined in \cref{eq:cores-paper-C}. At each recorded time,
the maximum bond length determines whether the chain has reached either
core. Between the cores, the chain retains its previous label. An excursion that
returns to the same core therefore does not count as a transition; the
label changes only when the opposite core is reached.

For each residence \(\alpha\) with an observed entry, the record contains its
observed duration \(D_\alpha^*\) and whether it ends in a switch to the
opposite core. The forward rate \(k_{\mathrm i\to\mathrm f}^{\mathrm{core},*}\)
is the number of intact-to-fractured switches \(N_{\mathrm i\to\mathrm f}\)
divided by the accumulated intact-labelled time
\(\mathcal T_{\mathrm i}^{\mathrm{LC},*}\), as in
\cref{eq:core-rate-normalization-paper}; the reverse rate
\(k_{\mathrm f\to\mathrm i}^{\mathrm{core},*}\) is obtained analogously. SI Secs.~S9.3--S9.4 explain how intervals
cut by the recording boundaries are treated and compare the rates across
core depths. This calculation uses the same equilibrium trajectories as
the ACF estimator, so their sampling errors are correlated. The residence estimator provides a
complementary measure based on observed state changes, normalized by the
last-core population.

\subsubsection{Nonlinear barrier-top reactive flux}
\label{sec:barrier-top-method}

The Bennett--Chandler reactive-flux method
\citep{Bennett1977,Chandler1978} quantifies how dynamical recrossing
modifies the fracture rate predicted by transition-state theory.
For \(N=7\)--12, 15, and 20 at coexistence, trajectories are launched
with one tagged bond at the adiabatic dividing surface
\(s^{*}=d_{N}^{*}=l_{\mathrm{ad}}^{\ddagger,*}\)
(panel c of \cref{fig:rate-workflows}). The remaining bonds are sampled
jointly from their conditional equilibrium distribution, subject to
the fixed total chain length and \(l_i^*<s^*\).
Initial particle velocities follow the Maxwell--Boltzmann distribution
conditioned on zero total momentum, obtained by removing the
center-of-mass component. Each configuration is launched with this
velocity vector and its negative, retaining both signs of the
tagged-bond velocity.
The trajectories evolve under Langevin dynamics with the full
Morse forces, yielding the transmission coefficient that corrects
the equilibrium TST rate.

Let \(h_{\mathrm f,d_N^*}=\mathbf1[\max_jl_{j}^{*}>d_{N}^{*}]\) indicate the
instantaneous fractured state. The transmission function is
\begin{equation}
  \kappa_{N,s^*\mid d_N^*}(t^*)
  =\frac{\left\langle
      v_{l}^{*}(0)h_{\mathrm f,d_N^*}[\bm r^{*}(t^*)]
    \right\rangle_{\Sigma_{s^*},\mathrm{tag}}}
    {\sqrt{c\Tstar/\pi}},\qquad v_{l}^{*}=\frac{dl^{*}}{dt^*}.
  \label{eq:transmission-paper}
\end{equation}
The average includes surface configurations, initial velocities and
thermostat noise. Each trajectory contributes its signed initial tagged
velocity when the chain is fractured at the observation time, and zero
otherwise. The denominator is the positive Maxwell crossing factor.
For \(s^{*}=d_{N}^{*}\), we write \(\kappa_{N,d_N^*\mid d_N^*}=\kappa_N\).

After the rapid recrossing transient, transitions in both directions relax
the finite intact and fractured populations. This regime is described by a
two-state model with constant directional rates: the departure probability
per unit time does not depend on residence age. At coexistence, equal
forward and reverse rates \(k_N^*\) give population decay
\(e^{-2k_N^*t^*}\). Defining the slow-regime transmission amplitude as
\(\kappa_{0,N}=k_N^*/k_N^{\mathrm{TST},*}\) gives the fitting form
\begin{equation}
  \kappa_N(t^*)=\kappa_{0,N}
   \exp\!\left[-2\kappa_{0,N}k_N^{\mathrm{TST},*}t^*\right].
  \label{eq:finite-pop-paper}
\end{equation}
SI Sec.~S10.2 derives this relation from the two-state transition
probability and its connection to reactive flux. The amplitude
\(\kappa_N^{\mathrm{nl}}=\kappa_{0,N}\) is fitted over \(3\le t^*\le10\)
at every chain length, holding the deterministic TST rate fixed.
Substitution into \cref{eq:rf-rate-paper} gives the simulated RF rate.
The amplitude describes the regime after the initial transient, rather
than the microscopic value at \(t^*=0\). For longer chains, the decay
is barely resolved and the fitted curve approaches a plateau; the
shortest chains retain small systematic departures from the exponential.
Sampling uncertainty and sensitivity to the fitting window are described
in SI Sec.~S10.2.

A separate \(N=10\) control at
\(\lambda=\CanonicalNtenFixedSurfaceLambda\) compares the adiabatic
saddle, site-PMF saddle and an outer launch surface while keeping
the physical states and population normalization fixed. For this control,
the time-dependent rates are averaged over \(4\le t^*\le8\), without imposing
the amplitude--decay relation in \cref{eq:finite-pop-paper} on the
displaced-surface correlations. This control shows that the final RF rate
is insensitive to the tested launch surfaces within the comparison
uncertainty at this state (SI Sec.~S10.3).

%% file: figures/rate_workflow_schematic/caption.tex
\caption{Rate estimators at \(N=10\). (a) Virial-tension trajectory and pooled
ACF with its exponential fit; the dark trajectory line is a moving-average
visual guide. The translucent strip records the instantaneous state
(red: fractured; green: intact). (b) Maximum bond length along the same trajectory, with intact
and fractured cores, instantaneous divider \(d_N^*\), and last-core label
(bottom strip). (c) Conditioned site PMF relative to its intact minimum,
with the RF launch surface \(s^*=d_N^*\) and opposite initial velocities. The plotted ordinate is
\([G_N^{\mathrm{site},*}(l^*)-G_N^{\mathrm{site},*}(l_{\mathrm i}^{\mathrm{site},*})]/T^*\).}

%% file: sections/04_results.tex
\section{Results}
\label{sec:results}

\input{sections/04_midpoint_and_barrier_results}
\input{sections/05_rate_results}
\input{sections/06_committor_and_crossover}

%% file: sections/04_midpoint_and_barrier_results.tex
\subsection{Validating equilibrium simulations against constrained thermodynamics}
\label{sec:thermodynamic-validation-paper}

\revphaseseven{The first comparison concerns the equilibrium state populations and
intact-chain tension from simulation with the predictions of constrained
thermodynamics. Constant-extension Langevin simulations sample five extensions
around coexistence for each \(N=4,\ldots,12\) and \(15\).
The target fractured probabilities are about 10\%, 25\%, 50\%, 75\% and
90\% (see SI Table~S8 and Sec.~S9.1).}

\begin{figure}[!tbp]
  \centering
  \includegraphics[width=8.5cm]{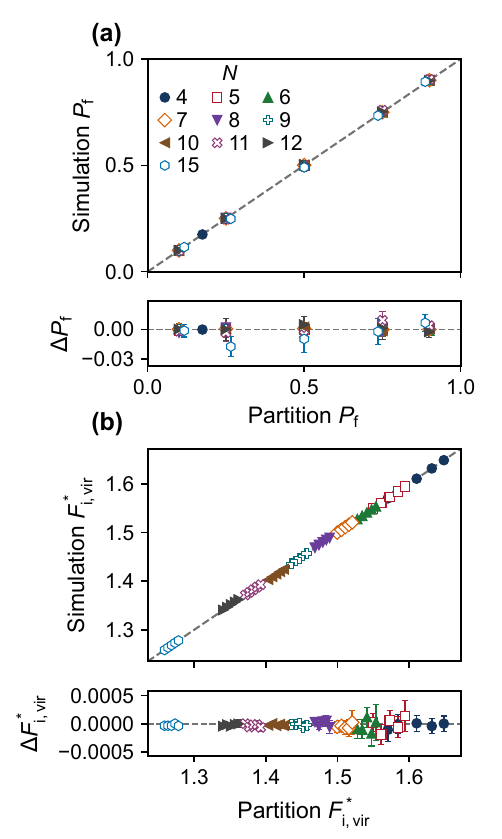}
  \input{figures/equilibrium_validation/caption_main}
  \label{fig:thermodynamic-comparison-paper}
\end{figure}

\revphaseseven{The simulations show excellent agreement with the deterministic
calculations: both the fractured probability and the mean tension within the
intact state closely follow the predictions
(\cref{fig:thermodynamic-comparison-paper}). Across all 50 states, the
root-mean-square differences are
\(\ScientificValue{\RegisteredCombinedOccupancyRmse}\) in
fractured probability and
\(\ScientificValue{\RegisteredCombinedVirialRmse}\) in reduced
intact-state virial tension. Agreement in both observables shows that the
simulations capture the relative statistical weights of the intact and
fractured states as well as the mechanical response of intact chains.
SI Sec.~S9.1 describes the equilibrium sampling and uncertainty estimation;
SI Sec.~S3.3 gives the complementary deterministic force-identity check.}

\subsection{Coexistence scaling and free-energy landscape}
\label{sec:thermodynamic-results}

\subsubsection{Scaled coexistence extension}
\label{sec:midpoint-results}

The coexistence extension, denoted by \(\lambda_{1/2}\), is the mean stretch
at which intact and fractured states have equal equilibrium populations. In simulations,
this condition becomes
\begin{equation}
  P_{\mathrm f}^{\mathrm{sim}}
  \!\left(\lambda_{1/2}^{\mathrm{sim}}\right)=\frac12.
  \label{eq:simulation-midpoint-paper}
\end{equation}
Linear interpolation of the sampled logit
\(g=\ln[P_{\mathrm f}/(1-P_{\mathrm f})]\) to zero estimates
\(\lambda_{1/2}^{\mathrm{sim}}\).

\begin{figure}[!tbp]
  \centering
  \includegraphics[width=8.5cm]{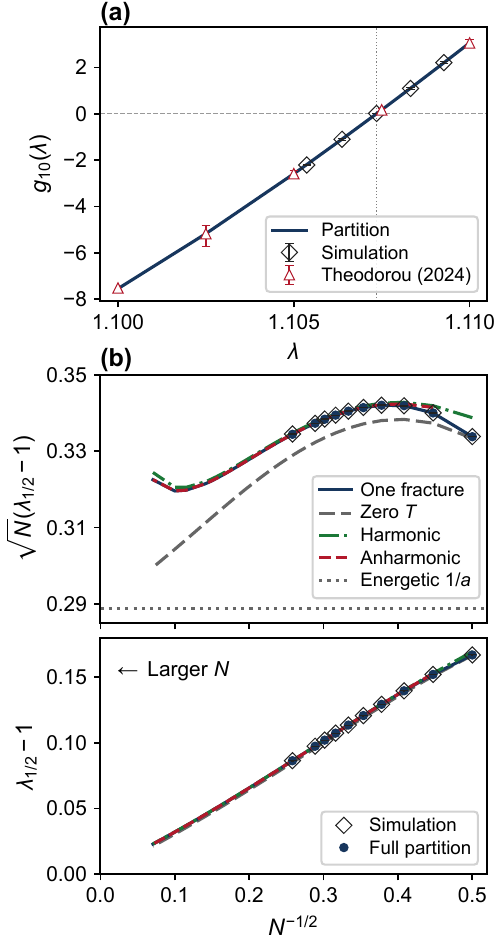}
  \input{figures/midpoint_scaling/caption}
  \label{fig:scaled-midpoint-paper}
\end{figure}

The logit comparison at \(N=10\) in \cref{fig:scaled-midpoint-paper}(a)
shows excellent agreement between the present simulations, deterministic
constrained-partition calculations and the results of \citet{Theodorou2024}.
As the chain is stretched, the logit increases through zero, marking the
point at which the two states are equally populated.

The chain-length dependence of \(\lambda_{1/2}\) is examined by comparing
the full partition calculation with a one-fracture approximation and
successive energetic, harmonic and anharmonic descriptions.
At the full-partition midpoint, the total fractured and intact weights are
equal, \(Z_{\mathrm f}^{*}=Z_{N,0}^{*}\), where \(Z_{N,0}^{*}\equiv Z_{\mathrm i}^{*}\).
Restricting the fractured state to configurations with one fractured bond
gives \(Z_{N,1}^{*}=Z_{N,0}^{*}\). Here \(Z_{N,1}^{*}\) includes all \(N\) possible fracture sites
(\cref{eq:midpoint-paper,eq:multiplicity-paper,eq:one-fracture-partition-paper}).

At zero temperature, the intact and stable one-fracture configurations
coexist when their energies are equal. Writing their energy difference as
\(\Delta E_{N}^{*}(\lambda)=E_{\mathrm f}^{*}(\lambda)-E_{\mathrm i}^{*}(\lambda)\),
the energetic midpoint \(\lambda_0(N)\) satisfies
\begin{equation}
  \Delta E_{N}^{*}\!\left(\lambda_0(N)\right)=0.
  \label{eq:midpoint-zero-temperature-paper}
\end{equation}
For a strongly extended fractured bond, this gives the energetic scaling
\begin{equation}
  \lambda_0(N)-1
  \simeq-\frac{1}{a}\ln\!\left(1-N^{-1/2}\right)
  =\frac{1}{a\sqrt N}+\frac{1}{2aN}
   +O(N^{-3/2}).
  \label{eq:midpoint-energetic-paper}
\end{equation}
At finite temperature, the choice of fracture site and thermal vibrations
also contribute to the statistical weights. The constrained harmonic
approximation includes these entropic contributions through
\begin{equation}
\begin{gathered}
 \Phi_{\mathrm H}(\lambda;N)
 =\frac{\Delta E_{N}^{*}(\lambda)}{\Tstar}-\ln N
 -\frac12\ln\!\left[\frac{D_{\mathrm i}^{*}(\lambda)}
 {D_{\mathrm f}^{*}(\lambda)}\right],\\
 \Phi_{\mathrm H}\!\left(\lambda_{\mathrm H}(N);N\right)=0,
\end{gathered}
  \label{eq:midpoint-harmonic-paper}
\end{equation}
where \(D_{b}^{*}(\lambda)\) is the constrained Hessian determinant evaluated
at the stationary configuration of basin \(b\) (SI Sec.~S5.2).

Including the leading anharmonic correction of the Morse potential gives
\begin{equation}
\begin{gathered}
 \Phi_{\mathrm{AH}}(\lambda;N)
 =\Phi_{\mathrm H}(\lambda;N)
 -\Tstar\left[\mathcal C_{1,\mathrm f}^{*}(\lambda)
 -\mathcal C_{1,\mathrm i}^{*}(\lambda)\right],\\
 \Phi_{\mathrm{AH}}\!\left(\lambda_{\mathrm{AH}}(N);N\right)=0.
\end{gathered}
  \label{eq:midpoint-anharmonic-paper}
\end{equation}
The coefficients \(\mathcal C_{1,\mathrm i}^{*}\) and
\(\mathcal C_{1,\mathrm f}^{*}\) describe the leading anharmonic corrections
to the intact and fractured basin log-partition functions. They are obtained
from a constrained Laplace expansion about the corresponding stationary
configurations. SI Sec.~S5.3 derives these coefficients; Secs.~S5.1--S5.2
give the stationary energies and harmonic weights.

Figure~\ref{fig:scaled-midpoint-paper}(b) compares these predictions with
the simulation midpoints. Full-partition and simulation results cover
\(N=4,\ldots,12,15\), while the deterministic one-fracture results extend
to \(N=200\). The lower plot shows that the excess stretch
\(\lambda_{1/2}-1\) decreases as the chain grows. Multiplying it by
\(\sqrt N\) in the upper plot makes departures from the leading energetic
law more visible. Both plots use \(N^{-1/2}\) on the horizontal axis, so larger
chains lie to the left.

The first-anharmonic prediction closely follows the numerical one-fracture
midpoint, differing by at most \(1.8\times10^{-5}\) for \(9\le N\le50\).
At the larger plotted sizes, the two curves are nearly indistinguishable.

The scaled midpoint can therefore rise while the excess stretch continues to
fall; this rise alone does not indicate an increasing multiple-fracture
contribution. The energetic scaling requires \(\Tstar\sqrt N\ll1\), so
agreement over a finite range cannot establish the large-\(N\) limit at
fixed temperature.

In addition, the fracture-count analysis shows that multiple-fracture
contributions remain negligible at the examined coexistence states for
\(N=4\)--50. SI Sec.~S3.4 gives their weights and the corresponding
linearized midpoint shifts.

\subsubsection{Growth and curvature of the site barrier}
\label{sec:barrier-results}

The site free-energy barrier and its local shape are then examined as
functions of chain length at coexistence. Let \(l_{\mathrm i}^{\mathrm{site},*}\) and
\(l_{\mathrm s}^{\mathrm{site},*}\) denote the intact minimum and neighboring maximum of
\(G_{N}^{\mathrm{site},*}\). Using this reduced profile and the reduced
bond coordinate \(l^{*}\), define
\[
\begin{gathered}
 B_N=\frac{G_{N}^{\mathrm{site},*}(l_{\mathrm s}^{\mathrm{site},*})
             -G_{N}^{\mathrm{site},*}(l_{\mathrm i}^{\mathrm{site},*})}{\Tstar},\\
 \mathcal K_{\alpha}^{*}=\left.\frac{d^2G_{N}^{\mathrm{site},*}}{d\bigl(l^{*}\bigr)^{2}}\right|_{l^{*}=l_{\alpha}^{\mathrm{site},*}},
 \quad \alpha\in\{\mathrm i,\mathrm s\}.
\end{gathered}
\]
The barrier \(B_N\) controls the exponential factor in the KGH rate;
the curvatures \(\mathcal K_{\mathrm i}^{*}>0\) and \(\mathcal K_{\mathrm s}^{*}<0\)
describe the local well stiffness and barrier shape and enter the KGH prefactor
(\cref{eq:kgh-definitions-paper,eq:kgh-rate-paper}). Deterministic,
harmonic and first-anharmonic site PMFs are compared with the adiabatic energy profile,
using the same finite-temperature midpoint extension for every method.
SI Fig.~S4 also compares the positions of the intact minimum, barrier
maximum and fractured minimum of the adiabatic and site-PMF profiles.

Fixing the tagged bond at \(l^{*}\) leaves the other \(N-1\) bonds with mean
length \(z^{*}=(N\lambda-l^{*})/(N-1)\). Expanding their energy about equal sharing
and integrating their constrained Gaussian fluctuations gives
\begin{equation}
  G_{N,\mathrm H}^{\mathrm{site},*}(l^{*})
  =u^{*}(l^{*})+(N-1)u^{*}(z^{*})
   +\frac{\Tstar}{2}(N-2)\ln u^{(2),*}(z^{*}),
  \label{eq:harmonic-site-pmf-paper}
\end{equation}
up to an additive constant. The first two terms give the adiabatic energy;
the logarithmic term accounts for reservoir entropy. The leading
anharmonic correction is
\begin{equation}
  \frac{G_{N}^{\mathrm{site},*}}{\Tstar}
  =\frac{G_{N,\mathrm H}^{\mathrm{site},*}}{\Tstar}
   -\Tstar\mathcal C_{1,N-1}^{*}(z^{*})+O({\Tstar}^2),
  \label{eq:anharmonic-site-pmf-paper}
\end{equation}
where \(\mathcal C_{1,N-1}^{*}\) describes the first correction to the Gaussian
reservoir integral. SI Sec.~S6.2 derives these expressions and the resulting
barriers and curvatures. For comparison, the entropy-free energetic barrier is
\begin{equation}
  B_N^{\mathrm E}(\lambda)
  =\frac{E_{\mathrm{ad}}^{(N-1),*}(l_{\mathrm s}^{\mathrm{ad},*};N,\lambda)
          -E_{\mathrm{ad}}^{(N-1),*}(l_{\mathrm i}^{\mathrm{ad},*};N,\lambda)}{\Tstar},
  \label{eq:energetic-barrier-paper}
\end{equation}
where \(l_{\mathrm i}^{\mathrm{ad},*}\) and \(l_{\mathrm s}^{\mathrm{ad},*}\)
are the intact minimum and saddle of the adiabatic profile (SI Sec.~S6.1). This reference
uses the same finite-temperature midpoint extension and energy scale as
the PMF comparison.

\begin{figure}[!tbp]
  \centering
  \includegraphics[width=8.5cm]{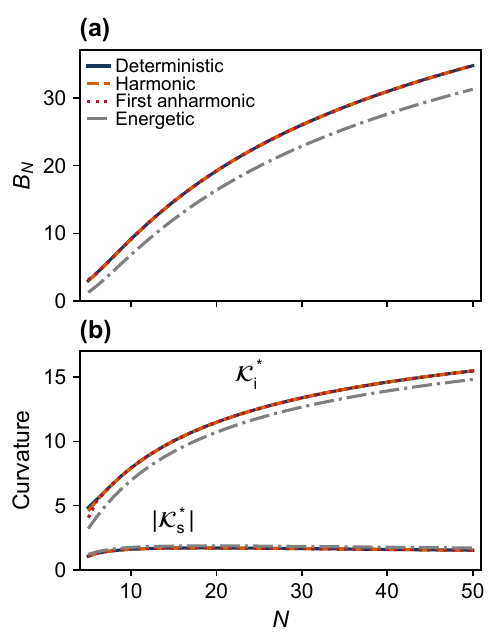}
  \input{figures/midpoint_barriers/curvature_preview/caption}
  \label{fig:barrier-paper}
\end{figure}

Figure~\ref{fig:barrier-paper}(a) shows that the harmonic and anharmonic
approximations closely follow the deterministic barrier. For \(N=7\)--50,
the maximum absolute harmonic error in the dimensional site barrier is
\(0.0607\,k_{\mathrm B}T\); the
first-anharmonic correction reduces it to \(0.0138\,k_{\mathrm B}T\).
The energetic barrier is lower, showing the contribution of reservoir
entropy. The barrier grows nonlinearly: its increment per added bond
falls from about \(1.26\,k_{\mathrm B}T\) between \(N=7\) and 8 to
\(0.343\,k_{\mathrm B}T\) between \(N=49\) and 50.

Panel~(b) shows the positive intact curvature
\(\mathcal K_{\mathrm i}^{*}\) and the magnitude of the negative
barrier curvature \(|\mathcal K_{\mathrm s}^{*}|\).
The intact well becomes progressively stiffer, whereas the barrier
curvature varies more weakly. The harmonic approximation captures
both trends. The first-anharmonic curvatures include both the
displacement of each extremum and the change in the profile's
second derivative, although agreement does not improve uniformly
(SI Fig.~S3). For the deterministic rate estimates presented below,
the first-anharmonic approximation retains the harmonic prefactor and
corrects only the barrier height.

%% file: figures/equilibrium_validation/caption_main.tex
\caption{Constrained-partition predictions compared with constant-extension
  equilibrium simulations at five extensions for each
  \(N=4,\ldots,12\) and \(15\).
  (a) Fractured probability \(P_{\mathrm f}\).
  (b) Intact-restricted virial tension \(F_{\mathrm i,vir}^*\).
  Dashed diagonals indicate equality; the lower plots show simulation minus
  theory at the same \(N\) and extension. Error bars show 95\% confidence intervals.
  Color and symbol
  identify \(N\). Sampling is described in SI Sec.~S9.1.}

%% file: figures/midpoint_scaling/caption.tex
\caption{Coexistence extension. (a) \(N=10\) logits from the present
  simulations, constrained partitions and \citet{Theodorou2024}; zero logit
  denotes equal populations. Simulation bars show 95\% sampling intervals;
  literature bars are transformed standard deviations (the first point has
  no finite lower logit bound). (b) The same midpoints shown as
  \(\sqrt N(\lambda_{1/2}-1)\) above and \(\lambda_{1/2}-1\) below,
  sharing \(N^{-1/2}\); larger chains lie to the left. Simulation and
  full-partition points cover \(N=4,\ldots,12,15\); one-fracture and
  stationary predictions extend through \(N=200\). The dotted line is the
  energetic reference \(1/a\). The anharmonic \(N=4\) point is omitted
  because it belongs to a different root branch.}

%% file: figures/midpoint_barriers/curvature_preview/caption.tex
\caption{Site-PMF barriers and curvatures at coexistence, \(T=450\,\mathrm K\).
(a) Barrier height \(B_N=\Delta G_N^{\mathrm{site}}/(k_{\mathrm B}T)\).
(b) Intact-minimum curvature \(\mathcal K_{\mathrm i}^*\) and barrier-curvature
magnitude \(|\mathcal K_{\mathrm s}^*|\), with energy in units of \(D_e\) and
bond length in units of \(l_0\). Colors and line styles identify the same
methods in both panels. Anharmonic curvatures include the first-order
displacement of the minimum and barrier position. All methods use common midpoint extensions:
full-partition values for \(N=5\)--12 and one-fracture values for
\(N=13\)--50. SI Sec.~S6 gives the derivation and Figure~S3 the differences
from the deterministic calculation.}

%% file: sections/05_rate_results.tex
\subsection{From free-energy barriers to fracture rates}
\label{sec:rate-results}

This section compares simulated and deterministic fracture rates,
beginning with the \(N=10\) literature benchmark, and then examines
their chain-length dependence at coexistence and their extension
dependence away from coexistence.

\subsubsection{Comparison of rate estimates at
\texorpdfstring{\(N=10\)}{N=10}}
\label{sec:theodorou-validation-main}

\begin{figure}[!tbp]
  \centering
  \includegraphics[width=8.5cm]{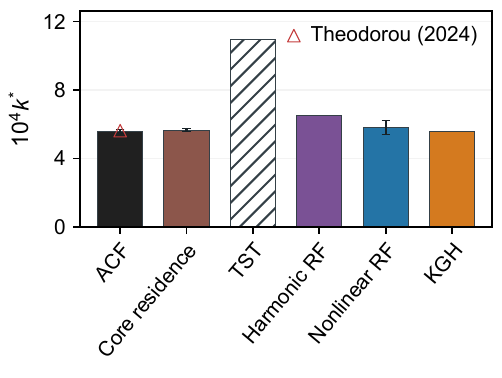}
  \input{figures/rate_comparison_n10/caption}
  \label{fig:n10-rate-comparison-paper}
\end{figure}

Figure~\ref{fig:n10-rate-comparison-paper} compares the present fracture rates with the results of
\citet{Theodorou2024} for \(N=10\), using the same potential, temperature,
friction and instantaneous state divider. At coexistence,
\(\lambda_{1/2}=1.1073539\), the present tension-ACF estimate is
\(5.5855\times10^{-4}\), compared with
\(5.6204\times10^{-4}\) obtained by interpolating the published ACF rates.
These reduced forward rates differ by only 0.62\%, showing excellent
agreement between the two applications of the ACF method.

The core-residence estimate differs from the ACF estimate by about 1\%.
It counts completed transitions between intact and fractured cores per
unit time carrying the origin-core label, whereas ACF infers the rate
from tension relaxation and instantaneous state populations. Their close
agreement therefore supports a consistent switching rate despite the
different state definitions and estimation procedures. The estimates use
the same equilibrium trajectories; SI Sec.~S9.4 compares them across
chain lengths and core definitions.

The barrier-based predictions illustrate the role of recrossing.
Bare TST counts outward crossing attempts without accounting for
recrossing and gives nearly twice the ACF rate. The harmonic dynamical
correction reduces this excess, but the corrected rate remains about
16\% higher. Nonlinear RF, which follows the full Morse dynamics,
and the site-PMF KGH prediction are much closer to the ACF estimate.
Thus the equilibrium crossing flux sets the attempt rate, while the
subsequent dynamics determine how much of that flux contributes to
transitions.

The excellent agreement between equilibrium estimates
(tension ACF and core residence), nonlinear reactive flux and
KGH theory supports a consistent fracture rate at this coexistence
state. These approaches connect equilibrium fluctuations,
completed transitions, barrier-crossing dynamics and free-energy
landscapes through distinct routes.

\subsubsection{Absolute rates at coexistence}
\label{sec:activated-rates}

\begin{figure}[!tbp]
  \centering
  \includegraphics[width=8.5cm]{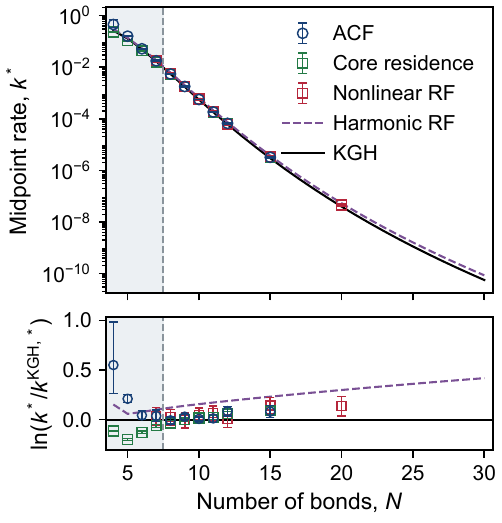}
  \input{figures/coexistence_rates/caption}
  \label{fig:absolute-rates-paper}
\end{figure}

Chain lengths are compared at their coexistence extensions
\(\lambda_{1/2}(N)\), where intact and fractured states have equal
equilibrium populations. Figure~\ref{fig:absolute-rates-paper} shows
fracture rates from equilibrium simulations using the ACF and
core-residence methods, from barrier-top reactive-flux simulations,
and from deterministic harmonic-RF and KGH predictions. The equilibrium
simulations span \(N=4\) to 15, and the barrier-top simulations extend
to \(N=20\).

The rate decreases approximately exponentially over the shorter-chain
range, but the decline becomes less steep as the chain grows. The nonlinear
reactive-flux results support this gentler decline through \(N=20\), in close
agreement with the KGH prediction, which continues the trend to larger
chain lengths. Slower switching makes the intact and fractured states increasingly
long lived. At fixed \(N\), stretching sufficiently beyond
\(\lambda_{1/2}(N)\) favors the fractured state, and reverse
transitions can become too rare to observe, making fracture
practically irreversible on the observation timescale.

This kinetic trend follows the free-energy landscape described in
\cref{sec:barrier-results}. The increasing site barrier suppresses
access to the transition region, while its diminishing increment per
added bond accounts for the gentler decline predicted for larger chains.
The transmission coefficient decreases much more slowly than the rate
(SI Fig.~S11): the nonlinear RF amplitude falls from about 0.62 to 0.40
between \(N=7\) and 20, while the corresponding TST rate decreases by
more than five orders of magnitude. Thus the growing free-energy barrier
primarily controls the slowing of fracture, whereas recrossing reduces the
absolute rate but contributes much less to its chain-length dependence.
The KGH expression allows these contributions to be separated explicitly.

The shaded region bounded by the dashed line in
Figure~\ref{fig:absolute-rates-paper} identifies the shorter chains below
the operational boundary \(N\ge8\) for approximately two-state kinetics
(\cref{sec:crossover-results-paper}). In this region, discrepancies between
rate estimates are appreciable, consistent with recrossing and memory
of the preceding motion. For larger chains, the simulations and KGH
predictions show exceptionally close agreement across several orders of
magnitude in rate. The site-PMF barrier and chain-dependent KGH prefactor
capture both the magnitude and chain-length dependence without fitting a
prefactor to the simulated rates. Harmonic RF remains systematically
higher: it includes a non-unit transmission coefficient but treats
recrossing within a harmonic approximation.
Explicit numerical rate comparisons for each case are reported in SI Sec.~S11.

\begin{figure}[!tbp]
 \centering
 \includegraphics[width=8.5cm]{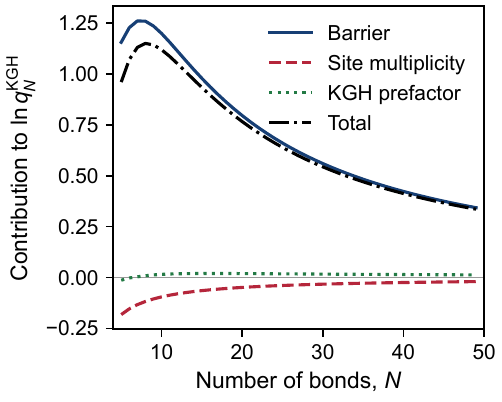}
 \input{figures/rate_decomposition/caption}
 \label{fig:rate-decomposition-paper}
\end{figure}

Given the overall agreement between KGH predictions and simulated rates,
the KGH expression is used to separate the contributions to the chain-length
dependence. For adjacent chains at their respective coexistence extensions
and the same temperature, \cref{eq:kgh-rate-paper} gives
\begin{equation}
 q_N^{\mathrm{KGH}}
 \equiv\frac{k_{1/2}^{\mathrm{KGH},*}(N)}{k_{1/2}^{\mathrm{KGH},*}(N+1)}
 =\underbrace{\frac{N}{N+1}}_{\text{sites}}
  \underbrace{\frac{A_{N}^{\mathrm{KGH},*}}{A_{N+1}^{\mathrm{KGH},*}}}_{\text{prefactor}}
  \underbrace{e^{B_{N+1}-B_N}}_{\text{barrier}}.
 \label{eq:rate-decomposition-paper}
\end{equation}
Here \(q_N^{\mathrm{KGH}}>1\) means that the longer chain fractures more
slowly; each barrier \(B_N\) and prefactor \(A_{N}^{\mathrm{KGH},*}\) is evaluated
at that chain's coexistence extension. Taking logarithms makes the three
contributions additive, as shown in \cref{fig:rate-decomposition-paper}.
The barrier increment dominates the slowing, while the greater
number of possible fracture sites partly offsets it and the
prefactor contribution remains small over the examined range.
As the barrier increment decreases, the rate reduction per added
bond becomes less pronounced, consistent with the gentler decline
at larger \(N\) in Figure~\ref{fig:absolute-rates-paper}.

\subsubsection{Extension dependence and activation length}
\label{sec:pointwise-kgh-paper}
\label{sec:activation-length-paper}

Away from coexistence, simulated fracture rates and KGH predictions are
examined as functions of extension. For each \(N=7,\ldots,12\) and \(15\),
rates are extracted from equilibrium simulations at five extensions corresponding to
fractured populations of approximately 10\%, 25\%, 50\%, 75\% and 90\%.
Evaluating the KGH barrier, curvatures and prefactor at these same extensions
gives excellent agreement with the tension-ACF rates throughout the sampled
range (SI Fig.~S12).

A simple description of the force dependence follows by retaining the
first-order change in the logarithm of the rate. This gives a local
Zhurkov-like expression \citep{Zhurkov1965,Suhir2013},
\begin{equation}
 k_{N}^{*}(F_{\mathrm i,\mathrm{vir}}^{*})\approx A_{Z,N}^{*}
 \exp\!\left(\frac{F_{\mathrm i,\mathrm{vir}}^{*}\ell_{Z,N}^*}{\Tstar}\right),
 \label{eq:local-zhurkov-paper}
\end{equation}
where \(F_{\mathrm i,\mathrm{vir}}^{*}\) is the intact-restricted virial
tension, \(A_{Z,N}^{*}\) is a rate coefficient, and the reduced activation length
is
\begin{equation}
 \ell_{Z,N}^*=\Tstar
 \left[\frac{\partial\ln k_{N}^{*}}{\partial F_{\mathrm i,\mathrm{vir}}^{*}}\right]_{1/2}.
 \label{eq:activation-length-definition-paper}
\end{equation}
The derivative is taken at fixed chain length and temperature and evaluated
at coexistence, denoted by the subscript \(1/2\); thus, \(\ell_{Z,N}^*\)
measures the local response near coexistence and can vary with \(N\).

Given the overall agreement with the simulated rates, these coefficients are
evaluated from the KGH model alone. Expanding its logarithmic rate and matching the
coexistence value gives (SI Sec.~S11.3)
\begin{equation}
 A_{Z,N}^{*}=k_{N,1/2}^{\mathrm{KGH},*}
 \exp\!\left[-\frac{F_{\mathrm i,\mathrm{vir},N,1/2}^{*}\ell_{Z,N}^*}{\Tstar}\right],
 \qquad \ell_{Z,N}^*=\ell_{Z,N}^{\mathrm{KGH},*}.
 \label{eq:local-zhurkov-intercept-paper}
\end{equation}
Here \(k_{N,1/2}^{\mathrm{KGH},*}\) and \(F_{\mathrm i,\mathrm{vir},N,1/2}^{*}\) are the KGH rate
and intact tension at coexistence. The coefficient \(A_{Z,N}^{*}\) is the
extrapolated zero-force intercept of the local exponential and differs from
the KGH prefactor \(A_{N}^{\mathrm{KGH},*}\).
Since \(k_{N}^{\mathrm{KGH},*}=N A_{N}^{\mathrm{KGH},*}e^{-B_N}\), the activation
length separates into barrier and prefactor contributions,
\(\ell_{Z,N}^{\mathrm{KGH},*}=\ell_{B,N}^*+\ell_{A,N}^*\), where
\begin{equation}
 \ell_{B,N}^*
 =-\Tstar\left[\frac{\partial B_N}{\partial F_{\mathrm i,\mathrm{vir}}^{*}}\right]_{1/2},
 \qquad
 \ell_{A,N}^*
 =\Tstar\left[\frac{\partial\ln A_{N}^{\mathrm{KGH},*}}
 {\partial F_{\mathrm i,\mathrm{vir}}^{*}}\right]_{1/2}.
 \label{eq:activation-length-components-paper}
\end{equation}
The dimensionless barrier \(B_N\) is the site-PMF barrier in units of thermal
energy; the constant site multiplicity \(N\) contributes no force derivative.

\begin{figure}[!tbp]
  \centering
  \includegraphics[width=8.5cm]{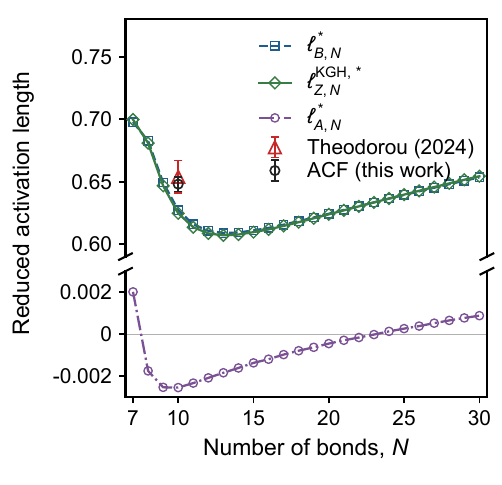}
  \input{figures/activation_lengths/caption.tex}
  \label{fig:activation-length-paper}
\end{figure}

Figure~\ref{fig:activation-length-paper} shows that the total activation
length decreases initially, reaches a shallow minimum around \(N=13\),
and then increases gradually toward \(N=30\). It closely follows the barrier
contribution, whereas the prefactor correction is much smaller. The local
sensitivity of fracture to tension is therefore controlled mainly by the
changing free-energy barrier.
For \(N=10\), the activation length \(\ell_Z^*=0.654\pm0.013\) reported by
\citet{Theodorou2024} is close to the present equilibrium-ACF estimate,
\(\ell_{Z,10}^*\simeq0.648\), obtained from the rate--tension slope across
the five sampled extensions (\cref{fig:activation-length-paper}).
SI Sec.~S11.3 describes the theoretical derivation and the simulation-based
activation-length estimate.

%% file: figures/rate_comparison_n10/caption.tex
\caption{Reduced forward fracture rates near coexistence at \(N=10\).
  Tension ACF and barrier-based predictions use instantaneous states;
  core residence uses last-core labels with \(\eta=0.5\). Hatched TST
  assumes unit transmission. The open red triangle gives the
  ACF rate of \citet{Theodorou2024}, interpolated in log-rate between the
  Table V entries at \(\lambda=1.1050\) and 1.1075 to
  \(\lambda_{1/2}=1.1073539\); no rate uncertainty is reported there.
  The core-residence error bar shows the approximate 95\% Poisson interval
  \(k_{\mathrm i\to f}^{\mathrm{core},*}\pm1.96\,\mathrm{SE}(k_{\mathrm i\to f}^{\mathrm{core},*})\), based on transition counts.
  ACF and nonlinear RF show separate 95\% bootstrap intervals, obtained by
  resampling equilibrium trajectories and launch configurations, respectively.
  These are individual-method intervals, not a paired comparison.}

%% file: figures/coexistence_rates/caption.tex
\caption{\revphaseseven{Fracture rates at coexistence and comparison with KGH theory.
  The upper panel shows tension-ACF, core-residence (\(\eta=0.5\)) and nonlinear-RF
  estimates as symbols, with KGH and harmonic-RF predictions as solid and dashed
  curves. The lower panel shows \(\ln(k^*/k^{\mathrm{KGH},*})\).
  Shading and the divider at \(N=7.5\) mark the region below the operational
  two-state boundary (\cref{sec:crossover-results-paper}). Error bars are 95\%
  bootstrap intervals for ACF and nonlinear RF, and approximate 95\% Poisson
  intervals for core residence, \(k^*\pm1.96\,\mathrm{SE}(k^*)\).}}

%% file: figures/rate_decomposition/caption.tex
\caption{Barrier, fracture-site and prefactor contributions to the logarithm of
the adjacent-chain KGH rate ratio (\cref{eq:rate-decomposition-paper}).
The barrier increment, \(B_{N+1}-B_N\), site multiplicity,
\(\ln[N/(N+1)]\), and prefactor,
\(\ln[A_N^{\mathrm{KGH},*}/A_{N+1}^{\mathrm{KGH},*}]\), sum to the total.
Each chain is evaluated at its coexistence extension at the same temperature.}

%% file: figures/activation_lengths/caption.tex
\caption{KGH activation length and its barrier and prefactor contributions
near coexistence. The broken vertical axis shows the barrier and total lengths
in the upper range and magnifies the much smaller prefactor contribution in
the lower range; both use the same units. The red triangle at \(N=10\) shows \citet{Theodorou2024},
\(\ell_Z^*=0.654\pm0.013\), with the quoted uncertainty. The black circle
is the present ACF estimate; its error bar shows \(\pm\) one standard error
from the regression, with force values treated as fixed. Numerical evaluation and precision
are discussed in SI Sec.~S11.3.}

%% file: sections/06_committor_and_crossover.tex
\subsection{Locating the transition state through committor analysis}
\label{sec:committor-crossover-results}
\label{sec:mechanistic-diagnostics}
\label{sec:committor-results-paper}

A useful transition coordinate should locate the transition and distinguish
configurations likely to fracture from those likely to return to the intact
state. Tagged-bond length is tested using the configuration committor \citep{PetersTrout2006}; i.e., the
probability that a trajectory launched from particle positions \(\bm r\)
reaches the fractured core before the intact core,
\begin{equation}
  q_{\mathrm{comm}}(\bm r)
  =\Pr_{\bm v,\mathcal W}\!\left(\tau_{\mathrm f}<\tau_{\mathrm i}\mid\bm r\right).
  \label{eq:configuration-committor-paper}
\end{equation}
The two cores are regions inside the intact and fractured basins, and
\(\tau_{\mathrm i}\) and \(\tau_{\mathrm f}\) are their first-entry times.
The probability averages over Maxwell-distributed initial velocities
\(\bm v\) with zero total momentum and subsequent Langevin noise
\(\mathcal W\). Fracture at any bond counts as reaching the fractured core.
The core boundaries and sampling are specified in SI Sec.~S12.

Different configurations can share the same reduced tagged-bond length
\(l^{*}\). Averaging over their conditional equilibrium distribution,
with the remaining bonds initially restricted to the intact reservoir,
gives the projected committor
\begin{equation}
  \overline q(l^{*})
  =\operatorname E_{\bm r\mid l^*}
    \!\left[q_{\mathrm{comm}}(\bm r)\right],
  \qquad \overline q(l_{1/2}^{*})=\tfrac12.
  \label{eq:projected-committor-paper}
\end{equation}
The tagged length is fixed when preparing a configuration and all bonds evolve
during the ensuing trajectory. At \(l_{1/2}^{*}\), the fracture probability
averaged over configurations is one half. However, individual configurations at the
same tagged length can have different probabilities because
the remaining bonds are arranged differently. For \(N=6,8,10\) at coexistence,
the crossing is located before independent configurations are sampled there
to measure this variation.

\begin{figure}[htbp]
  \centering
  \includegraphics[width=8.5cm]{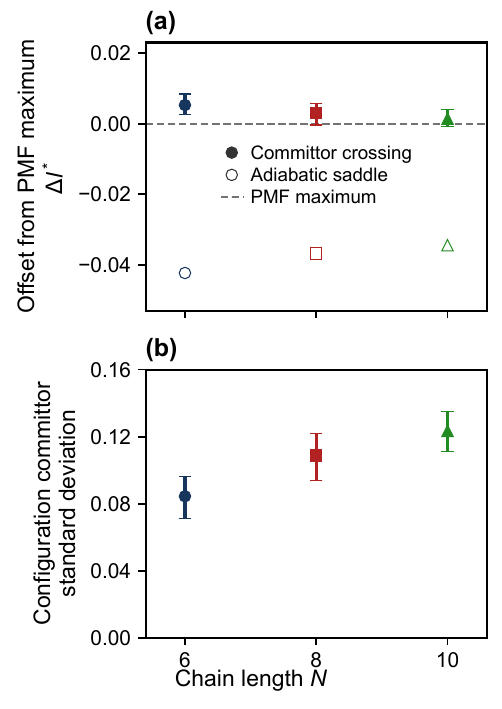}
  \input{figures/committor/caption_main}
  \label{fig:committor-results-paper}
\end{figure}

According to Fig.~\ref{fig:committor-results-paper}a the projected
crossing lies close to the finite-temperature site-PMF
maximum for all three chains and substantially farther from the adiabatic
saddle. The measured offsets
\(\widehat l_{1/2}^{*}-l_{\mathrm s}^{\mathrm{site},*}\) are approximately
\(\RegisteredCommittorPmfOffsetSix\),
\(\RegisteredCommittorPmfOffsetEight\), and
\(\RegisteredCommittorPmfOffsetTen\) for \(N=6,8,10\), respectively,
with each comparison made at the same state; therefore, this close
correspondence supports the site PMF as a useful description of the
projected dynamical transition.

The variation between configuration committors at fixed tagged
length is quantified by their probability-scale standard deviation,
\begin{equation}
  \sigma_q(l^{*})=\left\{
    \operatorname{Var}_{\bm r\mid l^{*}}
    \!\left[q_{\mathrm{comm}}(\bm r)\right]
  \right\}^{1/2}.
  \label{eq:committor-dispersion-paper}
\end{equation}
Repeated launches from each configuration produce success fractions whose
spread reflects both configuration differences and finite-shooting
variability. A hierarchical binomial model accounts for the configuration,
velocity and noise sampling levels to estimate \(\sigma_q\)
(SI Sec.~S12). At the selected crossings, the estimated SDs are
\(\RegisteredCommittorSdSix\), \(\RegisteredCommittorSdEight\), and
\(\RegisteredCommittorSdTen\) for \(N=6,8,10\)
(Fig.~\ref{fig:committor-results-paper}b). Tagged-bond length therefore
locates the average transition well, but does not fully determine an
individual configuration's fracture probability: different arrangements
of the remaining bonds give different probabilities at the same tagged length.

\subsection{Two-state accuracy and return-excursion timescales}
\label{sec:crossover-results-paper}

A chain may cross a dividing surface several times before completing a
transition between intact and fractured states. The analysis examines when this motion
can be described by two constant directional rates, and how the duration of
unsuccessful excursions compares with the time over which the state
populations relax. The calculations cover the sampled coexistence states
for \(N=6\)--10 at \(T=450\,\mathrm K\).

The states are identified using intact and fractured cores---regions lying
inside their respective basins. Each trajectory carries the label of the
most recently visited core until it reaches the opposite core. At an
observation lag \(\tau_{\mathrm{lag}}\), transition counts give the fractions of
intact-start and fractured-start observations whose endpoints carry the
opposite label. These transition probabilities determine the corresponding
two-state rates (SI Secs.~S13.1--S13.2).

One trajectory group provides the calibrated reference rates
\(k_{\mathrm{if}}^{\mathrm{core},0}\) and \(k_{\mathrm{fi}}^{\mathrm{core},0}\) and their sum defines the
population-relaxation time,
\begin{equation}
  k_{\mathrm{eff}}^{\mathrm{core},0}=k_{\mathrm{if}}^{\mathrm{core},0}+k_{\mathrm{fi}}^{\mathrm{core},0},\qquad
  \tau_{\mathrm{pop}}=\frac{1}{k_{\mathrm{eff}}^{\mathrm{core},0}}.
  \label{eq:crossover-clock-paper}
\end{equation}
In the two-state model, an initial deviation from the equilibrium population
decays exponentially with this time constant. The quantity
\(\tau_{\mathrm{pop}}\) compares observation times across chain lengths;
its calibration is described in SI Sec.~S13.3 and its values are listed in
Table~S20.

\begin{figure}[!tbp]
  \centering
  \includegraphics[width=8.5cm]{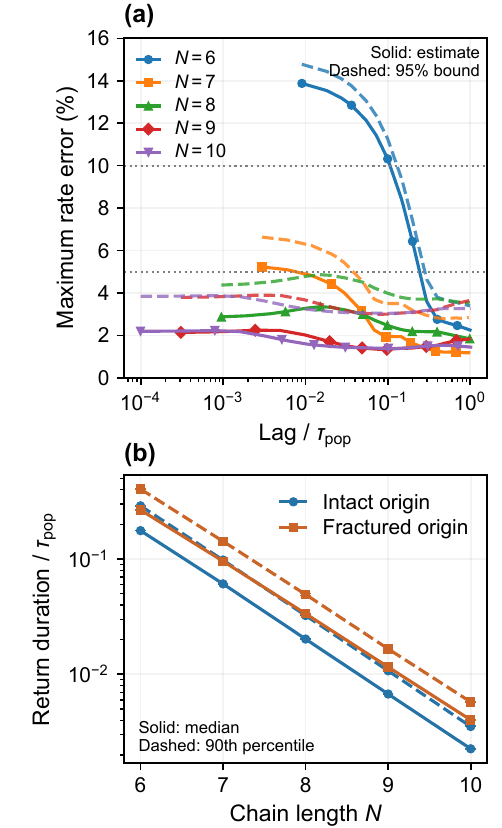}
  \caption{Two-state accuracy and failed-excursion timescales at the sampled
  coexistence states (\(N=6\)--10, \(T=450\,\mathrm K\)).
  (a) Measured maximum absolute relative discrepancy \(E_N(\tau_{\mathrm{lag}})\)
  (solid; Eq.~\eqref{eq:crossover-error-paper}, SI Eq.~(S204)) and
  sampling-uncertainty bound \(U_N(\tau_{\mathrm{lag}})\) (dashed),
  plotted against \(\tau_{\mathrm{lag}}/\tau_{\mathrm{pop}}\).
  \(U_N\) has approximate 95\% simultaneous coverage within
  each \(N\) (SI Sec.~S13.4). Dotted lines mark 5\% and 10\% discrepancies.
  (b) Median (solid) and 90th-percentile (dashed) durations of completed
  failed excursions divided by \(\tau_{\mathrm{pop}}\), using the
  adiabatic counting surface and primary fixed cores. Symbols distinguish
  intact and fractured origins. Bars are pointwise 95\% whole-trajectory
  bootstrap intervals, retaining dependence between duration and clock.
  Lines in (b) guide the eye.}
  \label{fig:kinetic-crossover-paper}
\end{figure}

\begin{samepage}
Independent validation trajectories test whether these reference rates
describe switching across different observation lags and state definitions.
At each lag, the measured discrepancy is
\begin{equation}
  E_N(\tau_{\mathrm{lag}})=
  \max_{\nu,c_{\mathrm{an}}}\left|
  \frac{\widehat k_{\nu}^{\mathrm{core},\mathrm{validation},c_{\mathrm{an}}}(\tau_{\mathrm{lag}})}
       {k_{\nu}^{\mathrm{core},0}}-1\right|,
  \label{eq:crossover-error-paper}
\end{equation}
\end{samepage}
with \(\nu=\mathrm{if}\) or \(\mathrm{fi}\) denoting intact-to-fractured or
reverse switching,
and \(c_{\mathrm{an}}\) the corresponding core definition and observation interval. The maximum
includes both directions and all nine tested conditions, including the
primary condition used for calibration; thus, a small \(E_N\) indicates
agreement for every tested condition. Figure~\ref{fig:kinetic-crossover-paper}a
shows the measured discrepancy and its sampling-uncertainty bound \(U_N\);
the latter has approximate 95\% simultaneous coverage within each chain
length (SI Sec.~S13.4).

With sampling uncertainty included, the results support broad intervals
of agreement within 5\% for \(N=7\)--10 while for \(N=8\)--10, agreement already
holds at the shortest tested lag, \(4\,\mathrm{fs}\). For \(N=6\),
individual lags satisfy the 5\% criterion, but a broad interval is supported
only at 10\%. The interval definition and numerical details are given
in SI Sec.~S13.4 and Table~S21.

Failed excursions are then examined: trajectories leave one core, reach
the adiabatic dividing surface, and return to their original core before
reaching the opposite one. Figure~\ref{fig:kinetic-crossover-paper}b shows
their median and 90th-percentile durations divided by
\(\tau_{\mathrm{pop}}\). The median durations increase modestly from
\(\RegisteredCrossoverMedianSixI/\RegisteredCrossoverMedianSixF\,\mathrm{fs}\)
for intact/fractured origins at \(N=6\) to
\(\RegisteredCrossoverMedianTenI/\RegisteredCrossoverMedianTenF\,\mathrm{fs}\)
at \(N=10\), whereas the population time increases much faster, from
\(\RegisteredCrossoverTauSix\) to
\(\RegisteredCrossoverTauTen\,\mathrm{fs}\). Failed excursions therefore occupy a
progressively smaller fraction of the population-relaxation time; these
ratios quantify the separation between excursion and population timescales.
SI Sec.~S13.5 explains the duration and uncertainty estimates; Figure~S15
shows the physical durations and their dependence on the counting surface
at fixed cores.

%% file: figures/committor/caption_main.tex
\caption{Committor analysis at coexistence for \(N=6,8,10\)
at \(450\,\mathrm K\).
(a) Projected half-committor crossings \(\widehat l_{1/2}^*\) (filled symbols)
and adiabatic saddles \(l_{\mathrm{ad}}^{\ddagger,*}\) (open symbols), relative to the site-PMF maximum:
\(\Delta l^*=l^*-l_{\mathrm s}^{\mathrm{site},*}\), with \(l^*=l/l_0\).
The dashed line marks the PMF maximum.
(b) Standard deviation of configuration committors at each estimated crossing.
Error bars show 95\% sampling intervals with the shooting lengths held fixed
(SI Sec.~S12). Colors and marker shapes identify the chain lengths in both panels.}

%% file: sections/07_conclusions.tex
\section{Conclusions}
\label{sec:conclusions}

This study examines fracture and healing in periodic one-dimensional Morse
chains at fixed extension, building on the reversible-chain model of
\citet{Theodorou2024} and motivated by strongly stretched, approximately
collinear strands near failure. The extension constraint couples a
stretching bond to the remaining chain, while the intact-reservoir site PMF
describes first fracture by excluding configurations already fractured
elsewhere that contribute to the unrestricted tagged-bond PMF. Equilibrium
surface probability and the
constrained-coordinate kinetic metric determine the attempt flux; dynamical
recrossing then corrects this flux to obtain the transition rate.

Equilibrium simulations and direct numerical evaluation of the constrained
partition function give practically identical intact and fractured
probabilities, \(P_{\mathrm i}\) and \(P_{\mathrm f}\), and mean intact-state
tensions across the sampled states. Partition models allowing one or
multiple fractured bonds show negligible multiple-fracture weight near
coexistence for \(N=4\)--50 at the studied temperature, supporting the
one-fracture approximation over this range.

Fracture rates were estimated using deterministic transition-state theory (TST),
harmonic reactive-flux theory and Kramers--Grote--Hynes (KGH) theory;
deterministic TST flux with nonlinear barrier-top simulations;
and tension-autocorrelation and core-residence analyses of equilibrium
constant-extension trajectories. KGH, nonlinear reactive flux and the
equilibrium estimates show excellent agreement over several orders of
magnitude where two-state kinetics is accurate, without fitting a kinetic
prefactor to measured rates. Bare TST and harmonic reactive flux remain
higher, demonstrating the importance of the dynamical treatment in
connecting equilibrium thermodynamics to fracture kinetics.

Increasing chain length lowers the coexistence stretch while raising the
site-PMF barrier, whose growth dominates the slowing of fracture; additional
fracture sites partly compensate, while transmission varies much more weakly
than the rate. The diminishing barrier increment explains the gentler rate
decline supported by nonlinear reactive flux through \(N=20\). Slow switching
at coexistence also has a practical consequence: under continued stretching
on a finite loading timescale, a long chain may remain intact beyond
coexistence before fracturing. At these larger extensions the fractured state
is favored, and healing can become too rare to observe. Fracture can thus be
practically irreversible even though the underlying model permits healing.

The site-PMF KGH model also agrees closely with equilibrium simulations
across the sampled extensions away from coexistence, providing a compact
route from barriers, curvatures and prescribed dynamics to rates. Near
coexistence, linearizing its logarithmic rate yields a Zhurkov-type tension
dependence with a chain-dependent activation length, dominated by the
barrier derivative with only a small prefactor correction. The local
theoretical derivative and the simulation slope over the sampled extension
interval connect the landscape to the force response within the
fixed-extension ensemble.

Recrossing and relaxation tests clarify when constant rates adequately
describe switching. Independent trajectory comparisons support broad intervals of rate
agreement and motivate the operational \(N\ge8\) convention under the
studied conditions. Failed returns occupy a decreasing fraction of the
population-relaxation time as chains grow, complementing the direct rate
test and clarifying when a two-state description becomes useful despite
repeated crossings of the dividing surface.

Committor calculations measure the probability of reaching the fractured
core before the intact core and test whether tagged-bond length predicts
this outcome. The projected half-probability crossings lie close to
site-PMF maxima, supporting their role in locating the transition region,
but configurations with the same tagged length retain different fracture
probabilities because the remaining bonds are arranged differently.
Bond length therefore identifies the average transition without fully
determining individual outcomes.

These findings offer a basis for parametrizing mesoscopic fracture models
with state- and chain-length-dependent rates. Natural next steps include
tension-controlled chains \citep{PuthurSebastian2002,Ghosh2010},
finite-compliance loading devices connecting fixed-extension and
fixed-tension limits \citep{BucheRimsza2023}, and fracture in bulk polymer
networks. These extensions would clarify how loading conditions, interchain
interactions and load redistribution influence metastability, free-energy
landscapes, rates and the two-state crossover, with the present
fixed-extension model providing a reference limit.

%% file: si_sections/notation_table.tex
\section{Notation}
\label{si:global-notation}

Stars identify reduced quantities; physical quantities are unstarred.
The dimensionless groups \(a,c,\lambda\),
probabilities, ratios, counts and geometric matrices remain unstarred.
Physical and reduced counterparts share a row where useful. The reference
scales are \(l_0\) for length, \(D_{\mathrm e}\) for energy and \(\xi^{-1}\)
for time; temperature is reduced as \(T^*=k_{\mathrm B}T/D_{\mathrm e}\).
Physical simulation settings and selected thermal-unit comparisons retain
their stated units. Section~\ref{si:notation} gives the state and unit
conventions in detail.

Table~\ref{si:tab:notation} groups Latin symbols alphabetically by base
letter, followed by Greek symbols in Greek alphabetical order, then
indices and modifiers. Local reuses of a letter are identified by their
scope. Definitions at first use remain authoritative for their stated
ensemble, coordinate and estimator.

\begingroup
\small
\setlength{\tabcolsep}{5pt}
\renewcommand{\arraystretch}{1.08}
\begin{longtable}{@{}>{\raggedright\arraybackslash}p{\dimexpr0.36\linewidth-\tabcolsep\relax}>{\raggedright\arraybackslash}p{\dimexpr0.64\linewidth-\tabcolsep\relax}@{}}
\caption{Notation for the main article and supplementary information.}
\label{si:tab:notation}\\
\toprule
Symbol & Description \\
\midrule
\endfirsthead
\multicolumn{2}{l}{\tablename~\thetable\ (continued)}\\
\toprule
Symbol & Description \\
\midrule
\endhead
\midrule
\multicolumn{2}{r}{\textit{Continued on next page}}\\
\endfoot
\bottomrule
\endlastfoot
\multicolumn{2}{@{}l}{\textbf{Latin symbols}}\\*
\addlinespace[3pt]
\(a\) & Morse shape parameter, \(a=\alpha l_0\); dimensionless. \\ 
\addlinespace[3pt]
\(a_0,\allowbreak\ a_1^*\) & Intercept and reduced-tension slope in \(\ln k_j^{\mathrm{ACF},*}=a_0+a_1^*F_{\mathrm i,\mathrm{vir},j}^*\); \(\ell_Z^*=T^*a_1^*\) (Sec.~\ref{si:activation-length}). \\ 
\addlinespace[3pt]
\(A\) & Metropolis acceptance probability, \(\min\{1,e^{-\beta\Delta U}\}\) (Sec.~\ref{si:sec:crossover-trajectories}). \\ 
\addlinespace[3pt]
\(A_k\) & One specified set of \(k\) fractured-bond indices in the count-resolved partition (Sec.~\ref{si:fracture-count-resolution}). \\ 
\addlinespace[3pt]
\(A_N^{\mathrm{KGH},*},\allowbreak\ A_{Z,N}^*\) & Reduced one-site KGH prefactor and local Zhurkov rate coefficient, respectively. The latter includes the coexistence barrier and force shift (Sec.~\ref{si:activation-length}). \\ 
\addlinespace[3pt]
\(A_3^*,\allowbreak\ A_4^*\) & Cubic and quartic terms after displacement rescaling in the constrained Laplace expansion (Sec.~\ref{si:anharmonic-midpoint}). \\
\addlinespace[3pt]
\(A_+^*,\allowbreak\ \mathcal A^*\) & Growing-mode amplitude and linear Langevin drift matrix in reduced variables (Sec.~\ref{si:harmonic-transmission}). Here \(A_+^*\) differs from the endpoint prefactor abbreviated \(A_\pm^*\) in Sec.~\ref{si:activation-length}. \\ 
\addlinespace[3pt]
\(B\) & Geometric map from independent to all bond displacements in Sec.~\ref{si:anharmonic-midpoint}; periodic particle-to-bond incidence matrix in Sec.~\ref{si:harmonic-transmission}. Both are dimensionless. \\
\addlinespace[3pt]
\(B\) & Number of bootstrap samples (Sec.~\ref{si:sec:crossover-uncertainty}). \\ 
\addlinespace[3pt]
\(B_{cv}(u,w)\) & Binomial probability of the observed shooting count conditional on the two latent effects (Sec.~\ref{si:sec:committor-dispersion}). \\ 
\addlinespace[3pt]
\(B_N,\allowbreak\ B_N^{\mathrm E}\) & Site-PMF barrier in thermal units, \(B_N=\Delta G_N^{\mathrm{site},*}/T^*=\Delta G_N^{\mathrm{site}}/(k_{\mathrm B}T)\); the unstarred \(\Delta G_N^{\mathrm{site}}\) is physical. \(B_N^{\mathrm E}\) uses the adiabatic energy barrier instead. Neither includes site multiplicity. \\ 
\addlinespace[3pt]
\(\mathcal B_{\mathrm i},\allowbreak\ \mathcal B_{\mathrm f}\) & Instantaneous intact and fractured basins, separated by \(d_N^*\) (Sec.~\ref{si:sec:crossover-trajectories}). \\ 
\addlinespace[3pt]
\(c\) & Dimensionless dynamics parameter, \(D_{\mathrm e}/(m\xi^2l_0^2)\); distinct from the configuration index \(c\) in Sec.~\ref{si:sec:committor}. \\ 
\addlinespace[3pt]
\(c_{\mathrm{an}}\) & Crossover analysis condition \((\eta,\Delta t_{\mathrm{obs}})\); \(c_{\mathrm{an},0}\) is the primary condition (Sec.~\ref{si:sec:crossover-trajectories}). \\ 
\addlinespace[3pt]
\(c_b^*(\eta)\) & Reduced core boundary between minimum \(l_b^{\mathrm{ad},*}\) and divider \(d_N^*\). Physical boundaries are \(l_0c_b^*\); \(b=\mathrm i,\mathrm f\). \\ 
\addlinespace[3pt]
\(C,\allowbreak\ C_{ab}^{\mathcal D,c_{\mathrm{an}}}\) & Number of sampled configurations in Sec.~\ref{si:sec:committor-sampling}; counts of lagged state-label pairs from \(a\) to \(b\) in Sec.~\ref{si:sec:crossover-rate-estimation}. \\ 
\addlinespace[3pt]
\(C_{FF}(t^*),\allowbreak\ C_0\) & Normalized tension autocorrelation and amplitude of its slow exponential component (Sec.~\ref{si:acf-method}); both dimensionless. \\ 
\addlinespace[3pt]
\(C_{\mathrm{if}}(t^*)\) & Equilibrium state correlation \(\langle h_{\mathrm i}(0)h_{\mathrm f}(t^*)\rangle\) (Sec.~\ref{si:rf-estimation}). \\ 
\addlinespace[3pt]
\(C_{\mathrm{tag}}^*,\allowbreak\ C_{\mathrm{site}}^*\) & Arbitrary additive constants in reduced tagged and site PMFs. \\ 
\addlinespace[3pt]
\(C^*,\allowbreak\ C_q^*,\allowbreak\ C_{VA}^*\) & Scaled bond-displacement covariance in Sec.~\ref{si:anharmonic-midpoint}; surface-position covariance and tagged-velocity/growing-amplitude covariance in Sec.~\ref{si:harmonic-transmission}. \\
\addlinespace[3pt]
\(\mathcal C_{1,b}^*,\allowbreak\ \mathcal C_{1,M}^*\) & First constrained Laplace correction for basin \(b\) or an equal-sharing reservoir of \(M\) bonds; \(\ln Z_b^*=\ln Z_b^{\mathrm H,*}+T^*\mathcal C_{1,b}^*+O((T^*)^2)\). \\ 
\addlinespace[3pt]
\(\mathcal C_{\mathrm i},\allowbreak\ \mathcal C_{\mathrm f}\) & Intact and fractured target cores defined by the maximum bond length and \(c_b^*\); sets, not energies or covariances. \\ 
\addlinespace[3pt]
\(d_N^*,\allowbreak\ d^*\) & Reduced adiabatic saddle and instantaneous state divider, also written \(l_{\mathrm{ad}}^{\ddagger,*}\); \(d^*\) suppresses fixed \(N,\lambda\). The site-PMF reservoir uses this fixed cutoff. \\ 
\addlinespace[3pt]
\(d_N^{(1),*}\) & Derivative \(d d_N^*/d\lambda\) of the moving divider (Sec.~\ref{si:moving-divider-force}); its argument is stretch rather than bond length. \\ 
\addlinespace[3pt]
\(D_{\mathrm e}\) & Physical Morse depth per bond and reference energy; \(N_{\mathrm A}D_{\mathrm e}\) is the molar depth reported in the parameter table. \\ 
\addlinespace[3pt]
\(D_b^*,\allowbreak\ D_{\mathrm{res}}^*\) & Determinants of the reduced constrained Hessians at basin minimum \(b\) and in the equal-sharing reservoir (Secs.~\ref{si:midpoint-approximations}, \ref{si:site-barriers}). \\ 
\addlinespace[3pt]
\(D_\alpha^*,\allowbreak\ D_b^*\) & Observed duration of retained residence \(\alpha\), and full residence duration in state \(b\), measured in reduced time (Sec.~\ref{si:core-method}); distinct from the Hessian determinant \(D_b^*\) in S5. \\ 
\addlinespace[3pt]
\(D_e,\allowbreak\ D_{(j)},\allowbreak\ \widehat D_p\) & Physical duration of excursion \(e\), ordered duration \(j\), and estimated quantile at probability \(p\) (Sec.~\ref{si:sec:crossover-excursions}); reported in fs. Italic \(e\) labels an episode, unlike roman \(\mathrm e\) in the Morse depth. \\ 
\addlinespace[3pt]
\(\mathcal D\) & Calibration or validation trajectory set (Sec.~\ref{si:sec:crossover-rate-estimation}). \\ 
\addlinespace[3pt]
\(e_\alpha\) & Residence exit indicator: one for a recorded switch and zero for terminal right-censoring (Sec.~\ref{si:core-method}). \\ 
\addlinespace[3pt]
\(\bm e_j\) & Unit vector selecting bond \(j\) (Sec.~\ref{si:harmonic-transmission}). \\ 
\addlinespace[3pt]
\(E\) & Orthonormal basis removing uniform particle translation (Sec.~\ref{si:harmonic-transmission}); dimensionless. \\ 
\addlinespace[3pt]
\(E_b^*,\allowbreak\ \Delta E_N^*\) & Reduced energy at stationary basin configuration \(b\) and fractured-minus-intact difference, \(\Delta E_N^*=E_{\mathrm f}^*-E_{\mathrm i}^*\). Superscript \((2)\) on \(E^{(2),*}\) identifies the quadratic expansion term. \\ 
\addlinespace[3pt]
\(E_{\mathrm{ad}}^{(N-1),*},\allowbreak\ E_{\mathrm{ad}}^{(1),*}\) & Reduced equal-sharing and one-bond reservoir branches, both referenced to the uniform-chain energy \(Nu^*(\lambda)\); \(E_{\mathrm{env}}^*\) is their lower envelope. \\ 
\addlinespace[3pt]
\(E_{\mathrm{ad,abs}}^*\) & Absolute reduced equal-sharing energy, \(u^*(l^*)+(N-1)u^*(z^*)\). Its physical counterpart is \(U_{\mathrm{ad,abs}}^{(N-1)}=D_{\mathrm e}E_{\mathrm{ad,abs}}^*\). \\ 
\addlinespace[3pt]
\(E_{\mathrm{tag}}^{\mathrm{pot},*},\allowbreak\ U_{\mathrm{tag}}^{\mathrm{pot}}\) & Conditional mean chain potential energy at fixed tagged length, in reduced and physical units, respectively; \(E_{\mathrm{tag}}^{\mathrm{pot},*}=U_{\mathrm{tag}}^{\mathrm{pot}}/D_{\mathrm e}\) (Sec.~\ref{si:sec:conditional-potential-energy}). \\ 
\addlinespace[3pt]
\(E_N(\tau_{\mathrm{lag}})\) & Maximum absolute relative validation-rate discrepancy across directions and conditions (Sec.~\ref{si:sec:crossover-validation}); dimensionless, not an energy. \\ 
\addlinespace[3pt]
\(f^*(l^*)\) & Reduced single-bond tension, \(u^{(1),*}(l^*)\); \(f_b^*\) is its common value at stationary configuration \(b\). \\ 
\addlinespace[3pt]
\(F,\allowbreak\ F^*\) & Physical and reduced chain tension, \(F^*=Fl_0/D_{\mathrm e}\). \\ 
\addlinespace[3pt]
\(F_{b,\mathrm{vir}}^*,\allowbreak\ F_{b,\mathrm{conj}}^*\) & Basin-averaged reduced virial tension and conjugate tension including the moving-divider contribution (Sec.~\ref{si:moving-divider-force}). The intact virial tension is the activation-force coordinate. \\ 
\addlinespace[3pt]
\(F_{rj}^*,\allowbreak\ \bar F^*,\allowbreak\ \delta F^*\) & Reduced tension sample \(j\) of trajectory \(r\), pooled mean, and fluctuation about the equilibrium mean (Sec.~\ref{si:acf-method}). \\ 
\addlinespace[3pt]
\(g_N(\lambda),\allowbreak\ g_1(\lambda)\) & Population logit \(\ln(P_{\mathrm f}/P_{\mathrm i})\), and one-fracture log-weight ratio \(\ln(Z_{N,1}^*/Z_{N,0}^*)\) (Sec.~\ref{si:thermodynamics}). A bare \(g\) suppresses the fixed chain index. \\ 
\addlinespace[3pt]
\(G_N^{\mathrm{tag},*},\allowbreak\ G_N^{\mathrm{site},*}\) & Reduced unrestricted tagged-bond and intact-reservoir site potentials of mean force (PMFs). Only the latter excludes competing reservoir fractures; additive energy zeros are arbitrary. \\ 
\addlinespace[3pt]
\(h_b,\allowbreak\ h_{\mathrm f,d_N^*}\) & Instantaneous basin indicators; \(h_{\mathrm f,d_N^*}=\mathbf1[\max_jl_j^*>d_N^*]\) makes the fixed divider explicit. Indicators remain unstarred. \\ 
\addlinespace[3pt]
\(h\) & Interpolated order-statistic rank \(1+(n-1)p\) for duration quantiles (Sec.~\ref{si:sec:crossover-excursions}). \\ 
\addlinespace[3pt]
\(h_-^*\) & Negative eigenvalue of the reduced internal particle Hessian at the saddle (Sec.~\ref{si:harmonic-transmission}). \\ 
\addlinespace[3pt]
\(H^*,\allowbreak\ H_{\mathrm{int}}^*\) & Reduced Hessian: independent bond coordinates in Sec.~\ref{si:midpoint-approximations}, particle coordinates in Sec.~\ref{si:harmonic-transmission}; \(H_{\mathrm{int}}^*=E^{\mathsf T}H^*E\) removes translation. \\ 
\addlinespace[3pt]
\(I\) & Identity matrix of the required dimension. \\ 
\addlinespace[3pt]
\(J_{\mathrm{if}}^*,\allowbreak\ J_{\mathrm i\to\mathrm f}^{\mathrm{core},*}\) & Directional transition frequencies per total reduced observation time, for instantaneous and core-defined states, respectively; rates divide these currents by the corresponding origin population. \\ 
\addlinespace[3pt]
\(J_{N,s^*}^{+,*},\allowbreak\ J_{N,s^*\mid d_N^*}^*\) & Reduced positive crossing frequency at launch surface \(s^*\), and signed finite-time current with fixed state divider \(d_N^*\) (Sec.~\ref{si:rf-framework}). \\ 
\addlinespace[3pt]
\(k,\allowbreak\ k^*;\allowbreak\ k_N,\allowbreak\ k_N^*\) & Physical and reduced rates, \(k^*=k/\xi\); physical rates have inverse-time units. Subscript \(N\) specifies the chain fracture rate; method and state modifiers distinguish its estimators. \\ 
\addlinespace[3pt]
\(k_{\mathrm B}\) & Boltzmann constant; \(k_{\mathrm B}T\) is the physical thermal energy used in selected comparisons. \\ 
\addlinespace[3pt]
\(k_{\mathrm i\to\mathrm f}^*,\allowbreak\ k_{\mathrm f\to\mathrm i}^*,\allowbreak\ k_{\mathrm{eff}}^*\) & Reduced fracture, healing and total population-decay rates for instantaneous states; \(k_{\mathrm{eff}}^*=k_{\mathrm i\to\mathrm f}^*+k_{\mathrm f\to\mathrm i}^*\). \\ 
\addlinespace[3pt]
\(k_{b\to\bar b}^{\mathrm{core},*},\allowbreak\ k_\nu^{\mathrm{core}}\) & Reduced count/exposure residence rates (Sec.~\ref{si:core-method}) and physical finite-lag core rates (Sec.~\ref{si:sec:crossover-rate-estimation}). Their estimators differ; both use last-core labels. \\ 
\addlinespace[3pt]
\(k_{\mathrm{eff}}^{\mathrm{core}},\allowbreak\ k_\nu^{\mathrm{core},0}\) & Physical sum of directional core rates and calibrated directional reference rates. Superscript \(0\) denotes calibration; \(k_{\mathrm{eff}}^{\mathrm{core},0}\) defines the population clock (Sec.~\ref{si:sec:crossover-validation}). \\ 
\addlinespace[3pt]
\(k_{N,s^*\mid d_N^*}^*(t^*),\allowbreak\ \bar k_{s^*\mid d^*}^*\) & Reduced time-dependent RF rate and its direct window average. At fixed \(N,d^*\), subscripts may be suppressed as \(k_{s^*}^*\), \(\bar k_{s^*}^*\); \(k_{\mathrm{TST}}^*,k_{\mathrm{RF}}^*\) identify methods. \\ 
\addlinespace[3pt]
\(K(\bm l^*)\) & Instantaneous number of bonds above the divider; an unstarred integer count (Sec.~\ref{si:thermodynamics}). \\ 
\addlinespace[3pt]
\(K^*\) & Diagonal matrix of reduced single-bond curvatures, \(\operatorname{diag}(u_j^{(2),*})\), before enforcing the sum constraint (Sec.~\ref{si:midpoint-approximations}). \\ 
\addlinespace[3pt]
\(K_{N,s^*\mid d_N^*}^*\) & Reduced population-normalized surface factor \(J_{N,s^*}^{+,*}/P_{\mathrm i}(d_N^*)\); \(K_{s^*}^*\) abbreviates it at fixed \(N,d_N^*\). At \(s^*=d_N^*\), it equals \(k_N^{\mathrm{TST},*}\). \\ 
\addlinespace[3pt]
\(\mathcal K_{\mathrm i}^*,\allowbreak\ \mathcal K_{\mathrm s}^*\) & Second derivative of the reduced site PMF with respect to \(l^*\), at its intact minimum (positive) and saddle (negative). \(\mathcal K_\alpha^*\) denotes either; \(\mathcal K_{\mathrm{AH}}^*\) includes the first anharmonic correction. \\ 
\addlinespace[3pt]
\(l_j,\allowbreak\ l_j^*;\allowbreak\ l,\allowbreak\ l^*\) & Physical and reduced bond lengths, \(l_j^*=l_j/l_0\). Without a bond index, \(l\) denotes the tagged bond. The italic letter \(l\) distinguishes bonds from particle coordinates \(x_j\). \\ 
\addlinespace[3pt]
\(l_0\) & Physical equilibrium length of an isolated Morse bond and reference length. \\ 
\addlinespace[3pt]
\(\bm l^*,\allowbreak\ \bm l_{-j}^*\) & Vector of reduced bond lengths, and spectator vector excluding tagged bond \(j\); \(d\bm l^*=\prod_jdl_j^*\). \\ 
\addlinespace[3pt]
\(l_b^{\mathrm{ad},*},\allowbreak\ l_b^{\mathrm{site},*}\) & Reduced landmarks of the adiabatic or site-PMF profile: \(b=\mathrm i,\mathrm f\) for minima and \(\mathrm s\) for a saddle. In particular, \(l_{\mathrm i}^{\mathrm{ad},*}=\lambda\), while \(l_{\mathrm s}^{\mathrm{site},*}\) need not equal \(d_N^*\). \\ 
\addlinespace[3pt]
\(l_{\mathrm{short}}^*,\allowbreak\ l_{\mathrm{long}}^*\) & Common short-bond and long-bond lengths of a nonuniform stationary configuration (Sec.~\ref{si:stationary-morse}). \\ 
\addlinespace[3pt]
\(l_{1/2}^*,\allowbreak\ \widehat l_{1/2}^*\) & Projected half-committor crossing and its estimate; \(\overline q(l_{1/2}^*)=1/2\). Shooting lengths are \(l_s^*\), with surface index \(s\) (Sec.~\ref{si:sec:committor-crossing}). \\ 
\addlinespace[3pt]
\(l_{\max}^*(t^*),\allowbreak\ l_{\max}^*(t)\) & Maximum reduced bond length along a trajectory, evaluated at reduced time in residence analysis or physical time in crossover analysis. In integration-domain formulas, \(l_{\min}^*,l_{\max}^*\) instead denote numerical bounds (Sec.~\ref{si:sec:numerical-evaluation}). \\ 
\addlinespace[3pt]
\(\ell_{Z,N}^*,\allowbreak\ \ell_{B,N}^*,\allowbreak\ \ell_{A,N}^*\) & Reduced activation length and its barrier and prefactor contributions; \(\ell_{Z,N}^{\mathrm{KGH},*}=\ell_{B,N}^*+\ell_{A,N}^*\). Physical activation lengths are divided by \(l_0\); script \(\ell\) distinguishes them from bond lengths (Sec.~\ref{si:activation-length}). \\ 
\addlinespace[3pt]
\(L,\allowbreak\ L^*\) & Physical and reduced total chain length, \(L^*=L/l_0=N\lambda\). \\ 
\addlinespace[3pt]
\(L\) & Orthonormal basis within the local dividing surface in Sec.~\ref{si:harmonic-transmission}; a dimensionless matrix, not the chain length. \\ 
\addlinespace[3pt]
\(\mathcal L(\mu,s_{\mathrm c},s_{\mathrm v})\) & Hierarchical binomial likelihood with configuration and velocity effects integrated out (Sec.~\ref{si:sec:committor-dispersion}). \\ 
\addlinespace[3pt]
\(m,\allowbreak\ m_{\mathrm{eff}}\) & Physical particle mass and symmetric-path effective mass; \(N_{\mathrm A}m\) is the tabulated molar mass and \(m_{\mathrm{eff}}/m=\rho_N\). \\ 
\addlinespace[3pt]
\(M\) & Number of reservoir bonds, \(M=N-1\), in the site-barrier expansion (Sec.~\ref{si:site-barriers}). \\ 
\addlinespace[3pt]
\(\bm M,\allowbreak\ M_{l^*}\) & Particle mass matrix \(\bm M=mI\), and dimensional kinetic metric \(M_{l^*}=ml_0^2/2\) for a reduced bond coordinate. The star inside its subscript identifies the coordinate, not a reduced metric. \\ 
\addlinespace[3pt]
\(M_0\) & Maxwell velocity probability measure conditioned on zero total momentum (Sec.~\ref{si:sec:committor-definition}). \\ 
\addlinespace[3pt]
\(n_r,\allowbreak\ n_{\mathrm{conf}},\allowbreak\ n_s,\allowbreak\ n_{\mathrm f,s}\) & Recorded samples in trajectory \(r\); RF configurations; resolved shots and fracture-first shots at shooting surface \(s\). \(n_{\mathrm f,cv}\) counts successes within a configuration/velocity group. \\ 
\addlinespace[3pt]
\(n_{\mathrm{return},a}(s),\allowbreak\ n_{\mathrm{pass},a}(s)\) & Completed failed-return and successful-passage counts from origin core \(a\), qualifying at counting surface \(s\) (Sec.~\ref{si:sec:crossover-excursions}). \\ 
\addlinespace[3pt]
\(\bm n\) & Normal to the tagged local dividing surface in internal coordinates (Sec.~\ref{si:harmonic-transmission}); dimensionless. \\ 
\addlinespace[3pt]
\(N,\allowbreak\ N_{b\to\bar b}\) & Number of bonds (also particles) in the periodic chain; completed core switches from \(b\) to the opposite core \(\bar b\), respectively. \\ 
\addlinespace[3pt]
\(N_{\mathrm A}\) & Avogadro constant, relating molar mass and energy to particle-level quantities. \\ 
\addlinespace[3pt]
\(O\) & Orthogonal matrix diagonalizing the constrained bond Hessian (Sec.~\ref{si:midpoint-approximations}). \\ 
\addlinespace[3pt]
\(p_N^{\mathrm{tag},*},\allowbreak\ p_v^*\) & Probability densities with respect to reduced tagged length \(dl^*\) and reduced velocity \(dv^*\), respectively; their stars specify the measures. \\ 
\addlinespace[3pt]
\(p_{cv},\allowbreak\ p_{\mathrm f}(t)\) & Phase-space fracture probability for configuration/velocity group \((c,v)\) in Sec.~\ref{si:sec:committor-dispersion}; time-dependent fractured last-core population in Sec.~\ref{si:sec:crossover-rate-estimation}. \\ 
\addlinespace[3pt]
\(P_{\mathrm i},\allowbreak\ P_{\mathrm f},\allowbreak\ P_{\mathrm f}^{\mathrm{sim}}\) & Instantaneous equilibrium basin probabilities and sampled fractured fraction. \(P_{\mathrm i}(d_N^*)\) makes the state divider explicit. \\ 
\addlinespace[3pt]
\(P_{ab},\allowbreak\ P_{\mathrm f\mid\mathrm i}\) & Conditional endpoint transition probabilities. \(P_{ab}(\tau_{\mathrm{lag}})\) uses last-core labels in S13; \(P_{\mathrm f\mid\mathrm i}(t^*)\) uses instantaneous states in S10. \\ 
\addlinespace[3pt]
\(\widehat P_{\mathrm{return},a}(s)\) & Estimated return probability among completed qualifying excursions from core \(a\); censored episodes are excluded (Sec.~\ref{si:sec:crossover-excursions}). \\ 
\addlinespace[3pt]
\(q_\alpha^*,\allowbreak\ \bm q^*\) & Reduced normal-mode coordinates in Sec.~\ref{si:midpoint-approximations}, or internal particle displacements in Sec.~\ref{si:harmonic-transmission}; distinct from committor probabilities. \\ 
\addlinespace[3pt]
\(q_{\mathrm{comm}}(\bm r),\allowbreak\ q_c\) & Configuration committor after averaging initial velocities and subsequent noise; \(q_c=q_{\mathrm{comm}}(\bm r_c)\). Hats denote finite-shot estimates. \\ 
\addlinespace[3pt]
\(\overline q(l^*),\allowbreak\ \overline q_{\mathrm{fit}},\allowbreak\ q(u)\) & Projected configuration-averaged committor; hierarchical-model mean; and model probability after averaging velocity effects at fixed configuration effect \(u\) (Sec.~\ref{si:sec:committor-dispersion}). \\ 
\addlinespace[3pt]
\(q_N^{\mathrm{KGH}}\) & Adjacent-chain rate ratio \(k_{1/2}^{\mathrm{KGH},*}(N)/k_{1/2}^{\mathrm{KGH},*}(N+1)\), with each chain at its own coexistence stretch; dimensionless. \\ 
\addlinespace[3pt]
\(Q_{\mathrm{comm}}(\bm r,\bm v),\allowbreak\ \widehat Q_{\mathrm{comm},cv}\) & Phase-space committor with positions and velocities fixed, averaging subsequent noise; its empirical estimate for group \((c,v)\). \\ 
\addlinespace[3pt]
\(Q_\partial^*\) & Divider derivative \((\partial Z_{\mathrm i}^*/\partial d_N^*)_{L^*}\), the reduced boundary partition weight (Sec.~\ref{si:moving-divider-force}). \\ 
\addlinespace[3pt]
\(\bm r,\allowbreak\ \bm r^*\) & Physical particle configuration \((x_1,\ldots,x_N)\) and its reduced counterpart \(\bm r^*=\bm r/l_0\); \(\delta\bm r^*\) denotes reduced displacements from the saddle. \\ 
\addlinespace[3pt]
\(r_+^*,\allowbreak\ r_-^*\) & Growing and decaying rates of the unstable harmonic mode in reduced time (Sec.~\ref{si:harmonic-transmission}). \\ 
\addlinespace[3pt]
\(R_j(t)\) & Physical Langevin thermal force on particle \(j\) at physical time \(t\). \\ 
\addlinespace[3pt]
\(R,\allowbreak\ R_{cv}\) & Number of noise realizations per velocity vector, and resolved shot count in group \((c,v)\) (Sec.~\ref{si:sec:committor-sampling}). \\ 
\addlinespace[3pt]
\(R_{\ge2/1}\) & Ratio of partition weights with at least two fractured bonds to the exactly-one-fracture weight (Sec.~\ref{si:thermodynamics}). \\ 
\addlinespace[3pt]
\(\mathcal R_b\) & Retained residences carrying core label \(b\) (Sec.~\ref{si:core-method}). \\ 
\addlinespace[3pt]
\(s,\allowbreak\ s^*\) & Physical and reduced surface coordinates, \(s^*=s/l_0\). RF launch surfaces use \(s^*\); physical excursion-counting surfaces are \(s_{\mathrm{ad}},s_{\mathrm{PMF}},s_{\mathrm{out}}\), with \(s_{\mathrm{out}}=2s_{\mathrm{PMF}}-s_{\mathrm{ad}}\). \\ 
\addlinespace[3pt]
\(s_{\mathrm c},\allowbreak\ s_{\mathrm v}\) & Standard deviations of configuration and velocity latent effects on the logit scale (Sec.~\ref{si:sec:committor-dispersion}); not physical lengths or velocities. \\ 
\addlinespace[3pt]
\(s_{\nu c_{\mathrm{an}}}(\tau_{\mathrm{lag}})\) & Bootstrap standard deviation of the relative rate discrepancy (Sec.~\ref{si:sec:crossover-uncertainty}); dimensionless. \\ 
\addlinespace[3pt]
\(S_b(t^*)\) & Survival probability for a full residence in core label \(b\) beyond reduced time \(t^*\) (Sec.~\ref{si:core-method}). \\ 
\addlinespace[3pt]
\(t,\allowbreak\ t^*\) & Physical and reduced time, \(t^*=\xi t\); \(\Delta t\) and \(\Delta t_{\mathrm{obs}}\) are integration and observation intervals, with starred forms multiplied by \(\xi\). \\ 
\addlinespace[3pt]
\(t_e^{\mathrm{out}},\allowbreak\ t_e^{\mathrm{in}}\) & Last sampled time in the origin core and first subsequent sampled core arrival for excursion \(e\); physical times (Sec.~\ref{si:sec:crossover-excursions}). \\ 
\addlinespace[3pt]
\(T,\allowbreak\ T^*\) & Physical bath temperature and reduced thermal energy, \(T^*=k_{\mathrm B}T/D_{\mathrm e}\). \\ 
\addlinespace[3pt]
\(\mathcal T_b^{\mathrm{LC},*}\) & Accumulated retained last-core exposure in reduced time, \(\sum_{\alpha\in\mathcal R_b}D_\alpha^*\) (Sec.~\ref{si:core-method}). \\ 
\addlinespace[3pt]
\(\mathcal T_{r,c_{\mathrm{an}}}(\tau_{\mathrm{lag}})\) & Set of eligible starting times for lagged label pairs of trajectory \(r\) under condition \(c_{\mathrm{an}}\) (Sec.~\ref{si:sec:crossover-rate-estimation}); not an accumulated duration. \\ 
\addlinespace[3pt]
\(U(l),\allowbreak\ u^*(l^*)\) & Physical Morse bond potential and its reduced form, \(u^*(l^*)=U(l_0l^*)/D_{\mathrm e}\). Derivatives \(u^{(n),*}\), including local \(u_j^{(n),*}\), use reduced bond length. \\ 
\addlinespace[3pt]
\(u_c,\allowbreak\ w_{cv}\) & Configuration and velocity latent effects in the committor probability model (Sec.~\ref{si:sec:committor-dispersion}); dimensionless logit shifts. Integration variables \(u,w\) have the same local meaning. \\ 
\addlinespace[3pt]
\(\bm u\) & Unit eigenvector of the unstable internal harmonic mode, oriented so that \(\bm n^{\mathsf T}\bm u>0\) (Sec.~\ref{si:harmonic-transmission}). \\ 
\addlinespace[3pt]
\(\mathcal U^*,\allowbreak\ \Delta U\) & Reduced total chain energy \(\mathcal U^*=\sum_j u^*(l_j^*)\), and physical energy change in a Metropolis proposal, respectively. Physical chain energy is \(D_{\mathrm e}\mathcal U^*\). \\ 
\addlinespace[3pt]
\(U_N(\tau_{\mathrm{lag}})\) & Upper curve for the maximum relative rate discrepancy, including approximate simultaneous sampling uncertainty within each \(N\) (Sec.~\ref{si:sec:crossover-uncertainty}); dimensionless. \\ 
\addlinespace[3pt]
\(v_j,\allowbreak\ \bm v,\allowbreak\ \widetilde v_j\) & Physical particle velocities, velocity vector, and Gaussian draws before subtracting their mean. Units are length/time (Sec.~\ref{si:sec:committor-sampling}). \\ 
\addlinespace[3pt]
\(v_l^*,\allowbreak\ v^*,\allowbreak\ \bar v_+^*\) & Reduced tagged velocity \(dl^*/dt^*\), its integration variable, and positive-part Maxwell mean \(\bar v_+^*=\sqrt{cT^*/\pi}\) (Sec.~\ref{si:tst-flux}). \\ 
\addlinespace[3pt]
\(\bm v^*,\allowbreak\ V^*\) & Reduced internal velocity \(d\bm q^*/dt^*\) and initial tagged velocity \(V^*=\bm n^{\mathsf T}\bm v^*(0)\) in the harmonic calculation (Sec.~\ref{si:harmonic-transmission}). \\ 
\addlinespace[3pt]
\(V\) & Number of velocity vectors per sampled configuration (Sec.~\ref{si:sec:committor-sampling}); distinct from reduced velocity \(V^*\). \\ 
\addlinespace[3pt]
\(w(l^*),\allowbreak\ w_{\mathrm i},\allowbreak\ w_{\theta^*}\) & Boltzmann bond weight \(e^{-u^*(l^*)/T^*}\), its intact-restricted form, and exponentially tilted form \(w(l^*)e^{\theta^*l^*}\). These weights are dimensionless. \\ 
\addlinespace[3pt]
\(w_{\mathrm i}^*\) & Effective reduced width of the intact site-PMF well, obtained by integrating its minimum-referenced Boltzmann weight below \(d_N^*\) (Sec.~\ref{si:midpoint-rate-comparison}). \\ 
\addlinespace[3pt]
\(\mathcal W,\allowbreak\ \bm W^*\) & Thermostat randomness in committor probabilities; vector of standard Wiener processes in reduced time in the harmonic Langevin model, respectively. \\ 
\addlinespace[3pt]
\(x_j,\allowbreak\ x_j^*\) & Physical and reduced scalar particle coordinates, \(x_j^*=x_j/l_0\); collected in \(\bm r\) and \(\bm r^*\). \\ 
\addlinespace[3pt]
\(y,\allowbreak\ y_0,\allowbreak\ y_j\) & Morse exponential coordinates: \(y=e^{-a(l_{\mathrm{long}}^*-1)}\), \(y_0=e^{-a(\lambda-1)}\), and \(y_j=e^{-a(l_j^{(b),*}-1)}\) at stationary configuration \(b\). \\ 
\addlinespace[3pt]
\(Y,\allowbreak\ Y_{cvr}\) & Binary fracture-first outcome, \(\mathbf1(\tau_{\mathrm f}<\tau_{\mathrm i})\), with sampling indices \(c,v,r\) when needed (Sec.~\ref{si:sec:committor-sampling}). \\ 
\addlinespace[3pt]
\(z^*,\allowbreak\ z_j\) & Reduced mean reservoir length \((N\lambda-l^*)/(N-1)\) in S6; reduced saddle displacement \(l^*-l_{\mathrm s}^{\mathrm{site},*}\) in S8.5. In S3.3, \(z_j=l_j^*/\lambda\) is a scaled integration coordinate. \\
\addlinespace[3pt]
\(z,\allowbreak\ z_{\mathrm{LC}}(t)\) & Short-bond Morse exponential \(z=e^{-a(l_{\mathrm{short}}^*-1)}=1-y\) in S3.3; last-visited-core label in S13. A bare \(z\) also serves as a generic logit argument in S12. \\ 
\addlinespace[3pt]
\(Z_N^*,\allowbreak\ Z_N^{b,*},\allowbreak\ Z_{N,k}^*\) & Reduced configurational partition densities: unrestricted, basin-restricted, and exactly \(k\) fractured bonds. \(Z_{N,0}^*=Z_N^{\mathrm i,*}\); \(Z_{N,1}^*\) includes all candidate sites. Kinetic and translation factors are removed. \\ 
\addlinespace[3pt]
\(Z_{N,\mathrm{site}}^{\mathrm f,*},\allowbreak\ Z_{N-1}^{\mathrm i,*}\) & One specified fractured-site weight and intact reservoir partition density. The latter retains its stated total length and cutoff; superscript H denotes a harmonic approximation. \\ 
\addlinespace[3pt]
\(Z_{\mathrm{boot}}^{(b)},\allowbreak\ Z_{\mathrm{boot},0.95}\) & Maximum standardized discrepancy in bootstrap sample \(b\), and its 95th-percentile threshold (Sec.~\ref{si:sec:crossover-uncertainty}); dimensionless statistics, not partition densities. \\ 
\addlinespace[3pt]
\multicolumn{2}{@{}l}{\textbf{Greek symbols}}\\*
\addlinespace[3pt]
\(\alpha\) & Physical inverse Morse width; \(a=\alpha l_0\). The same letter is a local mode, extremum or residence index when explicitly indexed. \\ 
\addlinespace[3pt]
\(\beta,\allowbreak\ \beta^*\) & Physical and reduced inverse thermal energies, \(\beta=(k_{\mathrm B}T)^{-1}\), \(\beta^*=D_{\mathrm e}\beta=1/T^*\). \\ 
\addlinespace[3pt]
\(\gamma_+^*\) & Positive reduced growth rate at the KGH barrier, \(\sqrt{1/4+(\omega_{\mathrm s}^*)^2}-1/2\) (Sec.~\ref{si:kgh-factor}). \\ 
\addlinespace[3pt]
\(\widehat\Gamma_m,\allowbreak\ \widehat\Gamma_m^*\) & Estimated physical and reduced tension covariances at integer lag \(m\); \(\widehat\Gamma_m^*=(l_0/D_{\mathrm e})^2\widehat\Gamma_m\) (Sec.~\ref{si:acf-method}). \\ 
\addlinespace[3pt]
\(\delta(\cdot),\allowbreak\ \delta_{ij}\) & Dirac delta distribution and Kronecker delta, respectively; the latter is one for \(i=j\) and zero otherwise. \\ 
\addlinespace[3pt]
\(\delta_{\mathrm{dom}}^*,\allowbreak\ \delta_{\mathrm{prof}}^*\) & Reduced long-bond integration-domain and site-profile margins (Sec.~\ref{si:sec:numerical-evaluation}). \\ 
\addlinespace[3pt]
\(\delta_{\nu c_{\mathrm{an}}}(\tau_{\mathrm{lag}})\) & Relative validation-rate discrepancy from the corresponding calibrated reference (Sec.~\ref{si:sec:crossover-validation}); dimensionless. \\ 
\addlinespace[3pt]
\(\Delta^*\) & Difference \(r_+^*-r_-^*\) between reduced harmonic growth rates (Sec.~\ref{si:harmonic-transmission}). \\ 
\addlinespace[3pt]
\(\epsilon_i^*,\allowbreak\ \bm\epsilon^*\) & Independent reduced bond displacements in the constrained Gaussian calculation; reservoir displacements in the site-PMF expansion (Secs.~\ref{si:midpoint-approximations}, \ref{si:site-barriers}). \\ 
\addlinespace[3pt]
\(\zeta_j^*\) & Thermally rescaled reduced bond displacement, \(\delta l_j^*=\sqrt{T^*}\,\zeta_j^*\), in the Laplace expansion (Sec.~\ref{si:anharmonic-midpoint}). \\
\addlinespace[3pt]
\(\eta\) & Dimensionless core fraction: zero at the basin minimum and one at the saddle; the primary cores use \(\eta=0.5\). \\ 
\addlinespace[3pt]
\(\theta^*,\allowbreak\ \Theta(\cdot)\) & Tilt conjugate to reduced length in the convolution, and Heaviside step function, respectively. \(\Theta(y)=1\) for \(y\ge0\) and zero otherwise. \\ 
\addlinespace[3pt]
\(\kappa_{N,s^*\mid d_N^*}(t^*),\allowbreak\ \kappa_N(t^*)\) & Dimensionless signed RF transmission; the second abbreviates coincident launch and state surfaces, \(s^*=d_N^*\). Bars denote direct time-window averages; a star in a surface index does not star the probability ratio. \\ 
\addlinespace[3pt]
\(\kappa_{0,N},\allowbreak\ \kappa_N^{\mathrm{nl}},\allowbreak\ \kappa_N^{\mathrm{harm}}\) & Slow-regime fitted transmission amplitude (\(\kappa_N^{\mathrm{nl}}=\kappa_{0,N}\)) and harmonic RF coefficient; distinct from microscopic time-zero transmission. \\ 
\addlinespace[3pt]
\(\kappa_{\mathrm{eff}}\) & Rate divided by the TST rate at the same state, \(k^*/k^{\mathrm{TST},*}\); not generally equal to the local KGH factor \(\chi_N\). \\ 
\addlinespace[3pt]
\(\lambda\) & Dimensionless mean stretch, \(L/(Nl_0)=L^*/N\); unstarred. \\ 
\addlinespace[3pt]
\(\lambda_{1/2},\allowbreak\ \lambda_{1/2}^{(1)}\) & Full thermodynamic midpoint and exactly-one-fracture midpoint, respectively; \(\lambda_{1/2}^{\mathrm{sim}}\) is estimated from simulation populations. \\ 
\addlinespace[3pt]
\(\lambda_0,\allowbreak\ \lambda_{\mathrm H},\allowbreak\ \lambda_{\mathrm{AH}}\) & Zero-temperature energetic, harmonic and first-anharmonic midpoint estimates. \(\lambda_\pm=\lambda_{1/2}\pm\delta\lambda\) are finite-difference endpoints; \(\Delta\lambda_{\mathrm{lin}}\) is the linearized multiple-fracture midpoint shift. \\ 
\addlinespace[3pt]
\(\mu,\allowbreak\ \mu_l\) & Central logit in the hierarchical committor model; physical tagged relative-length mass \(\mu_l=m/2\), respectively. Their scopes are S12.4 and S8.1. \\ 
\addlinespace[3pt]
\(\nu\) & Transition direction, \(\mathrm{if}\) or \(\mathrm{fi}\), in crossover rate comparisons. \\ 
\addlinespace[3pt]
\(\xi\) & Physical Langevin friction rate; inverse time, with reference time \(\xi^{-1}\). \\ 
\addlinespace[3pt]
\(\pi_{l^*},\allowbreak\ \Pi_b^{\mathrm{LC}}\) & Conditional configuration probability measure at tagged length \(l^*\), and stationary last-core population, respectively. Both remain unstarred; the star in \(\pi_{l^*}\) identifies its conditioning coordinate. \\ 
\addlinespace[3pt]
\(\rho_N\) & Dimensionless symmetric-path mass ratio \(m_{\mathrm{eff}}/m=N(N+1)/[12(N-1)]\). \\ 
\addlinespace[3pt]
\(\rho_\alpha^*\) & Eigenvalues of the reduced constrained bond Hessian (Sec.~\ref{si:midpoint-approximations}). \\ 
\addlinespace[3pt]
\(\rho_{N,s^*}^*,\allowbreak\ \rho_a(d\bm l^*)\) & Unnormalized reduced tagged launch-surface density in S8.1; normalized basin-restricted initial probability measure in S13.1, respectively. \\ 
\addlinespace[3pt]
\(\sigma_v^*,\allowbreak\ \sigma_V^*,\allowbreak\ \sigma_A^*,\allowbreak\ \sigma_u^*\) & Reduced standard deviations: Maxwell tagged velocity, initial harmonic tagged velocity, growing amplitude, and initial displacement along the unstable mode, respectively (Sec.~\ref{si:reactive-flux}). \\ 
\addlinespace[3pt]
\(\sigma_q(l^*)\) & Probability-scale standard deviation of configuration committors at fixed tagged length, after velocity averaging; distinct from latent logit SD \(s_{\mathrm c}\). \\ 
\addlinespace[3pt]
\(\Sigma_{s^*}\) & Union of tagged launch-surface components: one bond at \(s^*\), all others shorter (Sec.~\ref{si:tst-flux}). \\ 
\addlinespace[3pt]
\(\tau_a\) & Physical first-entry time into core \(a\), with \(a=\mathrm i,\mathrm f\), in the committor definition. \\ 
\addlinespace[3pt]
\(\tau_{\mathrm{lag}},\allowbreak\ \tau_{\mathrm{pop}},\allowbreak\ \tau_{\mathrm{plan}}\) & Physical observation lag, calibrated population-relaxation time \(1/k_{\mathrm{eff}}^{\mathrm{core},0}\), and ACF-based planning time \(1/k_{\mathrm{eff}}^{\mathrm{ACF}}\). Crossover times retain fs units; these are distinct from reduced \(t^*\). \\ 
\addlinespace[3pt]
\(\phi_s,\allowbreak\ \Phi_{\mathrm H},\allowbreak\ \Phi_{\mathrm{AH}}\) & Zero-mean Gaussian density with SD \(s\) in S12.4; dimensionless harmonic and anharmonic midpoint residuals whose zeros define \(\lambda_{\mathrm H},\lambda_{\mathrm{AH}}\), respectively (main Sec.~IV~B~1). \\
\addlinespace[3pt]
\(\chi_N\) & Local KGH dynamical transmission factor \(\gamma_+^*/\omega_{\mathrm s}^*\); dimensionless. \\ 
\addlinespace[3pt]
\(\omega_\alpha,\allowbreak\ \omega_\alpha^*\) & Physical and reduced KGH well/barrier frequencies, \(\omega_\alpha^*=\omega_\alpha/\xi\), \(\alpha=\mathrm i,\mathrm s\). The barrier frequency uses the magnitude of the negative curvature. \\ 
\addlinespace[3pt]
\multicolumn{2}{@{}l}{\textbf{Subscripts, superscripts and other notation}}\\*
\addlinespace[3pt]
\({}^*\) & Explicitly reduced quantity. Stars inside an index identify a reduced coordinate; they do not change the units of the enclosing object, e.g. \(M_{l^*}\). \\ 
\addlinespace[3pt]
\(j,k,p,n,m,\alpha\) & Local integer indices: bonds/particles, recursion size or derivative order, sample/lag number, mode or residence. In S3, \(k\) is a fracture count; in S8.5, \(p\) is a particle index; in S13.5, \(p\) is a quantile probability. Scope is stated where introduced. \\
\addlinespace[3pt]
\(\mathrm i,\mathrm f,\mathrm s;\allowbreak\ a,b,\bar b\) & Intact, fractured and saddle labels. Local \(a,b\) select states, and \(\bar b\) is the opposite core; they are not the Morse shape parameter in this use. \\ 
\addlinespace[3pt]
\(c,v,r;\allowbreak\ a,\sigma\) & Configuration, velocity and noise indices in S12; configuration and antithetic sign \(\sigma=\pm1\) in S10. Scalar \(r\) also indexes equilibrium trajectories, unlike configuration vector \(\bm r\). \\ 
\addlinespace[3pt]
\({}_{1/2},\allowbreak\ {}^{(1)},\allowbreak\ {}^\ddagger\) & Evaluation at coexistence, or half-committor probability when attached to \(l^*\); exactly-one-fracture approximation when attached to \(\lambda_{1/2}\); saddle marker. On adiabatic energies, \((1)\) labels the one-bond reservoir branch. \\ 
\addlinespace[3pt]
\(\mathrm{ad},\mathrm{abs},\mathrm{env},\mathrm{tag},\mathrm{site},\mathrm{pot}\) & Adiabatic construction, absolute energy, lower envelope, unrestricted tagged ensemble, intact-reservoir site ensemble, and potential-energy average, respectively. \\ 
\addlinespace[3pt]
\(\mathrm H,\mathrm{AH},\mathrm{TST},\mathrm{KGH},\mathrm{RF},\mathrm{ACF},\mathrm{RT}\) & Harmonic, first-anharmonic, transition-state theory, Kramers--Grote--Hynes, reactive flux, autocorrelation function, and residence time. PMF denotes potential of mean force; \(\mathrm{harm}\), \(\mathrm{nl}\) and \(\mathrm{sim}\) denote harmonic, nonlinear and simulation estimates. \(\mathrm E\), \(\mathrm{approx}\) and \(\mathrm{det}\) identify energetic, approximate and deterministic references. \\ 
\addlinespace[3pt]
\(\mathrm{vir},\mathrm{conj},\mathrm{core},\mathrm{LC},\mathrm{eff}\) & Virial, conjugate, core-defined, last-core-labelled, and effective or summed rate, respectively, as specified by the quantity. \(\mathrm{int}\) denotes internal modes. \\ 
\addlinespace[3pt]
\(\mathrm{PMF},\mathrm{out},\allowbreak\ s^*\mid d_N^*\) & PMF and outer surface labels; the vertical bar separates launch coordinate from the fixed state divider. Suppressing common \(N,d_N^*\) in S10 leaves the same definitions. \\ 
\addlinespace[3pt]
\(\widehat{(\cdot)},\allowbreak\ \overline{(\cdot)},\allowbreak\ \widetilde{(\cdot)}\) & Estimate; specified ensemble or time average; and locally defined preprocessing. Tildes mark raw Gaussian velocity draws or an isotonic fitted committor profile in S12. \\ 
\addlinespace[3pt]
\({}^{(n),*},\allowbreak\ {}',\allowbreak\ \dot{(\cdot)}\) & Reduced profile derivative of order \(n\) with respect to \(l^*\); a prime denotes differentiation in its stated argument, except proposed bond lengths in S13. A dot on a physical coordinate denotes differentiation with respect to physical time. \(d_N^{(1),*}\) differentiates in \(\lambda\). \\ 
\addlinespace[3pt]
\(\delta,\allowbreak\ \Delta,\allowbreak\ {}_\pm\) & Perturbation, fluctuation, grid increment or difference as locally defined; \(\pm\) marks paired finite-difference endpoints in S11. \(\Delta l^*\) is the reduced integration-grid spacing. In the committor figure, \(\Delta l^*=l^*-l_{\mathrm s}^{\mathrm{site},*}\) instead denotes the offset from the PMF maximum. \\ 
\addlinespace[3pt]
\({}^{(b)},\allowbreak\ {}^0,\allowbreak\ {}_{0.95}\) & Bootstrap sample index in S13; calibrated reference; and 95th-percentile threshold, respectively. A superscript \((b)\) on stationary bond lengths in S5 instead labels the basin. \\ 
\addlinespace[3pt]
\({}^{\mathsf T},\allowbreak\ \bm1,\allowbreak\ \mathbf1[\mathcal A]\) & Matrix transpose, all-ones vector, and indicator of condition \(\mathcal A\). Bold symbols denote vectors or the explicitly defined mass matrix; other matrices retain their local styling. \(O(\cdot)\) denotes asymptotic order; \(\mathbb R\) denotes the real numbers. \\ 
\addlinespace[3pt]
\(\langle\cdot\rangle,\allowbreak\ \Pr,\allowbreak\ \operatorname E,\allowbreak\ \operatorname{Var},\allowbreak\ \operatorname{Cov}\) & Specified ensemble average, probability, expectation, variance and covariance; conditioning labels identify the ensemble. \(\mathcal N\) and Binomial denote Gaussian and binomial laws; SE and SD denote standard error and standard deviation. \\ 
\addlinespace[3pt]
\(\operatorname{logit},\allowbreak\ \operatorname{logit}^{-1}\) & \(\operatorname{logit}p=\ln[p/(1-p)]\), with inverse \((1+e^{-z})^{-1}\); probability and logit variables are unstarred. \\ 
\addlinespace[3pt]
\end{longtable}
\endgroup

%% file: si_sections/notation_and_estimands.tex
\section{Notation and state conventions}
\label{si:notation}

Abbreviations: autocorrelation function (ACF), transition-state theory (TST),
reactive flux (RF), and Kramers--Grote--Hynes (KGH).

\subsection{Reduced variables and state definition}

For \(N\) periodic Morse bonds, the reduced lengths \(l_{j}^{*}=l_{j}/l_{0}\)
and mean stretch \(\lambda=L/(Nl_{0})\) satisfy
\begin{equation}
  \sum_{j=1}^{N}l_{j}^{*}=N\lambda\equiv L^{*}=\frac{L}{l_0}.
  \label{si:eq:constraint}
\end{equation}
Using Morse depth \(\De\), inverse width \(\alpha\), equilibrium length
\(l_{0}\), and friction rate \(\xi\), we define \(a=\alpha l_{0}\) and
\begin{equation}
  \begin{gathered}
    u^{*}(l^{*})=\frac{U(l_{0}l^{*})}{\De}=\left[1-e^{-a(l^{*}-1)}\right]^2,\\
    \Tstar=\frac{k_{\mathrm B}T}{\De},\qquad
    \Fstar=\frac{Fl_{0}}{\De},\qquad
    t^*=\xi t,\qquad k^*=\frac{k}{\xi}.
  \end{gathered}
  \label{si:eq:units}
\end{equation}
Here \(U(l)\) is the physical bond potential, \(F\) the chain tension,
and \(k\) a dimensional rate. The particle-level mass \(m\) and bond
energy \(\De\), also used in \(c=\De/(m\xi^2l_{0}^{2})\), equal the tabulated
molar quantities divided by Avogadro's constant. Stars mark reduced
quantities; dimensional quantities are unstarred.
The dimensionless groups \(a,c,\lambda\), probabilities,
counts and ratios remain unstarred. Physical simulation settings and
thermal-unit comparisons retain their stated units.

Particle configurations are \(\bm r=(x_1,\ldots,x_N)\), with
\(x_j^*=x_j/l_0\) and \(\bm r^*=(x_1^*,\ldots,x_N^*)\).
The bond coordinates obey \(l_j^*=x_j^*-x_{j-1}^*\), with periodic closure.
For reduced potentials and profiles, a superscript \((n),*\) denotes
an \(n\)th derivative with respect to the reduced bond coordinate; for example,
\(u^{(n),*}(l^*)=d^n u^*(l^*)/d(l^*)^n\).
Physical reference scales, such as \(l_0\), \(\De\), \(m\) and \(\xi\),
remain dimensional.

The two state classifications defined in main-text Sec.~II~B also apply here.
The adiabatic path has intact minimum \(l_{\mathrm i}^{\mathrm{ad},*}=\lambda\),
fractured minimum \(l_{\mathrm f}^{\mathrm{ad},*}\), and saddle
\(d_{N}^{*}(\lambda)\equiv l_{\mathrm{ad}}^{\ddagger,*}(N,\lambda)\) (Sec.~\ref{si:stationary-morse}).

\textit{Instantaneous classification.}
For \(\bm l^{*}=(l_{1}^{*},\ldots,l_{N}^{*})\), the equilibrium basin indicators are
\begin{equation}
 h_{\mathrm f}(\bm l^{*};N,\lambda)=
 \mathbf 1\!\left[\max_jl_{j}^{*}>d_{N}^{*}(\lambda)\right],
 \qquad h_{\mathrm i}=1-h_{\mathrm f},
 \label{si:eq:state}
\end{equation}
where \(\mathbf 1[\mathcal A]\) is one if \(\mathcal A\) holds and zero
otherwise. Thus the chain is classified as fractured if any bond fractures.

\textit{Core classification.}
For \(0<\eta<1\), the boundaries are
\[
 c_{b}^{*}(\eta)=l_{b}^{\mathrm{ad},*}+\eta(d_{N}^{*}-l_{b}^{\mathrm{ad},*}),
 \qquad b\in\{\mathrm i,\mathrm f\}.
\]
The intact core has \(\max_jl_{j}^{*}\le c_{\mathrm i}^{*}(\eta)\); the fractured
core has \(\max_jl_{j}^{*}\ge c_{\mathrm f}^{*}(\eta)\). Unless stated otherwise,
\(\eta=0.5\). A trajectory retains the label of its most recently visited
core until it reaches the opposite core; its label is undefined before
the first core visit.

The first-fracture site PMF restricts reservoir bonds below the full-chain
divider \(d_{N}^{*}\). The separate geometric launch-surface restriction used in the
RF control is specified in Sec.~\ref{si:barrier-top}; it does not change the equilibrium
basin definition above.

%% file: si_sections/constrained_thermodynamics.tex
\section{Constrained equilibrium thermodynamics}
\label{si:thermodynamics}

\subsection{Partition densities and convolution}

The constrained partition density of main-text Sec.~II~C,
\begin{equation}
 Z_{N}^{*}(L^{*})=\int_{(0,\infty)^N}
 e^{-\sum_j u^{*}(l_{j}^{*})/\Tstar}
 \delta\!\left(\sum_jl_{j}^{*}-L^{*}\right)d\bm l^{*},
 \label{si:eq:partition}
\end{equation}
can be evaluated by separating one bond from the remaining reservoir.

The bond-coordinate measure is the same as that in \citet{Theodorou2024}
(Eq.~(11) therein). The dimensionless partition used here differs from the scaled
configurational integral \(Z_{\mathrm T}\) in that work's Eq.~(15), which uses bond lengths
scaled by the total length \(L\). Theodorou's potential is lower than ours by
\(\De\) per bond, multiplying the Boltzmann weight by \(e^{N/\Tstar}\).
Changing the bond-length scale from \(l_0\) to \(L\) contributes
\((l_0/L)^N\) from the integration measure and \(L/l_0\) from the
fixed-length delta function. Together these factors give
\[
 Z_{\mathrm T}
 =e^{N/\Tstar}\left(\frac{l_0}{L}\right)^{N-1}Z_N^*.
\]
The full partition in that formulation contains the compensating factor
\(L^{N-1}\). A configurational average is an observable-weighted integral
divided by the partition density. Both acquire the same conversion factor,
which cancels in their ratio. The two formulations therefore give identical
normalized configurational averages for the same physical parameters and
restrictions.

With \(w(l^{*})=e^{-u^{*}(l^{*})/\Tstar}\), fix the first bond at \(l^{*}\):
\[
\begin{aligned}
 Z_n^*(L^*)&=\int_0^{L^*} w(l^*)\Bigg[
 \int_{(0,\infty)^{n-1}}\prod_{j=2}^{n}w(l_j^*)\,
 \delta\!\left(\sum_{j=2}^{n}l_j^*-(L^*-l^*)\right)
 \prod_{j=2}^{n}dl_j^*\Bigg]dl^*.
\end{aligned}
\]
The bracketed integral is \(Z_{n-1}^*(L^*-l^*)\). For a single bond, the
constraint fixes its length, giving \(Z_1^*(L^*)=w(L^*)\). Hence
\begin{equation}
 Z_{1}^{*}(L^{*})=w(L^{*}),\qquad
 Z_{n}^{*}(L^{*})=\int_0^{L^{*}} w(l^{*})Z_{n-1}^{*}(L^{*}-l^{*})\,dl^{*}.
 \label{si:eq:convolution}
\end{equation}
Each step assigns length \(l^{*}\) to one bond and \(L^{*}-l^{*}\) to the others.
For \(Z_{N}^{\mathrm i,*}\), replace \(w(l^{*})\) by
\(w(l^{*})\Theta(d_{N}^{*}-l^{*})\), keeping the full-chain divider
\(d_{N}^{*}(\lambda)\) fixed throughout the recursion. Then
\begin{equation}
 P_{\mathrm i}=\frac{Z_{N}^{\mathrm i,*}}{Z_{N}^{*}},\qquad
 P_{\mathrm f}=1-P_{\mathrm i},\qquad
 g_N(\lambda)=\ln\frac{P_{\mathrm f}}{P_{\mathrm i}}.
 \label{si:eq:prob-logit}
\end{equation}
The midpoint is the bracketed root \(g_N=0\), with the partition densities
and divider reevaluated at each trial extension.

\subsection{Numerical evaluation and convergence}
\label{si:sec:numerical-evaluation}

The constrained convolution in \cref{si:eq:convolution} is evaluated on a
uniform bond-length grid, using direct logarithmic summation or an
exponentially tilted fast Fourier transform (FFT).

\subsubsection{Direct logarithmic convolution}
\label{si:num-direct-logarithmic}

The probability and PMF calculations use direct discrete convolution.
On a uniform grid with spacing \(\Delta l^*\), the recurrence in
\cref{si:eq:convolution} becomes
\[
 Z_n^*(L^*)\approx\Delta l^*\sum_\alpha
 w_\alpha Z_{n-1}^*(L^*-l_\alpha^*).
\]
Here \(\alpha\) indexes the grid points, \(L^*\) lies on the corresponding
total-length grid, and \(w_\alpha\) is the bond weight at \(l_\alpha^*\).
At an intact boundary, this weight includes the fraction of the cell below
\(d_N^*\). The sum is evaluated logarithmically:
\[
 \ln Z_n^*(L^*)\approx\ln\Delta l^*
 +\operatorname{LSE}_\alpha\!\left[
 \ln w_\alpha+\ln Z_{n-1}^*(L^*-l_\alpha^*)\right].
\]
For logarithmic terms \(A_\alpha\), the log-sum-exp operation is
\[
 \operatorname{LSE}_\alpha(A_\alpha)
 =A_{\max}+\ln\sum_\alpha e^{A_\alpha-A_{\max}},
 \qquad A_{\max}=\max_\alpha A_\alpha.
\]
Subtracting the largest term avoids exponentiating very small absolute
weights, retaining partition densities that would underflow in direct
arithmetic. Zero terms have logarithm \(-\infty\); if every term is zero,
the density is zero. Off-grid values are obtained by piecewise cubic Hermite
interpolation of the logarithmic density (log-PCHIP)
\citep{FritschButland1984}, without extrapolating beyond its finite support.
This is also the method used for main-text Figure~2(b).
Local fourth-degree fits to the PMF, rather than second derivatives of
log-PCHIP, determine the extremum curvatures.

\subsubsection{Exponentially tilted FFT convolution}
\label{si:num-tilted-fft}

For the one-fracture midpoint and restricted site-barrier sequences,
exponential tilting improves the numerical representation of small target
densities. The tilted weights are convolved by FFT, retaining their scales
logarithmically; removing the tilt recovers the original partition densities:
\begin{equation}
  w_{\theta^{*}}(l^{*})=w(l^{*})e^{\theta^{*} l^{*}},\qquad
  Z_{n}^{*}(L^{*})=e^{-\theta^{*} L^{*}}Z_{n,\theta}^{*}(L^{*}).
  \label{si:eq:tilt}
\end{equation}
At fixed total length, the product of the added factors is
\(\prod_j e^{\theta^*l_j^*}=e^{\theta^*L^*}\), independent of the arrangement
of the bonds. Removing this factor restores the original ensemble.
On the grid, let \(w_\alpha^{\mathrm i}\) be the Boltzmann weight at
\(l_\alpha^*\), multiplied by the fraction of that cell below \(d_N^*\).
The tilt is chosen by solving
\[
 \frac{\sum_\alpha l_\alpha^*w_\alpha^{\mathrm i}e^{\theta^*l_\alpha^*}}
      {\sum_\alpha w_\alpha^{\mathrm i}e^{\theta^*l_\alpha^*}}
 =\bar l_{\mathrm{target}}^*,
\]
where \(\alpha\) indexes the grid points and
\(\bar l_{\mathrm{target}}^*\) is the desired mean of the tilted weight.
A bracketed scalar root solve determines \(\theta^*\).
For the one-fracture midpoint, the targets are \(\lambda\) for the intact
chain and the short-bond length at the fractured adiabatic minimum for its
reservoir. For the site PMF, several target means are used; at each tagged
length \(l^*\), the nearest to \((N\lambda-l^*)/(N-1)\) is selected.
These means choose the numerical representation; the reservoir bonds remain
jointly integrated rather than fixed at their mean lengths.

The FFT accelerates the same discrete convolution. Let
\(\bm w_\theta\) and \(\bm Z_{n,\theta}^*\) denote the arrays of tilted
bond weights and partition densities on their respective grids. Before
numerical rescaling and noise filtering, their convolution is
\[
 \bm Z_{n,\theta}^*
 =\Delta l^*\,\mathcal F^{-1}\!\left[
 \mathcal F(\bm Z_{n-1,\theta}^*)\odot
 \mathcal F(\bm w_\theta)\right],
\]
where \(\mathcal F\) is the discrete Fourier transform and \(\odot\)
denotes elementwise multiplication. The arrays are zero-padded to a common
transform length at least equal to the sum of their lengths minus one;
retaining that many entries gives linear convolution without wraparound.
This identity evaluates the discrete sum, which approximates the original
constrained integral.
Each discrete convolution is evaluated with SciPy's
\texttt{scipy.signal.fftconvolve} \citep{Virtanen2020} in
\texttt{mode="full"} and multiplied by \(\Delta l^{*}\). Each result is
divided by its maximum, and the accumulated scale is retained logarithmically.

For a query between grid points, linear interpolation of the scaled tilted
density and addition of the accumulated logarithmic scale recover
\(\ln Z_{n,\theta}^*(L^*)\). Removing the tilt then gives
\[
 \ln Z_n^*(L^*)=\ln Z_{n,\theta}^*(L^*)-\theta^*L^*,
\]
the logarithmic form of \cref{si:eq:tilt}. Thus this calculation
interpolates the density itself, whereas the direct logarithmic method
above uses log-PCHIP. Negative values and positive values at or below the
numerical noise threshold are set to zero; queries at or below that
threshold are treated as unavailable.

\subsubsection{Choice of numerical method}
\label{si:num-method-choice}

Table~\ref{si:tab:partition-methods} summarizes the numerical choices for
the published calculations. Both algorithms evaluate the corresponding
constrained partition integrals. The bond restrictions define the physical
ensemble; exponential tilting changes the numerical representation without
changing those restrictions. The energetic, harmonic and
anharmonic midpoint approximations are separate analytical calculations
(Sec.~\ref{si:coexistence-midpoint}).

\begin{table}[H]
  \centering
  \small
  \setlength{\tabcolsep}{4pt}
  \renewcommand{\arraystretch}{1.15}
  \caption{Numerical methods for the published constrained-partition calculations.}
  \label{si:tab:partition-methods}
  \begin{tabularx}{\textwidth}{@{}
    >{\hsize=1.02\hsize\linewidth=\hsize\raggedright\arraybackslash}X
    >{\hsize=0.85\hsize\linewidth=\hsize\raggedright\arraybackslash}X
    >{\hsize=1.13\hsize\linewidth=\hsize\raggedright\arraybackslash}X@{}}
    \toprule
    Calculation & Numerical method & Reason for the choice \\
    \midrule
    Production equilibrium populations and full-partition coexistence midpoints
    & Direct logarithmic convolution
    & Retains small densities, including unrestricted tails. \\
    \addlinespace
    Tagged and site PMFs in main-text Figure~2
    & Direct logarithmic convolution
    & Retains small densities and gives positive off-grid values through log-PCHIP. \\
    \addlinespace
    One-fracture midpoint sequence in main-text Figure~5
    & Tilted FFT for the reservoir, followed by long-bond integration
    & Extends the chain-length sequence efficiently; the restricted calculation passed the checks below. \\
    \addlinespace
    Numerical site-PMF barriers, curvatures and landmarks associated with main-text Figure~6
    & Tilted FFT convolution
    & Evaluates restricted reservoir densities efficiently; accuracy was checked by refinement. \\
    \addlinespace
    Fracture-count decomposition (Sec.~\ref{si:fracture-count-resolution})
    & Tilted FFT convolution
    & Evaluates count-resolved weights efficiently; their sum was checked against independent logarithmic convolution at the states reported below. \\
    \bottomrule
  \end{tabularx}
\end{table}

\subsubsection{Numerical benchmarks and convergence}
\label{si:num-benchmarks}

Numerical accuracy at the tested states was assessed through independent
comparisons and grid- and domain-refinement checks.
The reported differences are numerical comparisons rather than rigorous
absolute-error bounds.

For the direct logarithmic method, values at grid points were checked against
a separate convolution recurrence that sums positive weights in extended
precision, constructing every order independently of the logarithmic
calculation. Direct quadrature checked the off-grid probabilities, PMFs
and derived quantities.

For the tilted fracture-count calculation in
Sec.~\ref{si:fracture-count-resolution}, the summed weights at the tested
grid-aligned states for \(N=4\)--12 were compared with an independent full
logarithmic convolution. The maximum absolute differences are
\(1.54\times10^{-5}\) in \(\ln Z_{N}^{*}\), corresponding to approximately
\(0.00154\%\) in partition density, and \(7.52\times10^{-6}\) in the
basin probabilities. The following grid and domain checks concern the
one-fracture, restricted site-PMF and fracture-count calculations.

For the one-fracture midpoint, reservoir lengths range from
\(l_{\min}^{*}=0.5\) to \(d_{N}^{*}\), and the long-bond integration extends to
\begin{equation}
 \color{revisionphasetwo}
 l_{\max}^{*}=\max\!\left(l_{\mathrm f}^{\mathrm{ad},*}+\delta_{\mathrm{dom}}^{*},
                         d_{N}^{*}+\delta_{\mathrm{dom}}^{*}\right).
 \label{si:eq:midpoint-domain}
\end{equation}
Here \(l_{\mathrm f}^{\mathrm{ad},*}\) is the fractured adiabatic
minimum. The calculations use \(\Delta l^{*}=0.001\) and \(\delta_{\mathrm{dom}}^{*}=0.8\); Table~\ref{si:tab:midpoint-convergence}
compares grid spacings and margins \(\delta_{\mathrm{dom}}^{*}=0.5,0.8,1.1\). The one-fracture and
site-PMF upper limits are rounded up to the next grid point. The all-count
calculation uses the same lower bound and the fixed-sum upper bound
\(N\lambda-(N-1)l_{\min}^{*}\).

\begin{table}[H]
  \centering
  \small
  \color{revisionphasetwo}
  \caption{\revphaseseven{One-fracture midpoint grid and long-bond-domain checks.
  The grid column compares \(\Delta l^{*}=0.002\) with the
  \(\Delta l^{*}=0.001,\delta_{\mathrm{dom}}^{*}=1.1\) reference; the domain column is the range across
  \(\delta_{\mathrm{dom}}^{*}=0.5,0.8,1.1\) at \(\Delta l^{*}=0.001\).}}
  \label{si:tab:midpoint-convergence}
  \input{publication_tables/midpoint_convergence/table.tex}\end{table}

Halving the all-count grid spacing from \(0.001\) to \(0.0005\) changes
the midpoint by \(3.20\times10^{-7}\), \(3.20\times10^{-8}\),
\(4.01\times10^{-9}\), and \(1.84\times10^{-9}\) for
\(N=5,10,20,50\), respectively.

For the site PMF, the tagged coordinate spans
\begin{equation}
 \color{revisionphasetwo}
 l^{*}\in\left[0.5,
 \max\!\left(l_{\mathrm f}^{\mathrm{ad},*}+\delta_{\mathrm{prof}}^{*},
              l_{\mathrm{ad}}^{\ddagger,*}+0.55\right)\right].
 \label{si:eq:barrier-domain}
\end{equation}
The calculations use \(\delta_{\mathrm{prof}}^{*}=0.40\) and test \(\delta_{\mathrm{prof}}^{*}=0.25,0.40,0.70\).
For \(N=5,10,20,50\), halving \(\Delta l^*\) from \(0.001\) to \(0.0005\)
changes the barrier by at most \(1.36\times10^{-5}k_{\mathrm B}T\) and the
intact or saddle curvature by at most \(0.0062\%\)
(Table~\ref{si:tab:barrier-convergence}).

\begin{table}[H]
  \centering
  \small
  \color{revisionphasetwo}
  \caption{\revphaseseven{Site-PMF barrier and curvature convergence. The listed barriers use
  \(\Delta l^{*}=0.001\). Grid differences compare \(\Delta l^{*}=0.001\) with
  \(\Delta l^{*}=0.0005\); the last column is the largest relative change of the
  intact or saddle curvature.  The barrier was unchanged across all three
  profile margins at each fixed grid spacing.}}
  \label{si:tab:barrier-convergence}
  \input{publication_tables/site_barrier_convergence/table.tex}\end{table}

Tail sensitivity is checked by lowering \(l_{\min}^{*}\)
from 0.5 to 0.4 and increasing the long-bond margin from 0.8 to 1.2 at the
\(N=7,10,12\) midpoints, keeping \(\Delta l^{*}=0.001\).
Table~\ref{si:tab:phase3-tail-controls} reports the relative partition
changes \(\lvert Z_{\mathrm{base}}^{*}/Z_{\mathrm{ext}}^{*}-1\rvert\), where
``base'' and ``ext'' denote the baseline and extended domains, together
with the midpoint and barrier shifts.

\begin{table}[H]
  \centering
  \small
  \color{revisionphasethree}
  \caption{\revphaseseven{Integration-domain and tail sensitivity.
  ``Site low'' lowers the reservoir integration bound from 0.5 to 0.4 in the
  tagged-site fractured integral; ``site high'' extends its long-bond margin
  from 0.8 to 1.2. ``Count'' denotes the complete count-resolved partition.  The last two columns give the induced
  one-fracture midpoint and refined site-barrier shifts.}}
  \label{si:tab:phase3-tail-controls}
  \input{publication_tables/integration_controls/table.tex}\end{table}

Across these domain checks, the relative partition changes remain below
\(10^{-5}\). The midpoint and barrier change by at most
\(1.42\times10^{-11}\) and \(5.73\times10^{-7}k_{\mathrm B}T\), respectively.

\subsection{Moving-divider force identity}
\label{si:moving-divider-force}

At fixed temperature, changing \(\lambda\) changes the chain length and
moves the boundary used to classify the chain. The conjugate tension includes
both effects. Write \(Z_{b}^{*}=Z_{N}^{b,*}(L^{*};d_{N}^{*})\), with
\(L^{*}=N\lambda\) and \(b\in\{\mathrm i,\mathrm f\}\).

The divider is the saddle of the equal-sharing adiabatic path. At fixed
\(\sum_jl_{j}^{*}=N\lambda\), transferring length between any two bonds must
leave the energy unchanged to first order at a stationary configuration.
Thus all bonds carry equal tension. For one long bond \(l_{\mathrm{long}}^{*}\)
and \(N-1\) short bonds \(l_{\mathrm{short}}^{*}\), this requires
\(u^{(1),*}(l_{\mathrm{long}}^{*})=u^{(1),*}(l_{\mathrm{short}}^{*})\). For the Morse potential,
\begin{equation}
 u^{(1),*}(l^{*})=2a e^{-a(l^{*}-1)}[1-e^{-a(l^{*}-1)}].
 \label{si:eq:morse-bond-tension}
\end{equation}
Set \(y=e^{-a(l_{\mathrm{long}}^{*}-1)}\) and
\(z=e^{-a(l_{\mathrm{short}}^{*}-1)}\). Substitution into
Eq.~\eqref{si:eq:morse-bond-tension} gives the equal-tension condition
\[
 \begin{aligned}
 z(1-z)-y(1-y)&=0\\
 \Longrightarrow\quad (z-y)-(z^2-y^2)&=0\\
 \Longrightarrow\quad (z-y)(1-z-y)&=0.
 \end{aligned}
\]
The nonuniform configuration has \(z\ne y\), hence \(z=1-y\), with
\(0<y<1/2\) selecting the long bond. Taking logarithms gives
\[
 \ln y=-a(l_{\mathrm{long}}^{*}-1),\qquad
 \ln(1-y)=-a(l_{\mathrm{short}}^{*}-1),
\]
so
\begin{equation}
 l_{\mathrm{long}}^{*}=1-\frac{\ln y}{a},\qquad
 l_{\mathrm{short}}^{*}=1-\frac{\ln(1-y)}{a}.
 \label{si:eq:stationary-bond-lengths}
\end{equation}
Substituting Eq.~\eqref{si:eq:stationary-bond-lengths} into the
total-length constraint yields
\[
 N\lambda=l_{\mathrm{long}}^{*}+(N-1)l_{\mathrm{short}}^{*}
 =N-\frac{\ln y+(N-1)\ln(1-y)}{a}.
\]
Division by \(N\) gives \(\lambda(y)\). At the saddle, the long-bond
length defines the divider \(d_{N}^{*}\), so
\begin{equation}
 \lambda(y)=1-\frac{\ln y+(N-1)\ln(1-y)}{aN},
 \qquad d_{N}^{*}=1-\frac{\ln y}{a},
 \label{si:eq:lambda-y}
\end{equation}
with the saddle selected by \(1/N<y<1/2\). Differentiating both expressions
in Eq.~\eqref{si:eq:lambda-y} gives
\begin{equation}
 \frac{d\lambda}{dy}=\frac{Ny-1}{aNy(1-y)},\qquad
 \frac{d d_{N}^{*}}{dy}=-\frac{1}{ay},
 \label{si:eq:stationary-branch-derivatives}
\end{equation}
whose ratio gives the boundary speed,
\begin{equation}
 d_{N}^{(1),*}(\lambda)\equiv\frac{d d_{N}^{*}}{d\lambda}
 =\frac{d d_N^*/dy}{d\lambda/dy}
 =\frac{N(1-y)}{1-Ny}.
 \label{si:eq:divider-speed}
\end{equation}
To relate the conjugate tension to the sampled observable, define the
basin-averaged virial tension
\begin{equation}
 F_{b,\mathrm{vir}}^*
 =\frac{1}{N\lambda}\left\langle\sum_{j=1}^{N}l_{j}^{*} u^{(1),*}(l_{j}^{*})\right\rangle_b,
 \label{si:eq:basin-virial-tension}
\end{equation}
where \(\langle\cdot\rangle_b\) denotes an equilibrium average conditioned on
basin \(b\). Recall the conjugate-tension definition in main-text Sec.~II~C,
Eq.~(12):
\begin{equation}
 F_{b,\mathrm{conj}}^*
 =-\frac{\Tstar}{N}\frac{d}{d\lambda}
 \ln Z_b^*\!\left(N\lambda,d_N^*(\lambda)\right).
 \label{si:eq:basin-conjugate-tension}
\end{equation}
This total derivative follows the changing length and divider at fixed
\(N\) and temperature. To separate their contributions, set
\(l_j^*=\lambda z_j\) and
\(\widetilde d_N^*=d_N^*/\lambda\). Thus \(z_j=l_j^*/\lambda\) measures
each bond length relative to the mean, and dividing
\(\sum_jl_j^*=N\lambda\) by \(\lambda\) gives \(\sum_jz_j=N\).
The basin indicators in these scaled coordinates are
\[
 \widetilde h_{\mathrm i}(\bm z;\widetilde d_N^*)
 =\prod_{j=1}^{N}\Theta(\widetilde d_N^*-z_j),\qquad
 \widetilde h_{\mathrm f}=1-\widetilde h_{\mathrm i}.
\]
The Dirac delta obeys the scaling identity
\[
 \delta(a y)=\frac{\delta(y)}{|a|},\qquad a\ne0.
\]
Since \(\lambda>0\), the length constraint therefore contributes
\(\delta[\lambda(\sum_jz_j-N)]=\lambda^{-1}\delta(\sum_jz_j-N)\).
Each coordinate differential is \(dl_j^*=\lambda\,dz_j\), so their product
contributes \(\lambda^N\). Combining this with the delta factor gives
the measure factor \(\lambda^{N-1}\):
\begin{equation}
 \begin{aligned}
 Z_b^*(N\lambda,d_N^*)&=\lambda^{N-1}
 \widetilde Z_b^*(\lambda,\widetilde d_N^*),\\
 \widetilde Z_b^*(\lambda,\widetilde d_N^*)&=
 \int_{(0,\infty)^N}
 e^{-\sum_j u^*(\lambda z_j)/\Tstar}
 \widetilde h_b(\bm z;\widetilde d_N^*)
 \delta\!\left(\sum_jz_j-N\right)d\bm z.
 \end{aligned}
 \label{si:eq:scaled-basin-partition}
\end{equation}
Taking logarithms of Eq.~\eqref{si:eq:scaled-basin-partition} separates
the explicit measure factor from the scaled integral:
\begin{equation}
 \ln Z_b^*(N\lambda,d_N^*)
 =(N-1)\ln\lambda+\ln\widetilde Z_b^*(\lambda,\widetilde d_N^*).
 \label{si:eq:scaled-log-partition}
\end{equation}
At fixed \(N\), the first term in Eq.~\eqref{si:eq:scaled-log-partition}
contributes
\[
 \frac{d}{d\lambda}\big[(N-1)\ln\lambda\big]=\frac{N-1}{\lambda}.
\]
The remaining integral still depends on the bond energies and the scaled
divider. Holding \(\widetilde d_N^*\) fixed keeps the constraint and basin
indicator independent of \(\lambda\). Differentiating its Boltzmann factor
at fixed \(\bm z\) gives
\[
 \frac{\partial}{\partial\lambda}
 e^{-\sum_j u^*(\lambda z_j)/\Tstar}
 =-\frac{1}{\Tstar}\sum_j z_j u^{(1),*}(\lambda z_j)
 e^{-\sum_j u^*(\lambda z_j)/\Tstar}.
\]
Integrating this expression in Eq.~\eqref{si:eq:scaled-basin-partition}
and dividing by \(\widetilde Z_b^*\) converts the derivative to a basin average:
\begin{equation}
 \left.\frac{\partial\ln\widetilde Z_b^*}{\partial\lambda}
 \right|_{\widetilde d_N^*}
 =-\frac{1}{\Tstar}\left\langle
 \sum_j z_j u^{(1),*}(\lambda z_j)\right\rangle_b
 =-\frac{N}{\Tstar}F_{b,\mathrm{vir}}^*.
 \label{si:eq:scaled-partition-energy-response}
\end{equation}
The last equality uses \(z_j=l_j^*/\lambda\) and the virial-tension
definition in Eq.~\eqref{si:eq:basin-virial-tension}.
All derivatives retain fixed \(N\) and temperature.

Along the chosen divider \(d_N^*(\lambda)\), differentiating
\(\widetilde d_N^*=d_N^*/\lambda\) by the quotient rule gives
\begin{equation}
 \frac{d\widetilde d_N^*}{d\lambda}
 =\frac{\lambda d_N^{(1),*}-d_N^*}{\lambda^2}
 =\frac{d_N^{(1),*}}{\lambda}-\frac{d_N^*}{\lambda^2}.
 \label{si:eq:scaled-divider-speed}
\end{equation}
To evaluate the integral's response to this motion, hold \(\lambda\) fixed,
and hence \(L^*=N\lambda\). The measure factor is then constant, while
changing \(\widetilde d_N^*\) changes \(d_N^*=\lambda\widetilde d_N^*\).
Differentiating Eq.~\eqref{si:eq:scaled-log-partition} with respect to
this boundary therefore gives
\begin{equation}
 \left.\frac{\partial\ln\widetilde Z_b^*}{\partial\widetilde d_N^*}
 \right|_{\lambda}
 =\lambda\left.\frac{\partial\ln Z_b^*}{\partial d_N^*}\right|_{L^*}.
 \label{si:eq:scaled-partition-boundary-response}
\end{equation}
Here the partial boundary derivative changes the divider at fixed total
length; it does not stretch the chain. Differentiating
Eq.~\eqref{si:eq:scaled-log-partition} along the moving divider and
substituting Eqs.~\eqref{si:eq:scaled-partition-energy-response}--%
\eqref{si:eq:scaled-partition-boundary-response} gives
\begin{equation}
 \begin{aligned}
 \frac{d\ln Z_b^*}{d\lambda}
 &=\frac{N-1}{\lambda}
 +\left.\frac{\partial\ln\widetilde Z_b^*}{\partial\lambda}
 \right|_{\widetilde d_N^*}
 +\left.\frac{\partial\ln\widetilde Z_b^*}{\partial\widetilde d_N^*}
 \right|_{\lambda}\frac{d\widetilde d_N^*}{d\lambda}\\
 &=\frac{N-1}{\lambda}-\frac{N}{\Tstar}F_{b,\mathrm{vir}}^*
 +\left(d_N^{(1),*}-\frac{d_N^*}{\lambda}\right)
 \left.\frac{\partial\ln Z_b^*}{\partial d_N^*}\right|_{L^*}.
 \end{aligned}
 \label{si:eq:total-partition-response}
\end{equation}
Multiplying Eq.~\eqref{si:eq:total-partition-response} by \(-\Tstar/N\)
and using Eq.~\eqref{si:eq:basin-conjugate-tension} gives
\begin{equation}
 F_{b,\mathrm{conj}}^*
 =F_{b,\mathrm{vir}}^*-\frac{\Tstar(N-1)}{N\lambda}
 +\frac{\Tstar}{N}\left(\frac{d_{N}^{*}}{\lambda}-d_{N}^{(1),*}\right)
 \left.\frac{\partial\ln Z_{b}^{*}}{\partial d_{N}^{*}}\right|_{L^{*}}.
 \label{si:eq:conjugate-virial-force}
\end{equation}
{\setlength{\emergencystretch}{1em}
For the unrestricted ensemble, the distinction between the free-energy
derivative and the potential-virial tension agrees with
\citet{Theodorou2024}, Eqs.~(20)--(21); the basin-boundary contribution follows
from the state definitions used here.\par}

The second term in Eq.~\eqref{si:eq:conjugate-virial-force} is common to
both basins and cancels in the difference
between their conjugate tensions. The last term accounts for boundary motion relative to the
uniform stretching of the chain.

To evaluate this boundary response, hold \(L^*\) fixed and differentiate the
intact-bond restrictions:
\[
 \frac{\partial}{\partial d_N^*}\prod_{j=1}^{N}\Theta(d_N^*-l_j^*)
 =\sum_{j=1}^{N}\delta(d_N^*-l_j^*)
   \prod_{k\ne j}\Theta(d_N^*-l_k^*).
\]
In each term, integration over \(l_j^*\) fixes that bond at \(d_N^*\), giving
its Boltzmann weight \(e^{-u^*(d_N^*)/\Tstar}\). The other bonds remain intact
and share length \(L^*-d_N^*\), so their integral is
\(Z_{N-1}^{\mathrm i,*}(L^*-d_N^*;d_N^*)\). All \(N\) terms are identical
by cyclic symmetry. The resulting boundary response is
\begin{equation}
 Q_{\partial}^{*}(L^{*},d_{N}^{*})=\left.\frac{\partial Z_{\mathrm i}^{*}}{\partial d_{N}^{*}}\right|_{L^{*}}
 =N e^{-u^{*}(d_{N}^{*})/\Tstar}Z_{N-1}^{\mathrm i,*}(L^{*}-d_{N}^{*};d_{N}^{*}).
 \label{si:eq:surface-density}
\end{equation}
Thus \(Q_{\partial}^{*}\) is a partition weight per unit boundary displacement.
Increasing \(d_N^*\) transfers this weight from the fractured to the intact
basin, giving opposite signs:
\(\partial_{d_N^*}Z_{\mathrm i}^*=Q_{\partial}^*\) and
\(\partial_{d_N^*}Z_{\mathrm f}^*=-Q_{\partial}^*\).

Using the basin probabilities in Eq.~\eqref{si:eq:prob-logit}, the common
total partition cancels in their ratio:
\begin{equation}
 \frac{P_{\mathrm f}}{P_{\mathrm i}}
 =\frac{Z_{\mathrm f}^*/Z_N^*}{Z_{\mathrm i}^*/Z_N^*}
 =\frac{Z_{\mathrm f}^*}{Z_{\mathrm i}^*},\qquad
 g_N(\lambda)=\ln\frac{P_{\mathrm f}}{P_{\mathrm i}}
 =\ln Z_{\mathrm f}^*-\ln Z_{\mathrm i}^*.
 \label{si:eq:basin-logit-ratio}
\end{equation}
Differentiating Eq.~\eqref{si:eq:basin-logit-ratio} at fixed \(N\) and
temperature, and using Eq.~\eqref{si:eq:basin-conjugate-tension}, gives
\begin{equation}
 \begin{aligned}
 g_N'(\lambda)
 &=\frac{d}{d\lambda}\ln\frac{P_{\mathrm f}}{P_{\mathrm i}}
 =\frac{d\ln Z_{\mathrm f}^*}{d\lambda}
  -\frac{d\ln Z_{\mathrm i}^*}{d\lambda}\\
 &=\frac{N}{\Tstar}
  \left(F_{\mathrm i,conj}^*-F_{\mathrm f,conj}^*\right).
 \end{aligned}
 \label{si:eq:conjugate-identity}
\end{equation}
To make the boundary contribution explicit, divide the two boundary
derivatives given by Eq.~\eqref{si:eq:surface-density} and its opposite
fractured-basin response by their respective partition weights:
\begin{equation}
 \left.\frac{\partial\ln Z_{\mathrm i}^{*}}{\partial d_{N}^{*}}\right|_{L^{*}}
 =\frac{Q_{\partial}^{*}}{Z_{\mathrm i}^{*}},\qquad
 \left.\frac{\partial\ln Z_{\mathrm f}^{*}}{\partial d_{N}^{*}}\right|_{L^{*}}
 =-\frac{Q_{\partial}^{*}}{Z_{\mathrm f}^{*}}.
 \label{si:eq:basin-log-boundary-response}
\end{equation}
Substituting Eq.~\eqref{si:eq:basin-log-boundary-response} into the
difference of the two force decompositions in
Eq.~\eqref{si:eq:conjugate-virial-force} gives
\begin{equation}
 \begin{aligned}
 F_{\mathrm i,conj}^*-F_{\mathrm f,conj}^*
 &=F_{\mathrm i,vir}^*-F_{\mathrm f,vir}^*\\
 &\quad+\frac{\Tstar}{N}\left(\frac{d_{N}^{*}}{\lambda}-d_{N}^{(1),*}\right)
 Q_{\partial}^{*}\left(\frac1{Z_{\mathrm i}^{*}}+\frac1{Z_{\mathrm f}^{*}}\right).
 \end{aligned}
 \label{si:eq:conjugate-tension-difference}
\end{equation}
Substituting Eq.~\eqref{si:eq:conjugate-tension-difference} into
Eq.~\eqref{si:eq:conjugate-identity} cancels
\(\Tstar/N\) in the boundary term and yields
\begin{equation}
 g_N'(\lambda)=\frac{N}{\Tstar}
 \left(F_{\mathrm i,vir}^*-F_{\mathrm f,vir}^*\right)
 +\left(\frac{d_{N}^{*}}{\lambda}-d_{N}^{(1),*}\right)Q_{\partial}^{*}
  \left(\frac1{Z_{\mathrm i}^{*}}+\frac1{Z_{\mathrm f}^{*}}\right).
 \label{si:eq:moving-divider}
\end{equation}
This identity is checked by comparing partition-derived logit differences
with the integral of its right-hand side along \(\lambda\) at fixed \(N\).

\begin{figure}[H]
  \centering
  \includegraphics[width=16cm]{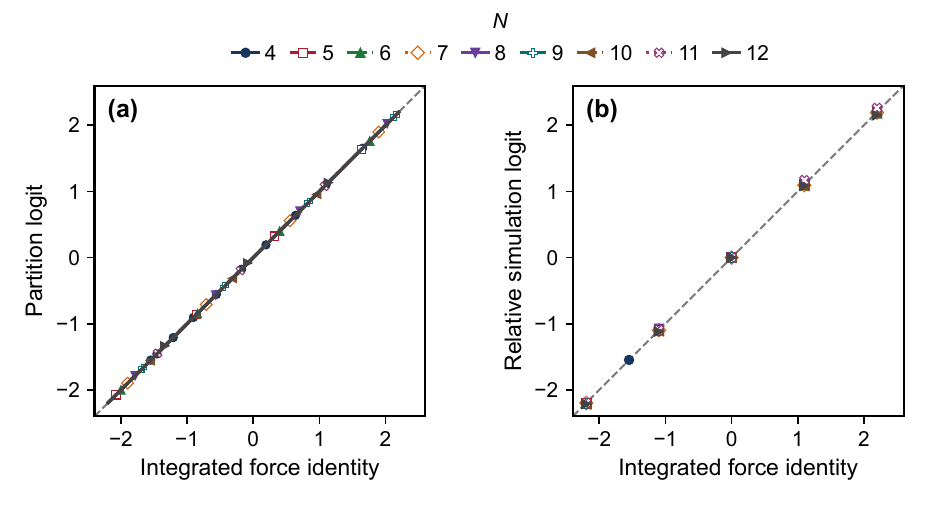}
  \input{figures/moving_divider_identity/caption}
  \label{si:fig:thermodynamic-validation}
\end{figure}

Figure~\ref{si:fig:thermodynamic-validation} compares the partition logit
and the relative simulation logit with the integrated force identity for
\(N=4\)--12. Across \CanonicalMovingDividerDensePoints{} partition points,
the maximum absolute difference is
\(\CanonicalMovingDividerIntegratedResidual\); the simulation comparison
uses 45 sampled states. Subtracting the simulation logit's interpolated
value at the partition midpoint makes the comparison sensitive to variation
with stretch.

\subsection{Fracture-count resolution}
\label{si:fracture-count-resolution}

\begin{figure}[H]
  \centering
  \includegraphics[width=16cm]{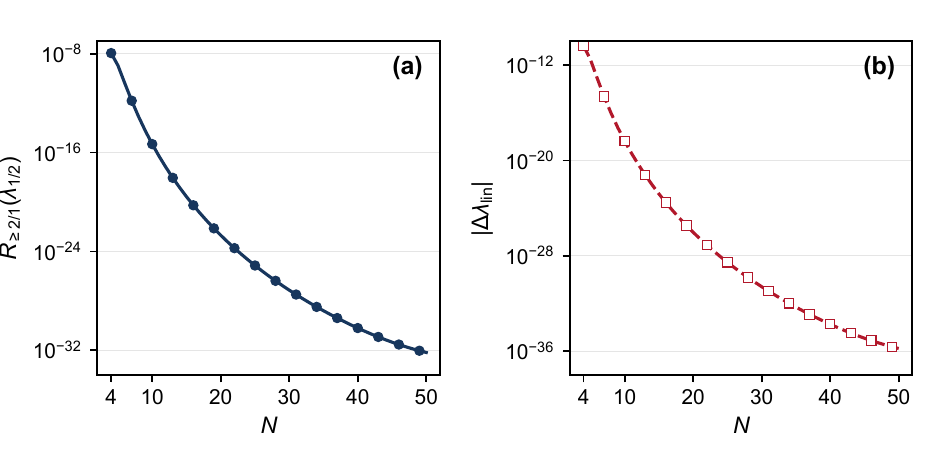}
  \input{figures/fracture_counts/caption}
  \label{si:fig:fracture-counts}
\end{figure}

The one-fracture approximation near coexistence is assessed by separating the
partition into contributions with different numbers of fractured bonds.
Using the same full-chain divider \(d_{N}^{*}(\lambda)\), the instantaneous count is
\begin{equation}
 K(\bm l^{*})=\#\{j:l_{j}^{*}>d_{N}^{*}(\lambda)\}.
\end{equation}
Let \(A_k\) be one specified set of \(k\) bond indices. The bonds in this set lie above
\(d_{N}^{*}\), while all others lie at or below it. Since the bonds are equivalent,
the \(\binom Nk\) choices of this set give the count-resolved partition
\begin{align}
 Z_{N,k}^{*}(L^{*})
 &=\binom Nk\int e^{-\sum_j u^{*}(l_{j}^{*})/\Tstar}
 \prod_{j\in A_k}\mathbf1(l_{j}^{*}>d_{N}^{*})
 \prod_{j\notin A_k}\mathbf1(l_{j}^{*}\le d_{N}^{*})
 \delta\!\left(\sum_jl_{j}^{*}-L^{*}\right)d\bm l^{*} ,
 \label{si:eq:count-partition}
\end{align}
where \(L^{*}=N\lambda\). Thus \(Z_{\mathrm i}^{*}=Z_{N,0}^{*}\) and
\(Z_{\mathrm f}^{*}=\sum_{k=1}^{N}Z_{N,k}^{*}\). The full midpoint
\(\lambda_{1/2}\) satisfies \(Z_{\mathrm f}^{*}=Z_{N,0}^{*}\), whereas the
one-fracture midpoint \(\lambda_{1/2}^{(1)}\) satisfies \(Z_{N,1}^{*}=Z_{N,0}^{*}\).
For each \(N=4\)--50 at the baseline temperature, the
calculation evaluated count-resolved weights using the tilted FFT method of
\cref{si:sec:numerical-evaluation} and located both midpoints by numerically
solving their respective weight equalities.
The relative weight omitted by the latter approximation is
\begin{equation}
 R_{\ge2/1}(\lambda)=\frac{\sum_{k=2}^{N}Z_{N,k}^{*}}{Z_{N,1}^{*}}.
 \label{si:eq:multiple-fracture-ratio}
\end{equation}
At \(\lambda_{1/2}\), this ratio is at most
\(1.11702269\times10^{-8}\), with the maximum at \(N=4\)
(\cref{si:fig:fracture-counts}a).

To estimate the resulting midpoint shift, define
\(g_1(\lambda)=\ln[Z_{N,1}^{*}/Z_{N,0}^{*}]\). Factoring the fractured
weight gives the full population logit exactly:
\[
 \begin{aligned}
 Z_{\mathrm f}^{*}
 &=Z_{N,1}^{*}+\sum_{k=2}^{N}Z_{N,k}^{*}
 =Z_{N,1}^{*}[1+R_{\ge2/1}(\lambda)],\\
 \ln\frac{P_{\mathrm f}}{P_{\mathrm i}}
 &=\ln\frac{Z_{\mathrm f}^{*}}{Z_{N,0}^{*}}
 =g_1(\lambda)+\ln[1+R_{\ge2/1}(\lambda)].
 \end{aligned}
\]
Let \(\delta\lambda=\lambda_{1/2}-\lambda_{1/2}^{(1)}\). Since
\(g_1(\lambda_{1/2}^{(1)})=0\), expanding \(g_1\) about this midpoint
and evaluating the small additional-weight correction there gives
\[
 0\simeq g_1'(\lambda_{1/2}^{(1)})\,\delta\lambda
 +\ln[1+R_{\ge2/1}(\lambda_{1/2}^{(1)})].
\]
The change in this correction across the small shift enters at higher
order. Thus, to first order in the omitted weight,
\[
 \lambda_{1/2}-\lambda_{1/2}^{(1)}\simeq\Delta\lambda_{\mathrm{lin}}
 =-\frac{\ln[1+R_{\ge2/1}(\lambda_{1/2}^{(1)})]}{g_1'(\lambda_{1/2}^{(1)})}.
\]
Because \(\ln(1+R)=R+O(R^2)\), retaining the logarithm changes only
higher-order terms in this estimate.
The slope is evaluated by a centered difference with stretch increment
\(10^{-6}\). It is positive at the tested midpoints, so including additional
fractured configurations shifts coexistence to a slightly smaller stretch.
The estimated shift with the largest magnitude is
\(-4.10263\times10^{-11}\), at \(N=4\).
The direct differences between the separately calculated roots are
numerically distinguishable only for \(N=4\)--6; for larger chains, the
separation is too small to resolve by subtracting the roots.
Figure~\ref{si:fig:fracture-counts}b therefore shows the linearized estimate
throughout.
These results support the one-fracture approximation near coexistence for
\(N=4\)--50 at this temperature.
Additional benchmarks away from coexistence (not shown) indicate that
single-fracture configurations dominate at the attainable tested states
with \(P_{\mathrm f}=0.01,0.10,0.90,0.99\) for \(N=4,10,15,50\).

%% file: publication_tables/midpoint_convergence/table.tex
\begin{tabular}{rccc}
  \toprule
  \(N\) & \(\lambda_{1/2}^{(1)}\) at \(\Delta l^*=0.001\) & \(|\Delta\lambda|_{\mathrm{grid}}\) & \(\operatorname{range}(\lambda)_{\mathrm{domain}}\) \\
  \midrule
  4 & 1.166915565452 & \(5.20\times10^{-7}\) & 0 \\
  10 & 1.107353865574 & \(6.51\times10^{-8}\) & 0 \\
  20 & 1.073935237972 & \(1.90\times10^{-8}\) & \(2.90\times10^{-13}\) \\
  50 & 1.045470112426 & \(3.68\times10^{-9}\) & \(2.61\times10^{-9}\) \\
  \bottomrule
\end{tabular}

%% file: publication_tables/site_barrier_convergence/table.tex
\begin{tabular}{rccc}
  \toprule
  \(N\) & \(\Delta G_N^{\mathrm{site},*}/T^*\) & \(|\Delta(\beta^* G^*)|_{\mathrm{grid}}\) & max. relative curvature change \\
  \midrule
  5 & 3.008461636 & \(1.36\times10^{-5}\) & \(6.18\times10^{-5}\) \\
  10 & 9.146891647 & \(5.99\times10^{-6}\) & \(5.83\times10^{-5}\) \\
  20 & 19.206523140 & \(1.94\times10^{-6}\) & \(9.59\times10^{-6}\) \\
  50 & 34.719013397 & \(1.00\times10^{-6}\) & \(7.75\times10^{-6}\) \\
  \bottomrule
\end{tabular}

%% file: publication_tables/integration_controls/table.tex
\begin{tabular}{rcccccc}
  \toprule
  \(N\) & intact & site low & site high & count & \(\Delta\lambda_{1/2}\) & \(\Delta(\beta^* G^{\ddagger,*})\) \\
  \midrule
  7 & 0 & \(7.13\times10^{-11}\) & 0 & \(3.57\times10^{-11}\) & \(-3.53\times10^{-14}\) & \(-5.73\times10^{-7}\) \\
  10 & \(4.55\times10^{-13}\) & \(1.24\times10^{-8}\) & 0 & \(6.19\times10^{-9}\) & \(1.42\times10^{-11}\) & \(-2.89\times10^{-8}\) \\
  12 & \(4.55\times10^{-13}\) & \(2.30\times10^{-8}\) & 0 & \(1.15\times10^{-8}\) & \(-5.96\times10^{-12}\) & \(5.07\times10^{-8}\) \\
  \bottomrule
\end{tabular}

%% file: figures/moving_divider_identity/caption.tex
\caption{Moving-divider force identity for \(N=4\)--12.
(a) Partition logit compared with the integrated right-hand side of
\cref{si:eq:moving-divider}, anchored at the interpolated partition midpoint.
(b) Simulation logit relative to its interpolated value at that same midpoint,
compared with the force integral. Dashed diagonals indicate equality.
Bars show approximate propagated replica standard errors, neglecting
covariance with the subtracted reference.}

%% file: figures/fracture_counts/caption.tex
\caption{Contributions from multiple fractured bonds near coexistence at the
baseline temperature, for \(N=4\)--50. (a) Their partition weight relative to
configurations with one fractured bond, evaluated at \(\lambda_{1/2}\).
(b) Magnitude of the linearized midpoint shift, \(\Delta\lambda_{\mathrm{lin}}\).
Both quantities decrease with chain length over the tested range.
Lines include every calculated size; markers are shown at every third size.}

%% file: si_sections/conditional_equilibrium_profiles.tex
\section{Adiabatic energies and conditional equilibrium profiles}
\label{si:conditional-profiles}

This section gives the constructions used in main-text Sec.~II~D and Figure~2.
The tagged bond has reduced length \(l^{*}\); the remaining \(N-1\) bonds form a
reservoir with total length \(L^{*}-l^{*}\), where \(L^{*}=N\lambda\).

\subsection{Adiabatic constructions}
\label{si:sec:adiabatic-profiles}

The two adiabatic branches prescribe the reservoir configuration
\citep{Theodorou2024}. Equal
sharing assigns each reservoir bond a length \((L^{*}-l^{*})/(N-1)\); the complementary
branch holds \(N-2\) bonds at their isolated minimum \(l_{j}^{*}=1\), leaving one
bond of length \(L^{*}-l^{*}-(N-2)\). Their reduced energies, relative to the uniform
chain, are
\[
 \begin{aligned}
 E_{\mathrm{ad}}^{(N-1),*}(l^{*})
 &=u^{*}(l^{*})+(N-1)u^{*}\!\left(\frac{L^{*}-l^{*}}{N-1}\right)-Nu^{*}(\lambda),\\
 E_{\mathrm{ad}}^{(1),*}(l^{*})
 &=u^{*}(l^{*})+u^{*}\!\left[L^{*}-l^{*}-(N-2)\right]+(N-2)u^{*}(1)-Nu^{*}(\lambda).
 \end{aligned}
\]
The lower envelope takes the smaller of these two energies at each \(l^{*}\).
It compares the prescribed configurations, rather than minimizing over every
possible reservoir arrangement. Thermal profiles instead average or integrate
over the compatible configurations, as described below.

\subsection{Tagged and intact-reservoir PMFs}
\label{si:sec:profile-pmfs}

The two PMFs follow from the constrained Boltzmann distribution. For
positive bond lengths \(\bm l^{*}=(l_{1}^{*},\ldots,l_{N}^{*})\) at total length \(L^{*}=N\lambda\),
the configuration probability measure is
\begin{equation}
 \frac{1}{Z_{N}^{*}(L^{*})}e^{-\sum_{j=1}^N u^{*}(l_{j}^{*})/\Tstar}
 \delta\!\left(\sum_{j=1}^N l_{j}^{*}-L^{*}\right)\prod_{j=1}^N dl_{j}^{*}.
 \label{si:eq:conditional-boltzmann-measure}
\end{equation}
Fixing the tagged coordinate in
Eq.~\eqref{si:eq:conditional-boltzmann-measure} at \(l_1^*=l^*\) and
integrating over the other bonds, before division by \(Z_N^*\), gives the
unnormalized tagged-coordinate density
\begin{equation}
 e^{-u^{*}(l^{*})/\Tstar}
 \int_{(0,\infty)^{N-1}}e^{-\sum_{j=2}^N u^{*}(l_{j}^{*})/\Tstar}
 \delta\!\left(\sum_{j=2}^N l_{j}^{*}-(L^{*}-l^{*})\right)\prod_{j=2}^N dl_{j}^{*}
 =e^{-u^{*}(l^{*})/\Tstar}Z_{N-1}^{*}(L^{*}-l^{*}).
 \label{si:eq:tagged-coordinate-weight}
\end{equation}
The integral of Eq.~\eqref{si:eq:tagged-coordinate-weight} over
\(0<l^*<L^*\) is \(Z_N^*(L^*)\), so division by this partition density normalizes \(p_{N}^{\mathrm{tag},*}(l^{*}\mid L^{*})\). Taking
\(-\Tstar\ln p_{N}^{\mathrm{tag},*}+C_{\mathrm{tag}}^{*}\) gives the tagged PMF.

For the site ensemble, restrict every reservoir bond to \(l_{j}^{*}\le d_{N}^{*}\)
by inserting \(\prod_{j=2}^N\Theta(d_N^*-l_j^*)\) in the integral of
Eq.~\eqref{si:eq:tagged-coordinate-weight}. The full-chain divider \(d_{N}^{*}\) is held fixed as \(l^{*}\) varies. The resulting
weight and PMF are
\begin{equation}
 \begin{gathered}
 e^{-u^{*}(l^{*})/\Tstar}Z_{N-1}^{\mathrm i,*}(L^{*}-l^{*};d_{N}^{*}),
 \\
 G_{N}^{\mathrm{site},*}(l^{*};L^{*})
 =u^{*}(l^{*})-\Tstar\ln Z_{N-1}^{\mathrm i,*}(L^{*}-l^{*};d_{N}^{*})+C_{\mathrm{site}}^{*}.
 \end{gathered}
 \label{si:eq:restricted-site-weight-and-pmf}
\end{equation}
The normalization over \(l^{*}\) is independent of \(l^{*}\) and is absorbed into
\(C_{\mathrm{site}}^{*}\).
Integrating the weight in Eq.~\eqref{si:eq:restricted-site-weight-and-pmf}
only over \(d_N^*<l^*<L^*\) includes only configurations with a fractured
tagged bond and an intact reservoir:
\begin{equation}
 Z_{N,\mathrm{site}}^{\mathrm f,*}(L^*)
 =\int_{d_N^*}^{L^*}e^{-u^*(l^*)/\Tstar}
   Z_{N-1}^{\mathrm i,*}(L^*-l^*;d_N^*)\,dl^*.
 \label{si:eq:one-fracture-site-integral}
\end{equation}
This is the one-fracture integral in main-text Sec.~II~C.

\subsection{Conditional potential energy}
\label{si:sec:conditional-potential-energy}

The conditional potential energy in main-text Figure~2 is the thermal mean
of the total chain potential energy when the tagged bond has length \(l^{*}\).
Writing \(\langle\cdot\rangle_{L^{*}-l^{*}}\) for the unrestricted reservoir's
constrained Boltzmann average, its definition is
\[
 E_{\mathrm{tag}}^{\mathrm{pot},*}(l^{*};L^{*})\equiv \frac{U_{\mathrm{tag}}^{\mathrm{pot}}(l^{*};L^{*})}{D_e}
 =u^{*}(l^{*})+\left\langle\sum_{j=2}^N u^{*}(l_{j}^{*})\right\rangle_{L^{*}-l^{*}}.
\]
Differentiating the reservoir Boltzmann weight brings down minus its potential
energy. Thus, with \(\beta^*=1/\Tstar\), the same average can be calculated as
\[
 E_{\mathrm{tag}}^{\mathrm{pot},*}(l^{*};L^{*})
 =u^{*}(l^{*})-\left.\frac{\partial}{\partial\beta^*}
 \ln Z_{N-1}^{*}(L^{*}-l^{*};\beta^*)\right|_{L^{*},l^{*}}.
\]
The first term is the tagged bond energy; the derivative gives the mean
reservoir potential energy at the fixed remaining length. The derivative is evaluated
by a centered difference with relative step \(10^{-4}\) in \(\beta^*\), using
the same unrestricted convolution as the tagged PMF. This thermal average
contains no kinetic energy and differs from both a minimum energy and a
PMF. Figure~2 shows the same thermal-unit ratio,
\(E_{\mathrm{tag}}^{\mathrm{pot},*}/\Tstar
=U_{\mathrm{tag}}^{\mathrm{pot}}/(k_{\mathrm B}T)\), shifted to zero at
\(l^{*}=\lambda\), independently of the PMFs.

%% file: si_sections/coexistence_midpoint.tex
\section{Coexistence extension and its chain-length dependence}
\label{si:derivations}
\label{si:coexistence-midpoint}

\subsection{Stationary configurations and the zero-temperature midpoint}
\label{si:stationary-morse}

The stationary configurations are parameterized by the equal-tension
construction in Sec.~\ref{si:moving-divider-force},
Eq.~\eqref{si:eq:lambda-y}. We recall that \(l_{\mathrm{long}}^{*}\) is the
long-bond length, \(l_{\mathrm{short}}^{*}\) is the common length of the other
\(N-1\) bonds, and \(y=e^{-a(l_{\mathrm{long}}^{*}-1)}\), with
\(e^{-a(l_{\mathrm{short}}^{*}-1)}=1-y\). We now determine their constrained
stability and compare their energies with that of the uniform chain.

The two nonuniform branches are distinguished by their stability under the
length constraint. Define \(u_{\mathrm{long}}^{(2),*}=u^{(2),*}(l_{\mathrm{long}}^{*})\) and
\(u_{\mathrm{short}}^{(2),*}=u^{(2),*}(l_{\mathrm{short}}^{*})\).
Differentiating the bond tension in Eq.~\eqref{si:eq:morse-bond-tension}
and using the lengths in Eq.~\eqref{si:eq:stationary-bond-lengths} gives
\begin{equation}
 u_{\mathrm{long}}^{(2),*}=-2a^2y(1-2y),\qquad
 u_{\mathrm{short}}^{(2),*}=2a^2(1-y)(1-2y)>0.
 \label{si:eq:stationary-bond-curvatures}
\end{equation}
Let the long bond change by \(\delta l^{*}\), while each short bond changes
by \(-\delta l^{*}/(N-1)\), preserving the total length. Equal tensions cancel
the linear energy terms:
\[
 u^{(1),*}(l_{\mathrm{long}}^{*})\delta l^{*}
 +(N-1)u^{(1),*}(l_{\mathrm{short}}^{*})\left(-\frac{\delta l^{*}}{N-1}\right)=0.
\]
The quadratic terms, using the curvatures in
Eq.~\eqref{si:eq:stationary-bond-curvatures}, give
\begin{equation}
 \Delta E^{(2),*}
 =\frac12 u_{\mathrm{long}}^{(2),*}(\delta l^{*})^2
 +(N-1)\frac12 u_{\mathrm{short}}^{(2),*}\left(\frac{\delta l^{*}}{N-1}\right)^2
 =\frac12\left(u_{\mathrm{long}}^{(2),*}+\frac{u_{\mathrm{short}}^{(2),*}}{N-1}\right)(\delta l^{*})^2.
 \label{si:eq:stationary-path-quadratic-energy}
\end{equation}
Because \(u_{\mathrm{short}}^{(2),*}>0\), equal sharing minimizes the sum of the
short-bond squared displacements at fixed total displacement, and hence
their quadratic energy cost. Substituting
Eq.~\eqref{si:eq:stationary-bond-curvatures} into the coefficient in
Eq.~\eqref{si:eq:stationary-path-quadratic-energy} gives
\begin{equation}
 \begin{aligned}
 u_{\mathrm{long}}^{(2),*}+\frac{u_{\mathrm{short}}^{(2),*}}{N-1}
 &=2a^2(1-2y)\left(-y+\frac{1-y}{N-1}\right)\\
 &=\frac{2a^2(1-2y)(1-Ny)}{N-1}.
 \end{aligned}
 \label{si:eq:stationary-path-curvature}
\end{equation}
This coefficient has the sign of \(1-Ny\), since \(1-2y>0\) on \(0<y<1/2\).
The remaining \(N-2\) independent directions redistribute length among
short bonds and have positive curvature. For \(0<y<1/N\), the energy
therefore increases to quadratic order in every allowed direction, giving
a stable fractured minimum. For \(1/N<y<1/2\), transferring length between
the long bond and the short-bond reservoir lowers the energy, while the
other directions remain stable: this is a saddle with one unstable direction.

At \(y=1/N\), the coefficient in Eq.~\eqref{si:eq:stationary-path-curvature}
vanishes. The extension derivative in
Eq.~\eqref{si:eq:stationary-branch-derivatives} is
\[
 \frac{d\lambda}{dy}=\frac{Ny-1}{aNy(1-y)},
\]
so \(\lambda(y)\) has its minimum at the same point, where the stable and
saddle branches meet. At a prescribed extension where both branches
exist, solving Eq.~\eqref{si:eq:lambda-y} on the two intervals selects the
corresponding values of \(y\). Substituting them into the long-bond length
in Eq.~\eqref{si:eq:stationary-bond-lengths} gives the fractured-minimum coordinate
\(l_{\mathrm f}^{\mathrm{ad},*}\) from the stable branch and the divider
\(d_{N}^{*}\) from the saddle branch, used in the state definitions of
Sec.~\ref{si:notation}.

To compare the fractured and intact energies, both configurations must have
the same \(\lambda\). Substituting \(\lambda(y)\) from
Eq.~\eqref{si:eq:lambda-y} into \(y_0=e^{-a(\lambda-1)}\) gives
\[
 \ln y_0=-a[\lambda(y)-1]
 =\frac{\ln y+(N-1)\ln(1-y)}{N}.
\]
Exponentiating yields
\begin{equation}
 y_0(y)\equiv e^{-a[\lambda(y)-1]}
 =\left[y(1-y)^{N-1}\right]^{1/N}.
 \label{si:eq:uniform-stationary-coordinate}
\end{equation}
Here \(y_0\) in Eq.~\eqref{si:eq:uniform-stationary-coordinate} describes
each bond in the uniform chain at the same total extension as the
nonuniform configuration. From
\(u^{*}(l^{*})=[1-e^{-a(l^{*}-1)}]^2\), each uniform intact bond contributes
\((1-y_0)^2\). In the fractured configuration, the long bond contributes
\((1-y)^2\), while each short bond contributes
\([1-(1-y)]^2=y^2\). Summing gives the total configurational energies
in units of \(\De\):
\begin{equation}
 E_{\mathrm i}^{*}=N(1-y_0)^2,\qquad
 E_{\mathrm f}^{*}=(1-y)^2+(N-1)y^2=1-2y+Ny^2.
 \label{si:eq:stationary-energies}
\end{equation}
Equating the two energies in Eq.~\eqref{si:eq:stationary-energies} gives
the energetic midpoint equation,
\begin{equation}
 E_{\mathrm f}^*-E_{\mathrm i}^*
 =1-2y+Ny^2
 -N\left\{1-\left[y(1-y)^{N-1}\right]^{1/N}\right\}^{2}=0,
 \qquad 0<y<1/N,
 \label{si:eq:energetic-midpoint-root}
\end{equation}
where Eq.~\eqref{si:eq:uniform-stationary-coordinate} supplies \(y_0(y)\).
Substituting the root into Eq.~\eqref{si:eq:lambda-y} gives
\(\lambda_0(N)\).
In the implementation, this scalar root is found in \(\ln y\), so very
small long-bond exponential coordinates remain resolved. The finite-\(N\)
zero-temperature curves use this energy equality.

For large \(N\), the stable branch has \(y<1/N\), so
Eq.~\eqref{si:eq:stationary-energies} gives \(E_{\mathrm f}^{*}\to1\):
one bond approaches its dissociation energy while the short bonds relax
toward their minima. The energy equality in
Eq.~\eqref{si:eq:energetic-midpoint-root} therefore reduces to
\begin{equation}
 N(1-y_0)^2\simeq1,\qquad y_0\simeq1-N^{-1/2}.
 \label{si:eq:energetic-large-n-balance}
\end{equation}
Using Eq.~\eqref{si:eq:energetic-large-n-balance} in
\(\lambda_0-1=-\ln y_0/a\), from
Eq.~\eqref{si:eq:uniform-stationary-coordinate}, and expanding with
\(-\ln(1-z)=z+z^2/2+z^3/3+\cdots\), we obtain
\begin{equation}
 \lambda_0-1\simeq-\frac1a\ln(1-N^{-1/2})
 =\frac1{a\sqrt N}+\frac1{2aN}
  +\frac1{3aN^{3/2}}+\cdots.
 \label{si:eq:energetic-midpoint}
\end{equation}
Equation~\eqref{si:eq:energetic-midpoint} is a large-\(N\) approximation
to the root in Eq.~\eqref{si:eq:energetic-midpoint-root}. Its
neglected long-bond corrections are smaller than any power of \(N^{-1}\):
the constraint \(y(1-y)^{N-1}=y_0^N\), together with energy equality,
gives \(\ln y=-\sqrt N+O(1)\). Thus
\(\sqrt N(\lambda_0-1)\to1/a\), with the leading finite-size correction
\(1/(2a\sqrt N)\).

\subsection{Constrained harmonic midpoint}
\label{si:midpoint-approximations}

At finite temperature, a basin's statistical weight depends both on its
minimum energy and on the thermally accessible configurations around that
minimum. A broader basin can therefore compete with a deeper one. Equal
intact and fractured weights determine the midpoint; here the calculation approximates
the fractured weight by the contributions of the \(N\) one-fracture minima.
The reduced energy is expanded around the stationary configurations of
Sec.~\ref{si:stationary-morse}. Let \(l_{j}^{(b),*}\) denote the bond lengths
at minimum \(b\), where \(b=\mathrm i\) is the uniform intact minimum and
\(b=\mathrm f\) is a fractured minimum with one specified long bond.
The fixed total length constrains the displacements:
\begin{equation}
 \delta l_{j}^{*}=l_{j}^{*}-l_{j}^{(b),*},\qquad
 \sum_{j=1}^{N}\delta l_{j}^{*}=0.
 \label{si:eq:midpoint-displacement-constraint}
\end{equation}
Thus specifying \(N-1\) displacements fixes the last one. Define the minimum
energy and local derivatives
\[
 E_{b}^{*}=\sum_j u^{*}(l_{j}^{(b),*}),\qquad
 u_{j}^{(2),*}=u^{(2),*}(l_{j}^{(b),*}),\quad
 u_{j}^{(3),*}=u^{(3),*}(l_{j}^{(b),*}),\quad u_{j}^{(4),*}=u^{(4),*}(l_{j}^{(b),*}).
\]
At a stationary configuration, every \(u^{(1),*}(l_{j}^{(b),*})\) equals the same
tension \(f_{b}^{*}\). The linear term therefore vanishes by
Eq.~\eqref{si:eq:midpoint-displacement-constraint},
\(\sum_j u^{(1),*}(l_{j}^{(b),*})\delta l_{j}^{*}=f_{b}^{*}\sum_j\delta l_{j}^{*}=0\).
The energy expansion becomes
\begin{equation}
 E^{*}=E_{b}^{*}+\frac12\sum_j u_{j}^{(2),*}(\delta l_{j}^{*})^2
       +\frac16\sum_j u_{j}^{(3),*}(\delta l_{j}^{*})^3
       +\frac1{24}\sum_j u_{j}^{(4),*}(\delta l_{j}^{*})^4+\cdots.
 \label{si:eq:midpoint-energy-expansion}
\end{equation}
Keeping only the quadratic term in
Eq.~\eqref{si:eq:midpoint-energy-expansion} gives the harmonic approximation. We
extend the local displacement integral over the entire fixed-sum subspace,
which is appropriate when the thermal fluctuations are concentrated near
the minimum. With the same delta-function measure as the partition density
in Sec.~\ref{si:thermodynamics}, its weight is
\begin{equation}
 Z_{b}^{\mathrm H,*}=e^{-E_{b}^{*}/\Tstar}
 \int_{\mathbb R^N}
 \exp\!\left[-\frac{1}{2\Tstar}\sum_j u_{j}^{(2),*}(\delta l_{j}^{*})^2\right]
 \delta\!\left(\sum_j\delta l_{j}^{*}\right)\prod_jd(\delta l_{j}^{*}).
 \label{si:eq:harmonic-basin-integral}
\end{equation}

To evaluate Eq.~\eqref{si:eq:harmonic-basin-integral}, use the constraint in
Eq.~\eqref{si:eq:midpoint-displacement-constraint} to choose \(N-1\) independent
coordinates and eliminate the last displacement:
\begin{equation}
 \epsilon_{i}^{*}=\delta l_{i}^{*},\quad i=1,\ldots,N-1,\qquad
 \delta l_{N}^{*}=-\sum_{i=1}^{N-1}\epsilon_{i}^{*}.
 \label{si:eq:independent-bond-displacements}
\end{equation}
First separate the last bond's contribution to the quadratic term of
Eq.~\eqref{si:eq:midpoint-energy-expansion}:
\[
 E^{(2),*}=\frac12\sum_{j=1}^{N}u_{j}^{(2),*}(\delta l_{j}^{*})^2
 =\frac12\sum_{i=1}^{N-1}u_{i}^{(2),*}(\delta l_{i}^{*})^2
  +\frac12u_{N}^{(2),*}(\delta l_{N}^{*})^2.
\]
Here \(u_{N}^{(2),*}=u^{(2),*}(l_{N}^{(b),*})\) is the potential curvature of the eliminated
bond, evaluated at the chosen minimum. Its displacement compensates for
those of the other bonds. Substituting
Eq.~\eqref{si:eq:independent-bond-displacements} gives
\begin{equation}
 \begin{aligned}
 E^{(2),*}
 &=\frac12\sum_{i=1}^{N-1}u_{i}^{(2),*}\bigl(\epsilon_{i}^{*}\bigr)^{2}
   +\frac12u_{N}^{(2),*}\left(-\sum_{i=1}^{N-1}\epsilon_{i}^{*}\right)^2\\
 &=\frac12\sum_{i=1}^{N-1}u_{i}^{(2),*}\bigl(\epsilon_{i}^{*}\bigr)^{2}
   +\frac12u_{N}^{(2),*}\left(\sum_{i=1}^{N-1}\epsilon_{i}^{*}\right)^2.
 \end{aligned}
 \label{si:eq:constrained-quadratic-energy}
\end{equation}
In Eq.~\eqref{si:eq:harmonic-basin-integral}, the delta function selects this
same value of the last displacement:
\[
 \int_{-\infty}^{\infty}
 e^{-u_N^{(2),*}(\delta l_N^*)^2/(2\Tstar)}
 \delta\!\left(\delta l_N^*+\sum_{i=1}^{N-1}\epsilon_i^*\right)d(\delta l_N^*)
 =
 e^{-u_N^{(2),*}\left(\sum_{i=1}^{N-1}\epsilon_i^*\right)^2/(2\Tstar)}.
\]
The derivative of the delta-function argument with respect to
\(\delta l_N^*\) is one, so its reciprocal absolute value contributes no
additional factor. The remaining measure is
\(d\bm\epsilon^*=\prod_{i=1}^{N-1}d\epsilon_i^*\).

The constrained Hessian is the matrix of second derivatives of
Eq.~\eqref{si:eq:constrained-quadratic-energy}. Since each derivative of
\(\sum_j\epsilon_j^*\) with respect to an independent coordinate is one,
\[
 \frac{\partial E^{(2),*}}{\partial\epsilon_{i}^{*}}
 =u_{i}^{(2),*}\epsilon_{i}^{*}+u_{N}^{(2),*}\sum_{j=1}^{N-1}\epsilon_{j}^{*},\qquad
 H_{ij}^{*}=\frac{\partial^2E^{(2),*}}{\partial\epsilon_{i}^{*}\partial\epsilon_{j}^{*}}
 =u_{i}^{(2),*}\delta_{ij}+u_{N}^{(2),*},
\]
where \(\delta_{ij}\) is one for \(i=j\) and zero otherwise. In matrix form,
\begin{equation}
 H^{*}=\operatorname{diag}(u_{1}^{(2),*},\ldots,u_{N-1}^{(2),*})
   +u_{N}^{(2),*}\bm1_{N-1}\bm1_{N-1}^{\mathsf T},\qquad
 E^{(2),*}=\frac12\bigl(\bm\epsilon^{*}\bigr)^{\mathsf T}H^{*}\bm\epsilon^{*},
 \label{si:eq:constrained-hessian}
\end{equation}
Here \(\bm1_{N-1}\) is the column vector of \(N-1\) ones, so every entry of
its outer product \(\bm1_{N-1}\bm1_{N-1}^{\mathsf T}\) equals one. Every independent
displacement changes the eliminated bond, so its energy contributes
\(u_{N}^{(2),*}\) to every Hessian entry.

Because \(H^*\) in Eq.~\eqref{si:eq:constrained-hessian} is real and symmetric,
an orthogonal transformation
\(\bm\epsilon^{*}=O\bm q^{*}\) diagonalizes it:
\begin{equation}
 O^{\mathsf T}H^{*}O=\operatorname{diag}(\rho_{1}^{*},\ldots,\rho_{N-1}^{*}),
 \qquad E^{(2),*}=\frac12\sum_{\alpha=1}^{N-1}\rho_{\alpha}^{*} \bigl(q_{\alpha}^{*}\bigr)^{2}.
 \label{si:eq:constrained-normal-modes}
\end{equation}
Here \(\rho_{\alpha}^{*}\) are the Hessian eigenvalues. The transformation removes
cross terms and preserves integration volume because \(|\det O|=1\).
It also preserves the determinant. Using the diagonal matrix in
Eq.~\eqref{si:eq:constrained-normal-modes},
\begin{equation}
 \det H^*
 =\det(O^{\mathsf T}H^*O)
 =\det[\operatorname{diag}(\rho_1^*,\ldots,\rho_{N-1}^*)]
 =\prod_{\alpha=1}^{N-1}\rho_\alpha^*.
 \label{si:eq:hessian-eigenvalue-product}
\end{equation}
The first equality follows from
\(\det(O^{\mathsf T}H^*O)=\det(O)^2\det H^*\) and \(\det(O)^2=1\).
The Boltzmann factor from Eq.~\eqref{si:eq:constrained-normal-modes}
separates because the exponential of a sum is a product:
\[
 \exp\!\left[-\frac{1}{2\Tstar}
 \sum_{\alpha=1}^{N-1}\rho_\alpha^*(q_\alpha^*)^2\right]
 =\prod_{\alpha=1}^{N-1}
 \exp\!\left[-\frac{\rho_\alpha^*(q_\alpha^*)^2}{2\Tstar}\right].
\]
Each factor depends on just one coordinate. The one-dimensional Gaussian
integral \citep[App.~A.1.2]{PetersenPedersen2012} is
\begin{equation}
 \int_{-\infty}^{\infty}e^{-\rho^*(q^*)^2/(2\Tstar)}\,dq^*
 =\sqrt{\frac{2\pi\Tstar}{\rho^*}},\qquad \rho^*>0.
 \label{si:eq:one-dimensional-gaussian}
\end{equation}
Applying Eq.~\eqref{si:eq:one-dimensional-gaussian} to each coordinate of
the extended domain \(\mathbb R^{N-1}\), and then using
Eq.~\eqref{si:eq:hessian-eigenvalue-product}, gives
\begin{equation}
 \begin{aligned}
 \int_{\mathbb R^{N-1}}e^{-E^{(2),*}/\Tstar}\,d\bm\epsilon^{*}
 &=\prod_{\alpha=1}^{N-1}\int_{-\infty}^{\infty}
 e^{-\rho_{\alpha}^{*}(q_{\alpha}^{*})^{2}/(2\Tstar)}\,dq_{\alpha}^{*}\\
 &=\prod_{\alpha=1}^{N-1}\sqrt{\frac{2\pi\Tstar}{\rho_\alpha^*}}
 =\frac{(2\pi\Tstar)^{(N-1)/2}}
 {\sqrt{\prod_{\alpha=1}^{N-1}\rho_\alpha^*}}
 =\frac{(2\pi\Tstar)^{(N-1)/2}}{\sqrt{\det H^*}}.
 \end{aligned}
 \label{si:eq:constrained-gaussian-integral}
\end{equation}
Convergence requires \(\rho_{\alpha}^{*}>0\) for every direction, corresponding
to a stable constrained minimum. Each direction contributes a thermal width
proportional to \(\sqrt{\Tstar/\rho_\alpha^*}\). Their product in
Eq.~\eqref{si:eq:constrained-gaussian-integral} explains why a larger determinant,
and hence narrower fluctuations, reduces the basin weight.

To express this determinant in the bond curvatures, define the diagonal
part of Eq.~\eqref{si:eq:constrained-hessian} as
\begin{equation}
 A^*=\operatorname{diag}(u_1^{(2),*},\ldots,u_{N-1}^{(2),*}).
 \label{si:eq:diagonal-hessian-part}
\end{equation}
These curvatures are nonzero at the chosen minima, so the inverse is
\begin{equation}
 (A^*)^{-1}=\operatorname{diag}\!\left(
 \frac1{u_1^{(2),*}},\ldots,\frac1{u_{N-1}^{(2),*}}\right).
 \label{si:eq:inverse-diagonal-hessian-part}
\end{equation}
Using Eqs.~\eqref{si:eq:diagonal-hessian-part} and
\eqref{si:eq:inverse-diagonal-hessian-part}, factor \(A^*\) from the Hessian:
\[
 H^*=A^*
 \left[I+u_N^{(2),*}(A^*)^{-1}
 \bm1_{N-1}\bm1_{N-1}^{\mathsf T}\right],
\]
where \(I\) is the \((N-1)\)-dimensional identity matrix. The matrix in
brackets leaves vectors whose components sum to zero unchanged, giving
\(N-2\) eigenvalues equal to one. Acting on
\((A^*)^{-1}\bm1_{N-1}\) multiplies it by
\(1+u_N^{(2),*}\bm1_{N-1}^{\mathsf T}(A^*)^{-1}\bm1_{N-1}\),
the remaining eigenvalue. As in
Eq.~\eqref{si:eq:hessian-eigenvalue-product}, the determinant is their
product. Equation~\eqref{si:eq:diagonal-hessian-part} also gives
\(\det A^*=\prod_{i=1}^{N-1}u_i^{(2),*}\), so
\[
 \begin{aligned}
 \det H^*
 &=\det A^*\,
 \det\!\left[I+u_N^{(2),*}(A^*)^{-1}
 \bm1_{N-1}\bm1_{N-1}^{\mathsf T}\right]\\
 &=\left(\prod_{i=1}^{N-1}u_i^{(2),*}\right)
 \left[1+u_N^{(2),*}
 \bm1_{N-1}^{\mathsf T}(A^*)^{-1}\bm1_{N-1}\right].
 \end{aligned}
\]
Multiplication by the all-ones vectors sums the diagonal entries in
Eq.~\eqref{si:eq:inverse-diagonal-hessian-part}:
\[
 \bm1_{N-1}^{\mathsf T}(A^*)^{-1}\bm1_{N-1}
 =\sum_{i=1}^{N-1}\frac1{u_i^{(2),*}}.
\]
This gives the first determinant expression in
Eq.~\eqref{si:eq:constrained-gaussian}.

To include the eliminated bond in the same sum, write
\(1=u_N^{(2),*}/u_N^{(2),*}\) and factor \(u_N^{(2),*}\):
\[
 \begin{aligned}
 1+u_N^{(2),*}\sum_{i=1}^{N-1}\frac1{u_i^{(2),*}}
 &=u_N^{(2),*}
 \left[\frac1{u_N^{(2),*}}+\sum_{i=1}^{N-1}\frac1{u_i^{(2),*}}\right]\\
 &=u_N^{(2),*}\sum_{i=1}^N\frac1{u_i^{(2),*}}.
 \end{aligned}
\]
The bracket now contains the reciprocal curvatures of all \(N\) bonds.
The extracted curvature also completes the product:
\[
 \left(\prod_{i=1}^{N-1}u_i^{(2),*}\right)u_N^{(2),*}
 =\prod_{i=1}^N u_i^{(2),*}.
\]
Together these steps give the symmetric determinant in
Eq.~\eqref{si:eq:constrained-gaussian}. Substituting it
into Eq.~\eqref{si:eq:constrained-gaussian-integral}, and restoring the
minimum-energy factor from Eq.~\eqref{si:eq:harmonic-basin-integral}, gives
\begin{equation}
 \begin{aligned}
 \det H^{*}
 &=\left(\prod_{i=1}^{N-1}u_{i}^{(2),*}\right)
   \left(1+u_{N}^{(2),*}\sum_{i=1}^{N-1}\frac1{u_{i}^{(2),*}}\right)
 \\
 &=\left(\prod_{i=1}^{N}u_{i}^{(2),*}\right)
   \left(\sum_{i=1}^{N}\frac1{u_{i}^{(2),*}}\right)\equiv D_{b}^{*},\\
 Z_{b}^{\mathrm H,*}
 &=e^{-E_{b}^{*}/\Tstar}\frac{(2\pi\Tstar)^{(N-1)/2}}{\sqrt{D_{b}^{*}}}.
 \end{aligned}
 \label{si:eq:constrained-gaussian}
\end{equation}
The final determinant expression treats every bond identically: it does not
depend on which bond was eliminated. The subscript \(b\) identifies the
basin minimum at which the curvatures are evaluated. The exponential weights
its minimum energy, while the determinant factor accounts for fluctuations
around it. The Gaussian integral is evaluated exactly within the harmonic
approximation; the original basin integral is still approximated.
The relevant positive Hessian is \(H^{*}\), restricted to allowed fluctuations.
The unconstrained bond-curvature matrix is
\begin{equation}
 K^*=\operatorname{diag}(u_1^{(2),*},\ldots,u_N^{(2),*}).
 \label{si:eq:bond-curvature-matrix}
\end{equation}
It need not be positive. In particular, the long-bond curvature is negative
at the stable fractured minimum, as shown in Sec.~\ref{si:stationary-morse}.

For the intact state, all curvatures equal
\(u_{\mathrm i}^{(2),*}=u^{(2),*}(\lambda)\). The curvature product in
Eq.~\eqref{si:eq:constrained-gaussian} is therefore
\(\bigl(u_{\mathrm i}^{(2),*}\bigr)^N\), while its reciprocal sum is
\(N/u_{\mathrm i}^{(2),*}\). Multiplying gives
\begin{equation}
 D_{\mathrm i}^*
 =\bigl(u_{\mathrm i}^{(2),*}\bigr)^N\frac{N}{u_{\mathrm i}^{(2),*}}
 =N\bigl(u_{\mathrm i}^{(2),*}\bigr)^{N-1}.
 \label{si:eq:intact-harmonic-determinant}
\end{equation}
For a fractured minimum, the product contains one long-bond curvature and
\(N-1\) short-bond curvatures. The reciprocal sum contains the same
multiplicities, so Eq.~\eqref{si:eq:constrained-gaussian} gives
\[
 \begin{aligned}
 D_{\mathrm f}^*
 &=u_{\mathrm{long}}^{(2),*}
 \bigl(u_{\mathrm{short}}^{(2),*}\bigr)^{N-1}
 \left[\frac{1}{u_{\mathrm{long}}^{(2),*}}
       +\frac{N-1}{u_{\mathrm{short}}^{(2),*}}\right]\\
 &=\bigl(u_{\mathrm{short}}^{(2),*}\bigr)^{N-1}
 +(N-1)u_{\mathrm{long}}^{(2),*}
 \bigl(u_{\mathrm{short}}^{(2),*}\bigr)^{N-2}.
 \end{aligned}
\]
Factoring \(\bigl(u_{\mathrm{short}}^{(2),*}\bigr)^{N-2}\) yields the
first form in Eq.~\eqref{si:eq:fractured-harmonic-determinant} below.
To obtain its second form, use the Morse curvatures in
Eq.~\eqref{si:eq:stationary-bond-curvatures}:
\[
 \frac{u_{\mathrm{long}}^{(2),*}}{u_{\mathrm{short}}^{(2),*}}
 =\frac{-2a^2y(1-2y)}{2a^2(1-y)(1-2y)}
 =-\frac{y}{1-y}.
\]
Thus the remaining curvature factor simplifies to
\[
 \begin{aligned}
 u_{\mathrm{short}}^{(2),*}+(N-1)u_{\mathrm{long}}^{(2),*}
 &=u_{\mathrm{short}}^{(2),*}
 \left[1-\frac{(N-1)y}{1-y}\right]\\
 &=u_{\mathrm{short}}^{(2),*}\frac{1-y-(N-1)y}{1-y}
 =u_{\mathrm{short}}^{(2),*}\frac{1-Ny}{1-y}.
 \end{aligned}
\]
Substitution gives the two equivalent expressions
\begin{equation}
 D_{\mathrm f}^{*}
 =\bigl(u_{\mathrm{short}}^{(2),*}\bigr)^{N-2}[u_{\mathrm{short}}^{(2),*}+(N-1)u_{\mathrm{long}}^{(2),*}]
 =\bigl(u_{\mathrm{short}}^{(2),*}\bigr)^{N-1}\frac{1-Ny}{1-y}.
 \label{si:eq:fractured-harmonic-determinant}
\end{equation}
Here \(y\), \(u_{\mathrm{long}}^{(2),*}\) and \(u_{\mathrm{short}}^{(2),*}\) are those of
Sec.~\ref{si:stationary-morse}; \(D_{\mathrm f}^{*}>0\) on the stable fractured branch.
There are \(N\) equivalent choices of the long bond, each with the same
single-site weight. Using the determinants in
Eqs.~\eqref{si:eq:intact-harmonic-determinant} and
\eqref{si:eq:fractured-harmonic-determinant}, the basin-weight formula
in Eq.~\eqref{si:eq:constrained-gaussian} gives
\begin{equation}
 Z_{\mathrm i}^{\mathrm H,*}
 =\frac{(2\pi\Tstar)^{(N-1)/2}}{\sqrt{D_{\mathrm i}^{*}}}
 e^{-E_{\mathrm i}^{*}/\Tstar},\qquad
 N Z_{N,\mathrm{site}}^{\mathrm f,\mathrm H,*}
 =N\frac{(2\pi\Tstar)^{(N-1)/2}}{\sqrt{D_{\mathrm f}^{*}}}
 e^{-E_{\mathrm f}^{*}/\Tstar}.
 \label{si:eq:harmonic-basin-weights}
\end{equation}
Here the first weight is intact and the second sums all one-fracture sites.
The common temperature factor in Eq.~\eqref{si:eq:harmonic-basin-weights}
cancels in their ratio:
\begin{equation}
 \frac{N Z_{N,\mathrm{site}}^{\mathrm f,\mathrm H,*}}{Z_{\mathrm i}^{\mathrm H,*}}
 =N\sqrt{\frac{D_{\mathrm i}^{*}}{D_{\mathrm f}^{*}}}
   e^{-\Delta E_{N}^{*}/\Tstar},\qquad
 \Delta E_{N}^{*}=E_{\mathrm f}^{*}-E_{\mathrm i}^{*}.
 \label{si:eq:harmonic-basin-ratio}
\end{equation}
At the midpoint, the ratio in Eq.~\eqref{si:eq:harmonic-basin-ratio} equals one.
Taking its logarithm and rearranging gives the harmonic midpoint condition,
\begin{equation}
 \begin{aligned}
 0&=\ln N+\frac12\ln\frac{D_{\mathrm i}^*}{D_{\mathrm f}^*}
       -\frac{\Delta E_N^*}{\Tstar},\\
 \frac{\Delta E_{N}^{*}}{\Tstar}
 &=\ln N+\frac12\ln\frac{D_{\mathrm i}^{*}}{D_{\mathrm f}^{*}}.
 \end{aligned}
 \label{si:eq:harmonic-midpoint}
\end{equation}
Thus both the number of fracture sites and the volume of thermal fluctuations
shift coexistence away from equal minimum energies.

\subsection{First-anharmonic midpoint correction}
\label{si:anharmonic-midpoint}

The harmonic approximation of Sec.~\ref{si:midpoint-approximations}
neglects the asymmetry and higher-order shape
of the local energy landscape. The first anharmonic, or Laplace, correction
retains the leading change in basin weight from the cubic and quartic
terms of Eq.~\eqref{si:eq:midpoint-energy-expansion}. To evaluate these
terms, make the change of variables
\begin{equation}
 \zeta_j^*=\frac{\delta l_j^*}{\sqrt{\Tstar}},\qquad
 \delta l_j^*=\sqrt{\Tstar}\,\zeta_j^*,\qquad
 \sum_{j=1}^N\zeta_j^*=0.
 \label{si:eq:anharmonic-displacement-scaling}
\end{equation}
Substitution into the quadratic energy term of
Eq.~\eqref{si:eq:midpoint-energy-expansion} removes the explicit temperature
factor from the harmonic Boltzmann exponent:
\[
 \frac{E^{(2),*}}{\Tstar}
 =\frac1{2\Tstar}\sum_j u_j^{(2),*}(\delta l_j^*)^2
 =\frac1{2\Tstar}\sum_j u_j^{(2),*}
       (\sqrt{\Tstar}\,\zeta_j^*)^2
 =\frac12\sum_j u_j^{(2),*}(\zeta_j^*)^2.
\]
At a specified minimum, the rescaled fluctuations still depend on its
curvatures and the fixed-length constraint.
The variables \(\zeta_j^*\) have zero mean under the harmonic distribution,
but are correlated by the length constraint. Here
\(\langle\cdot\rangle_{\mathrm H}\) denotes an average under that normalized
constrained Gaussian distribution, with all derivatives evaluated at the
chosen minimum \(b\).

Recall from Eq.~\eqref{si:eq:independent-bond-displacements} that
\(\epsilon_i^*=\delta l_i^*\), for \(i=1,\ldots,N-1\), and
\(\delta l_N^*=-\sum_{i=1}^{N-1}\epsilon_i^*\).
Collecting these relations into vectors gives
\begin{equation}
 B=\begin{pmatrix}I_{N-1}\\-\bm1_{N-1}^{\mathsf T}\end{pmatrix},
 \qquad
 \delta\bm l^*=B\bm\epsilon^*.
 \label{si:eq:anharmonic-bond-map}
\end{equation}
The identity block copies the first \(N-1\) displacements; the bottom row
produces their negative sum, the displacement of the eliminated bond.
Dividing Eq.~\eqref{si:eq:anharmonic-bond-map} by \(\sqrt{\Tstar}\) and
using Eq.~\eqref{si:eq:anharmonic-displacement-scaling} gives
\[
 \bm\zeta^*=\frac{\delta\bm l^*}{\sqrt{\Tstar}}
 =B\,\frac{\bm\epsilon^*}{\sqrt{\Tstar}}.
\]

The separated normal modes of
Eq.~\eqref{si:eq:constrained-normal-modes} have zero means and variances
\(\Tstar/\rho_\alpha^*\), from the Gaussian factors in
Eq.~\eqref{si:eq:one-dimensional-gaussian}. Dividing each coordinate by
\(\sqrt{\Tstar}\) divides its variance by \(\Tstar\), giving
\[
 \left\langle
 \frac{q_\alpha^*}{\sqrt{\Tstar}}\,
 \frac{q_\beta^*}{\sqrt{\Tstar}}
 \right\rangle_{\mathrm H}
 =\frac{\delta_{\alpha\beta}}{\rho_\alpha^*}.
\]
Here \(\delta_{\alpha\beta}\) is the Kronecker delta; different normal modes
are uncorrelated. Transforming back with
\(\bm\epsilon^*=O\bm q^*\), as in
Eq.~\eqref{si:eq:constrained-normal-modes}, gives the covariance
\[
 \left\langle
 \frac{\bm\epsilon^*}{\sqrt{\Tstar}}\,
 \frac{(\bm\epsilon^*)^{\mathsf T}}{\sqrt{\Tstar}}
 \right\rangle_{\mathrm H}
 =O\left\langle
 \frac{\bm q^*}{\sqrt{\Tstar}}\,
 \frac{(\bm q^*)^{\mathsf T}}{\sqrt{\Tstar}}
 \right\rangle_{\mathrm H}O^{\mathsf T}
 =O\,\operatorname{diag}\!\left(\frac1{\rho_1^*},\ldots,
                              \frac1{\rho_{N-1}^*}\right)O^{\mathsf T}.
\]
Equation~\eqref{si:eq:constrained-normal-modes} also implies
\[
 H^*=O\,\operatorname{diag}(\rho_1^*,\ldots,\rho_{N-1}^*)O^{\mathsf T},
 \qquad
 (H^*)^{-1}
 =O\,\operatorname{diag}\!\left(\frac1{\rho_1^*},\ldots,
                              \frac1{\rho_{N-1}^*}\right)O^{\mathsf T},
\]
since \(O^{-1}=O^{\mathsf T}\). Thus the preceding covariance equals
\((H^*)^{-1}\). Define
\(C_{ij}^*=\langle\zeta_i^*\zeta_j^*\rangle_{\mathrm H}\) as the covariance
of the rescaled displacements of bonds \(i\) and \(j\). Using the rescaling
and reconstruction in Eqs.~\eqref{si:eq:anharmonic-displacement-scaling}
and \eqref{si:eq:anharmonic-bond-map}, and taking the fixed matrix \(B\)
outside the average, gives
\[
 \begin{aligned}
 C^*&=\langle\bm\zeta^*(\bm\zeta^*)^{\mathsf T}\rangle_{\mathrm H}
 =\left\langle
 B\frac{\bm\epsilon^*}{\sqrt{\Tstar}}\,
 \frac{(\bm\epsilon^*)^{\mathsf T}}{\sqrt{\Tstar}}B^{\mathsf T}
 \right\rangle_{\mathrm H}\\
 &=B\left\langle\frac{\bm\epsilon^*(\bm\epsilon^*)^{\mathsf T}}
 {\Tstar}\right\rangle_{\mathrm H}B^{\mathsf T}
 =B(H^*)^{-1}B^{\mathsf T}.
 \end{aligned}
\]

To obtain its entries explicitly, invert the diagonal-plus-rank-one
Hessian in Eq.~\eqref{si:eq:constrained-hessian}, using the diagonal matrix
and its inverse in Eqs.~\eqref{si:eq:diagonal-hessian-part} and
\eqref{si:eq:inverse-diagonal-hessian-part}:
\[
 (H^*)^{-1}
 =(A^*)^{-1}
 -\frac{u_N^{(2),*}(A^*)^{-1}\bm1_{N-1}\bm1_{N-1}^{\mathsf T}(A^*)^{-1}}
 {1+u_N^{(2),*}\bm1_{N-1}^{\mathsf T}(A^*)^{-1}\bm1_{N-1}}.
\]
Multiplying this expression by \(H^*\) verifies the inverse. Dividing its
second numerator and denominator by \(u_N^{(2),*}\), and using the
diagonal entries from Eq.~\eqref{si:eq:inverse-diagonal-hessian-part}, gives
\[
 [(H^*)^{-1}]_{ij}
 =\frac{\delta_{ij}}{u_i^{(2),*}}
 -\frac{1/[u_i^{(2),*}u_j^{(2),*}]}
        {\sum_{k=1}^N1/u_k^{(2),*}},
 \qquad i,j=1,\ldots,N-1.
\]
In particular, \(C_{iN}^*=\langle\zeta_i^*\zeta_N^*\rangle_{\mathrm H}\)
is the covariance of bond \(i\) with the eliminated bond, and
\(C_{NN}^*=\langle(\zeta_N^*)^2\rangle_{\mathrm H}\) is that bond's variance.
Combining the rescaling in Eq.~\eqref{si:eq:anharmonic-displacement-scaling}
with the reconstruction in Eq.~\eqref{si:eq:anharmonic-bond-map} gives
\(\zeta_N^*=-\sum_{j=1}^{N-1}\zeta_j^*\).
For \(i,j<N\), the identity block of \(B\) leaves the inverse-Hessian entries
unchanged, so \(C_{ij}^*=[(H^*)^{-1}]_{ij}\).
For a covariance involving the eliminated bond, its negative sum gives
\[
 \begin{aligned}
 C_{iN}^*
 &=-\sum_{j=1}^{N-1}[(H^*)^{-1}]_{ij}\\
 &=-\frac1{u_i^{(2),*}}
 +\frac{(1/u_i^{(2),*})\sum_{j=1}^{N-1}1/u_j^{(2),*}}
 {\sum_{k=1}^N1/u_k^{(2),*}}\\
 &=-\frac{1/[u_i^{(2),*}u_N^{(2),*}]}
 {\sum_{k=1}^N1/u_k^{(2),*}},
 \qquad i=1,\ldots,N-1.
 \end{aligned}
\]
The last step uses
\(\sum_{k=1}^N1/u_k^{(2),*}-\sum_{j=1}^{N-1}1/u_j^{(2),*}
=1/u_N^{(2),*}\).
For the eliminated bond's variance, the two negative sums multiply to
give a positive double sum. The diagonal term sums once over the first
\(N-1\) bonds, while the product term factors into a squared sum:
\[
 \begin{aligned}
 C_{NN}^*
 &=\sum_{i,j=1}^{N-1}[(H^*)^{-1}]_{ij}\\
 &=\sum_{i=1}^{N-1}\frac1{u_i^{(2),*}}
 -\frac{\left(\sum_{i=1}^{N-1}1/u_i^{(2),*}\right)^2}
 {\sum_{k=1}^N1/u_k^{(2),*}}\\
 &=\frac{(1/u_N^{(2),*})\sum_{i=1}^{N-1}1/u_i^{(2),*}}
 {\sum_{k=1}^N1/u_k^{(2),*}}
 =\frac1{u_N^{(2),*}}
 -\frac{1/(u_N^{(2),*})^2}{\sum_{k=1}^N1/u_k^{(2),*}}.
 \end{aligned}
\]
The final expressions for \(C_{iN}^*\) and \(C_{NN}^*\) have the same form
as the entries with \(i,j<N\). To write this result in matrix form, use
\(K^*\) from Eq.~\eqref{si:eq:bond-curvature-matrix}. Its nonzero diagonal
entries give
\[
 \begin{aligned}
 [(K^*)^{-1}]_{ij}&=\frac{\delta_{ij}}{u_i^{(2),*}},\\
 [(K^*)^{-1}\bm1]_i&=\frac1{u_i^{(2),*}},\\
 \bm1^{\mathsf T}(K^*)^{-1}\bm1&=\sum_{k=1}^N\frac1{u_k^{(2),*}},
 \end{aligned}
\]
where \(\bm1\) has \(N\) entries. The outer product in the numerator
below therefore has entries \(1/[u_i^{(2),*}u_j^{(2),*}]\).
Combining these results gives the covariance for every pair of bonds:
\begin{equation}
 \begin{aligned}
 C^*&=B(H^*)^{-1}B^{\mathsf T}
 =(K^*)^{-1}-
 \frac{(K^*)^{-1}\bm1\bm1^{\mathsf T}(K^*)^{-1}}
      {\bm1^{\mathsf T}(K^*)^{-1}\bm1},\\
 C_{ij}^*&\equiv\langle\zeta_i^*\zeta_j^*\rangle_{\mathrm H}
 =\frac{\delta_{ij}}{u_i^{(2),*}}
 -\frac{1/[u_i^{(2),*}u_j^{(2),*}]}
        {\sum_{k=1}^N1/u_k^{(2),*}},\qquad i,j=1,\ldots,N.
 \end{aligned}
 \label{si:eq:laplace-covariance}
\end{equation}
Combining the rescaling in Eq.~\eqref{si:eq:anharmonic-displacement-scaling}
with the covariance definition in Eq.~\eqref{si:eq:laplace-covariance} gives
\(\langle\delta l_i^*\delta l_j^*\rangle_{\mathrm H}=\Tstar C_{ij}^*\).
The covariance respects fixed length because its rows sum to zero:
\[
 \sum_{j=1}^N C_{ij}^*
 =\frac1{u_i^{(2),*}}
 -\frac{(1/u_i^{(2),*})\sum_{j=1}^N1/u_j^{(2),*}}
        {\sum_{k=1}^N1/u_k^{(2),*}}=0.
\]
It is derived from the stable constrained Hessian \(H^*\), even when the
unconstrained matrix \(K^*\) has a negative entry.

Next collect the cubic and quartic polynomials in the rescaled variables:
\begin{equation}
 A_3^*=\frac16\sum_{j=1}^N u_j^{(3),*}(\zeta_j^*)^3,\qquad
 A_4^*=\frac1{24}\sum_{j=1}^N u_j^{(4),*}(\zeta_j^*)^4.
 \label{si:eq:anharmonic-polynomials}
\end{equation}
Substituting Eq.~\eqref{si:eq:anharmonic-displacement-scaling} into the
energy expansion, Eq.~\eqref{si:eq:midpoint-energy-expansion}, and using
\(K^*\) from Eq.~\eqref{si:eq:bond-curvature-matrix} gives
\begin{equation}
 \begin{aligned}
 \frac{E^*-E_b^*}{\Tstar}
 &=\frac12\sum_j u_j^{(2),*}(\zeta_j^*)^2
 +\frac{\sqrt{\Tstar}}6\sum_j u_j^{(3),*}(\zeta_j^*)^3
 +\frac{\Tstar}{24}\sum_j u_j^{(4),*}(\zeta_j^*)^4+\cdots\\
 &=\frac12(\bm\zeta^*)^{\mathsf T}K^*\bm\zeta^*
   +\sqrt{\Tstar}A_3^*+\Tstar A_4^*+\cdots.
 \end{aligned}
 \label{si:eq:rescaled-anharmonic-energy}
\end{equation}
Thus the cubic and quartic contributions to the Boltzmann exponent are of
orders \(\sqrt{\Tstar}\) and \(\Tstar\), respectively. Factoring out the
harmonic Boltzmann weight leaves an exponential that can be expanded using
\(e^{-z}=1-z+z^2/2+\cdots\):
\begin{equation}
 \begin{aligned}
 e^{-\sqrt{\Tstar}A_3^*-\Tstar A_4^*+\cdots}
 &=1-(\sqrt{\Tstar}A_3^*+\Tstar A_4^*)
   +\tfrac12(\sqrt{\Tstar}A_3^*+\Tstar A_4^*)^2+\cdots\\
 &=1-\sqrt{\Tstar}A_3^*
   +\Tstar\left[\tfrac12(A_3^*)^2-A_4^*\right]
   +O((\Tstar)^{3/2}).
 \end{aligned}
 \label{si:eq:anharmonic-boltzmann-expansion}
\end{equation}
Dividing the local basin integral by its harmonic weight,
Eq.~\eqref{si:eq:constrained-gaussian}, converts this expansion into a
normalized Gaussian average. The harmonic distribution is unchanged when
all \(\zeta_j^*\) change sign, whereas \(A_3^*\) changes sign.
Hence \(\langle A_3^*\rangle_{\mathrm H}=0\), but
\(\langle(A_3^*)^2\rangle_{\mathrm H}\) survives. The terms of order
\((\Tstar)^{3/2}\) are also odd and average to zero. The first correction to
the basin weight is therefore
\begin{equation}
 \begin{aligned}
 \frac{Z_b^*}{Z_b^{\mathrm H,*}}
 &=1+\Tstar\mathcal C_{1,b}^*+O((\Tstar)^2),\\
 \mathcal C_{1,b}^*
 &=-\langle A_4^*\rangle_{\mathrm H}
   +\tfrac12\langle(A_3^*)^2\rangle_{\mathrm H}.
 \end{aligned}
 \label{si:eq:laplace-weight-correction}
\end{equation}
The cubic term thus contributes to the leading correction through its
square, at the same order as the quartic term.

These averages are calculated from the covariance in
Eq.~\eqref{si:eq:laplace-covariance}. For a zero-mean Gaussian distribution,
an even moment is the sum over all pairings of its factors, with each pair
replaced by its covariance. Four copies of \(\zeta_j^*\) have three
pairings. For three copies each of \(\zeta_i^*\) and \(\zeta_j^*\), a
pairing has either one or three cross pairs. With one cross pair, its two
factors can be chosen in \(3\times3=9\) ways; the remaining factors pair
within each index. With three cross pairs, the matching has \(3!=6\)
choices. Hence
\begin{equation}
 \begin{aligned}
 \langle(\zeta_j^*)^4\rangle_{\mathrm H}
 &=3(C_{jj}^*)^2,\\
 \langle(\zeta_i^*)^3(\zeta_j^*)^3\rangle_{\mathrm H}
 &=9C_{ii}^*C_{jj}^*C_{ij}^*+6(C_{ij}^*)^3.
 \end{aligned}
 \label{si:eq:laplace-gaussian-moments}
\end{equation}
Expanding the sums in Eq.~\eqref{si:eq:anharmonic-polynomials} before
averaging makes their numerical factors explicit:
\begin{equation}
 \begin{aligned}
 \langle A_4^*\rangle_{\mathrm H}
 &=\frac1{24}\sum_j u_j^{(4),*}\langle(\zeta_j^*)^4\rangle_{\mathrm H}
 =\frac18\sum_j u_j^{(4),*}(C_{jj}^*)^2,\\
 \tfrac12\langle(A_3^*)^2\rangle_{\mathrm H}
 &=\frac12\left(\frac16\right)^2
   \sum_{ij}u_i^{(3),*}u_j^{(3),*}
   \langle(\zeta_i^*)^3(\zeta_j^*)^3\rangle_{\mathrm H}\\
 &=\frac1{72}\sum_{ij}u_i^{(3),*}u_j^{(3),*}
   \left[9C_{ii}^*C_{jj}^*C_{ij}^*+6(C_{ij}^*)^3\right].
 \end{aligned}
 \label{si:eq:laplace-polynomial-averages}
\end{equation}
Substitution into Eq.~\eqref{si:eq:laplace-weight-correction} gives the
coefficient used in the calculation:
\begin{equation}
 \mathcal C_{1}^{*}=-\frac18\sum_ju_{j}^{(4),*}\bigl(C_{jj}^{*}\bigr)^{2}
 +\frac1{72}\sum_{ij}u_{i}^{(3),*}u_{j}^{(3),*}
 \left(9C_{ii}^{*}C_{jj}^{*}C_{ij}^{*}+6\bigl(C_{ij}^{*}\bigr)^{3}\right).
 \label{si:eq:laplace-c1}
\end{equation}
All sums run over the \(N\) bonds, and the basin index is implicit in the
derivatives and covariance. For the Morse potential, define
\(y_j=e^{-a(l_j^{(b),*}-1)}\), so \(dy_j/dl_j^*=-a y_j\).
Successive differentiation of \(u_j^{(2),*}=2a^2y_j(2y_j-1)\) gives
\begin{equation}
 \begin{aligned}
 u_j^{(3),*}
 &=-a y_j\frac{d}{dy_j}\left[2a^2y_j(2y_j-1)\right]
 =2a^3y_j(1-4y_j),\\
 u_j^{(4),*}
 &=-a y_j\frac{d}{dy_j}\left[2a^3y_j(1-4y_j)\right]
 =2a^4y_j(8y_j-1).
 \end{aligned}
 \label{si:eq:laplace-morse-derivatives}
\end{equation}
Equations~\eqref{si:eq:laplace-covariance},
\eqref{si:eq:laplace-c1} and \eqref{si:eq:laplace-morse-derivatives} are
evaluated separately for the intact configuration
\((y_0,\ldots,y_0)\) and the fractured configuration
\((y,1-y,\ldots,1-y)\), at the same extension
(Sec.~\ref{si:stationary-morse}). They give
\(\mathcal C_{1,\mathrm i}^*\) and \(\mathcal C_{1,\mathrm f}^*\).

Finally, convert the weight correction in
Eq.~\eqref{si:eq:laplace-weight-correction} into a correction to the
logarithm. Using \(\ln(1+z)=z+O(z^2)\),
\begin{equation}
 \begin{aligned}
 \ln Z_b^*-\ln Z_b^{\mathrm H,*}
 &=\ln\!\left[1+\Tstar\mathcal C_{1,b}^*+O((\Tstar)^2)\right]\\
 &=\Tstar\mathcal C_{1,b}^*+O((\Tstar)^2).
 \end{aligned}
 \label{si:eq:anharmonic-log-weight}
\end{equation}
Subtracting the intact log weight from the fractured one and including
the \(N\) equivalent fracture sites gives
\begin{equation}
 \begin{aligned}
 \ln\frac{N Z_{N,\mathrm{site}}^{\mathrm f,*}}{Z_{\mathrm i}^*}
 &=\ln\frac{N Z_{N,\mathrm{site}}^{\mathrm f,\mathrm H,*}}
              {Z_{\mathrm i}^{\mathrm H,*}}
   +\Tstar(\mathcal C_{1,\mathrm f}^*-\mathcal C_{1,\mathrm i}^*)
   +O((\Tstar)^2)\\
 &=\ln N+\frac12\ln\frac{D_{\mathrm i}^*}{D_{\mathrm f}^*}
   -\frac{\Delta E_N^*}{\Tstar}
   +\Tstar(\mathcal C_{1,\mathrm f}^*-\mathcal C_{1,\mathrm i}^*)
   +O((\Tstar)^2),
 \end{aligned}
 \label{si:eq:anharmonic-basin-ratio}
\end{equation}
where the second line uses the harmonic ratio,
Eq.~\eqref{si:eq:harmonic-basin-ratio}. At coexistence this logarithm is
zero, so rearrangement yields the first-anharmonic midpoint equation:
\begin{equation}
 \frac{\Delta E_{N}^{*}}{\Tstar}
 =\ln N+\frac12\ln\frac{D_{\mathrm i}^{*}}{D_{\mathrm f}^{*}}
 +\Tstar(\mathcal C_{1,\mathrm f}^{*}-\mathcal C_{1,\mathrm i}^{*})
 +O({\Tstar}^2).
 \label{si:eq:anharmonic-midpoint}
\end{equation}
For each trial \(y\), both configurations are evaluated at the same
\(\lambda(y)\). The midpoint is obtained by solving
Eq.~\eqref{si:eq:harmonic-midpoint}, or
Eq.~\eqref{si:eq:anharmonic-midpoint} truncated after the \(\Tstar\) term,
in \(\ln y\) on the stable branch.
If several roots are found, the implementation selects the one nearest the
zero-temperature root in \(\ln y\). The selected first-anharmonic
\(N=4\) root belongs to a different branch from the displayed sequence
and is omitted from Figure~5, as stated in its caption. These are
low-temperature expansions at fixed \(N\); their coefficients depend on
chain length, so their accuracy is not uniform as \(N\to\infty\) at
fixed temperature.

These calculations distinguish finite-temperature midpoint behavior from the
zero-temperature energetic limit. Agreement over the examined chain lengths
does not establish the asymptotic behavior at fixed temperature.

%% file: si_sections/site_barriers.tex
\section{Site free-energy barriers}
\label{si:site-barriers}

The site PMF describes a tagged bond with every other bond restricted below
the fixed full-chain divider \(d_{N}^{*}\) (Sec.~\ref{si:sec:profile-pmfs}).
We first derive the equal-sharing adiabatic energy, then include reservoir
fluctuations through harmonic and first-anharmonic approximations, and finally
compare their barriers and curvatures with numerical evaluations.
The profiles below use energy in units of \(D_e\), and derivatives are with
respect to the reduced bond length \(l^{*}\). The barrier errors reported
below retain their physical thermal-energy units. At fixed \(N\), extension
and temperature, define
\begin{equation}
 M=N-1,\qquad z^{*}=\frac{N\lambda-l^{*}}{M},\qquad
 \frac{dz^{*}}{dl^{*}}=-\frac1M.
 \label{si:eq:site-reservoir-mean}
\end{equation}
Here \(z^*\) is the mean length of the remaining bonds: length gained by
the tagged bond is lost by its reservoir.

\subsection{Adiabatic energy barrier and curvatures}
\label{si:energetic-site-barrier}

The equal-sharing branch in Sec.~\ref{si:sec:adiabatic-profiles} assigns
length \(z^*\) to each of the \(M\) reservoir bonds. Its total potential
energy is therefore the tagged-bond energy plus \(M\) identical reservoir
contributions, \(u^*(l^*)+Mu^*(z^*)\). This prescribed configuration supplies
an energetic reference without integrating over thermal fluctuations.
To differentiate along the path, use the length constraint in
Eq.~\eqref{si:eq:site-reservoir-mean}:
\[
 \frac{d}{dl^*}[Mu^*(z^*)]
 =M u^{(1),*}(z^*)\left(-\frac1M\right)
 =-u^{(1),*}(z^*).
\]
The tagged contribution differentiates to \(u^{(1),*}(l^*)\).
Differentiating both contributions once more gives
\[
 \frac{d}{dl^*}[u^{(1),*}(l^*)]=u^{(2),*}(l^*),\qquad
 \frac{d}{dl^*}[-u^{(1),*}(z^*)]
 =-u^{(2),*}(z^*)\left(-\frac1M\right)
 =\frac{u^{(2),*}(z^*)}{M}.
\]
Thus the energy and its two derivatives are
\begin{equation}
 \begin{aligned}
 E_{\mathrm{ad,abs}}^*(l^*)&=u^*(l^*)+Mu^*(z^*),\\
 E_{\mathrm{ad,abs}}^{(1),*}(l^*)&=u^{(1),*}(l^*)-u^{(1),*}(z^*),\\
 E_{\mathrm{ad,abs}}^{(2),*}(l^*)&=u^{(2),*}(l^*)+\frac{u^{(2),*}(z^*)}{M}.
 \end{aligned}
 \label{si:eq:energetic-site-profile}
\end{equation}
Stationarity requires equal tensions,
\(u^{(1),*}(l^*)=u^{(1),*}(z^*)\), as in
Sec.~\ref{si:stationary-morse}. The uniform intact minimum has
\(l_{\mathrm i}^{\mathrm{ad},*}=z^*=\lambda\); the saddle coordinate is
\(l_{\mathrm s}^{\mathrm{ad},*}=d_N^*\).
The second derivative in Eq.~\eqref{si:eq:energetic-site-profile} is
positive at the intact minimum and negative at the saddle, identifying
their stability along this length-preserving path.

The relative energy used in Sec.~\ref{si:sec:adiabatic-profiles} is
\(E_{\mathrm{ad}}^{(N-1),*}
=E_{\mathrm{ad,abs}}^*-Nu^*(\lambda)\).
At fixed \(\lambda\), the subtracted term is independent of \(l^*\);
it changes neither derivatives nor energy differences. The energetic
barrier is consequently
\begin{equation}
 B_N^{\mathrm E}(\lambda)
 =\frac{E_{\mathrm{ad,abs}}^*(l_{\mathrm s}^{\mathrm{ad},*})
       -E_{\mathrm{ad,abs}}^*(l_{\mathrm i}^{\mathrm{ad},*})}{\Tstar}.
 \label{si:eq:energetic-site-barrier}
\end{equation}
This reference is evaluated at the same finite-temperature midpoint
extension and temperature as the PMF comparisons below.

\subsection{Harmonic and anharmonic site-barrier approximations}
\label{si:site-barrier-approximations}

The site PMF includes the reservoir's thermal fluctuations. Its exact
profile follows from the restricted partition density derived in
Sec.~\ref{si:sec:profile-pmfs}:
\begin{equation}
 G_{N}^{\mathrm{site},*}(l^{*})
 =u^{*}(l^{*})-\Tstar\ln Z_{M}^{\mathrm i,*}(Mz^{*};d_{N}^{*})+\text{constant}.
 \label{si:eq:site-reservoir-profile}
\end{equation}
The divider \(d_N^*\) is held fixed as \(l^*\) varies. Approximating the
reservoir integral in Eq.~\eqref{si:eq:site-reservoir-profile} gives the
harmonic and first-anharmonic profiles below.

To approximate the reservoir integral, write its bond lengths as
\(z^{*}+\epsilon_{j}^{*}\), with \(\sum_{j=1}^M\epsilon_{j}^{*}=0\). Expansion about
equal sharing gives
\begin{equation}
 \begin{aligned}
 \sum_{j=1}^M u^{*}(z^{*}+\epsilon_{j}^{*})
 &=Mu^{*}(z^{*})+u^{(1),*}(z^{*})\underbrace{\sum_j\epsilon_j^*}_{0}\\
 &\quad+\frac{u^{(2),*}(z^{*})}2
 \sum_{j=1}^M\bigl(\epsilon_{j}^{*}\bigr)^{2}+\cdots.
 \end{aligned}
 \label{si:eq:site-reservoir-expansion}
\end{equation}
The length constraint cancels the linear term. Keeping the quadratic term
in Eq.~\eqref{si:eq:site-reservoir-expansion} gives the constrained
Gaussian problem of Sec.~\ref{si:midpoint-approximations}.
Equation~\eqref{si:eq:constrained-gaussian} gives the general harmonic
weight under a fixed-length constraint. Here it applies to the \(M=N-1\)
reservoir bonds; the fixed tagged bond is excluded from their fluctuations.
At equal sharing, every reservoir curvature is \(u^{(2),*}(z^*)\), so its
determinant becomes
\begin{equation}
 \begin{aligned}
 D_{\mathrm{res}}^*
 &=\left(\prod_{j=1}^{M}u^{(2),*}(z^*)\right)
   \left(\sum_{j=1}^{M}\frac1{u^{(2),*}(z^*)}\right)\\
 &=[u^{(2),*}(z^*)]^M\frac{M}{u^{(2),*}(z^*)}\\
 &=M[u^{(2),*}(z^*)]^{M-1}.
 \end{aligned}
 \label{si:eq:site-reservoir-determinant}
\end{equation}
The product contains \(M\) identical curvatures, and the sum contains
\(M\) identical reciprocal curvatures. This determinant supplies the
Gaussian fluctuation factor \((2\pi\Tstar)^{(M-1)/2}/\sqrt{D_{\mathrm{res}}^*}\)
for the \(M-1\) independent reservoir displacements.
Multiplying this factor by the Boltzmann weight of the equal-sharing
configuration gives the harmonic reservoir partition density needed in
Eq.~\eqref{si:eq:site-reservoir-profile}:
\begin{equation}
 \begin{aligned}
 Z_{M,\mathrm H}^{\mathrm i,*}(Mz^{*};d_{N}^{*})
 &\simeq e^{-Mu^{*}(z^{*})/\Tstar}
 \frac{(2\pi\Tstar)^{(M-1)/2}}{\sqrt{D_{\mathrm{res}}^*}}\\
 &=\frac{e^{-Mu^{*}(z^{*})/\Tstar}}{\sqrt M}
 \left(\frac{2\pi\Tstar}{u^{(2),*}(z^{*})}\right)^{(M-1)/2}.
 \end{aligned}
 \label{si:eq:harmonic-reservoir-weight}
\end{equation}
This local approximation requires \(u^{(2),*}(z^{*})>0\) and fluctuations
concentrated inside the intact reservoir, allowing the local integration
domain to be extended to the full constrained subspace. Taking the
logarithm of Eq.~\eqref{si:eq:harmonic-reservoir-weight} gives
\begin{equation}
 \ln Z_{M,\mathrm H}^{\mathrm i,*}
 =-\frac{Mu^{*}(z^{*})}{\Tstar}-\frac12\ln M
  +\frac{M-1}{2}\left[\ln(2\pi\Tstar)-\ln u^{(2),*}(z^{*})\right].
 \label{si:eq:harmonic-reservoir-log-weight}
\end{equation}
Multiplying Eq.~\eqref{si:eq:harmonic-reservoir-log-weight} by
\(-\Tstar\) and substituting into Eq.~\eqref{si:eq:site-reservoir-profile}
gives
\[
 \begin{aligned}
 u^*(l^*)-\Tstar\ln Z_{M,\mathrm H}^{\mathrm i,*}
 &=u^*(l^*)+Mu^*(z^*)+\frac{\Tstar}{2}\ln M\\
 &\quad-\frac{\Tstar}{2}(M-1)\ln(2\pi\Tstar)
 +\frac{\Tstar}{2}(M-1)\ln u^{(2),*}(z^*).
 \end{aligned}
\]
Absorbing the terms independent of \(l^{*}\) into the additive constant leaves
\begin{equation}
 G_{N,\mathrm H}^{\mathrm{site},*}(l^{*})
 =u^{*}(l^{*})+Mu^{*}(z^{*})+\frac{\Tstar}{2}(M-1)\ln u^{(2),*}(z^{*}).
 \label{si:eq:harmonic-site}
\end{equation}
Thus the reservoir contributes both its equal-sharing energy and an
entropic term. The first term in Eq.~\eqref{si:eq:harmonic-site}
differentiates directly to \(u^{(1),*}(l^*)\).
For the reservoir-energy term and the logarithmic factor in the entropic
term, the chain rule with \(dz^*/dl^*=-1/M\) from
Eq.~\eqref{si:eq:site-reservoir-mean} gives
\begin{equation}
 \begin{aligned}
 \frac{d}{dl^{*}}[Mu^{*}(z^{*})]
 &=M u^{(1),*}(z^*)\frac{dz^*}{dl^*}=-u^{(1),*}(z^{*}),\\
 \frac{d}{dl^{*}}\ln u^{(2),*}(z^{*})
 &=\frac{u^{(3),*}(z^*)}{u^{(2),*}(z^*)}\frac{dz^*}{dl^*}
 =-\frac1M\frac{u^{(3),*}(z^{*})}{u^{(2),*}(z^{*})}.
 \end{aligned}
 \label{si:eq:site-harmonic-chain-rule}
\end{equation}
Multiplying the second identity in Eq.~\eqref{si:eq:site-harmonic-chain-rule}
by \(\Tstar(M-1)/2\) and adding all three contributions gives the
profile's first derivative:
\begin{equation}
 G_{N,\mathrm H}^{(1),*}(l^*)
 =u^{(1),*}(l^*)-u^{(1),*}(z^*)
 -\frac{\Tstar}{2}\frac{M-1}{M}
  \frac{u^{(3),*}(z^*)}{u^{(2),*}(z^*)}.
 \label{si:eq:site-harmonic-first-derivative}
\end{equation}
Setting Eq.~\eqref{si:eq:site-harmonic-first-derivative} to zero gives
the harmonic stationary equation,
\begin{equation}
 0=u^{(1),*}(l^{*})-u^{(1),*}(z^{*})
 -\frac{\Tstar}{2}\frac{M-1}{M}\frac{u^{(3),*}(z^{*})}{u^{(2),*}(z^{*})}.
 \label{si:eq:site-stationary}
\end{equation}

To calculate the curvature, differentiate
Eq.~\eqref{si:eq:site-harmonic-first-derivative} once more.
Its first term gives \(u^{(2),*}(l^*)\); the reservoir-energy contribution
gives
\[
 \frac{d}{dl^*}[-u^{(1),*}(z^*)]
 =-u^{(2),*}(z^*)\left(-\frac1M\right)
 =\frac{u^{(2),*}(z^*)}{M}.
\]
For the entropic contribution, differentiating the ratio by the quotient
rule gives
\[
 \frac{d}{dl^*}\frac{u^{(3),*}(z^*)}{u^{(2),*}(z^*)}
 =-\frac1M\,
 \frac{u^{(4),*}(z^*)u^{(2),*}(z^*)-[u^{(3),*}(z^*)]^2}
 {[u^{(2),*}(z^*)]^2}.
\]
Multiplying this derivative by \(-\Tstar(M-1)/(2M)\), and adding
the other two contributions, gives
\begin{equation}
 G_{N,\mathrm H}^{(2),*}(l^{*})=u^{(2),*}(l^{*})+\frac{u^{(2),*}(z^{*})}M
 +\frac{\Tstar}{2}\frac{M-1}{M^2}
 \left[\frac{u^{(4),*}(z^{*})}{u^{(2),*}(z^{*})}
 -\left(\frac{u^{(3),*}(z^{*})}{u^{(2),*}(z^{*})}\right)^2\right].
 \label{si:eq:site-curvature}
\end{equation}
Solving Eq.~\eqref{si:eq:site-stationary} identifies the intact minimum
\(l_{\mathrm i,H}^{\mathrm{site},*}\) and neighboring maximum \(l_{\mathrm s,H}^{\mathrm{site},*}\),
where Eq.~\eqref{si:eq:site-curvature} is positive and negative,
respectively. Their free-energy difference defines the harmonic barrier
in thermal-energy units:
\begin{equation}
 B_{N,\mathrm H}
 =\frac{G_{N,\mathrm H}^{\mathrm{site},*}(l_{\mathrm s,H}^{\mathrm{site},*})
       -G_{N,\mathrm H}^{\mathrm{site},*}(l_{\mathrm i,H}^{\mathrm{site},*})}{\Tstar}.
 \label{si:eq:harmonic-site-barrier}
\end{equation}
The curvatures are the second derivatives at these same points.

The cubic and quartic terms in the reservoir energy give the first
constrained Laplace correction (Sec.~\ref{si:anharmonic-midpoint}).
Apply the scaled covariance in Eq.~\eqref{si:eq:laplace-covariance}
to the \(M=N-1\) reservoir bonds, excluding the fixed tagged bond.
With \(u_j^{(2),*}=u^{(2),*}(z^*)\) for every reservoir bond, its
denominator simplifies to
\[
 \sum_{j=1}^{M}\frac1{u_j^{(2),*}}
 =\frac{M}{u^{(2),*}(z^*)}.
\]
The constraint correction in each covariance entry is therefore
\[
 \frac{1/[u^{(2),*}(z^*)]^2}{M/u^{(2),*}(z^*)}
 =\frac1{M u^{(2),*}(z^*)}.
\]
Substituting these expressions into Eq.~\eqref{si:eq:laplace-covariance}
gives
\begin{equation}
 C_{ij}^*=\frac{\delta_{ij}}{u^{(2),*}(z^*)}
 -\frac{1/[u^{(2),*}(z^*)]^2}{M/u^{(2),*}(z^*)}
 =\frac{\delta_{ij}-1/M}{u^{(2),*}(z^*)},
 \qquad i,j=1,\ldots,M.
 \label{si:eq:site-reservoir-covariance}
\end{equation}
Here \(\delta_{ij}\) is the Kronecker delta. Diagonal entries are
\((M-1)/[M u^{(2),*}(z^*)]\), and off-diagonal entries are
\(-1/[M u^{(2),*}(z^*)]\). Each row sums to zero, as required by the fixed
reservoir length. All cubic and quartic derivatives are also identical
across reservoir bonds. The sums in Eq.~\eqref{si:eq:laplace-c1} therefore
reduce to
\begin{equation}
 \begin{aligned}
 \sum_j(C_{jj}^*)^2
 &=M\left[\frac{M-1}{M u^{(2),*}(z^*)}\right]^2
 =\frac{(M-1)^2}{M[u^{(2),*}(z^*)]^2},\\
 \sum_{ij}C_{ii}^*C_{jj}^*C_{ij}^*
 &=\left[\frac{M-1}{M u^{(2),*}(z^*)}\right]^2
 \sum_{ij}C_{ij}^*=0,\\
 \sum_{ij}(C_{ij}^*)^3
 &=M\left[\frac{M-1}{M u^{(2),*}(z^*)}\right]^3
 +M(M-1)\left[\frac{-1}{M u^{(2),*}(z^*)}\right]^3\\
 &=\frac{(M-1)(M-2)}{M[u^{(2),*}(z^*)]^3}.
 \end{aligned}
 \label{si:eq:site-reservoir-contractions}
\end{equation}
The last sum counts \(M\) diagonal and \(M(M-1)\) off-diagonal entries.
Substituting Eq.~\eqref{si:eq:site-reservoir-contractions} into
Eq.~\eqref{si:eq:laplace-c1} gives
\begin{equation}
 \mathcal C_{1,M}^{*}(z^{*})
 =-\frac{(M-1)^2}{8M}\frac{u^{(4),*}(z^{*})}{[u^{(2),*}(z^{*})]^2}
  +\frac{(M-1)(M-2)}{12M}\frac{[u^{(3),*}(z^{*})]^2}{[u^{(2),*}(z^{*})]^3}.
 \label{si:eq:site-reservoir-laplace-coefficient}
\end{equation}
To convert this coefficient into a profile correction, use
Eq.~\eqref{si:eq:laplace-weight-correction} for the reservoir and take
the logarithm, as in Eq.~\eqref{si:eq:anharmonic-log-weight}:
\begin{equation}
 \begin{aligned}
 \frac{Z_M^{\mathrm i,*}}{Z_{M,\mathrm H}^{\mathrm i,*}}
 &=1+\Tstar\mathcal C_{1,M}^*(z^*)+O((\Tstar)^2),\\
 \ln Z_M^{\mathrm i,*}-\ln Z_{M,\mathrm H}^{\mathrm i,*}
 &=\ln[1+\Tstar\mathcal C_{1,M}^*(z^*)+O((\Tstar)^2)]\\
 &=\Tstar\mathcal C_{1,M}^*(z^*)+O((\Tstar)^2).
 \end{aligned}
 \label{si:eq:site-reservoir-log-correction}
\end{equation}
Inserting Eq.~\eqref{si:eq:site-reservoir-log-correction} into
Eq.~\eqref{si:eq:site-reservoir-profile} gives
\begin{equation}
 \frac{G_{N}^{\mathrm{site},*}}{\Tstar}
 =\frac{G_{N,\mathrm H}^{\mathrm{site},*}}{\Tstar}
 -\Tstar\mathcal C_{1,M}^{*}(z^{*})+O({\Tstar}^2).
 \label{si:eq:site-anharmonic}
\end{equation}
Only the reservoir is expanded: the tagged-bond energy \(u^{*}(l^{*})\)
remains unapproximated. To derive the barrier and curvature corrections,
write the harmonic profile as
\(G_{\mathrm H}^{*}=G_{N,\mathrm H}^{\mathrm{site},*}\) and its leading
perturbation from Eq.~\eqref{si:eq:site-anharmonic} as
\begin{equation}
 \delta G^{*}(l^{*})=-{\Tstar}^2\mathcal C_{1,M}^{*}(z^{*}(l^{*})).
 \label{si:eq:site-profile-perturbation}
\end{equation}
Let a harmonic extremum \(l_{\mathrm H}^{*}\) shift to
\(l_{\mathrm H}^{*}+\delta l^{*}\).
At first order in this perturbation,
\begin{equation}
   \begin{aligned}
   &G_{\mathrm H}^*(l_{\mathrm H}^*+\delta l^*)
   +\delta G^*(l_{\mathrm H}^*+\delta l^*)\\
   &\quad=G_{\mathrm H}^*(l_{\mathrm H}^*)
   +\underbrace{G_{\mathrm H}^{(1),*}(l_{\mathrm H}^*)}_{0}\delta l^*
   +\delta G^*(l_{\mathrm H}^*)\\
   &\qquad+\frac12G_{\mathrm H}^{(2),*}(l_{\mathrm H}^*)(\delta l^*)^2
   +\delta G^{(1),*}(l_{\mathrm H}^*)\delta l^*+\cdots\\
   &\quad\simeq G_{\mathrm H}^*(l_{\mathrm H}^*)
   +\delta G^*(l_{\mathrm H}^*).
   \end{aligned}
   \label{si:eq:site-extremum-energy-correction}
  \end{equation}
Terms quadratic in the displacement, or multiplying it by the perturbation,
are higher order. Thus displacement of the extrema does not contribute
to the barrier at first order. Define
\(z_{\alpha,\mathrm H}^*=(N\lambda-l_{\alpha,\mathrm H}^{\mathrm{site},*})/M\)
for \(\alpha=\mathrm i,\mathrm s\).
Applying Eq.~\eqref{si:eq:site-extremum-energy-correction} at the
maximum and minimum gives the first-order barrier:
\begin{equation}
 \begin{aligned}
 B_{N,\mathrm{AH}}
 &\simeq\frac{
 [G_{\mathrm H}^*(l_{\mathrm s,H}^{\mathrm{site},*})
  +\delta G^*(l_{\mathrm s,H}^{\mathrm{site},*})]
 -[G_{\mathrm H}^*(l_{\mathrm i,H}^{\mathrm{site},*})
  +\delta G^*(l_{\mathrm i,H}^{\mathrm{site},*})]}{\Tstar}\\
 &=B_{N,\mathrm H}
 +\frac{\delta G^*(l_{\mathrm s,H}^{\mathrm{site},*})
       -\delta G^*(l_{\mathrm i,H}^{\mathrm{site},*})}{\Tstar}\\
 &=B_{N,\mathrm H}
 -\Tstar\bigl[\mathcal C_{1,M}^{*}(z_{\mathrm s,H}^{*})
                    -\mathcal C_{1,M}^{*}(z_{\mathrm i,H}^{*})\bigr].
 \end{aligned}
 \label{si:eq:anharmonic-site-barrier}
\end{equation}
The harmonic-profile difference gives \(B_{N,\mathrm H}\), defined
in Eq.~\eqref{si:eq:harmonic-site-barrier}. The final line substitutes
\(\delta G^*=-{\Tstar}^2\mathcal C_{1,M}^*\) from
Eq.~\eqref{si:eq:site-profile-perturbation}; division by \(\Tstar\)
leaves the factor \(-\Tstar\).

The curvature corrections do require the extremum displacement.
The harmonic extremum satisfies Eq.~\eqref{si:eq:site-stationary},
so \(G_{\mathrm H}^{(1),*}(l_{\mathrm H}^*)=0\).
The shifted extremum must instead satisfy the corrected stationary condition,
\begin{equation}
 0=G_{\mathrm H}^{(1),*}(l_{\mathrm H}^*+\delta l^*)
 +\delta G^{(1),*}(l_{\mathrm H}^*+\delta l^*).
 \label{si:eq:site-corrected-stationarity}
\end{equation}
Taylor-expanding Eq.~\eqref{si:eq:site-corrected-stationarity} about
\(l_{\mathrm H}^*\) gives
\[
 0\simeq \underbrace{G_{\mathrm H}^{(1),*}(l_{\mathrm H}^{*})}_{0}
 +G_{\mathrm H}^{(2),*}(l_{\mathrm H}^{*})\delta l^{*}
 +\delta G^{(1),*}(l_{\mathrm H}^{*}).
\]
Solving for the displacement yields
\begin{equation}
 \delta l^{*}=-\frac{\delta G^{(1),*}(l_{\mathrm H}^{*})}{G_{\mathrm H}^{(2),*}(l_{\mathrm H}^{*})}.
 \label{si:eq:site-extremum-shift}
\end{equation}
This expansion assumes an isolated extremum with nonzero harmonic curvature.
For the second derivative, Taylor expansion gives
\[
 G_{\mathrm H}^{(2),*}(l_{\mathrm H}^*+\delta l^*)
 =G_{\mathrm H}^{(2),*}(l_{\mathrm H}^*)
 +G_{\mathrm H}^{(3),*}(l_{\mathrm H}^*)\delta l^*+\cdots.
\]
Adding the perturbation's second derivative at the harmonic extremum
therefore gives the corrected curvature:
\begin{equation}
 \mathcal K_{\mathrm{AH}}^{*}\simeq
 G_{\mathrm H}^{(2),*}(l_{\mathrm H}^{*})+\delta G^{(2),*}(l_{\mathrm H}^{*})
 +G_{\mathrm H}^{(3),*}(l_{\mathrm H}^{*})\delta l^{*}.
 \label{si:eq:anharmonic-site-curvature}
\end{equation}
The last two terms describe the direct change in curvature and the change
caused by moving along the harmonic profile. Differentiating
Eq.~\eqref{si:eq:site-profile-perturbation} with
\(dz^*/dl^*=-1/M\) from Eq.~\eqref{si:eq:site-reservoir-mean}
gives the quantities used in Eqs.~\eqref{si:eq:site-extremum-shift}
and \eqref{si:eq:anharmonic-site-curvature}:
\begin{equation}
 \begin{aligned}
 \delta G^{(1),*}(l^{*})
 &=-{\Tstar}^2\frac{d\mathcal C_{1,M}^{*}}{dz^{*}}
 \left(-\frac1M\right)
 =\frac{{\Tstar}^2}{M}\frac{d\mathcal C_{1,M}^{*}}{dz^{*}},\\
 \delta G^{(2),*}(l^{*})
 &=\frac{{\Tstar}^2}{M}\frac{d^2\mathcal C_{1,M}^{*}}{d(z^{*})^{2}}
 \left(-\frac1M\right)
 =-\frac{{\Tstar}^2}{M^2}\frac{d^2\mathcal C_{1,M}^{*}}{d(z^{*})^{2}}.
 \end{aligned}
 \label{si:eq:site-perturbation-derivatives}
\end{equation}
These formulas apply at both extrema and define a first-order correction,
rather than a new minimization of the truncated profile. The barrier plots
show the magnitude of the negative curvature. The first-anharmonic rate
estimates retain the harmonic KGH prefactor; the curvature corrections shown
here are a separate comparison of the profiles.

\subsection{Numerical evaluation and comparisons}
\label{si:numerical-site-barriers}

\begin{figure}[H]
 \centering
 \includegraphics[width=16cm]{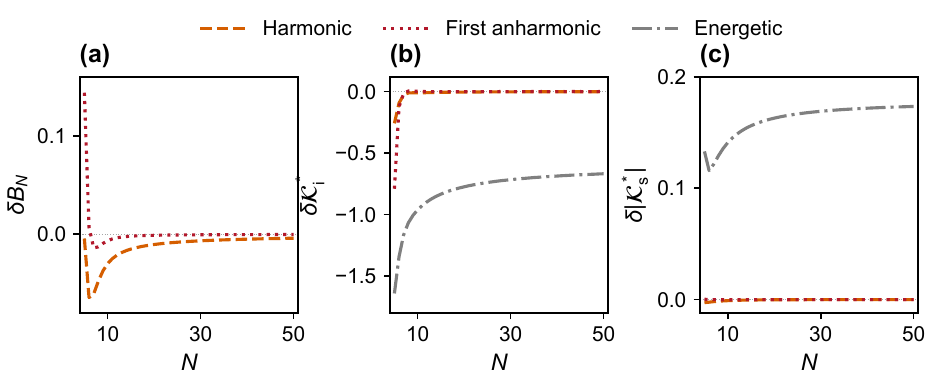}
 \input{figures/midpoint_barriers/curvature_preview/errors_caption}
 \label{si:fig:barrier-curvature-errors}
\end{figure}

For the deterministic reference, the restricted reservoir density is
evaluated by tilted convolution (Sec.~\ref{si:sec:numerical-evaluation}).
Local fourth-degree polynomials, fitted within \(0.03\) in \(l^{*}\) of
each grid extremum, refine the minimum and maximum and provide their
second derivatives. The fits use \(G_{N}^{\mathrm{site},*}/\Tstar\), so
multiplication by \(\Tstar\) converts their curvatures to the units
used above. Figure~\ref{si:fig:barrier-curvature-errors} shows the signed
differences from this reference. For \(N=7\)--50, the maximum absolute
barrier error decreases from \(0.0607\,k_{\mathrm B}T\) in the harmonic
approximation to \(0.0138\,k_{\mathrm B}T\) with the first correction.
Curvature improvement is not uniform: in particular, the correction moves
the \(N=5\) intact curvature further from the deterministic value.

\begin{figure}[H]
 \centering
 \includegraphics[width=16cm]{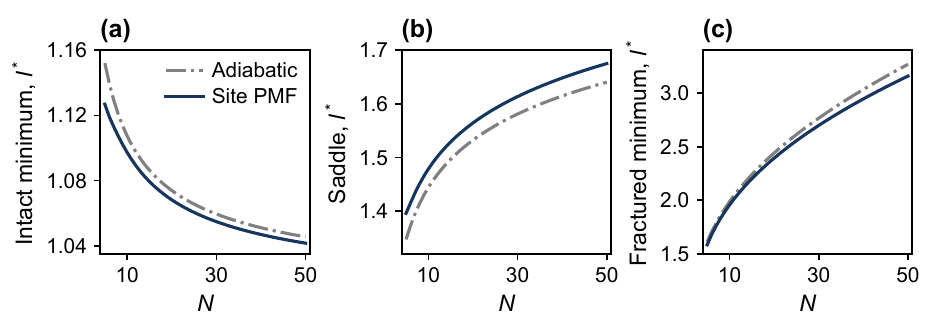}
 \input{figures/landscape_landmarks/caption}
 \label{si:fig:landscape-landmarks}
\end{figure}

Figure~\ref{si:fig:landscape-landmarks} compares the positions of the
intact minimum, barrier maximum and fractured minimum of the adiabatic
and site-PMF profiles at the same midpoint states. Reservoir entropy
shifts the barrier towards larger bond lengths and both minima towards
smaller lengths. The PMF extrema are refined with the same local
polynomial procedure used above.

%% file: figures/midpoint_barriers/curvature_preview/errors_caption.tex
\caption{Differences from deterministic site-PMF calculations at the same
midpoint extensions, \(T=450\,\mathrm K\): (a) barrier height,
\(\delta B_N=B_N^{\mathrm{approx}}-B_N^{\mathrm{det}}\);
(b) intact-minimum curvature,
\(\delta\mathcal K_{\mathrm i}^*=\mathcal K_{\mathrm i}^{\mathrm{approx},*}-\mathcal K_{\mathrm i}^{\mathrm{det},*}\);
(c) barrier-curvature magnitude,
\(\delta|\mathcal K_{\mathrm s}^*|=|\mathcal K_{\mathrm s}^{\mathrm{approx},*}|-|\mathcal K_{\mathrm s}^{\mathrm{det},*}|\).
Harmonic and first-anharmonic differences are shown in all panels; energetic
curvature differences are also shown in (b,c). Curvatures use energy in units
of \(D_e\) and length in units of \(l_0\). The midpoint extensions come from
the full partition for \(N=5\)--12 and the one-fracture partition for
\(N=13\)--50. All points are retained, including \(N=5\), where the
first-anharmonic correction worsens the intact-curvature estimate.}

%% file: figures/landscape_landmarks/caption.tex
\caption{Positions of the (a) intact minimum, (b) barrier maximum and
(c) fractured minimum of the adiabatic energy profile (gray dash-dot) and
site PMF (navy solid), at \(T=450\,\mathrm K\). Bond lengths are reduced, \(l^*=l/l_0\).
The curves locate the adiabatic and site-PMF extrema, respectively;
their saddles are \(l_{\mathrm{ad}}^{\ddagger,*}\) and \(l_{\mathrm s}^{\mathrm{site},*}\). Both profiles use the same finite-temperature midpoint extension
for each \(N\): full-partition values for \(N=5\)--12 and one-fracture
values for \(N=13\)--50. The three panels use different vertical scales.}

%% file: si_sections/langevin_dynamics.tex
\section{Langevin dynamics and numerical checks}
\label{si:langevin}

\subsection{Langevin dynamics}
\label{si:langevin-dynamics}

LAMMPS \citep{Thompson2022} integrates the Langevin equation given in main-text Sec.~III~B in
\texttt{real} units. The chain direction is periodic
(\texttt{boundary p f f}), with fixed box length \(L\).
The default parameters are listed in main-text Table~I:
\(T=450\,\mathrm K\), \(\xi=0.022\,\mathrm{fs}^{-1}\), and timestep
\(0.25\,\mathrm{fs}\). Velocity-Verlet integration uses \texttt{fix nve},
followed by \texttt{fix langevin} with damping time
\(1/\xi=45.454545\,\mathrm{fs}\).

The thermostat uses the chain-direction temperature computed by
\texttt{temp/partial 1 0 0}. Transverse velocities are set to zero, and
\texttt{fix setforce NULL 0.0 0.0} suppresses transverse forces, keeping
the motion one-dimensional. Initial velocities have zero total momentum;
subsequently, the center-of-mass momentum evolves under the thermostat,
without further momentum removal or the \texttt{zero yes} option.
Configuration and velocity preparation are specified with each simulation
method.

Periodic bond forces require access to the appropriate images of bonded
particles. The communication cutoff is set to
\(L/2+0.3\,\mathrm{\mathring A}\) using \texttt{comm\_modify cutoff}.
The neighbor list is checked every step
(\texttt{neigh\_modify every 1 delay 0 check yes cluster yes});
\texttt{cluster yes} checks the periodic geometry of bonded groups.

\subsection{Numerical checks}
\label{si:corrected-timestep}

Conservative forces and potential energies were checked against independent
Morse-potential calculations using minimum-image bond lengths.
Across 4,355,800 configurations from the timestep comparison below,
the maximum force discrepancy was
\(3.50\times10^{-12}\,\mathrm{kcal\,mol^{-1}\,\mathring A^{-1}}\).

Timestep sensitivity was assessed for \(N=10\) at
\(\lambda=1.107353897439319\), with the default temperature and friction.
The box length was \(17.05325002056551\,\mathrm{\mathring A}\), and
instantaneous fracture was defined by a maximum bond length above
\(2.223544999321842\,\mathrm{\mathring A}\). At this state, the partition
calculation predicts a fractured fraction of one half; halving its grid
spacing from 0.00025 to 0.000125 changes that fraction by
\(2.2\times10^{-8}\).
Each trajectory contains 2.5 ns of equilibration and 7.5 ns of production,
with observables recorded every 100 fs. Initial intact and fractured
configurations are equally represented, with the initially long bond
cycled through the chain sites. The comparison uses 2900 trajectories at each of
0.25 and 0.125 fs, together with 128 trajectories at 0.5 fs.

\begin{table}[htbp]
  \centering
  \small
  \caption{Fractured fractions at the three tested timesteps for
  \(N=10\). Parentheses give the lower and upper bounds of the
  95\% whole-trajectory bootstrap confidence intervals.
  The partition prediction is one half.}
  \label{si:tab:corrected-timestep-populations}
  {\color{revisionphasefifteen}
  \input{publication_tables/timestep_populations/table.tex}  }
\end{table}

Rates are obtained from the tension ACF as described in
Sec.~\ref{si:acf-method}, using the same fitting window
\(300\le t^*\le800\) at every timestep. For the two finer steps,
95\% intervals use 10,000 whole-trajectory bootstrap samples, stratified
by initial basin. Population and ACF estimates are recomputed together,
retaining the covariance between matched trajectories across timesteps.
The 0.5/0.25 fs comparison combines 100,000 independent pairs of saved
bootstrap draws from the two ensembles.

\begin{table}[htbp]
  \centering
  \small
  \caption{Relative differences in tension-ACF decay and directional
  rates, \(100(k_{\mathrm{larger\ step}}/k_{\mathrm{smaller\ step}}-1)\),
  at the common fitting window. Entries are percentages; parentheses give
  the lower and upper bounds of the 95\% confidence intervals.}
  \label{si:tab:corrected-timestep-rates}
  {\color{revisionphasefifteen}
  \input{publication_tables/timestep_rate_contrasts/table.tex}  }
\end{table}

Relative to 0.125 fs, the 0.25 fs estimates are higher by 0.82\% for the
ACF decay, 0.97\% for the fracture rate and 0.67\% for the return rate
(\cref{si:tab:corrected-timestep-rates}). Both finer-step fractured
fractions agree with one half within their 95\% intervals
(\cref{si:tab:corrected-timestep-populations}). The adopted timestep is
0.25 fs, half the 0.5 fs used by \citet{Theodorou2024}.

%% file: publication_tables/timestep_populations/table.tex
\begin{tabular}{rrc}
  \toprule
  Step / fs & Trajectories & \(P_{\mathrm f}^{\mathrm{sim}}\) \\
  \midrule
  0.500 & 128 & \(0.5063\;(0.4976,\,0.5150)\) \\
  0.250 & 2900 & \(0.5006\;(0.4987,\,0.5025)\) \\
  0.125 & 2900 & \(0.4999\;(0.4979,\,0.5018)\) \\
  \bottomrule
\end{tabular}

%% file: publication_tables/timestep_rate_contrasts/table.tex
\begin{tabular}{lcc}
  \toprule
  Quantity & 0.5 / 0.25 fs & 0.25 / 0.125 fs \\
  \midrule
  Fitted decay & \(-5.0\;(-8.0,\,-1.8)\) & \(0.82\;(-0.11,\,1.77)\) \\
  Intact to fractured & \(-3.9\;(-7.5,\,-0.2)\) & \(1.0\;(-0.1,\,2.1)\) \\
  Fractured to intact & \(-6.1\;(-9.3,\,-2.6)\) & \(0.7\;(-0.4,\,1.8)\) \\
  \bottomrule
\end{tabular}

%% file: si_sections/deterministic_rates.tex
\section{Rate theory and deterministic predictions}
\label{si:reactive-flux}

\subsection{Tagged-coordinate metric and surface flux}
\label{si:tst-flux}

A transition-state estimate counts outward crossings of the intact basin's
boundary at fixed \(N\), extension and temperature. Its two ingredients are
the equilibrium density at the surface and the thermal velocity towards
fracture. The derivation below obtains the velocity factor from particle
motion, combines it with the constrained surface weight, and divides the
resulting crossing frequency by the intact population.

The tagged length \(l_j=x_j-x_{j-1}\) changes through the relative motion
of two neighboring particles. For the reduced length \(l_j^*=l_j/l_0\),
the gradient with respect to the physical configuration \(\bm r\) has
entries \(+1/l_0\) and \(-1/l_0\) at these particles and zero elsewhere.
With equal masses \(m\), the particle mass matrix is \(\bm M=mI\), where
\(I\) is the identity matrix. The mass-weighted gradient and its inverse
are therefore
\begin{equation}
 \begin{aligned}
 \bigl(\nabla l_j^*\bigr)^{\mathsf T}\bm M^{-1}\nabla l_j^*
 &=\frac1m\left[\left(\frac1{l_0}\right)^2
 +\left(-\frac1{l_0}\right)^2\right]=\frac2{ml_0^2},\\
 M_{l^*}
 &\equiv\left[\bigl(\nabla l_j^*\bigr)^{\mathsf T}
 \bm M^{-1}\nabla l_j^*\right]^{-1}
 =\frac{ml_0^2}{2}.
 \end{aligned}
 \label{si:eq:tagged-metric}
\end{equation}
The coefficient \(M_{l^*}\) applies to the dimensionless length and
carries units of mass times length squared; the corresponding kinetic
mass for the physical bond length is \(\mu_l=m/2\).
Its connection to the velocity variance follows below.

Before imposing zero total momentum, the particle velocities are independent
Gaussians with mean zero and variance \(k_{\mathrm B}T/m\). Subtracting
their common mean conditions the velocities on zero total momentum.
Each particle then has variance \((k_{\mathrm B}T/m)(1-1/N)\), and
distinct particle velocities have covariance \(-k_{\mathrm B}T/(mN)\).
Together these statements give
\begin{equation}
 \operatorname{Cov}(\dot x_i,\dot x_j)
 =\frac{k_{\mathrm B}T}{m}\left(\delta_{ij}-\frac1N\right),
 \label{si:eq:maxwell-particle-covariance}
\end{equation}
where \(\delta_{ij}\) is the Kronecker delta. The negative covariance arises
from the zero-sum condition: one particle's velocity is linked to those
of the others. For two random variables \(A\) and \(B\),
\(\operatorname{Var}(A-B)=\operatorname{Var}(A)+\operatorname{Var}(B)
-2\operatorname{Cov}(A,B)\).
Applying this identity to the two neighboring velocities in
Eq.~\eqref{si:eq:maxwell-particle-covariance}, and using \(t^*=\xi t\),
gives
\begin{equation}
 \begin{aligned}
 v_l^*=\frac{dl_j^*}{dt^*}
 &=\frac{\dot x_j-\dot x_{j-1}}{\xi l_0},\\
 \bigl(\sigma_v^*\bigr)^2\equiv\operatorname{Var}(v_l^*)
 &=\frac{\operatorname{Var}(\dot x_j)+\operatorname{Var}(\dot x_{j-1})
 -2\operatorname{Cov}(\dot x_j,\dot x_{j-1})}{\xi^2l_0^2}\\
 &=\frac{2(k_{\mathrm B}T/m)(1-1/N)+2k_{\mathrm B}T/(mN)}
 {\xi^2l_0^2}\\
 &=\frac{2k_{\mathrm B}T}{m\xi^2l_0^2}
 =\frac{k_{\mathrm B}T}{\xi^2M_{l^*}}\\
 &=2\frac{\De}{m\xi^2l_0^2}\frac{k_{\mathrm B}T}{\De}
 =2c\Tstar.
 \end{aligned}
 \label{si:eq:tagged-velocity-variance}
\end{equation}
The \(1/N\) terms cancel, consistently with common-mean subtraction
leaving the relative velocity unchanged. The fourth line substitutes
\(M_{l^*}=ml_0^2/2\) from Eq.~\eqref{si:eq:tagged-metric};
the factor \(1/\xi^2\) converts its velocity variance from physical
to reduced time. The final line uses \(c=\De/(m\xi^2l_0^2)\) and
\(\Tstar=k_{\mathrm B}T/\De\) (Sec.~\ref{si:notation}).
The reduced tagged velocity has the normalized Gaussian density
\begin{equation}
 p_v^*(v^*)=
 \frac{\exp[-(v^*)^2/(2(\sigma_v^*)^2)]}
 {\sqrt{2\pi}\,\sigma_v^*},
 \qquad -\infty<v^*<\infty.
 \label{si:eq:tagged-maxwell-density}
\end{equation}
An outward crossing has \(v_l^*>0\). Its velocity factor is the average
of the positive part of the velocity, with \(\Theta\) denoting the
Heaviside function:
\begin{equation}
 \begin{aligned}
 \bar v_+^*&\equiv\left\langle v_l^*\Theta(v_l^*)\right\rangle
 =\int_0^\infty v^*p_v^*(v^*)\,dv^*\\
 &=\frac1{\sqrt{2\pi}\,\sigma_v^*}
 \int_0^\infty v^*e^{-(v^*)^2/(2(\sigma_v^*)^2)}\,dv^*\\
 &=\left[-\frac{\sigma_v^*}{\sqrt{2\pi}}
 e^{-(v^*)^2/(2(\sigma_v^*)^2)}\right]_0^\infty
 =\frac{\sigma_v^*}{\sqrt{2\pi}}
 =\sqrt{\frac{c\Tstar}{\pi}}.
 \end{aligned}
 \label{si:eq:maxwell-flux}
\end{equation}
The last equality uses Eq.~\eqref{si:eq:tagged-velocity-variance}.
This is the classical forward-velocity factor evaluated by
\citet[p.~110]{Eyring1935}, with reaction-coordinate mass \(m/2\) and
conversion to reduced variables.
This average uses the full velocity distribution in
Eq.~\eqref{si:eq:tagged-maxwell-density}; negative velocities contribute
zero. Since half the velocities are positive, conditioning on positive
velocity would instead give \(2\bar v_+^*\).

To construct the configurational factor, consider the surface where the
maximum bond length reaches a chosen value \(s^*\). It consists of
\(N\) equivalent tagged components:
\begin{equation}
 \Sigma_{s^*}=\bigcup_{j=1}^N
 \{l_j^*=s^*,\ l_i^*<s^*\ \text{for all }i\ne j\}.
 \label{si:eq:rf-launch-surface}
\end{equation}
The symbol \(\bigcup_{j=1}^N\) denotes the union of the \(N\) sets:
the surface includes configurations in which any one tagged bond has
length \(s^*\), with all other bonds shorter than \(s^*\).
Simultaneous ties have zero equilibrium measure. On one component, fixing
the tagged bond at \(s^*\) leaves length \(L^*-s^*\) for the reservoir,
where \(L^*=N\lambda\). Applying the restricted-weight construction in
Eq.~\eqref{si:eq:restricted-site-weight-and-pmf} at tagged length \(s^*\),
with reservoir cutoff \(s^*\), gives the unnormalized surface density
\begin{equation}
 \rho_{N,s^*}^*(L^*)
 =e^{-u^*(s^*)/\Tstar}
 Z_{N-1}^{\mathrm i,*}(L^*-s^*;s^*).
 \label{si:eq:rf-surface-density}
\end{equation}
Division by the full partition density \(Z_N^*\), as in the probability
measure of Eq.~\eqref{si:eq:conditional-boltzmann-measure}, normalizes it
per unit tagged length.

The canonical distribution separates positions and velocities, so the density
in Eq.~\eqref{si:eq:rf-surface-density} multiplies the Maxwell factor. A positive velocity \(v^*\)
sweeps a length \(v^*dt^*\) in time \(dt^*\); density times swept length
gives the crossing probability. Dividing by \(dt^*\), averaging over
positive velocities and summing the \(N\) surface components yields
\begin{equation}
 \begin{aligned}
 J_{N,s^*}^{+,*}
 &=N\frac{\rho_{N,s^*}^*(L^*)}{Z_N^*(L^*)}
 \int_0^\infty v^*p_v^*(v^*)\,dv^*\\
 &=N\bar v_+^*\frac{\rho_{N,s^*}^*(L^*)}{Z_N^*(L^*)}.
 \end{aligned}
 \label{si:eq:surface-density-flux}
\end{equation}
Thus \(J_{N,s^*}^{+,*}\) is the equilibrium number of outward crossings
per unit reduced time, counting all equivalent bonds once.

The physical intact basin is defined by the divider \(d_N^*\)
(Sec.~\ref{si:notation}), with population
\begin{equation}
 P_{\mathrm i}(d_N^*)=
 \frac{Z_N^{\mathrm i,*}(L^*;d_N^*)}{Z_N^*(L^*)}.
 \label{si:eq:tst-intact-population}
\end{equation}
A crossing frequency per intact chain is obtained by dividing
Eq.~\eqref{si:eq:surface-density-flux} by
Eq.~\eqref{si:eq:tst-intact-population}. The full partition cancels:
\begin{equation}
 \begin{aligned}
 K_{N,s^*\mid d_N^*}^*
 &=\frac{J_{N,s^*}^{+,*}}{P_{\mathrm i}(d_N^*)}\\
 &=\frac{N\bar v_+^*\rho_{N,s^*}^*(L^*)/Z_N^*(L^*)}
 {Z_N^{\mathrm i,*}(L^*;d_N^*)/Z_N^*(L^*)}\\
 &=N\bar v_+^*
 \frac{\rho_{N,s^*}^*(L^*)}{Z_N^{\mathrm i,*}(L^*;d_N^*)}.
 \end{aligned}
 \label{si:eq:exact-tst}
\end{equation}
The reservoir cutoff at the crossing surface is \(s^*\), while the
population in the denominator retains the fixed divider \(d_N^*\).
When \(s^*=d_N^*\), the surface is the actual intact--fractured boundary
and Eq.~\eqref{si:eq:exact-tst} gives the chain TST estimate:
\begin{equation}
 k_N^{\mathrm{TST},*}\equiv K_{N,d_N^*\mid d_N^*}^*
 =N\sqrt{\frac{c\Tstar}{\pi}}\,
 \frac{e^{-u^*(d_N^*)/\Tstar}
 Z_{N-1}^{\mathrm i,*}(L^*-d_N^*;d_N^*)}
 {Z_N^{\mathrm i,*}(L^*;d_N^*)}.
 \label{si:eq:production-tst-rate}
\end{equation}
This expression specializes the classical TST construction of
\citet[Eqs.~(4.7)--(4.8), (8.2)]{Chandler1978} to the constrained chain's
intact-basin boundary.
The partition densities retain the constrained equilibrium ensemble,
without equal-sharing or harmonic approximations. The TST assumption
is dynamical: each outward boundary crossing is taken to complete a
transition \citep[Eq.~(8.1)]{Chandler1978}. The following subsection accounts for
recrossing through the transmission coefficient.

\subsection{General reactive-flux framework}
\label{si:rf-framework}

TST counts every outward boundary crossing as a transition. Reactive flux
instead follows the subsequent motion and weights each initial crossing
by the state reached at the observation time. This accounts for trajectories
that cross and return.

The launch ensemble is the constrained equilibrium distribution on one
tagged component of \(\Sigma_{s^*}\), defined in
Eq.~\eqref{si:eq:rf-launch-surface}. The tagged bond has length \(s^*\);
the other bonds share \(L^*-s^*\), with their Boltzmann weights and lengths
restricted below \(s^*\). Initial velocities follow the full Maxwell
distribution conditioned on zero total momentum. The brackets
\(\langle\cdot\rangle_{\Sigma_{s^*},\mathrm{tag}}\) denote an average over
these configurations, velocities and independent subsequent Langevin
noise realizations. Both signs of the initial tagged velocity
\(v_l^*(0)\), defined in Eq.~\eqref{si:eq:tagged-velocity-variance}, enter
this average.

The physical states retain the divider \(d_N^*\), which need not coincide
with the launch length \(s^*\). Their instantaneous fractured-state
indicator is
\begin{equation}
 h_{\mathrm f,d_N^*}[\bm r^*(t^*)]
 =\mathbf1\!\left[\max_j l_j^*(t^*)>d_N^*\right].
 \label{si:eq:rf-fractured-indicator}
\end{equation}
It equals one if any bond is fractured at \(t^*\), and zero otherwise.
The tagged bond identifies the initial crossing coordinate; the later
fracture may occur at any bond. The indicator records the endpoint state,
including any intervening returns, rather than whether fracture has ever
occurred.

The signed transmission function is \citep[Eq.~(7.9)]{Chandler1978}
\begin{equation}
 \kappa_{N,s^*\mid d_N^*}(t^*)=
 \frac{\left\langle v_{l}^{*}(0)
 h_{\mathrm f,d_N^*}[\bm r^{*}(t^*)]\right\rangle_{\Sigma_{s^*},\mathrm{tag}}}
 {\bar v_+^*}.
 \label{si:eq:rf}
\end{equation}
The denominator is the positive Maxwell factor in
Eq.~\eqref{si:eq:maxwell-flux}. The numerator weights a fractured endpoint
by the signed initial velocity. To make the subtraction explicit, write
\(h_{\mathrm f}(t^*)\) for the indicator in
Eq.~\eqref{si:eq:rf-fractured-indicator}. Then
\begin{equation}
 \begin{aligned}
 \left\langle v_l^*(0)h_{\mathrm f}(t^*)\right\rangle_{\Sigma_{s^*},\mathrm{tag}}
 &=
 \left\langle v_l^*(0)\Theta(v_l^*(0))h_{\mathrm f}(t^*)\right\rangle_{\Sigma_{s^*},\mathrm{tag}}
 \\
 &\quad-
 \left\langle |v_l^*(0)|\Theta(-v_l^*(0))h_{\mathrm f}(t^*)\right\rangle_{\Sigma_{s^*},\mathrm{tag}}.
 \end{aligned}
 \label{si:eq:rf-signed-numerator}
\end{equation}
Positive launches contribute positively when fractured at the observation
time; negative launches reaching that same state subtract from them.
Either sign contributes zero when the chain is intact. Thus
\(\kappa\) is a normalized signed average, not a success fraction.

The positive surface flux in Eq.~\eqref{si:eq:surface-density-flux}
contains the normalized density, the Maxwell factor and all \(N\)
equivalent sites. Replacing its positive-velocity factor by the signed
average in Eq.~\eqref{si:eq:rf} gives the finite-time current. Dividing
by the intact population in Eq.~\eqref{si:eq:tst-intact-population}
then gives
\begin{equation}
 \begin{aligned}
 J_{N,s^*\mid d_N^*}^{*}(t^*)
 &=N\frac{\rho_{N,s^*}^*(L^*)}{Z_N^*(L^*)}
 \left\langle v_l^*(0)h_{\mathrm f}(t^*)\right\rangle_{\Sigma_{s^*},\mathrm{tag}}
 =J_{N,s^*}^{+,*}\kappa_{N,s^*\mid d_N^*}(t^*),\\
 k_{N,s^*\mid d_N^*}^{*}(t^*)
 &=\frac{J_{N,s^*\mid d_N^*}^{*}(t^*)}{P_{\mathrm i}(d_N^*)}
 =K_{N,s^*\mid d_N^*}^{*}\kappa_{N,s^*\mid d_N^*}(t^*).
 \end{aligned}
 \label{si:eq:fixed-state-rf-rate}
\end{equation}
The last equality uses Eq.~\eqref{si:eq:exact-tst}.
Site multiplicity is already included in \(K\).

When the launch surface is the physical boundary, \(s^*=d_N^*\), write
\(\kappa_{N,d_N^*\mid d_N^*}=\kappa_N\).
Equations~\eqref{si:eq:production-tst-rate} and
\eqref{si:eq:fixed-state-rf-rate} reduce to
\begin{equation}
 k_{\mathrm{RF}}^*(t^*)=k_N^{\mathrm{TST},*}\kappa_N(t^*).
 \label{si:eq:rf-boundary-rate}
\end{equation}
Immediately after launch, the endpoint lies on the fractured side
precisely when \(v_l^*(0)>0\). Hence
\(\kappa_N(0^+)=
\langle v_l^*(0)\Theta(v_l^*(0))\rangle/\bar v_+^*=1\),
by Eq.~\eqref{si:eq:maxwell-flux}, and the initial RF rate is the TST rate.
Later recrossing changes this value. The time-dependent quantity in
Eq.~\eqref{si:eq:rf-boundary-rate} is converted into a kinetic rate after
the initial transient; Sec.~\ref{si:rf-estimation} explains the treatment
of subsequent population relaxation.
The next subsection evaluates transmission with harmonic dynamics,
whereas Sec.~\ref{si:barrier-top} uses trajectories with the full Morse forces.

\subsection{Full-coordinate harmonic transmission}
\label{si:harmonic-transmission}

Near the saddle, the harmonic approximation separates motion into stable
fluctuations and one unstable mode that carries the chain away from the
dividing surface. Initial positions, velocities and thermal noise determine
its growing amplitude. The correlation between this amplitude and the
initial tagged-bond velocity gives the harmonic transmission coefficient,
which corrects the TST rate for recrossing. Replacing the potential by its
quadratic expansion gives linear Langevin dynamics, so the signed average
in Eq.~\eqref{si:eq:rf} can be evaluated from Gaussian statistics.
Its use as an activated-rate approximation assumes well-separated basins
and a barrier large compared with the thermal energy.

Let \(\delta\bm r^*\) denote particle displacements from the saddle of
Sec.~\ref{si:stationary-morse}, in units of \(l_0\).
The periodic incidence matrix \(B\) converts them into bond
displacements:
\((B\delta\bm r^*)_j=\delta x_j^*-\delta x_{j-1}^*=\delta l_j^*\).
At the saddle, the bond tensions are equal. The linear energy change
therefore vanishes because fixed total length imposes
\(\sum_j\delta l_j^*=0\). With
\(u_j^{(2),*}=u^{(2),*}(l_j^{\mathrm s,*})\), the quadratic contribution is
\[
 \frac12\sum_j u_j^{(2),*}(\delta l_j^*)^2
 =\frac12(B\delta\bm r^*)^{\mathsf T}
 \operatorname{diag}(u_j^{(2),*})(B\delta\bm r^*).
\]
Thus the dimensionless chain energy
\(\mathcal U^*=\sum_j u^*(l_j^*)\) has the expansion
\begin{equation}
 \mathcal U^{*}-\mathcal U_{\mathrm s}^{*}\simeq\tfrac12\bigl(\delta\bm r^{*}\bigr)^{\mathsf T}H^{*}\delta\bm r^{*},
 \qquad H^{*}=B^{\mathsf T}\operatorname{diag}(u_{j}^{(2),*})B.
 \label{si:eq:harmonic-hessian}
\end{equation}
Uniform translation changes no bond length and gives a zero mode of
\(H^*\). Remove it by choosing the columns of \(E\) as an orthonormal
basis perpendicular to the uniform vector \(\bm1\), whose entries are one:
\begin{equation}
 E^{\mathsf T}E=I,\qquad E^{\mathsf T}\bm1=0,\qquad
 \delta\bm r^*=E\bm q^*,\qquad
 H_{\mathrm{int}}^*=E^{\mathsf T}H^*E.
 \label{si:eq:harmonic-internal-coordinates}
\end{equation}
The \(N-1\) coordinates \(\bm q^*\) describe internal motion.
Substitution into Eq.~\eqref{si:eq:harmonic-hessian} gives the energy
\(\frac12(\bm q^*)^{\mathsf T}H_{\mathrm{int}}^*\bm q^*\).
For \(\bm v^*=d\bm q^*/dt^*\), the corresponding linear Langevin dynamics is
\begin{equation}
 d\begin{pmatrix}\bm q^{*}\\\bm v^{*}\end{pmatrix}
 =\underbrace{\begin{pmatrix}0&I\\-cH_{\mathrm{int}}^{*}&-I\end{pmatrix}}_{\mathcal A^{*}}
 \begin{pmatrix}\bm q^{*}\\\bm v^{*}\end{pmatrix}dt^*
 +\begin{pmatrix}0\\\sqrt{2c\Tstar}\,d\bm W^{*}\end{pmatrix},
 \label{si:eq:harmonic-langevin}
\end{equation}
where \(I\) is the identity matrix and \(\bm W^*\) contains \(N-1\)
independent standard Wiener processes. Each noise increment has variance
\(dt^*\). The two rows respectively give the position change and the
velocity change from the harmonic force, friction and thermal noise.
The matrices \(B,E\), the surface basis \(L\) below and unit vectors are
geometric and remain unstarred.

For tagged bond \(j\), Eq.~\eqref{si:eq:harmonic-internal-coordinates}
gives
\(\delta l_j^*=\bm e_j^{\mathsf T}BE\bm q^*=\bm n^{\mathsf T}\bm q^*\),
where \(\bm e_j\) selects that bond and
\(\bm n=E^{\mathsf T}B^{\mathsf T}\bm e_j\).
Fixing its length at the saddle defines the local surface
\(\bm n^{\mathsf T}\bm q^*=0\). Let the \(N-2\) columns of \(L\) form
an orthonormal basis tangent to it, so \(L^{\mathsf T}L=I\) and
\(\bm n^{\mathsf T}L=0\). Writing
\(\bm q^*=L\bm q_\parallel^*\), the harmonic surface distribution is
\begin{equation}
 \begin{aligned}
 p_\parallel^*(\bm q_\parallel^*)&\propto
 \exp\!\left[-\frac{(\bm q_\parallel^*)^{\mathsf T}
 (L^{\mathsf T}H_{\mathrm{int}}^*L)\bm q_\parallel^*}{2\Tstar}\right],\\
 \operatorname{Cov}(\bm q_\parallel^*)&=
 \Tstar(L^{\mathsf T}H_{\mathrm{int}}^*L)^{-1}.
 \end{aligned}
 \label{si:eq:harmonic-tangent-gaussian}
\end{equation}
As in Eq.~\eqref{si:eq:one-dimensional-gaussian}, each positive curvature
gives a Gaussian variance equal to temperature divided by curvature.
The tangential coordinates are integrated over \(\mathbb R^{N-2}\);
\(L^{\mathsf T}H_{\mathrm{int}}^*L\) must be positive definite for the
distribution to be normalizable. Transforming the covariance back with
\(\bm q^*=L\bm q_\parallel^*\) yields
\begin{equation}
 C_{q}^{*}=\Tstar L(L^{\mathsf T}H_{\mathrm{int}}^{*}L)^{-1}L^{\mathsf T}.
 \label{si:eq:harmonic-surface-covariance}
\end{equation}
Initial velocities are independent of positions. Reducing the particle
covariance in Eq.~\eqref{si:eq:maxwell-particle-covariance} and projecting
with \(E\) gives their internal covariance:
\begin{equation}
 \operatorname{Cov}(\bm v^*(0))
 =E^{\mathsf T}\!\left[c\Tstar
 \left(I-\frac{\bm1\bm1^{\mathsf T}}N\right)\right]E
 =c\Tstar I.
 \label{si:eq:harmonic-velocity-covariance}
\end{equation}
The last equality uses both orthogonality relations in
Eq.~\eqref{si:eq:harmonic-internal-coordinates}.
The subsequent thermostat noise is independent of these initial draws.

The internal Hessian has one negative eigenvalue \(h_-^*\).
Choose its unit eigenvector \(\bm u\) with
\(\bm n^{\mathsf T}\bm u>0\), so growth in the positive direction
lengthens the tagged bond. Projecting Eq.~\eqref{si:eq:harmonic-langevin}
onto this mode gives
\begin{equation}
 \begin{aligned}
 q_-^*&=\bm u^{\mathsf T}\bm q^*,\qquad
 v_-^*=\bm u^{\mathsf T}\bm v^*,\qquad
 W_u^*=\bm u^{\mathsf T}\bm W^*,\\
 dq_-^*&=v_-^*\,dt^*,\\
 dv_-^*&=(-ch_-^*q_-^*-v_-^*)\,dt^*
 +\sqrt{2c\Tstar}\,dW_u^*.
 \end{aligned}
 \label{si:eq:harmonic-unstable-mode}
\end{equation}
Because \(\bm u\) has unit norm, \(W_u^*\) is a standard Wiener process.
Without noise, substituting \(q_-^*\propto e^{r^*t^*}\) gives
\((r^*)^2+r^*+ch_-^*=0\). Its growing and decaying roots are
\begin{equation}
 r_{\pm}^{*}=\frac{-1\pm\sqrt{1-4ch_{-}^{*}}}2,
 \qquad \Delta^{*}=r_{+}^{*}-r_{-}^{*}.
 \label{si:eq:harmonic-growth}
\end{equation}
Here \(r_+^*>0\) and \(r_-^*<0\).
These equations specialize the linear saddle dynamics of
\citet[Eqs.~(9)--(15)]{Bartsch2005}. In the reduced variables used here, their
force, friction and drift matrices correspond to
\(\Omega=-cH_{\mathrm{int}}^*\), \(\Gamma=I\) and
\(A=\mathcal A^*\), respectively; their eigenvalues
\(\epsilon_{\mathrm u}\) and \(\epsilon_{\mathrm s}\) correspond to
\(r_+^*\) and \(r_-^*\).
The homogeneous solution is a sum of these two exponentials; its
coefficients are fixed by \(q_-^*(0)\) and \(v_-^*(0)\).
Adding the response to the thermal forcing in
Eq.~\eqref{si:eq:harmonic-unstable-mode} gives
\begin{equation}
 \begin{aligned}
 q_-^*(t^*)&=
 \frac{[v_-^*(0)-r_-^*q_-^*(0)]e^{r_+^*t^*}
 -[v_-^*(0)-r_+^*q_-^*(0)]e^{r_-^*t^*}}{\Delta^*}\\
 &\quad+\frac{\sqrt{2c\Tstar}}{\Delta^*}
 \int_0^{t^*}
 \left[e^{r_+^*(t^*-s^*)}-e^{r_-^*(t^*-s^*)}\right]dW_u^*(s^*).
 \end{aligned}
 \label{si:eq:harmonic-unstable-solution}
\end{equation}
The kernel in the integral is the displacement response to a unit velocity
impulse. Multiplying by \(e^{-r_+^*t^*}\) removes the growing exponential;
the terms containing the decaying root vanish as \(t^*\to\infty\).
Using the projections in Eq.~\eqref{si:eq:harmonic-unstable-mode} gives
the growing amplitude
\begin{equation}
 \begin{aligned}
 A_+^*&\equiv\lim_{t^*\to\infty}e^{-r_+^*t^*}q_-^*(t^*)\\
 &=\frac{
 \bm u^{\mathsf T}\bm v^*(0)-r_-^*\bm u^{\mathsf T}\bm q^*(0)
 +\sqrt{2c\Tstar}\int_0^\infty
 e^{-r_+^*s^*}\bm u^{\mathsf T}d\bm W^*(s^*)
 }{\Delta^*}.
 \end{aligned}
 \label{si:eq:harmonic-growing-amplitude}
\end{equation}
Here \(s^*\) is the reduced integration time.
In the phase-space eigenbasis of
\citet[Eq.~(32), p.~204102-8]{Bartsch2005},
the unstable coordinate is
\((v_-^*-r_-^*q_-^*)/\Delta^*\), with noise increment
\(\sqrt{2c\Tstar}\,dW_u^*/\Delta^*\).
Their transition-state coordinate is the negative of the future-noise
integral \citep[Eqs.~(22)--(23)]{Bartsch2005}.
Subtracting it from the initial unstable coordinate gives the
coefficient of the growing exponential in
Eq.~\eqref{si:eq:harmonic-growing-amplitude}.
Stable modes do not grow, so
\(\bm n^{\mathsf T}\bm q^*(t^*)\sim
(\bm n^{\mathsf T}\bm u)A_+^*e^{r_+^*t^*}\).
Thus \(A_+^*>0\) selects the fractured side of the local harmonic surface,
as in the product-side criterion of \citet[Sec.~IV~D]{Bartsch2005}.

To calculate the signed average, define the initial tagged velocity
and the initial unstable-coordinate variance:
\begin{equation}
 V^*=\bm n^{\mathsf T}\bm v^*(0)=v_l^*(0),\qquad
 (\sigma_u^*)^2=\operatorname{Var}(q_-^*(0))
 =\bm u^{\mathsf T}C_q^*\bm u.
 \label{si:eq:harmonic-launch-projections}
\end{equation}
Fixing the tagged length need not fix the projection along \(\bm u\),
so this position variance need not vanish.
The velocity, position and noise contributions in
Eq.~\eqref{si:eq:harmonic-growing-amplitude} are independent.
Equation~\eqref{si:eq:harmonic-velocity-covariance} gives
\(\operatorname{Var}(\bm u^{\mathsf T}\bm v^*(0))=c\Tstar\).
Independent Wiener increments have variance \(ds^*\), so the variance of
the weighted noise integral is the integral of its squared weight:
\begin{equation}
 \operatorname{Var}\!\left[
 \sqrt{2c\Tstar}\int_0^\infty e^{-r_+^*s^*}\,dW_u^*(s^*)\right]
 =2c\Tstar\int_0^\infty e^{-2r_+^*s^*}\,ds^*
 =\frac{c\Tstar}{r_+^*}.
 \label{si:eq:harmonic-noise-variance}
\end{equation}
The noise projection by \(1/\Delta^*\) and the resulting variance,
\(c\Tstar/[r_+^*(\Delta^*)^2]\), correspond to
\citet[Eqs.~(33)--(34), p.~204102-8]{Bartsch2005}.
Their \(\kappa\) denotes the noise amplitude, rather than a transmission
coefficient.
Only the initial-velocity contribution correlates with \(V^*\):
\[
 \operatorname{Cov}(\bm n^{\mathsf T}\bm v^*(0),
                    \bm u^{\mathsf T}\bm v^*(0))
 =\bm n^{\mathsf T}(c\Tstar I)\bm u
 =c\Tstar\,\bm n^{\mathsf T}\bm u.
\]
Adding the three variances and dividing this covariance by \(\Delta^*\)
therefore gives
\begin{equation}
 \begin{aligned}
 \bigl(\sigma_{A}^{*}\bigr)^{2}\equiv\operatorname{Var}(A_{+}^{*})
 &=\frac{c\Tstar+\bigl(r_{-}^{*}\bigr)^{2}\bigl(\sigma_{u}^{*}\bigr)^{2}+c\Tstar/r_{+}^{*}}{\bigl(\Delta^{*}\bigr)^{2}},\\
 C_{VA}^{*}\equiv\operatorname{Cov}(V^{*},A_{+}^{*})
 &=\frac{c\Tstar\,\bm n^{\mathsf T}\bm u}{\Delta^{*}},
 \\
 \bigl(\sigma_{V}^{*}\bigr)^{2}&=c\Tstar\,\bm n^{\mathsf T}\bm n.
 \end{aligned}
 \label{si:eq:harmonic-amplitude-moments}
\end{equation}
The surface normal is not normalized: each bond difference has two
entries, \(+1\) and \(-1\), and is perpendicular to translation, so
\(\bm n^{\mathsf T}\bm n=2\). The last line consequently reproduces
the tagged-velocity variance in Eq.~\eqref{si:eq:tagged-velocity-variance}.

Both \(V^*\) and \(A_+^*\) are centered Gaussian variables, since they
are linear combinations of Gaussian initial draws and noise.
Subtracting \(C_{VA}^*A_+^*/(\sigma_A^*)^2\) from \(V^*\) leaves a
centered Gaussian uncorrelated with, and therefore independent of,
\(A_+^*\). Their conditional mean is consequently
\begin{equation}
 \mathbb E[V^*\mid A_+^*]
 =\frac{C_{VA}^*}{(\sigma_A^*)^2}A_+^*.
 \label{si:eq:harmonic-gaussian-conditional-mean}
\end{equation}
First averaging at fixed \(A_+^*\), then over positive amplitudes,
evaluates the signed numerator:
\begin{equation}
 \begin{aligned}
 \left\langle V^*\mathbf1(A_+^*>0)\right\rangle
 &=\left\langle\mathbb E[V^*\mid A_+^*]\,
                 \mathbf1(A_+^*>0)\right\rangle\\
 &=\frac{C_{VA}^*}{(\sigma_A^*)^2}
       \left\langle A_+^*\Theta(A_+^*)\right\rangle\\
 &=\frac{C_{VA}^*}{(\sigma_A^*)^2}
       \frac{\sigma_A^*}{\sqrt{2\pi}}
 =\frac{C_{VA}^*}{\sqrt{2\pi}\,\sigma_A^*}.
 \end{aligned}
 \label{si:eq:harmonic-signed-gaussian-average}
\end{equation}
The penultimate step uses the positive Gaussian moment derived in
Eq.~\eqref{si:eq:maxwell-flux}, now with standard deviation \(\sigma_A^*\).
For \(V^*\), the same integral gives
\(\langle V^*\Theta(V^*)\rangle=\sigma_V^*/\sqrt{2\pi}\).
Dividing Eq.~\eqref{si:eq:harmonic-signed-gaussian-average} by this
Maxwell factor, as in Eq.~\eqref{si:eq:rf}, and using
Eq.~\eqref{si:eq:rf-boundary-rate} yields
\begin{equation}
 \kappa_N^{\mathrm{harm}}=\frac{C_{VA}^{*}}{\sigma_{V}^{*}\sigma_{A}^{*}},
 \qquad
 k_N^{\mathrm{harm\,RF},*}=k_N^{\mathrm{TST},*}\kappa_N^{\mathrm{harm}}.
 \label{si:eq:harmonic-transmission-result}
\end{equation}
The coefficient is exact for this local Gaussian, harmonic dynamical model;
its use for Morse-chain recrossing is an approximation.
The TST factor is calculated separately from the full equilibrium partition
functions in Eq.~\eqref{si:eq:production-tst-rate}.
No simulated trajectories or fitted kinetic parameters enter the harmonic
transmission calculation.

\subsection{Connection to Kramers--Langer theory}
\label{si:kramers-langer-connection}

The Gaussian reactive-flux calculation above and Langer's harmonic escape
current describe the same local saddle dynamics. The connection becomes
explicit when the fluctuation covariances are written as Hessian
determinants. In the preceding translation-free particle coordinates, define
\begin{equation}
 H_{\mathrm s}^*=H_{\mathrm{int}}^*,\qquad
 H_{\mathrm i}^*=
 \left.\nabla_{\bm q^*}\nabla_{\bm q^*}^{\mathsf T}\mathcal U^*
 \right|_{\mathrm i},\qquad
 H_\Sigma^*=L^{\mathsf T}H_{\mathrm s}^*L.
 \label{si:eq:langer-hessians}
\end{equation}
Here \(H_{\mathrm i}^*\) is the potential-energy Hessian at the intact
minimum, \(H_{\mathrm s}^*\) is its saddle counterpart, and \(H_\Sigma^*\)
contains the stable curvatures on the tagged launch surface. Common
translation modes have already been removed.

Let \(\widehat{\bm n}=\bm n/\|\bm n\|\) and
\(\alpha=\widehat{\bm n}^{\mathsf T}\bm u>0\).
Expressing \(H_{\mathrm s}^*\) in the orthonormal basis
\((\widehat{\bm n},L)\) separates the normal and tangential coordinates.
Block inversion rewrites the surface covariance in
Eq.~\eqref{si:eq:harmonic-surface-covariance}; the normal cofactor gives
the determinant ratio. Projecting onto the unstable eigenvector
\(H_{\mathrm s}^*\bm u=h_-^*\bm u\) then gives
\begin{equation}
 \begin{aligned}
 \frac{C_q^*}{\Tstar}
 &=(H_{\mathrm s}^*)^{-1}
 -\frac{(H_{\mathrm s}^*)^{-1}\widehat{\bm n}
        \widehat{\bm n}^{\mathsf T}(H_{\mathrm s}^*)^{-1}}
       {\widehat{\bm n}^{\mathsf T}(H_{\mathrm s}^*)^{-1}
        \widehat{\bm n}},\\
 \widehat{\bm n}^{\mathsf T}(H_{\mathrm s}^*)^{-1}\widehat{\bm n}
 &=\frac{\det H_\Sigma^*}{\det H_{\mathrm s}^*},\\
 \frac{(\sigma_u^*)^2}{\Tstar}
 &=\frac1{h_-^*}
   +\frac{\alpha^2}{(h_-^*)^2}
    \frac{|\det H_{\mathrm s}^*|}{\det H_\Sigma^*}.
 \end{aligned}
 \label{si:eq:langer-covariance-determinants}
\end{equation}
The last line uses the variance definition in
Eq.~\eqref{si:eq:harmonic-launch-projections} and the negative sign of
\(\det H_{\mathrm s}^*\).
The growth roots in Eq.~\eqref{si:eq:harmonic-growth} satisfy
\(r_+^*+r_-^*=-1\) and \(r_+^*r_-^*=ch_-^*\).
Substituting Eq.~\eqref{si:eq:langer-covariance-determinants} into the
amplitude variance of Eq.~\eqref{si:eq:harmonic-amplitude-moments}
and using these root identities gives
\begin{equation}
 \begin{aligned}
 \frac{(\Delta^*)^2(\sigma_A^*)^2}{\Tstar}
 &=\underbrace{c\!\left(1+\frac1{r_+^*}\right)
       +\frac{(r_-^*)^2}{h_-^*}}_{0}
   +\frac{(r_-^*)^2\alpha^2}{(h_-^*)^2}
    \frac{|\det H_{\mathrm s}^*|}{\det H_\Sigma^*}\\
 &=\frac{c^2\alpha^2}{(r_+^*)^2}
    \frac{|\det H_{\mathrm s}^*|}{\det H_\Sigma^*}.
 \end{aligned}
 \label{si:eq:langer-amplitude-reduction}
\end{equation}
The same moment equation gives
\(C_{VA}^*=c\Tstar\|\bm n\|\alpha/\Delta^*\) and
\((\sigma_V^*)^2=c\Tstar\|\bm n\|^2\).
Their substitution into Eq.~\eqref{si:eq:harmonic-transmission-result},
using Eq.~\eqref{si:eq:langer-amplitude-reduction}, yields
\begin{equation}
 \kappa_N^{\mathrm{harm}}
 =\frac{C_{VA}^*}{\sigma_V^*\sigma_A^*}
 =\frac{r_+^*}{\sqrt c}
 \sqrt{\frac{\det H_\Sigma^*}{|\det H_{\mathrm s}^*|}}.
 \label{si:eq:harmonic-langer-identity}
\end{equation}

To connect this coefficient to the escape rate, suppose the TST
thermodynamics is also evaluated harmonically. Let
\(\mathcal U_{\mathrm i}^*\) and \(\mathcal U_{\mathrm s}^*\) be the total
potential energies at the intact minimum and saddle, with
\(\Delta\mathcal U^*=\mathcal U_{\mathrm s}^*-\mathcal U_{\mathrm i}^*\).
Extending the quadratic integrals over their full Gaussian domains in
these internal coordinates gives the intact weight and one tagged-surface
weight:
\begin{equation}
 \begin{aligned}
 Z_{\mathrm i,H}^*
 &= e^{-\mathcal U_{\mathrm i}^*/\Tstar}
   \frac{(2\pi\Tstar)^{(N-1)/2}}{\sqrt{\det H_{\mathrm i}^*}},\\
 \rho_{\Sigma,\mathrm H}^*
 &=\frac{e^{-\mathcal U_{\mathrm s}^*/\Tstar}}{\|\bm n\|}
   \frac{(2\pi\Tstar)^{(N-2)/2}}{\sqrt{\det H_\Sigma^*}}.
 \end{aligned}
 \label{si:eq:langer-harmonic-weights}
\end{equation}
These are the Gaussian fluctuation factors of
Eq.~\eqref{si:eq:constrained-gaussian}, now in orthonormal particle
coordinates. Any common coordinate Jacobian cancels in their ratio.
The surface factor \(1/\|\bm n\|\) comes from fixing the tagged
displacement \(\bm n^{\mathsf T}\bm q^*=0\).
The positive Maxwell factor in Eq.~\eqref{si:eq:maxwell-flux} is
\(\bar v_+^*=\|\bm n\|\sqrt{c\Tstar/(2\pi)}\), which cancels this factor.
Applying the surface-weight/intact-weight construction of
Eq.~\eqref{si:eq:exact-tst} therefore gives
\begin{equation}
 k_{N,\mathrm H}^{\mathrm{TST},*}
 =N\bar v_+^*\frac{\rho_{\Sigma,\mathrm H}^*}{Z_{\mathrm i,H}^*}
 =\frac{N\sqrt c}{2\pi}
 \sqrt{\frac{\det H_{\mathrm i}^*}{\det H_\Sigma^*}}
 e^{-\Delta\mathcal U^*/\Tstar}.
 \label{si:eq:langer-harmonic-tst}
\end{equation}
Multiplying Eqs.~\eqref{si:eq:langer-harmonic-tst} and
\eqref{si:eq:harmonic-langer-identity} cancels both the surface
determinant and \(\sqrt c\):
\begin{equation}
 \begin{aligned}
 k_{N,\mathrm H}^{\mathrm{TST},*}\kappa_N^{\mathrm{harm}}
 &=
 \frac{N\sqrt c}{2\pi}
 \sqrt{\frac{\det H_{\mathrm i}^*}{\det H_\Sigma^*}}
 e^{-\Delta\mathcal U^*/\Tstar}
 \frac{r_+^*}{\sqrt c}
 \sqrt{\frac{\det H_\Sigma^*}{|\det H_{\mathrm s}^*|}}\\
 &=\frac{Nr_+^*}{2\pi}
 \sqrt{\frac{\det H_{\mathrm i}^*}{|\det H_{\mathrm s}^*|}}
 e^{-\Delta\mathcal U^*/\Tstar}.
 \end{aligned}
 \label{si:eq:langer-harmonic-rate}
\end{equation}
This is Langer's harmonic escape rate rewritten in the present reduced variables:
his Eq.~(3.22) supplies the harmonic basin normalization, while Eq.~(3.35)
combines it with saddle fluctuations and unstable growth
\citep{Langer1969}. The factor \(N\) counts the discrete multiplicity of
equivalent fracture sites. The quantity denoted \(|\kappa|\) by Langer
is a dimensional growth rate, equal here to \(\xi r_+^*\);
the transmission coefficient \(\kappa_N^{\mathrm{harm}}\) used here is dimensionless.

The article's harmonic-RF prediction in
Eq.~\eqref{si:eq:harmonic-transmission-result} retains the full-partition
TST factor of Eq.~\eqref{si:eq:production-tst-rate}.
Equation~\eqref{si:eq:langer-harmonic-rate} makes the additional harmonic
approximation to that thermodynamic factor.
Separate Gaussian and determinant evaluations of transmission for
\(N=4\)--30 differ by at most \(5.66\times10^{-15}\), supporting the
derivation and implementation within the harmonic Langevin model.
Accuracy for nonlinear Morse dynamics is assessed separately against
simulation.

\subsection{Symmetric-path KGH approximation}
\label{si:kgh-factor}

The Kramers--Grote--Hynes (KGH) approximation describes escape along one
coordinate, the tagged-bond length. The symmetric path specifies how the
particles move as this bond stretches; the site PMF supplies the equilibrium
free-energy profile along that coordinate. Together, the path mass and
the PMF determine the local well frequency, the unstable growth rate
and the fracture-rate estimate.

On this path, the tagged physical length \(l=l_0l^*\) changes while each
of the \(N-1\) reservoir bonds has length \((L-l)/(N-1)\).
At fixed \(L\), a change \(dl\) therefore shortens each reservoir bond
by \(dl/(N-1)\). Number the particles \(p=0,\ldots,N-1\) along the reservoir
segment. Their physical velocities satisfy
\(v_{p+1}-v_p=-\dot l/(N-1)\), where the dot here denotes differentiation
with respect to physical time. Thus
\(v_p=v_0-p\dot l/(N-1)\). Zero center-of-mass velocity requires
\[
 0=\sum_{p=0}^{N-1}v_p
 =Nv_0-\frac{\dot l}{N-1}\frac{N(N-1)}2,
 \qquad v_0=\frac{\dot l}{2}.
\]
The resulting velocities and kinetic energy are
\begin{equation}
 v_p=\frac{(N-1)/2-p}{N-1}\dot l,
 \qquad
 \frac m2\sum_{p=0}^{N-1}v_p^2
 =\frac{m_{\mathrm{eff}}}{2}\dot l^2.
 \label{si:eq:kgh-path-velocities}
\end{equation}
To evaluate the effective mass in
Eq.~\eqref{si:eq:kgh-path-velocities}, use the sums of consecutive integers
and their squares:
\[
 \begin{aligned}
 \sum_{p=0}^{N-1}\left(p-\frac{N-1}{2}\right)^2
 &=\sum_{p=0}^{N-1}p^2-(N-1)\sum_{p=0}^{N-1}p
   +\frac{N(N-1)^2}{4}\\
 &=\frac{N(N-1)(2N-1)}6-\frac{N(N-1)^2}2
   +\frac{N(N-1)^2}4
 =\frac{N(N^2-1)}{12}.
 \end{aligned}
\]
Substitution gives
\begin{equation}
 \rho_N=\frac{m_{\mathrm{eff}}}{m}
 =\frac{1}{(N-1)^2}\sum_{p=0}^{N-1}\left(p-\frac{N-1}{2}\right)^2
 =\frac{N(N+1)}{12(N-1)}.
 \label{si:eq:kgh-path-mass}
\end{equation}
Each particle experiences friction \(-m\xi v_p\).
Projection onto the same path gives the drag
\(-m\xi\sum_p(v_p/\dot l)^2\dot l=-m_{\mathrm{eff}}\xi\dot l\).
The friction per unit effective mass therefore remains \(\xi\).

This geometry specifies the kinetic model. Its free-energy profile is
the thermally averaged \(G_N^{\mathrm{site},*}\) of
Sec.~\ref{si:sec:profile-pmfs}, which integrates over reservoir
configurations rather than imposing equal sharing on the equilibrium
ensemble. With the reduced variables of Eq.~\eqref{si:eq:units}, the
deterministic part of the one-coordinate Langevin motion is
\begin{equation}
 \frac{d^2l^*}{d(t^*)^2}+\frac{dl^*}{dt^*}
 =-\frac{c}{\rho_N}G_N^{\mathrm{site},(1),*}(l^*).
 \label{si:eq:kgh-path-drift}
\end{equation}
The unit damping coefficient follows from \(t^*=\xi t\); the force
coefficient is \(c/\rho_N=D_e/(m_{\mathrm{eff}}\xi^2l_0^2)\).
Thermal fluctuations enter the escape problem through this Langevin
model and the PMF's equilibrium distribution.

Let \(l_{\mathrm i}^{\mathrm{site},*}\) and
\(l_{\mathrm s}^{\mathrm{site},*}\) denote the intact minimum and
maximum of the site PMF. Its local curvatures and dimensionless barrier are
\begin{equation}
 \begin{gathered}
 \mathcal K_{\mathrm i}^{*}=G_{N}^{\mathrm{site},(2),*}(l_{\mathrm i}^{\mathrm{site},*})>0,
 \qquad \mathcal K_{\mathrm s}^{*}=G_{N}^{\mathrm{site},(2),*}(l_{\mathrm s}^{\mathrm{site},*})<0,\\
 B_N=\frac{G_{N}^{\mathrm{site},*}(l_{\mathrm s}^{\mathrm{site},*})-G_{N}^{\mathrm{site},*}(l_{\mathrm i}^{\mathrm{site},*})}{\Tstar}.
 \end{gathered}
 \label{si:eq:kgh-pmf-parameters}
\end{equation}
The curvatures are obtained from local fourth-degree polynomial fits to
the computed PMF within \(0.03\) in reduced bond length of each grid
extremum, as described in Sec.~\ref{si:numerical-site-barriers}.
Each polynomial's second derivative is evaluated at its nearby stationary
point. For profiles stored as \(G_N^{\mathrm{site},*}/\Tstar\), multiplying
that derivative by \(\Tstar\) restores the curvature convention in
Eq.~\eqref{si:eq:kgh-pmf-parameters}.

Near either extremum, the first derivative vanishes, so the leading
free-energy change is quadratic. Writing
\(z_{\mathrm i}^*=l^*-l_{\mathrm i}^{\mathrm{site},*}\) and
\(z^*=l^*-l_{\mathrm s}^{\mathrm{site},*}\) gives
\[
 \begin{aligned}
 G_N^{\mathrm{site},*}(l_{\mathrm i}^{\mathrm{site},*}+z_{\mathrm i}^*)
 &\simeq G_N^{\mathrm{site},*}(l_{\mathrm i}^{\mathrm{site},*})
       +\tfrac12\mathcal K_{\mathrm i}^*(z_{\mathrm i}^*)^2,\\
 G_N^{\mathrm{site},*}(l_{\mathrm s}^{\mathrm{site},*}+z^*)
 &\simeq G_N^{\mathrm{site},*}(l_{\mathrm s}^{\mathrm{site},*})
       -\tfrac12|\mathcal K_{\mathrm s}^*|(z^*)^2.
 \end{aligned}
\]
Substituting these local forces into Eq.~\eqref{si:eq:kgh-path-drift}
identifies the frequencies
\begin{equation}
 \omega_{\mathrm i}^*=\sqrt{\frac{c\mathcal K_{\mathrm i}^{*}}{\rho_N}},
 \qquad
 \omega_{\mathrm s}^*=\sqrt{\frac{c|\mathcal K_{\mathrm s}^{*}|}{\rho_N}}.
 \label{si:eq:kgh-frequencies}
\end{equation}
The first is the undamped small-oscillation frequency in the well; the
second measures the magnitude of the unstable barrier curvature.
The corresponding local equations are
\begin{equation}
 \frac{d^2z_{\mathrm i}^*}{d(t^*)^2}
 +\frac{dz_{\mathrm i}^*}{dt^*}
 +(\omega_{\mathrm i}^*)^2z_{\mathrm i}^*=0,
 \qquad
 \frac{d^2z^*}{d(t^*)^2}
 +\frac{dz^*}{dt^*}
 -(\omega_{\mathrm s}^*)^2z^*=0.
 \label{si:eq:kgh-local-dynamics}
\end{equation}
At the barrier, substituting \(z^*\propto e^{\gamma_+^*t^*}\) into
Eq.~\eqref{si:eq:kgh-local-dynamics} gives
\begin{equation}
 (\gamma_+^*)^2+\gamma_+^*-(\omega_{\mathrm s}^*)^2=0,
 \qquad
 \gamma_+^*=\sqrt{\tfrac14+(\omega_{\mathrm s}^*)^2}-\tfrac12.
 \label{si:eq:kgh-growth-rate}
\end{equation}
Without friction, the growing solution has rate \(\omega_{\mathrm s}^*\).
The Markovian Grote--Hynes transmission factor is the ratio
\(\gamma_+^*/\omega_{\mathrm s}^*\)
\citep{Kramers1940,GroteHynes1980}.

The attempt frequency follows from the harmonic well population.
Extending the local quadratic approximation over the real line and
using the Gaussian integral in Eq.~\eqref{si:eq:one-dimensional-gaussian}
gives
\begin{equation}
 \begin{aligned}
 \int_{\mathrm{well}}e^{-G_N^{\mathrm{site},*}(l^*)/\Tstar}\,dl^*
 &\simeq
 e^{-G_N^{\mathrm{site},*}(l_{\mathrm i}^{\mathrm{site},*})/\Tstar}
 \int_{-\infty}^{\infty}
 e^{-\mathcal K_{\mathrm i}^*(z_{\mathrm i}^*)^2/(2\Tstar)}
 \,dz_{\mathrm i}^*\\
 &=
 e^{-G_N^{\mathrm{site},*}(l_{\mathrm i}^{\mathrm{site},*})/\Tstar}
 \sqrt{\frac{2\pi\Tstar}{\mathcal K_{\mathrm i}^*}}.
 \end{aligned}
 \label{si:eq:kgh-harmonic-well}
\end{equation}
Within this path model, the reduced velocity has Maxwell variance
\(c\Tstar/\rho_N\), so its positive-velocity average is
\(\sqrt{c\Tstar/(2\pi\rho_N)}\). Multiplying this factor by the surface
weight, dividing by Eq.~\eqref{si:eq:kgh-harmonic-well}, and including
the \(N\) equivalent fracture sites gives the harmonic path TST estimate
\begin{equation}
 k_{N,\mathrm{path}}^{\mathrm{TST},*}
 \simeq
 N\sqrt{\frac{c\Tstar}{2\pi\rho_N}}\,
 \frac{e^{-G_N^{\mathrm{site},*}(l_{\mathrm s}^{\mathrm{site},*})/\Tstar}}
 {e^{-G_N^{\mathrm{site},*}(l_{\mathrm i}^{\mathrm{site},*})/\Tstar}
  \sqrt{2\pi\Tstar/\mathcal K_{\mathrm i}^*}}
 =N\frac{\omega_{\mathrm i}^*}{2\pi}e^{-B_N}.
 \label{si:eq:kgh-path-tst}
\end{equation}
The equality uses the barrier and frequency definitions in
Eqs.~\eqref{si:eq:kgh-pmf-parameters} and
\eqref{si:eq:kgh-frequencies}. Thus the well curvature supplies the attempt
frequency \(\omega_{\mathrm i}^*/(2\pi)\), while friction reduces the
transmission by
\begin{equation}
 \chi_N=\frac{\gamma_{+}^{*}}{\omega_{\mathrm s}^*}
 =\frac{\sqrt{1/4+(\omega_{\mathrm s}^*)^2}-1/2}{\omega_{\mathrm s}^*},
 \qquad A_{N}^{\mathrm{KGH},*}=\frac{\omega_{\mathrm i}^*}{2\pi}\chi_N.
 \label{si:eq:kgh-prefactor}
\end{equation}
Multiplying Eq.~\eqref{si:eq:kgh-path-tst} by \(\chi_N\) yields
\begin{equation}
 k_N^{\mathrm{KGH},*}=N A_{N}^{\mathrm{KGH},*}e^{-B_N}.
 \label{si:eq:kgh-rate}
\end{equation}
KGH therefore combines site-PMF thermodynamics with harmonic well
normalization and one-coordinate dynamics. The full-partition TST
factor of Eq.~\eqref{si:eq:production-tst-rate} uses a different kinetic
metric and intact-population normalization. The barrier and curvatures
are evaluated at each state; no prefactor is fitted to simulation rates.

%% file: si_sections/equilibrium_simulation_rates.tex
\section{Rates from equilibrium simulations}
\label{si:simulation}

\subsection{Systems and trajectory design}
\label{si:equilibrium-design}

The equilibrium simulations cover \(N=4,\ldots,12\) and \(15\), with five extensions near
coexistence for each chain length. For \(N=5,\ldots,12\), the extensions
target constrained-partition fractured probabilities of 10\%, 25\%, 50\%,
75\% and 90\%; at \(N=4\), the lowest target is 17.5\%.
For \(N=15\), five equally spaced extensions around coexistence give
approximately 11.7\%, 26.6\%, 50.0\%, 73.6\% and 88.7\%.
Table~\ref{si:tab:trajectory-design} lists the actual simulated extensions.

Half the independent trajectories at
each state start from the intact stationary configuration and half from a
fractured stationary configuration. The initially long bond is distributed
cyclically among the chain sites. Initial particle velocities are Gaussian
at the specified temperature, with transverse components set to zero and
the center-of-mass velocity removed
(\texttt{velocity create}, \texttt{dist gaussian},
\texttt{velocity all zero linear}). Subsequent dynamics follows
Sec.~\ref{si:langevin-dynamics}.

The equilibration segment is discarded before collecting tension and
maximum-bond observations. Table~\ref{si:tab:trajectory-design} gives the
simulated extensions, trajectory allocations, trajectory lengths, recording
intervals and primary ACF fitting windows for all 50 states.

{\small
\setlength{\tabcolsep}{4pt}
\begin{longtable}{rrrrrrr}
  \caption{Equilibrium sampling at each simulated extension \(\lambda\).
  Trajectory counts and primary ACF fitting intervals are given for each state.
  Equilibration steps, production steps and recording stride on the first row
  of each \(N\) group apply to all five extensions in that group.
  The timestep is \(0.25\,\mathrm{fs}\); the recording stride is the number
  of timesteps between observations, and fitting intervals use reduced time \(t^*\).}
  \label{si:tab:trajectory-design}\\
  \toprule
  \input{publication_tables/equilibrium_sampling/table.tex}\end{longtable}
}

For the equilibrium comparison in main-text Figure~4, the fractured
probability is the fraction of all recorded observations classified as
fractured, and the intact-state virial tension is averaged over intact
observations only. Whole trajectories are resampled with replacement within
their initial-state groups, preserving each group's size and recomputing
both observables in each sample. The 2.5th and 97.5th percentiles give the
pointwise 95\% sampling intervals; these equilibrium estimates use all
trajectories, regardless of whether the kinetic fits succeed.

\subsection{Pooled tension autocorrelation}
\label{si:acf-method}

In the two-state description, the populations evolve according to
\begin{equation}
 \frac{dP_{\mathrm i}}{dt^*}
 =-k_{\mathrm i\to f}^*P_{\mathrm i}
 +k_{\mathrm f\to i}^*P_{\mathrm f},
 \qquad k_{\mathrm eff}^*=k_{\mathrm i\to f}^*+k_{\mathrm f\to i}^*.
 \label{si:eq:master}
\end{equation}
Using \(P_{\mathrm f}=1-P_{\mathrm i}\), a departure from equilibrium
decays as \(e^{-k_{\mathrm eff}^*t^*}\). Once fluctuations within each
basin have relaxed, the tension retains this slower contribution because
its mean differs between the two states.

At a fixed \((N,\lambda)\), let \(F_{rj}^{*}=F_{rj}l_0/\De\) be the reduced virial tension at
recorded sample \(j=0,\ldots,n_r-1\) of trajectory \(r\). All trajectories
are centered with the same pooled mean \(\bar F^{*}\). For integer lag
\(0\le m<\min_r n_r\), the mean and covariance estimate are
\begin{equation}
 \begin{aligned}
 \bar F^{*}&=\frac{\sum_r\sum_{j=0}^{n_r-1}F_{rj}^{*}}{\sum_r n_r},\\
 \widehat\Gamma_{m}^{*}&=
 \frac{\sum_r\sum_{j=0}^{n_r-m-1}
 (F_{rj}^{*}-\bar F^{*})(F_{r,j+m}^{*}-\bar F^{*})}{\sum_r(n_r-m)}.
 \end{aligned}
 \label{si:eq:pooled-acf}
\end{equation}
Each product pairs two observations from the same trajectory; the denominator
counts the available pairs. The covariance sums are evaluated by FFT with
zero-padding. The reduced covariance satisfies
\(\widehat\Gamma_m^*=(l_0/\De)^2\widehat\Gamma_m\), so the force
scale cancels from the normalized ACF. At lag \(t^*=m\Delta t_{\mathrm{obs}}^*\), where
\(\Delta t_{\mathrm{obs}}^*\) is the recording interval,
\(\widehat C_{FF}(t^*)=\widehat\Gamma_{m}^{*}/\widehat\Gamma_{0}^{*}\) estimates \citep{Theodorou2024}
\begin{equation}
 C_{FF}(t^*)=\frac{\langle\delta F^{*}(0)\delta F^{*}(t^*)\rangle}
 {\langle\bigl(\delta F^{*}\bigr)^{2}\rangle}
 \simeq C_0e^{-k_{\mathrm eff}^*t^*}.
 \label{si:eq:acf}
\end{equation}
Here \(\delta F^{*}\) denotes a fluctuation about the equilibrium mean.
The fitted amplitude \(C_0\) describes the slow component and need not
be one, although the complete normalized ACF equals one at zero lag.

Within the specified interval for each state,
\(\ln\widehat C_{FF}(t^*)=\ln C_0-k_{\mathrm eff}^*t^*\) is fitted by unweighted
linear least squares. The fit uses finite ACF values above \(10^{-8}\)
and requires at least three points and a negative slope. The state-specific
intervals are listed in Table~\ref{si:tab:trajectory-design}. Alternative fitting intervals
and the state-indicator ACF test the sensitivity of this inference. Figure~\ref{si:fig:acf-relaxation}
shows the ACFs and fitted intervals at sampled near-coexistence states.

The instantaneous fracture indicator \(h_{\mathrm f,rj}\) equals one
when the maximum reduced bond length exceeds \(d_{N}^{*}\), and zero otherwise.
Its pooled average,
\[
 P_{\mathrm f}^{\mathrm{sim}}
 =\frac{\sum_r\sum_{j=0}^{n_r-1}h_{\mathrm f,rj}}{\sum_r n_r},
\]
estimates the equilibrium fractured population. Detailed balance,
\(P_{\mathrm i}k_{\mathrm i\to f}^*=P_{\mathrm f}k_{\mathrm f\to i}^*\),
then gives the directional rates from the fitted decay rate \citep{Theodorou2024}
\begin{equation}
 k_{\mathrm i\to f}^*=P_{\mathrm f}^{\mathrm{sim}}k_{\mathrm eff}^*,
 \qquad
 k_{\mathrm f\to i}^*=(1-P_{\mathrm f}^{\mathrm{sim}})k_{\mathrm eff}^*.
 \label{si:eq:acf-rates}
\end{equation}
The measured population is used at every state, including those near
coexistence; setting it to one half would impose equal directional rates.

Sampling uncertainty is estimated with 5000 whole-trajectory bootstrap
samples. Trajectories are drawn with replacement within their initial-state
groups, preserving each group's size and every trajectory's linked
observations \citep{FieldWelsh2007}. The pooled mean, ACF, fitted decay and
population split are recomputed for each sample using the same fitting interval. The 2.5th and 97.5th
percentiles of successful fits give the 95\% sampling intervals, retaining
the covariance between population and decay estimates. These intervals do
not include sensitivity to the choice of fitting window.

Sensitivity is assessed by fitting both tension and state-indicator
ACFs over four intervals and comparing their decay rates with the primary
tension-ACF estimate. Besides the primary interval
\([t_{\min}^*,t_{\max}^*]\), the alternatives are
\([0.8t_{\min}^*,t_{\max}^*]\),
\([t_{\min}^*,1.2t_{\max}^*]\) and
\([1.2t_{\min}^*,1.2t_{\max}^*]\).
Short-chain estimates are more sensitive to these choices, particularly
at \(N=4\)--5. Table~\ref{si:tab:campaign-flags}
summarizes the states with larger sampling uncertainty, observable/window
dependence or an unsuccessful alternative fit; all remain in the reported
data. At \(N=4\), one alternative indicator-ACF fit does not give a
decaying exponential, and two bootstrap fits fail. The sampling intervals
above are conditional on successful fits. Interpolating rates to coexistence
does not remove sensitivity at the contributing sampled states.

\begin{table}[H]
\centering\small
\caption{Sampling uncertainty and observable/window sensitivity of the ACF rates.
Listed states have a relative 95\% interval half-width above 10\%, a decay-rate
change above 5\%, or an unsuccessful alternative fit. The third column is the
largest interval width divided by twice the rate estimate across the forward,
reverse and population-decay rates. The fourth is the largest absolute
relative change from the primary tension-ACF decay rate among successful
fits to either observable over the four intervals. Both are percentages.}
\label{si:tab:campaign-flags}
\begin{tabular}{rrrr}\toprule
\(N\) & \(\lambda\) & CI half-width (\%) & Observable/window change (\%)\\\midrule
\RegisteredCampaignFlagRows
\bottomrule\end{tabular}
\end{table}

\begin{figure}[p]
  \centering
  \includegraphics[width=16cm]{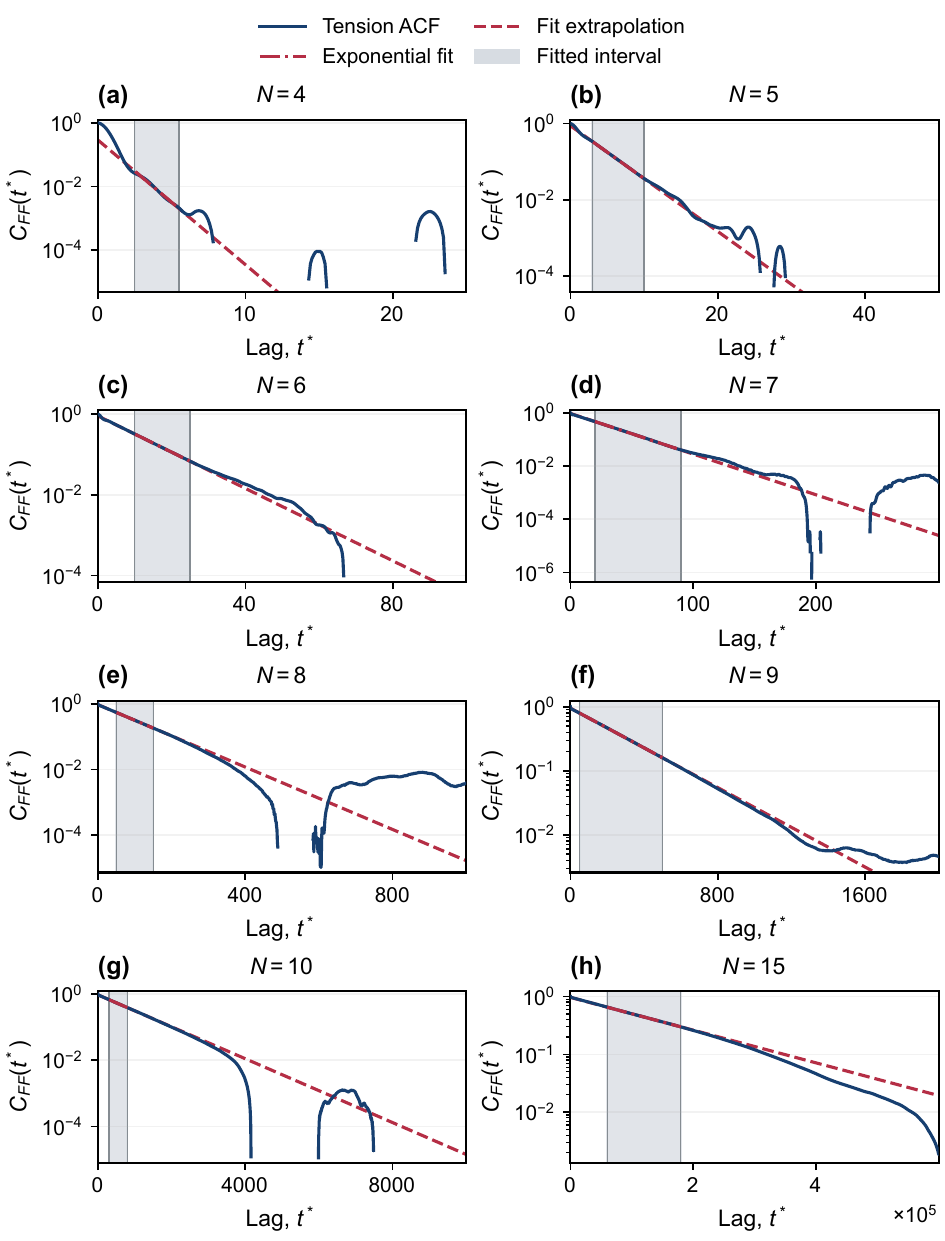}
  \input{figures/acf_relaxation/caption.tex}
\end{figure}

\FloatBarrier

\subsection{Last-core residence estimator}
\label{si:core-method}

The analysis uses the maximum-bond trace \(l_{\max}^{*}(t^*)=\max_jl_{j}^{*}(t^*)\) from the same
equilibrium trajectories as the tension ACF. Following the state convention
of Sec.~\ref{si:notation}, the core boundaries are
\begin{equation}
 c_{b}^{*}(\eta)=l_{b}^{\mathrm{ad},*}+\eta\bigl(d_{N}^{*}-l_{b}^{\mathrm{ad},*}\bigr),
 \qquad b\in\{\mathrm i,\mathrm f\}.
 \label{si:eq:residence-cores}
\end{equation}
Here \(l_{\mathrm i}^{\mathrm{ad},*}=\lambda\) and
\(l_{\mathrm f}^{\mathrm{ad},*}\) are the intact and fractured minima of the reduced profile
\(E_{\mathrm{ad}}^{(N-1),*}\), and \(d_{N}^{*}\) is its intervening saddle.
The intact core has \(l_{\max}^{*}\le c_{\mathrm i}^{*}\), and the fractured core has
\(l_{\max}^{*}\ge c_{\mathrm f}^{*}\). The primary analysis uses \(\eta=0.5\) and is repeated
with 0.25 and 0.75, placing the boundaries deeper within the basins and
closer to the saddle, respectively.

The recorded label changes when the trajectory first reaches the opposite
core and remains unchanged between cores. Consecutive samples with the
same label form a residence. Its start and end are the observed entry
times into successive opposite cores, so an excursion returning to the
same core remains part of that residence.

At the beginning of a record, the instantaneous divider sets the initial
label if the first sample lies between the cores. The initial
residence and its terminating switch are omitted because its entry time is unobserved.
Subsequent residences have observed entries. A residence still in progress
at the end contributes time up to the last recorded sample, but no completed
switch. A trajectory with no label change contributes neither count nor
time under this convention. Times are assigned at the recorded observations,
with the recording intervals given in Table~\ref{si:tab:trajectory-design}.

For each retained residence \(\alpha\), let \(D_\alpha^*\) be its observed
duration and \(e_\alpha=1\) if it ends in a recorded switch, or zero if
the record ends before a switch is observed. Let \(\mathcal R_b\) collect retained
residences with label \(b\), pooled over trajectories. For the opposite
label \(\bar b\), the exposure and transition count are
\[
 \mathcal T_{b}^{\mathrm{LC},*}=\sum_{\alpha\in\mathcal R_b}D_\alpha^*,
 \qquad N_{b\to\bar b}=\sum_{\alpha\in\mathcal R_b}e_\alpha.
\]
Dividing each transition count by its corresponding exposure gives the directional residence rates \citep{Schutte2011},
\begin{equation}
 k_{\mathrm i\to\mathrm f}^{\mathrm{core},*}
 =\frac{N_{\mathrm i\to\mathrm f}}{\mathcal T_{\mathrm i}^{\mathrm{LC},*}},
 \qquad
 k_{\mathrm f\to\mathrm i}^{\mathrm{core},*}
 =\frac{N_{\mathrm f\to\mathrm i}}{\mathcal T_{\mathrm f}^{\mathrm{LC},*}}.
 \label{si:eq:core-residence-rate}
\end{equation}
For an exponential residence-time model, these are the maximum-likelihood
estimates: completed residences contribute a transition and a duration,
while unfinished terminal residences contribute only their observed duration.

The rate comparisons use approximate Poisson counting errors,
\[
 \operatorname{SE}\!\left(k_{b\to\bar b}^{\mathrm{core},*}\right)
 \approx\frac{\sqrt{N_{b\to\bar b}}}{\mathcal T_b^{\mathrm{LC},*}},
 \qquad
 k_{b\to\bar b}^{\mathrm{core},*}\pm
 1.96\,\operatorname{SE}\!\left(k_{b\to\bar b}^{\mathrm{core},*}\right),
\]
where the second expression gives the approximate 95\% interval. This
counting convention differs from the ACF trajectory-bootstrap interval
and does not itself establish exponential residence statistics.

For a full residence duration \(D_b^*\) in state \(b\), the survival
probability \(S_b(t^*)=\Pr(D_b^*>t^*)\) is the probability of remaining
in that residence beyond \(t^*\). The Kaplan--Meier
construction \citep{KaplanMeier1958} estimates this survival probability,
omitting initial incomplete residences and retaining terminal
right-censoring as above. At each observed exit time, survival decreases by
the fraction exiting among the residences still at risk; censored residences
leave the risk set at the end of their observed duration.
A constant directional rate predicts
\[
 S_b(t^*)=\exp[-k_{b\to\bar b}^{\mathrm{core},*}t^*].
\]
\begin{figure}[p]
  \centering
  \includegraphics[width=16cm]{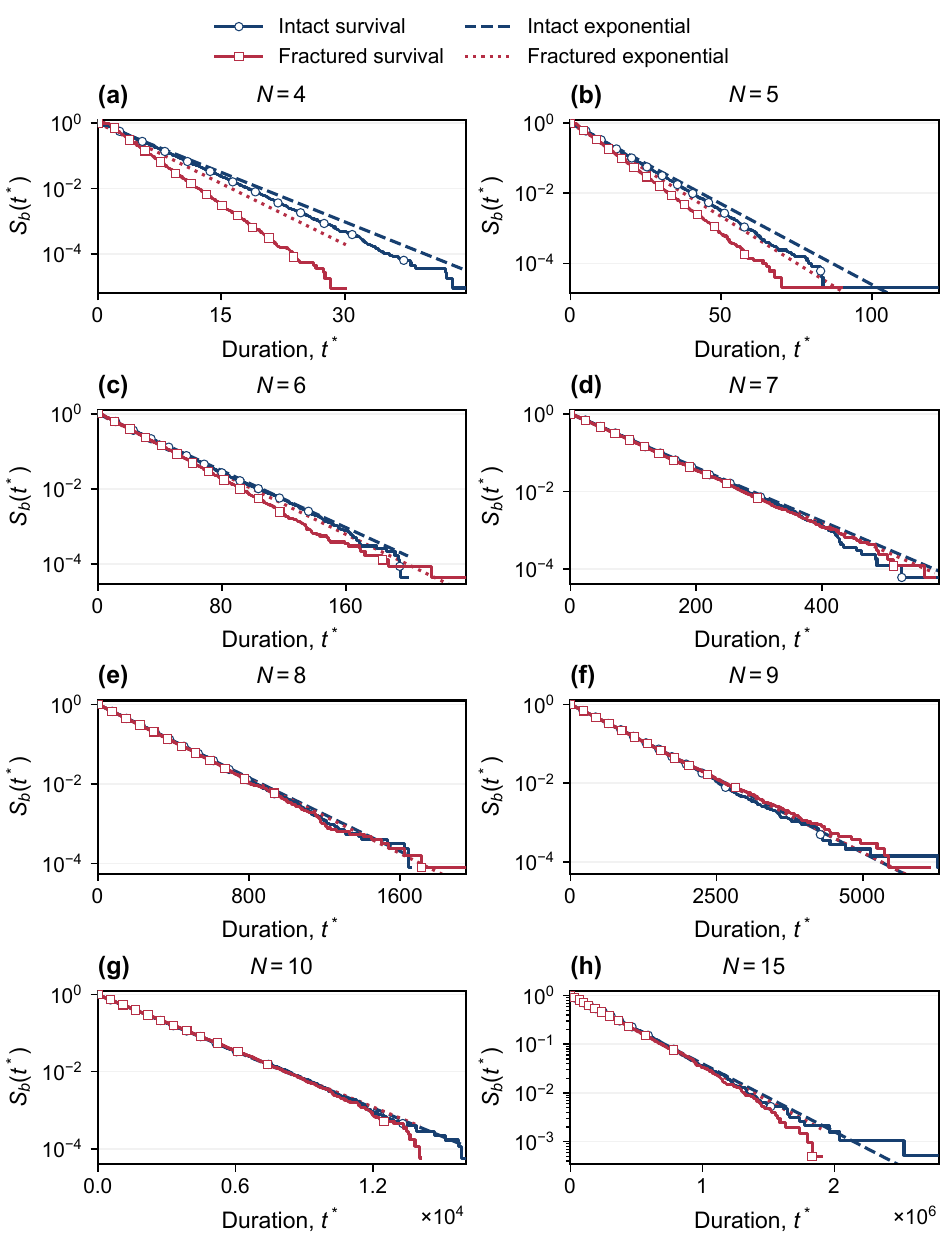}
  \input{figures/core_residence_survival/caption.tex}
\end{figure}

\FloatBarrier

The exponential comparisons use the count/exposure rates in
\cref{si:eq:core-residence-rate}, not ACF rates or separate fits to survival.
With negligible censoring, \(1/k_{b\to\bar b}^{\mathrm{core},*}\)
approximates the mean residence duration, but matching that mean does not
determine the distribution's shape. The negative slope
\(-d\ln S_b/dt^*\) is the departure rate conditional on having remained
in state \(b\) until \(t^*\); a changing slope reveals dependence on
residence age.

In Figure~\ref{si:fig:residence-survival}, at the sampled near-coexistence
state for \(N=4\), survival initially
exceeds the corresponding exponential and subsequently falls below it.
Thus one constant rate does not describe the entire residence distribution;
a steeper tail should not simply replace the overall count/exposure rate.
Finite travel between cores and memory of the entry configuration are
plausible contributors, but this comparison alone does not establish either
mechanism. The extreme tails also contain few observations.

\subsection{Comparison with the ACF estimate}
\label{si:core-acf-comparison}
The residence estimator uses time carrying a last-core label; the ACF rate
in \cref{si:eq:acf-rates} uses the instantaneous population and the slow
tension decay. For each \(N\), the directly sampled rates are compared at the
same extension nearest coexistence. The two estimates share trajectories
and therefore have correlated sampling errors. Their ratios are reported
as point estimates; joint-bootstrap intervals were not calculated.

\begin{figure}[htbp]
  \centering
  \includegraphics[width=16cm]{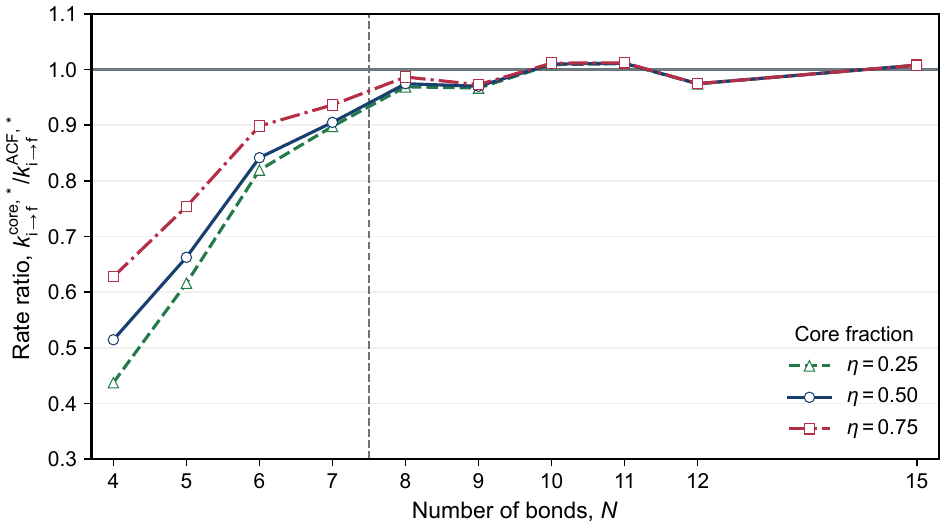}
  \caption{Directly sampled core-residence/ACF rate ratios at the same
  near-coexistence extension for each \(N=4,\ldots,12,15\), with
  \(\eta=0.25,0.50,0.75\). The horizontal line marks equal rates.
  The dashed divider at \(N=7.5\) indicates the operational \(N\geq8\)
  convention for robust two-state kinetics under the studied conditions
  (Sec.~\ref{si:sec:crossover-uncertainty}). Ratios are point estimates.}
  \label{si:fig:core-residence-acf}
\end{figure}

For \(\eta=0.5\), the core/ACF ratio spans
\RegisteredActivatedCoreMin--\RegisteredActivatedCoreMax over
\(N=7,\ldots,12,15\).
Figure~\ref{si:fig:core-residence-acf} also compares the three core fractions,
all evaluated on the same trajectories. Figures~\ref{si:fig:acf-relaxation}
and~\ref{si:fig:residence-survival} show the ACFs and residence survival
at sampled near-coexistence states.

%% file: publication_tables/equilibrium_sampling/table.tex
\(N\) & \(\lambda\) & Trajectories & \shortstack{Equilibration\\steps} & \shortstack{Production\\steps} & \shortstack{Recording\\stride} & \shortstack{ACF fit interval\\in \(t^*\)} \\
\midrule
\endfirsthead
\multicolumn{7}{c}{\tablename~\thetable\ (continued)} \\
\toprule
\(N\) & \(\lambda\) & Trajectories & \shortstack{Equilibration\\steps} & \shortstack{Production\\steps} & \shortstack{Recording\\stride} & \shortstack{ACF fit interval\\in \(t^*\)} \\
\midrule
\endhead
\midrule
\multicolumn{7}{r}{Continued on next page} \\
\endfoot
\bottomrule
\endlastfoot
4 & 1.162521 & 96 & \(10\,000\) & \(2\,000\,000\) & \(20\) & \(3\text{--}5.5\) \\*
 & 1.163354 & 80 &  &  &  & \(3\text{--}5.5\) \\*
 & 1.166917 & 80 &  &  &  & \(2.5\text{--}5.5\) \\*
 & 1.170934 & 80 &  &  &  & \(2.5\text{--}5.5\) \\*
 & 1.174530 & 112 &  &  &  & \(3\text{--}7\) \\
\addlinespace[5pt]
5 & 1.146069 & 96 & \(10\,000\) & \(2\,000\,000\) & \(20\) & \(3\text{--}7\) \\*
 & 1.148977 & 80 &  &  &  & \(3\text{--}7\) \\*
 & 1.152084 & 80 &  &  &  & \(3\text{--}10\) \\*
 & 1.154965 & 80 &  &  &  & \(3\text{--}10\) \\*
 & 1.157555 & 112 &  &  &  & \(2\text{--}10\) \\
\addlinespace[5pt]
6 & 1.135026 & 96 & \(500\,000\) & \(2\,000\,000\) & \(40\) & \(5\text{--}10\) \\*
 & 1.137388 & 96 &  &  &  & \(10\text{--}25\) \\*
 & 1.139610 & 96 &  &  &  & \(10\text{--}25\) \\*
 & 1.141662 & 96 &  &  &  & \(10\text{--}25\) \\*
 & 1.143582 & 128 &  &  &  & \(10\text{--}25\) \\
\addlinespace[5pt]
7 & 1.125874 & 96 & \(500\,000\) & \(4\,000\,000\) & \(40\) & \(20\text{--}40\) \\*
 & 1.127636 & 96 &  &  &  & \(20\text{--}40\) \\*
 & 1.129294 & 96 &  &  &  & \(20\text{--}90\) \\*
 & 1.130869 & 96 &  &  &  & \(20\text{--}90\) \\*
 & 1.132377 & 160 &  &  &  & \(20\text{--}90\) \\
\addlinespace[5pt]
8 & 1.118008 & 128 & \(200\,000\) & \(8\,000\,000\) & \(200\) & \(40\text{--}100\) \\*
 & 1.119395 & 112 &  &  &  & \(50\text{--}150\) \\*
 & 1.120725 & 112 &  &  &  & \(50\text{--}150\) \\*
 & 1.122008 & 160 &  &  &  & \(50\text{--}150\) \\*
 & 1.123257 & 288 &  &  &  & \(50\text{--}150\) \\
\addlinespace[5pt]
9 & 1.111231 & 224 & \(200\,000\) & \(16\,000\,000\) & \(400\) & \(50\text{--}150\) \\*
 & 1.112388 & 176 &  &  &  & \(50\text{--}500\) \\*
 & 1.113512 & 192 &  &  &  & \(50\text{--}500\) \\*
 & 1.114608 & 272 &  &  &  & \(50\text{--}500\) \\*
 & 1.115680 & 496 &  &  &  & \(50\text{--}500\) \\
\addlinespace[5pt]
10 & 1.105368 & 192 & \(10\,000\,000\) & \(30\,000\,000\) & \(400\) & \(200\text{--}800\) \\*
 & 1.106372 & 160 &  &  &  & \(300\text{--}1000\) \\*
 & 1.107354 & 400 &  &  &  & \(300\text{--}800\) \\*
 & 1.108319 & 256 &  &  &  & \(200\text{--}1500\) \\*
 & 1.109268 & 448 &  &  &  & \(500\text{--}1500\) \\
\addlinespace[5pt]
11 & 1.100259 & 176 & \(4\,000\,000\) & \(40\,000\,000\) & \(2\,000\) & \(700\text{--}2000\) \\*
 & 1.101155 & 144 &  &  &  & \(1000\text{--}4000\) \\*
 & 1.102034 & 512 &  &  &  & \(1500\text{--}4500\) \\*
 & 1.102903 & 208 &  &  &  & \(2500\text{--}5000\) \\*
 & 1.103759 & 368 &  &  &  & \(3000\text{--}5000\) \\
\addlinespace[5pt]
12 & 1.095770 & 192 & \(30\,000\,000\) & \(70\,000\,000\) & \(4\,000\) & \(3000\text{--}8000\) \\*
 & 1.096583 & 160 &  &  &  & \(4000\text{--}9000\) \\*
 & 1.097387 & 512 &  &  &  & \(5000\text{--}10000\) \\*
 & 1.098180 & 240 &  &  &  & \(5000\text{--}10000\) \\*
 & 1.098962 & 432 &  &  &  & \(5000\text{--}10000\) \\
\addlinespace[5pt]
15 & 1.085155884 & 128 & \(400\,000\,000\) & \(2\,000\,000\,000\) & \(20\,000\) & \(60000\text{--}180000\) \\*
 & 1.085755884 & 128 &  &  &  & \(60000\text{--}180000\) \\*
 & 1.086355884 & 128 &  &  &  & \(60000\text{--}180000\) \\*
 & 1.086955884 & 128 &  &  &  & \(60000\text{--}180000\) \\*
 & 1.087555884 & 128 &  &  &  & \(60000\text{--}180000\) \\

%% file: figures/acf_relaxation/caption.tex
\caption{Tension autocorrelation at sampled near-coexistence states for
\(N=4,5,6,7,8,9,10,15\) (panels a--h). The respective extensions are
\(1.166917\), \(1.152084\), \(1.139610\), \(1.129294\), \(1.120725\),
\(1.113512\), \(1.107354\), and \(1.0863558839711749\).
Blue curves show the tension ACF; red dash-dotted segments show its fitted
exponential within the shaded intervals, and red dashes extrapolate the same
fit. Nonpositive ACF values are omitted on the logarithmic axes, leaving gaps;
all positive observations and the full lag ranges are retained.}
\label{si:fig:acf-relaxation}

%% file: figures/core_residence_survival/caption.tex
\caption{Intact and fractured residence survival for \(N=4,5,6,7,8,9,10,15\)
(panels a--h), at the same near-coexistence states as
Figure~\ref{si:fig:acf-relaxation}, with \(\eta=0.5\).
\(S_b(t^*)\) is the probability that a residence lasts longer than \(t^*\).
Steps show Kaplan--Meier estimates; dashed and dotted exponentials use the
corresponding directional core-residence count/exposure rates, without fitting
survival curves. Initial incomplete residences are omitted and terminal
residences are right-censored. Zero survival endpoints are omitted on the
logarithmic axes; all positive observations and full duration ranges are retained.}
\label{si:fig:residence-survival}

%% file: si_sections/barrier_top_simulations.tex
\section{Rates from barrier-top simulations}
\label{si:barrier-top}

\subsection{Surface ensembles and trajectory design}
\label{si:rf-trajectory-design}

At coexistence, the adiabatic surface \(s^{*}=d_{N}^{*}\) is sampled for
\(N=7\)--12, 15, and 20. One tagged bond is fixed at \(s^{*}\), while
all other bonds lie below \(s^{*}\) and share the remaining length
\(N\lambda-s^{*}\). This is one of the equivalent components of
\(\Sigma_{s^*}\) in \cref{si:eq:rf-launch-surface}. The remaining bond lengths
are drawn sequentially using the restricted partition functions: each
possible length is weighted by its Boltzmann factor and the partition
function of the bonds still to be drawn, with the remaining total length
updated after each draw. Multiplying the cumulative sums of reduced bond lengths by \(l_0\) gives the
physical particle positions in the periodic box.

Each chain length uses 2048 independently sampled surface configurations.
For each configuration, Gaussian particle velocities are drawn at temperature
\(T\), their mean is subtracted, and the pair \(+\bm v,-\bm v\) is launched.
These antithetic launches share positions but have opposite velocities;
their subsequent thermostat noise is independent. The resulting 4096
trajectories follow the Langevin dynamics of Sec.~\ref{si:langevin-dynamics},
using \texttt{fix nve} and \texttt{fix langevin} without resetting the
prepared velocities. The timestep is \(0.25\,\mathrm{fs}\), each trajectory
lasts 1500 fs, and observations are recorded every 1 fs. The equilibrium
timestep checks are given in Sec.~\ref{si:corrected-timestep}.
The displaced-surface \(N=10\) control uses the same preparation principle;
its surfaces and larger sampling allocation are specified in
Sec.~\ref{si:sec:fixed-surface-control}.

\subsection{Transmission and rate estimation}
\label{si:rf-estimation}

For configuration \(a\) and velocity sign \(\sigma=\pm1\), let
\(v_{l,a\sigma}^{*}(0)\) be the initial reduced tagged velocity and
\(h_{\mathrm f,a\sigma}(t^*)=\mathbf1[\max_jl_{j,a\sigma}^{*}(t^*)>d_{N}^{*}]\)
the corresponding trajectory's instantaneous fractured-state indicator. With
\(n_{\mathrm{conf}}=2048\), the trajectory estimate of
Eq.~\eqref{si:eq:rf} is
\begin{equation}
 \widehat\kappa_N(t^*)=
 \frac{1}{2n_{\mathrm{conf}}\bar v_+^*}
 \sum_{a=1}^{n_{\mathrm{conf}}}\sum_{\sigma=\pm1}
 v_{l,a\sigma}^{*}(0)h_{\mathrm f,a\sigma}(t^*),
 \qquad \bar v_+^*=\sqrt{c\Tstar/\pi}.
 \label{si:eq:rf-trajectory-average}
\end{equation}
The factor \(2n_{\mathrm{conf}}\) counts all launches. Both velocity
signs enter the same average, and a trajectory that has returned to the
intact state contributes zero at that observation time. The denominator
is the analytical Maxwell factor in Eq.~\eqref{si:eq:maxwell-flux}.
Because \(v_{l,a,-}^{*}(0)=-v_{l,a,+}^{*}(0)\), each configuration's
contribution to the velocity sum is
\[
 v_{l,a,+}^{*}(0)
 \bigl[h_{\mathrm f,a,+}(t^*)-h_{\mathrm f,a,-}(t^*)\bigr].
\]
Thus the pair contributes a signed difference of endpoint indicators,
weighted by the initial tagged velocity, rather than a fracture count.

The regime after the rapid recrossing transient is described by a model
in which the chain switches between intact and fractured states with constant
rates \(k_{\mathrm i\to f}^*\) and \(k_{\mathrm f\to i}^*\). A constant
rate means that the departure probability per unit time does not depend
on residence age. This is the same basic kinetic framework as the crossover
analysis (Sec.~\ref{si:sec:kinetic-crossover}), but RF uses the instantaneous
divider \(d_{N}^{*}\), whereas crossover uses last-visited-core labels. Their
populations and inferred rates therefore need not coincide.

Write \(h_{\mathrm f}=\mathbf1[\max_jl_{j}^{*}>d_{N}^{*}]\) and
\(h_{\mathrm i}=1-h_{\mathrm f}\), with equilibrium populations
\(P_{\mathrm i}=\langle h_{\mathrm i}\rangle\) and
\(P_{\mathrm f}=\langle h_{\mathrm f}\rangle\). The equilibrium state
correlation is
\begin{equation}
 C_{\mathrm{if}}(t^*)
 =\langle h_{\mathrm i}(0)h_{\mathrm f}(t^*)\rangle
 =P_{\mathrm i}P_{\mathrm f\mid\mathrm i}(t^*).
 \label{si:eq:rf-state-correlation}
\end{equation}
The indicator product equals one when the chain starts intact and is
fractured at the later observation. Its average is the joint probability
of those two endpoint states. The last equality follows from conditional
probability: \(P_{\mathrm f\mid\mathrm i}(t^*)\) is the probability of
being fractured at \(t^*\), given an intact initial state. Several
transitions may occur between these endpoints; this is not a first-passage
probability.

At equilibrium, shifting both observation times by \(t_0^*\) leaves
this correlation unchanged. Differentiating with respect to that shift
and evaluating at \(t_0^*=0\) gives
\[
 \begin{aligned}
 0&=\frac{d}{dt_0^*}
 \langle h_{\mathrm i}(t_0^*)h_{\mathrm f}(t_0^*+t^*)\rangle\\
 &=\langle\dot h_{\mathrm i}(0)h_{\mathrm f}(t^*)\rangle
 +\langle h_{\mathrm i}(0)\dot h_{\mathrm f}(t^*)\rangle.
 \end{aligned}
\]
The second term is the time derivative of
Eq.~\eqref{si:eq:rf-state-correlation}. Using
\(h_{\mathrm i}=1-h_{\mathrm f}\) therefore yields
\begin{equation}
 \frac{dC_{\mathrm{if}}}{dt^*}
 =-\langle\dot h_{\mathrm i}(0)h_{\mathrm f}(t^*)\rangle
 =\langle\dot h_{\mathrm f}(0)h_{\mathrm f}(t^*)\rangle.
 \label{si:eq:rf-stationary-correlation}
\end{equation}
Dots denote differentiation with respect to reduced time. The derivative
of the indicator counts signed crossings of its boundary: positive on
entry into the fractured state and negative on exit.

Let \(J_{N}^{+,*}\equiv J_{N,d_N^*}^{+,*}\) denote the equilibrium positive
crossing flux in Eq.~\eqref{si:eq:surface-density-flux}. The finite-time
current in Eq.~\eqref{si:eq:fixed-state-rf-rate} becomes the signed
boundary-crossing average in Eq.~\eqref{si:eq:rf-stationary-correlation}
when \(s^*=d_N^*\). Using the surface density
\(\rho_{N,d_N^*}^*(L^*)\) from Eq.~\eqref{si:eq:rf-surface-density},
the common configurational factor can be shown explicitly:
\[
 \begin{aligned}
 \langle\dot h_{\mathrm f}(0)h_{\mathrm f}(t^*)\rangle
 &=N\frac{\rho_{N,d_N^*}^*(L^*)}{Z_N^*(L^*)}
 \left\langle v_l^*(0)h_{\mathrm f}(t^*)
 \right\rangle_{\Sigma_{d_N^*},\mathrm{tag}}\\
 &=N\frac{\rho_{N,d_N^*}^*(L^*)}{Z_N^*(L^*)}
 \bar v_+^*\kappa_N(t^*)
 =J_N^{+,*}\kappa_N(t^*).
 \end{aligned}
\]
The second line substitutes the signed average from Eq.~\eqref{si:eq:rf}.
Dividing by \(P_{\mathrm i}\), and using
\(k_N^{\mathrm{TST},*}=J_N^{+,*}/P_{\mathrm i}\) from
Eqs.~\eqref{si:eq:exact-tst} and \eqref{si:eq:production-tst-rate}, gives
\begin{equation}
 \frac{1}{P_{\mathrm i}}\frac{dC_{\mathrm{if}}}{dt^*}
 =\frac{J_N^{+,*}}{P_{\mathrm i}}\kappa_N(t^*)
 =k_N^{\mathrm{TST},*}\kappa_N(t^*).
 \label{si:eq:rf-stationary-flux}
\end{equation}
This stationary flux identity is exact for the chosen equilibrium ensemble
when the launch surface \(s^{*}=d_{N}^{*}\) is the boundary defining the physical
states. Site multiplicity is already included in the chain TST factor.
The exponential time dependence requires the additional two-state approximation.

In the two-state model, the intact fraction
\(1-P_{\mathrm f\mid\mathrm i}\) supplies probability to the fractured
state at rate \(k_{\mathrm i\to f}^*\), while the fractured fraction
loses it at rate \(k_{\mathrm f\to i}^*\). This gain--loss balance gives
\begin{equation}
 \begin{aligned}
 \frac{dP_{\mathrm f\mid\mathrm i}}{dt^*}
 &=k_{\mathrm i\to f}^*(1-P_{\mathrm f\mid\mathrm i})
 -k_{\mathrm f\to i}^*P_{\mathrm f\mid\mathrm i}\\
 &=k_{\mathrm i\to f}^*-k_{\mathrm{eff}}^{*}P_{\mathrm f\mid\mathrm i},
 \qquad k_{\mathrm{eff}}^{*}=k_{\mathrm i\to f}^*+k_{\mathrm f\to i}^*.
 \end{aligned}
 \label{si:eq:rf-two-state-balance}
\end{equation}
The initial condition is \(P_{\mathrm f\mid\mathrm i}(0)=0\), since the
chain starts intact. Moving \(k_{\mathrm{eff}}^{*}P_{\mathrm f\mid\mathrm i}\)
to the left of Eq.~\eqref{si:eq:rf-two-state-balance} and multiplying by
\(e^{k_{\mathrm{eff}}^{*}t^*}\) gives the derivative of a product.
Integration from zero to \(t^*\) then gives
\[
 \begin{aligned}
 \frac{d}{dt^*}\left[e^{k_{\mathrm{eff}}^{*}t^*}
 P_{\mathrm f\mid\mathrm i}(t^*)\right]
 &=k_{\mathrm i\to f}^*e^{k_{\mathrm{eff}}^{*}t^*},\\
 e^{k_{\mathrm{eff}}^{*}t^*}P_{\mathrm f\mid\mathrm i}(t^*)
 &=\frac{k_{\mathrm i\to f}^*}{k_{\mathrm{eff}}^{*}}
 \left(e^{k_{\mathrm{eff}}^{*}t^*}-1\right).
 \end{aligned}
\]
Dividing by the exponential gives the solution; its long-time limit
identifies \(P_{\mathrm f}=k_{\mathrm i\to f}^*/k_{\mathrm{eff}}^{*}\):
\begin{equation}
 P_{\mathrm f\mid\mathrm i}(t^*)=P_{\mathrm f}
 \bigl(1-e^{-k_{\mathrm{eff}}^{*}t^*}\bigr),
 \qquad P_{\mathrm f}k_{\mathrm{eff}}^{*}=k_{\mathrm i\to f}^*.
 \label{si:eq:rf-two-state-probability}
\end{equation}
Substituting Eq.~\eqref{si:eq:rf-two-state-probability} into
Eq.~\eqref{si:eq:rf-state-correlation} and differentiating gives
\begin{equation}
 \begin{aligned}
 \frac{1}{P_{\mathrm i}}\frac{dC_{\mathrm{if}}}{dt^*}
 &=P_{\mathrm f}k_{\mathrm{eff}}^{*}e^{-k_{\mathrm{eff}}^{*}t^*}
 =k_{\mathrm i\to f}^*e^{-k_{\mathrm{eff}}^{*}t^*},\\
 \kappa_N(t^*)
 &=\frac{k_{\mathrm i\to f}^*}{k_N^{\mathrm{TST},*}}
 e^{-k_{\mathrm{eff}}^{*}t^*}.
 \end{aligned}
 \label{si:eq:rf-two-state-transmission}
\end{equation}
The second line uses Eq.~\eqref{si:eq:rf-stationary-flux} to divide out
the TST rate. It identifies the slow-regime amplitude as
\begin{equation}
 \kappa_{0,N}=\frac{k_{\mathrm i\to f}^*}{k_N^{\mathrm{TST},*}}.
 \label{si:eq:rf-slow-amplitude}
\end{equation}
Detailed balance equates the two equilibrium directional currents,
\(P_{\mathrm i}k_{\mathrm i\to f}^*=P_{\mathrm f}k_{\mathrm f\to i}^*\).
At coexistence, \(P_{\mathrm i}=P_{\mathrm f}=1/2\), so the directional
rates coincide. Calling their common value \(k_N^*\), and using
Eq.~\eqref{si:eq:rf-slow-amplitude}, gives
\begin{equation}
 k_{\mathrm i\to f}^*=k_{\mathrm f\to i}^*=k_N^*,
 \qquad k_{\mathrm{eff}}^{*}=2k_N^*,
 \qquad k_N^*=\kappa_{0,N}k_N^{\mathrm{TST},*}.
 \label{si:eq:rf-coexistence-rate}
\end{equation}
Substituting these rates into Eq.~\eqref{si:eq:rf-two-state-transmission}
yields the fitting model
\begin{equation}
 \kappa_N(t^*)=\kappa_{0,N}
 \exp\!\left[-2\kappa_{0,N}k_N^{\mathrm{TST},*}t^*\right].
 \label{si:eq:finite-pop}
\end{equation}
The time-dependent RF rate is given by Eq.~\eqref{si:eq:rf-boundary-rate},
\(k_{\mathrm{RF}}^*(t^*)=k_N^{\mathrm{TST},*}\kappa_N(t^*)\).
After the initial recrossing transient, if \(2k_N^*t^*\ll1\) throughout
the observation window, population relaxation is negligible and
\(k_{\mathrm{RF}}^*(t^*)\approx k_N^*\). A direct average over this
plateau therefore approximates the rate obtained from the fitted amplitude.

The factor of two comes from forward plus reverse population relaxation,
not from the antithetic velocity pair. The fitted amplitude describes the extrapolated
slow-regime model and is not the microscopic transmission at time zero.

Equation~\eqref{si:eq:finite-pop} is fitted by unweighted least squares over
\(3\le t^*\le10\) at 33 evenly selected recorded times, holding
\(k_N^{\mathrm{TST},*}\) fixed. Writing these times as \(t_m^*\),
\(m=1,\ldots,33\), the fitted amplitude minimizes the squared deviations
between the trajectory average in
Eq.~\eqref{si:eq:rf-trajectory-average} and the model:
\begin{equation}
 \widehat\kappa_{0,N}
 =\underset{\kappa_{0,N}>0}{\operatorname{arg\,min}}
 \sum_{m=1}^{33}
 \left[\widehat\kappa_N(t_m^*)
 -\kappa_{0,N}e^{-2\kappa_{0,N}k_N^{\mathrm{TST},*}t_m^*}\right]^2.
 \label{si:eq:rf-amplitude-fit}
\end{equation}
All selected times have equal weight. One fitted parameter therefore
determines both the amplitude and the model decay rate. The reported
transmission \(\kappa_N^{\mathrm{nl}}\) is the fitted amplitude
\(\kappa_{0,N}\), and multiplication by the TST
factor in Eq.~\eqref{si:eq:rf-coexistence-rate} gives the RF rate.
Sensitivity is evaluated over six additional
windows: \(2\!:\!8\), \(2\!:\!10\), \(3\!:\!8\), \(4\!:\!8\),
\(4\!:\!10\), and \(4\!:\!12\).

The transmission decreases most visibly for the shortest chains. For
\(N\ge10\), it varies little over the fitting interval, so the fit mainly
determines its amplitude; the decay rate follows from the assumed two-state
relation rather than being independently resolved. The exponential model predicts eventual
population decay and captures the overall relaxation, but does not reproduce
every feature of the shorter-chain curves.

Sampling intervals use 5000 whole-configuration bootstrap samples.
Each resamples \(n_{\mathrm{conf}}\) configurations with replacement,
retaining both antithetic trajectories and all their linked times.
The curve in Eq.~\eqref{si:eq:rf-trajectory-average} is recomputed and
Eq.~\eqref{si:eq:rf-amplitude-fit} refitted with the TST factor fixed;
multiplication by that factor gives the corresponding RF rate.
The 2.5th and 97.5th percentiles give 95\% intervals for transmission
and rate, quantifying sampling uncertainty for the adopted model and
fitting window. Across the six alternative windows, fitted
amplitudes differed from the primary estimates by less than 0.66\%.

\begin{figure}[tbp]
  \centering
  \includegraphics[width=\linewidth]{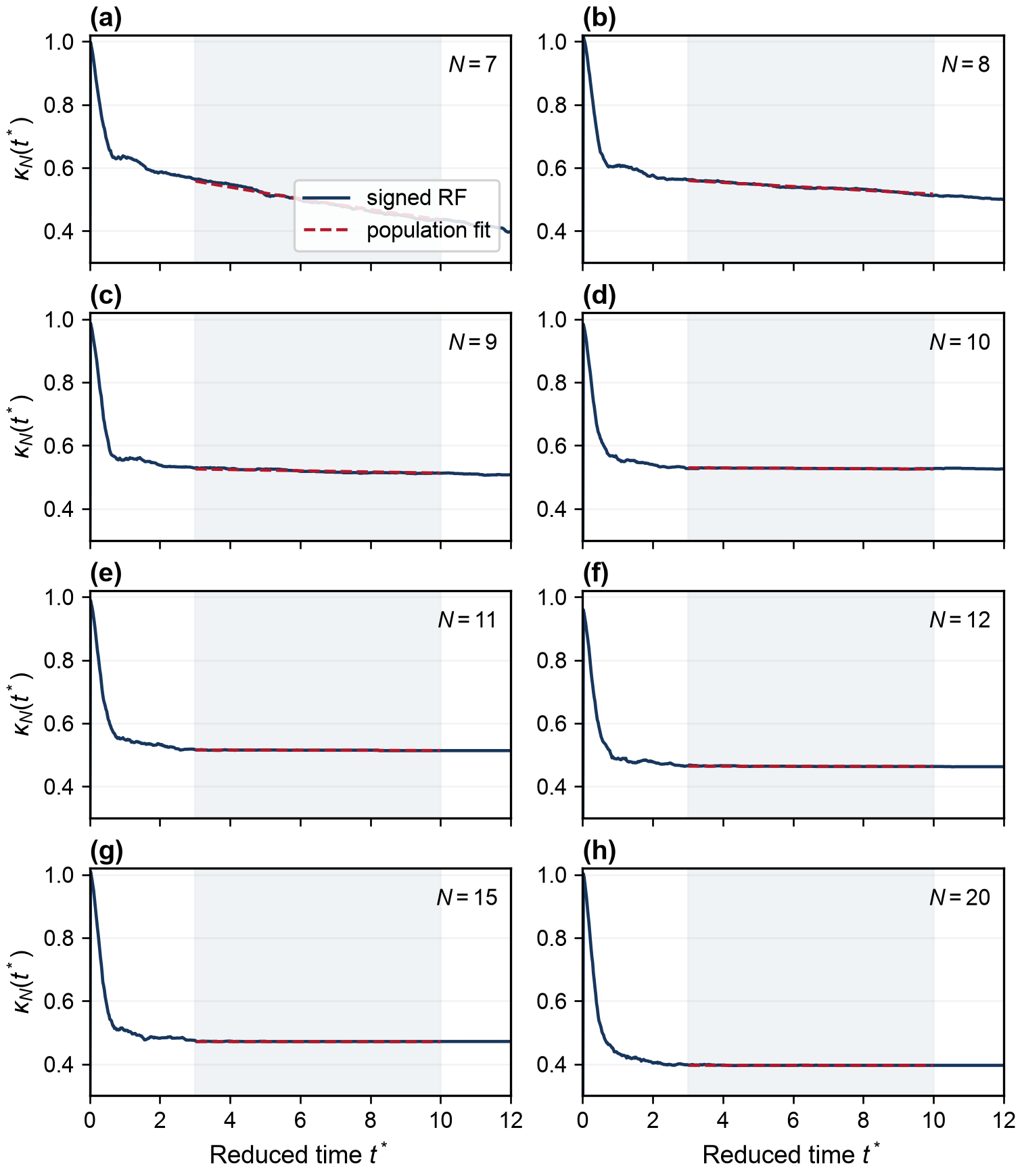}
  \caption{Nonlinear transmission functions for the eight core RF
  lengths.  Dashed curves show the common finite-population fit over
  \(3\le t^*\le10\).}
  \label{si:fig:rf-curves}
\end{figure}

\begin{table}[H]
  \centering
  \small
  \caption{Nonlinear transmission at the simulated coexistence extensions,
  using the common finite-population estimator fitted over \(3\le t^*\le10\).
  Parentheses give the lower and upper bounds of the 95\% whole-configuration
  bootstrap confidence intervals.}
  \label{si:tab:transmission}
  \input{publication_tables/reactive_flux_transmission/table.tex}\end{table}

\FloatBarrier
\subsection{\revphasefourteen{Fixed-state surface control at N=10}}
\label{si:sec:fixed-surface-control}

The RF rate's sensitivity to the launch surface is tested by varying the
launch coordinate while keeping the physical states fixed. At \(N=10\)
and \(\lambda=\CanonicalNtenFixedSurfaceLambda\), the reduced launch coordinates are
\begingroup\color{revisionphasefourteen}
\begin{align}
 d^{*}&=\CanonicalNtenFixedSurfaceAdiabaticCoordinate,
 &s_{\mathrm{PMF}}^{*}&=\CanonicalNtenFixedSurfacePMFCoordinate,\\
 s_{\mathrm{out}}^{*}&=2s_{\mathrm{PMF}}^{*}-d^{*}
 =\CanonicalNtenFixedSurfaceOuterCoordinate.
 \label{si:eq:fixed-surface-coordinates}
\end{align}
\endgroup
Here \(d^{*}=d_{N}^{*}\) is the adiabatic saddle, \(s_{\mathrm{PMF}}^{*}\) is the
finite-temperature site-PMF saddle, and \(s_{\mathrm{out}}^{*}\) lies equally
far beyond the PMF saddle. At each launch surface, one tagged bond is fixed
at \(s^{*}\) and the remaining bonds lie below \(s^{*}\), as in
\cref{si:eq:rf-launch-surface}. All three calculations use the same
fractured-state indicator \(h_{\mathrm f,d^*}=\mathbf1[\max_jl_{j}^{*}>d^{*}]\)
and intact population \(P_{\mathrm i}(d^{*})=1/2\). Changing \(s^{*}\) therefore
changes the launch ensemble and its positive flux, while the criterion
for counting a trajectory as fractured remains the same.

For displaced launch surfaces, the direct window average is used without
imposing the amplitude--decay relation derived in \cref{si:eq:finite-pop}
for coincident launch and state boundaries.

Each surface uses 40\,960 independently sampled configurations, with two
antithetic trajectories per configuration, giving 245\,760 trajectories in
total. The timestep is \(0.25\,\mathrm{fs}\), the duration is \(t^*=33\),
and observations are recorded every \(\Delta t^*=0.022\). For each surface,
the transmission is averaged over the recorded times in \(4\le t^*\le8\).
At this fixed \(N\) and \(d^*\), write
\(J_{s^*}^{+,*}=J_{N,s^*}^{+,*}\) and
\(K_{s^*}^{*}=K_{N,s^*\mid d^*}^{*}\). The resulting rate is
\[
 \bar k_{s^*\mid d^*}^{*}=K_{s^*}^{*}\bar\kappa_{s^*\mid d^*},
 \qquad K_{s^*}^{*}=\frac{J_{s^*}^{+,*}}{P_{\mathrm i}(d^{*})},
\]
where the overbar denotes this arithmetic time average and \(K_{s^*}^{*}\)
is the deterministic surface factor defined in \cref{si:eq:exact-tst}.
Thus the comparison combines the change in flux with the change in
transmission (\cref{si:tab:surface-control}).

Sampling uncertainty is estimated with 20\,000 whole-configuration bootstrap
samples, retaining each antithetic pair and all its observation times.
Configurations are resampled independently for each surface. Each bootstrap
sample gives three mean rates and their pairwise ratios; the 2.5th and
97.5th percentiles give the 95\% intervals. Log rate ratios are also reported,
for which zero corresponds to equal rates (\cref{si:tab:fixed-surface-pairs}).

\begin{table}[H]
  \centering
  \small
  \setlength{\tabcolsep}{4pt}
  \color{revisionphasefourteen}
  \caption{\revphasefourteen{Three-surface fixed-state control.
  \(K_{s^*}^{*}=J_{s^*}^{+,*}/P_{\mathrm i}(d^{*})\) is deterministic. Parentheses give independent-surface, whole-configuration
  bootstrap 95\% intervals for transmission and rate averaged over \(4\!:\!8\).}}
  \label{si:tab:surface-control}
  \input{publication_tables/reactive_flux_surface_rates/table.tex}\end{table}

\begingroup\color{revisionphasefourteen}
Moving from the adiabatic to the PMF launch surface lowers the surface
flux but increases transmission. These changes nearly compensate:
\begin{equation}
  \frac{K_{\mathrm{PMF}}^*}{K_{\mathrm{ad}}^*}
  =\CanonicalNtenFixedSurfacePMFOverAdiabaticFactorRatio,
  \qquad
  \frac{\bar\kappa_{\mathrm{PMF}\mid d^*}}{\bar\kappa_{\mathrm{ad}\mid d^*}}
  =\CanonicalNtenFixedSurfacePMFOverAdiabaticTransmissionRatio,
  \qquad
  \frac{\bar k_{\mathrm{PMF}\mid d^*}^{*}}{\bar k_{\mathrm{ad}\mid d^*}^{*}}
  =\CanonicalNtenFixedSurfacePMFOverAdiabaticRateRatio.
  \label{si:eq:surface-factor-ratios}
\end{equation}
\endgroup

\begin{table}[H]
  \centering
  \small
  \color{revisionphasefourteen}
  \caption{\revphasefourteen{Pairwise rate comparisons for the
  \(4\!:\!8\) mean.  Each interval uses independent resampling of the two
  surface ensembles.}}
  \label{si:tab:fixed-surface-pairs}
  \input{publication_tables/reactive_flux_surface_comparisons/table.tex}\end{table}

\begin{figure}[H]
  \centering
  \includegraphics[width=\linewidth]{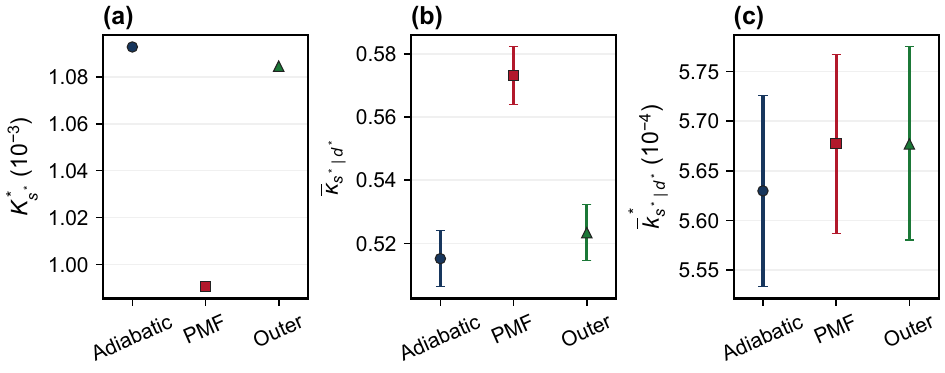}
  \caption{\revphasefourteen{Fixed-state \(N=10\) control.  The panels show
  the deterministic surface factor, direct transmission, and resulting rate
  for the adiabatic, PMF, and outer launch surfaces. The transmission and rate
are averaged over \(4\le t^*\le8\), with the state divider fixed at \(d^*\). Bars on stochastic
  quantities are independent-surface, whole-configuration 95\% intervals.}}
  \label{si:fig:surface-invariance}
\end{figure}

\begin{figure}[p]
  \centering
  \includegraphics[width=\linewidth]{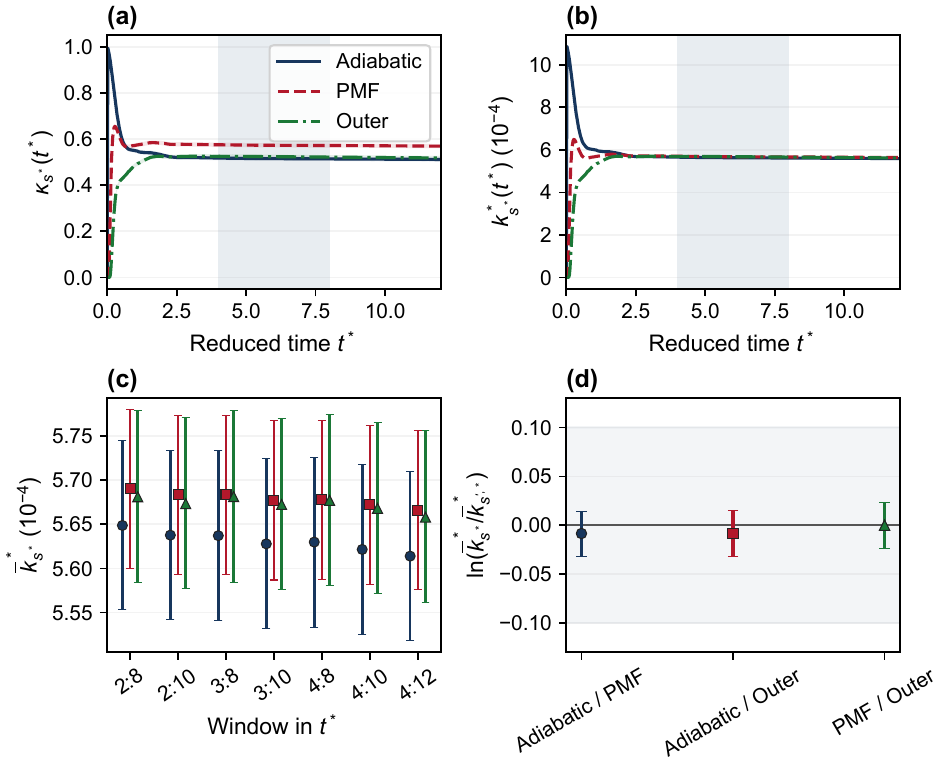}
  \caption{\revphasefourteen{Time curves and window sensitivity for the
  three fixed-state ensembles. The labels suppress the common state-divider
condition \(d^*\): \(k_{s^*}^*(t^*)=k_{s^*\mid d^*}^*(t^*)\) and
\(\kappa_{s^*}(t^*)=\kappa_{s^*\mid d^*}(t^*)\); overbars denote window averages.
Upper panels show transmission and rate;
  shading marks \(4\le t^*\le8\).  Lower panels show neighboring direct-window
  estimates and independent-surface 95\% log-rate-ratio intervals.  The
  shaded \(\pm0.10\) band marks the practical-agreement reference; its
  containment criterion uses 90\% intervals, while the displayed bars are
  95\% intervals.}}
  \label{si:fig:fixed-surface-windows}
\end{figure}

The largest difference between the three central rates is 0.85\%.
Every pairwise 95\% log-ratio interval includes zero, with half-widths
below 0.03; the corresponding 90\% intervals lie within \(\pm0.10\).
The largest change in a central rate across neighboring averaging windows
is \(\CanonicalNtenFixedSurfaceMaximumWindowShiftPercent\%\).
The final rate is therefore insensitive to the tested launch surfaces
within the comparison uncertainty at this state. The compensation between
flux and transmission is shown in \cref{si:fig:surface-invariance};
\cref{si:fig:fixed-surface-windows} shows the time dependence and the
stability of the window averages. Control and production estimates retain
separate sampling uncertainties.

\FloatBarrier

%% file: publication_tables/reactive_flux_transmission/table.tex
\begin{tabular}{rrc}
  \toprule
  \(N\) & \(\lambda\) & \(\kappa_N^{\mathrm{nl}}\) (95\% interval) \\
  \midrule
  7 & 1.129293228424449 & \(0.620\;(0.566,\,0.677)\) \\
  8 & 1.1207228320216314 & \(0.579\;(0.535,\,0.624)\) \\
  9 & 1.1135092525657033 & \(0.531\;(0.490,\,0.572)\) \\
  10 & 1.1073538970587944 & \(0.532\;(0.494,\,0.570)\) \\
  11 & 1.1020345665628868 & \(0.516\;(0.477,\,0.555)\) \\
  12 & 1.09738589408922 & \(0.464\;(0.427,\,0.502)\) \\
  15 & 1.0863558914897524 & \(0.472\;(0.433,\,0.511)\) \\
  20 & 1.0739352379713902 & \(0.396\;(0.359,\,0.436)\) \\
  \bottomrule
\end{tabular}

%% file: publication_tables/reactive_flux_surface_rates/table.tex
\begin{tabular}{lccc}
  \toprule
  Surface & \(10^3K_{s^*}^{*}\) & \(\bar\kappa_{s^*\mid d^*}\) (95\% interval) & \(10^4\bar k_{s^*\mid d^*}^{*}\) (95\% interval) \\
  \midrule
  Adiabatic & 1.092686 & \(0.5152\;(0.5064,\,0.5240)\) & \(5.630\;(5.534,\,5.726)\) \\
  PMF & 0.990639 & \(0.5731\;(0.5640,\,0.5822)\) & \(5.678\;(5.587,\,5.767)\) \\
  Outer & 1.084523 & \(0.5235\;(0.5145,\,0.5325)\) & \(5.677\;(5.580,\,5.775)\) \\
  \bottomrule
\end{tabular}

%% file: publication_tables/reactive_flux_surface_comparisons/table.tex
\begin{tabular}{lcc}
  \toprule
  Pair & Rate ratio (95\% interval) & 95\% log-ratio interval \\
  \midrule
  PMF/Adiabatic & \(1.008\;(0.986,\,1.032)\) & \((-0.014,\,0.032)\) \\
  Outer/Adiabatic & \(1.008\;(0.985,\,1.033)\) & \((-0.015,\,0.032)\) \\
  Outer/PMF & \(1.000\;(0.977,\,1.024)\) & \((-0.024,\,0.023)\) \\
  \bottomrule
\end{tabular}

%% file: si_sections/fracture_rate_results.tex
\section{Fracture-rate comparisons and chain-length dependence}
\label{si:fracture-rate-results}

\subsection{Midpoint rates and transmission factors}
\label{si:midpoint-rate-comparison}

For the KGH model, the effective coefficient in
Table~\ref{si:tab:effective-transmission} differs from the local dynamical
factor \(\chi_N\). To see why, use the site PMF \(G_{N}^{\mathrm{site},*}(l^{*})\), with
all spectator bonds restricted below the fixed divider \(d_{N}^{*}\), and let
\(l_{\mathrm i}^{\mathrm{site},*}\) and \(l_{\mathrm s}^{\mathrm{site},*}\) denote its intact minimum and
barrier maximum. The effective intact-well width is
\[
 w_{\mathrm i}^{*}=\int_{l^{*}<d_{N}^{*}}
 \exp\!\left[-\frac{G_{N}^{\mathrm{site},*}(l^{*})-G_{N}^{\mathrm{site},*}(l_{\mathrm i}^{\mathrm{site},*})}{T^*}\right]dl^{*},
\]
where the integral covers the allowed tagged-coordinate domain below
\(d_{N}^{*}\). Since \(l^{*}\) is dimensionless, so is \(w_{\mathrm i}^{*}\).
Integrating the restricted tagged weight over this domain gives the
intact partition weight. The weight at \(d_{N}^{*}\) divided by this intact
weight, together with the positive Maxwell crossing factor \(\bar v_+^*\) defined in
\cref{si:eq:maxwell-flux}, therefore gives
\[
 \begin{aligned}
 k_N^{\mathrm{TST},*}
 &=\frac{N\bar v_+^*}{w_{\mathrm i}^{*}}
   e^{-[G_{N}^{\mathrm{site},*}(d_{N}^{*})-G_{N}^{\mathrm{site},*}(l_{\mathrm i}^{\mathrm{site},*})]/T^*},\\
 k_N^{\mathrm{KGH},*}
 &=N\frac{\omega_{\mathrm i}^*}{2\pi}\chi_N
   e^{-[G_{N}^{\mathrm{site},*}(l_{\mathrm s}^{\mathrm{site},*})-G_{N}^{\mathrm{site},*}(l_{\mathrm i}^{\mathrm{site},*})]/T^*}.
 \end{aligned}
\]
The first expression rewrites the equilibrium TST rate; the second uses
the KGH approximation of Sec.~\ref{si:kgh-factor}. Taking their ratio cancels the
site multiplicity and the common reference free energy:
\[
 \kappa_{\mathrm{eff},N}^{\mathrm{KGH}}
 =\frac{k_N^{\mathrm{KGH},*}}{k_N^{\mathrm{TST},*}}
 =\chi_N\frac{\omega_{\mathrm i}^*w_{\mathrm i}^{*}}
                  {2\pi\bar v_+^*}
 \exp\!\left[-\frac{G_{N}^{\mathrm{site},*}(l_{\mathrm s}^{\mathrm{site},*})-G_{N}^{\mathrm{site},*}(d_{N}^{*})}{T^*}\right].
\]
Thus the effective transmission ratio includes the local dynamical factor, the
well-normalization factor, and the free-energy difference between the
PMF maximum and the basin divider. It need not equal \(\chi_N\).

\begin{table}[H]
  \centering
  \scriptsize
  \setlength{\tabcolsep}{3pt}
  \caption{Reduced forward rates near coexistence for \(N=4,\ldots,30\).
  Multiply all rate entries by the factor in the second column.
  Parentheses give 95\% intervals: whole-trajectory bootstrap for ACF,
  whole-configuration bootstrap for nonlinear RF, and approximate Poisson
  intervals for core residence (\(k^{*}\pm1.96\,\mathrm{SE}(k^*)\)).
  Core residence uses \(\eta=0.5\). Dashes denote unavailable values.}
  \label{tab:activated-rates-paper}
  \input{publication_tables/coexistence_rates/table.tex}
\end{table}

\begin{table}[H]
  \centering
  \scriptsize
  \setlength{\tabcolsep}{3pt}
  \caption{Effective transmission coefficients \(\kappa_{\mathrm{eff}}=k^{*}/k^{\mathrm{TST},*}\)
  near coexistence, normalized by the deterministic TST chain rate at each
  method's reporting state. Core residence uses \(\eta=0.5\).
  Parentheses give 95\% intervals with the conventions in
  Table~\ref{tab:activated-rates-paper} and the TST normalization held fixed.
  Dashes denote unavailable values.}
  \label{si:tab:effective-transmission}
  \input{publication_tables/effective_transmission/table.tex}
\end{table}

\begin{figure}[htbp]
  \centering
  \includegraphics[width=8.5cm]{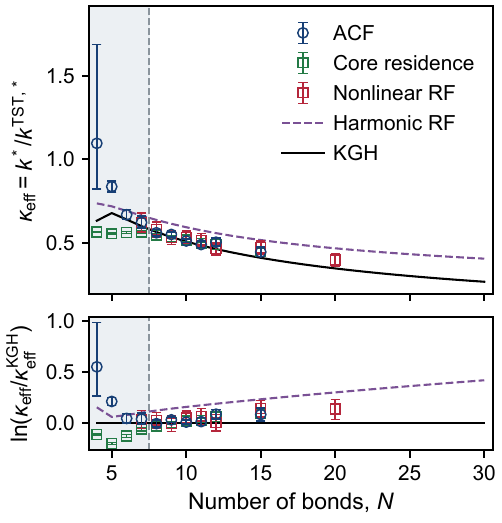}
  \input{figures/effective_transmission/caption}
  \label{si:fig:transmission-n}
\end{figure}
\FloatBarrier

\subsection{Fracture rates away from coexistence}
\label{si:off-coexistence-rates}

\begin{figure}[H]
  \centering
  \includegraphics[width=8.5cm]{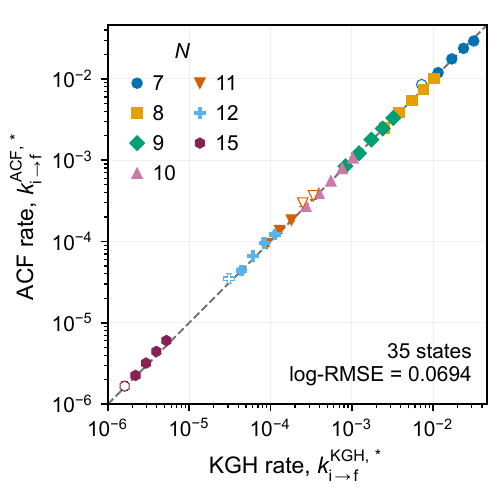}
  \input{figures/off_coexistence_rates/caption}
  \label{fig:pointwise-kgh-paper}
\end{figure}

The KGH rate is evaluated separately at five sampled extensions for each
\(N=7,\ldots,12\) and \(15\), using the site-PMF barrier, curvatures and
prefactor at that state. The 35 predictions agree closely with the tension-ACF
estimates, with logarithmic RMSE \(\CanonicalPointwiseKghRmse\)
(\cref{fig:pointwise-kgh-paper}).

\FloatBarrier

\subsection{Local force dependence and activation length}
\label{si:activation-length}

The activation length measures how rapidly the fracture rate changes with
tension near coexistence. At fixed \(N\) and \(T^*\), use the
intact-restricted virial tension \(F_{\mathrm i,\mathrm{vir}}^{*}\)
as the force coordinate and write
\(\Delta F_{\mathrm i,\mathrm{vir}}^{*}=F_{\mathrm i,\mathrm{vir}}^{*}-F_{\mathrm i,\mathrm{vir},N,1/2}^{*}\). A first-order Taylor expansion gives
\[
 \ln k_N^*(F_{\mathrm i,\mathrm{vir}}^{*})=
 \ln k_{N,1/2}^*+
 \left[\frac{\partial\ln k_N^*}{\partial F_{\mathrm i,\mathrm{vir}}^{*}}\right]_{1/2}
 \Delta F_{\mathrm i,\mathrm{vir}}^{*}+O\!\left((\Delta F_{\mathrm i,\mathrm{vir}}^{*})^2\right).
\]
Defining the reduced activation length as
\(\ell_{Z,N}^*=T^*[\partial\ln k_N^*/\partial F_{\mathrm i,\mathrm{vir}}^{*}]_{1/2}\)
and exponentiating the linear approximation yields
\[
 \begin{aligned}
 k_N^*(F_{\mathrm i,\mathrm{vir}}^{*})&\approx k_{N,1/2}^*
 \exp\!\left[\frac{\Delta F_{\mathrm i,\mathrm{vir}}^{*}\ell_{Z,N}^*}{T^*}\right]
 =A_{Z,N}^{*}\exp\!\left[\frac{F_{\mathrm i,\mathrm{vir}}^{*}\ell_{Z,N}^*}{T^*}\right],\\
 A_{Z,N}^{*}&=k_{N,1/2}^*
 \exp\!\left[-\frac{F_{\mathrm i,\mathrm{vir},N,1/2}^{*}\ell_{Z,N}^*}{T^*}\right].
 \end{aligned}
\]
Thus the local Zhurkov form follows from linearizing the logarithmic rate;
its coefficient \(A_{Z,N}^{*}\) ensures that the approximation passes through
the coexistence rate. It is an extrapolated intercept, not a separately
measured zero-force rate.

For KGH, \(k_N^{\mathrm{KGH},*}=N A_{N}^{\mathrm{KGH},*}e^{-B_N}\), where
\(A_{N}^{\mathrm{KGH},*}=\omega_{\mathrm i}^*\chi_N/(2\pi)\)
and \(B_N\) is the site-PMF barrier divided by thermal energy. Taking the
logarithm and differentiating separates the barrier and prefactor contributions:
\[
 \ln k_N^{\mathrm{KGH},*}=\ln N+\ln A_{N}^{\mathrm{KGH},*}-B_N,
 \qquad
 \frac{\partial\ln k_N^{\mathrm{KGH},*}}{\partial F_{\mathrm i,\mathrm{vir}}^{*}}
 =\frac{\partial\ln A_{N}^{\mathrm{KGH},*}}{\partial F_{\mathrm i,\mathrm{vir}}^{*}}
  -\frac{\partial B_N}{\partial F_{\mathrm i,\mathrm{vir}}^{*}}.
\]
The site multiplicity is constant at fixed \(N\), so it contributes no force
derivative. The reduced activation lengths are defined by
\(\ell_{B,N}^*=\ell_{B,N}/l_0\),
\(\ell_{A,N}^*=\ell_{A,N}/l_0\), and
\(\ell_{Z,N}^{\mathrm{KGH},*}=\ell_N^{\mathrm{KGH}}/l_0\).
Multiplying the logarithmic-rate derivatives by \(T^*\) gives
\begin{align}
 \ell_{B,N}^*
 &=-T^*\left[\frac{\partial B_N}{\partial F_{\mathrm i,\mathrm{vir}}^{*}}\right]_{1/2},\\
 \ell_{A,N}^*
 &=T^*\left[\frac{\partial\ln A_{N}^{\mathrm{KGH},*}}{\partial F_{\mathrm i,\mathrm{vir}}^{*}}\right]_{1/2},\\
 \ell_{Z,N}^*=\ell_{Z,N}^{\mathrm{KGH},*}
 &=\ell_{B,N}^*+\ell_{A,N}^*.
 \label{si:eq:activation-lengths}
\end{align}
The resulting logarithmic rate is
\begin{equation}
 \ln k_N^{\mathrm{KGH},*}(F_{\mathrm i,\mathrm{vir}}^{*})
 \approx\ln[N A_{N,1/2}^{\mathrm{KGH},*}]-B_{N,1/2}
 +\frac{\ell_{Z,N}^{\mathrm{KGH},*}}{T^*}
   (F_{\mathrm i,\mathrm{vir}}^{*}-F_{\mathrm i,\mathrm{vir},N,1/2}^{*}).
 \label{si:eq:connected-zhurkov}
\end{equation}
Substituting \(k_{N,1/2}^{\mathrm{KGH},*}
=N A_{N,1/2}^{\mathrm{KGH},*}e^{-B_{N,1/2}}\) into the expression for
\(A_{Z,N}^{*}\) above gives the main-text coefficient. Unlike
\(A_{N}^{\mathrm{KGH},*}\), it includes the coexistence barrier, site multiplicity
and force shift.

For the theoretical estimate, the derivatives are evaluated by central
differences about coexistence, using extension increments
\(\delta\lambda\leq0.001\). The PMF, its curvatures, the KGH
prefactor and the intact tension at
\(\lambda_\pm=\lambda_{1/2}\pm\delta\lambda\) are computed. Denote these endpoint
values by \(B_\pm\), \(A_{\pm}^{*}\) and \(F_{\mathrm i,\mathrm{vir},\pm}^{*}\).
Applying the chain rule through \(\lambda\) cancels the common
finite-difference denominator \(2\delta\lambda\), giving
\[
 \ell_{B,N}^*\approx
 -T^*\frac{B_+-B_-}{F_{\mathrm i,\mathrm{vir},+}^{*}-F_{\mathrm i,\mathrm{vir},-}^{*}},
 \qquad
 \ell_{A,N}^*\approx
 T^*\frac{\ln(A_{+}^{*}/A_{-}^{*})}{F_{\mathrm i,\mathrm{vir},+}^{*}-F_{\mathrm i,\mathrm{vir},-}^{*}}.
\]
Numerical checks confirmed that the total activation length is insensitive
to the tested discretization choices.

For the simulation estimate at \(N=10\), exponential fits to the tension
ACF first provide rates at five sampled extensions. A separate linear regression then estimates their force dependence,
\[
 \ln k_{j}^{\mathrm{ACF},*}=a_{0}+a_{1}^{*} F_{\mathrm i,\mathrm{vir},j}^{*},
 \qquad \ell_Z^*=T^*a_{1}^{*},
 \qquad j=1,\ldots,5,
\]
where \(F_{\mathrm i,\mathrm{vir},j}^{*}\) is the intact-state tension at sampled extension \(j\),
\(a_{0}\) is the intercept and \(a_{1}^{*}\) is the slope with respect
to reduced tension. No slope is shared across
chain lengths. This regression measures the slope over the sampled
extension interval, whereas KGH provides a local derivative at coexistence.
The ACF error bar represents one regression standard error with force coordinates
held fixed, not a 95\% interval or a full propagation of uncertainty.

%% file: publication_tables/coexistence_rates/table.tex
\begin{tabular}{rrcccrr}
  \toprule
  \(N\) & Factor & ACF & Core (\(\eta=0.5\)) & Nonlinear RF & Harmonic RF & KGH \\
  \midrule
  4 & \(10^{-1}\) & \(4.5\;(3.4,\,7.0)\) & \(2.321\;(2.307,\,2.334)\) & \textemdash & \(3.028\) & \(2.602\) \\
  5 & \(10^{-1}\) & \(1.599\;(1.537,\,1.666)\) & \(1.0598\;(1.0505,\,1.0691)\) & \textemdash & \(1.373\) & \(1.296\) \\
  6 & \(10^{-2}\) & \(5.18\;(4.96,\,5.44)\) & \(4.362\;(4.307,\,4.418)\) & \textemdash & \(5.358\) & \(4.962\) \\
  7 & \(10^{-2}\) & \(1.76\;(1.67,\,1.87)\) & \(1.596\;(1.572,\,1.620)\) & \(1.76\;(1.61,\,1.92)\) & \(1.879\) & \(1.700\) \\
  8 & \(10^{-3}\) & \(5.45\;(5.26,\,5.64)\) & \(5.312\;(5.221,\,5.403)\) & \(5.63\;(5.21,\,6.07)\) & \(6.196\) & \(5.491\) \\
  9 & \(10^{-3}\) & \(1.790\;(1.740,\,1.842)\) & \(1.737\;(1.709,\,1.766)\) & \(1.73\;(1.59,\,1.86)\) & \(1.998\) & \(1.739\) \\
  10 & \(10^{-4}\) & \(5.59\;(5.48,\,5.70)\) & \(5.641\;(5.559,\,5.722)\) & \(5.81\;(5.39,\,6.23)\) & \(6.481\) & \(5.545\) \\
  11 & \(10^{-4}\) & \(1.832\;(1.760,\,1.908)\) & \(1.852\;(1.816,\,1.888)\) & \(1.93\;(1.78,\,2.08)\) & \(2.150\) & \(1.811\) \\
  12 & \(10^{-5}\) & \(6.65\;(6.34,\,6.98)\) & \(6.48\;(6.32,\,6.64)\) & \(6.13\;(5.64,\,6.63)\) & \(7.367\) & \(6.109\) \\
  13 & \(10^{-5}\) & \textemdash & \textemdash & \textemdash & \(2.618\) & \(2.139\) \\
  14 & \(10^{-6}\) & \textemdash & \textemdash & \textemdash & \(9.668\) & \(7.784\) \\
  15 & \(10^{-6}\) & \(3.19\;(3.00,\,3.40)\) & \(3.22\;(3.09,\,3.35)\) & \(3.39\;(3.11,\,3.67)\) & \(3.711\) & \(2.945\) \\
  16 & \(10^{-6}\) & \textemdash & \textemdash & \textemdash & \(1.479\) & \(1.158\) \\
  17 & \(10^{-7}\) & \textemdash & \textemdash & \textemdash & \(6.118\) & \(4.723\) \\
  18 & \(10^{-7}\) & \textemdash & \textemdash & \textemdash & \(2.620\) & \(1.996\) \\
  19 & \(10^{-8}\) & \textemdash & \textemdash & \textemdash & \(11.60\) & \(8.720\) \\
  20 & \(10^{-8}\) & \textemdash & \textemdash & \(4.50\;(4.07,\,4.95)\) & \(5.302\) & \(3.933\) \\
  21 & \(10^{-8}\) & \textemdash & \textemdash & \textemdash & \(2.496\) & \(1.828\) \\
  22 & \(10^{-9}\) & \textemdash & \textemdash & \textemdash & \(12.09\) & \(8.744\) \\
  23 & \(10^{-9}\) & \textemdash & \textemdash & \textemdash & \(6.013\) & \(4.295\) \\
  24 & \(10^{-9}\) & \textemdash & \textemdash & \textemdash & \(3.066\) & \(2.164\) \\
  25 & \(10^{-9}\) & \textemdash & \textemdash & \textemdash & \(1.601\) & \(1.117\) \\
  26 & \(10^{-10}\) & \textemdash & \textemdash & \textemdash & \(8.552\) & \(5.893\) \\
  27 & \(10^{-10}\) & \textemdash & \textemdash & \textemdash & \(4.664\) & \(3.177\) \\
  28 & \(10^{-10}\) & \textemdash & \textemdash & \textemdash & \(2.595\) & \(1.747\) \\
  29 & \(10^{-11}\) & \textemdash & \textemdash & \textemdash & \(14.72\) & \(9.797\) \\
  30 & \(10^{-11}\) & \textemdash & \textemdash & \textemdash & \(8.494\) & \(5.593\) \\
  \bottomrule
\end{tabular}

%% file: publication_tables/effective_transmission/table.tex
\begin{tabular}{rcccrr}
  \toprule
  \(N\) & ACF & Core (\(\eta=0.5\)) & Nonlinear RF & Harmonic RF & KGH \\
  \midrule
  4 & \(1.10\;(0.82,\,1.69)\) & \(0.5639\;(0.5606,\,0.5672)\) & \textemdash & \(0.7360\) & \(0.6323\) \\
  5 & \(0.836\;(0.804,\,0.871)\) & \(0.5540\;(0.5492,\,0.5589)\) & \textemdash & \(0.7185\) & \(0.6780\) \\
  6 & \(0.668\;(0.639,\,0.701)\) & \(0.5618\;(0.5546,\,0.5690)\) & \textemdash & \(0.6905\) & \(0.6395\) \\
  7 & \(0.622\;(0.588,\,0.661)\) & \(0.5627\;(0.5543,\,0.5712)\) & \(0.620\;(0.566,\,0.677)\) & \(0.6626\) & \(0.5995\) \\
  8 & \(0.560\;(0.541,\,0.580)\) & \(0.5457\;(0.5364,\,0.5551)\) & \(0.579\;(0.535,\,0.624)\) & \(0.6369\) & \(0.5645\) \\
  9 & \(0.550\;(0.534,\,0.566)\) & \(0.5333\;(0.5247,\,0.5420)\) & \(0.531\;(0.490,\,0.572)\) & \(0.6138\) & \(0.5341\) \\
  10 & \(0.511\;(0.501,\,0.522)\) & \(0.5162\;(0.5087,\,0.5237)\) & \(0.532\;(0.494,\,0.570)\) & \(0.5931\) & \(0.5075\) \\
  11 & \(0.490\;(0.470,\,0.510)\) & \(0.4950\;(0.4854,\,0.5045)\) & \(0.516\;(0.477,\,0.555)\) & \(0.5746\) & \(0.4838\) \\
  12 & \(0.503\;(0.480,\,0.528)\) & \(0.490\;(0.478,\,0.503)\) & \(0.464\;(0.427,\,0.502)\) & \(0.5579\) & \(0.4626\) \\
  13 & \textemdash & \textemdash & \textemdash & \(0.5428\) & \(0.4434\) \\
  14 & \textemdash & \textemdash & \textemdash & \(0.5290\) & \(0.4259\) \\
  15 & \(0.445\;(0.418,\,0.474)\) & \(0.448\;(0.429,\,0.467)\) & \(0.472\;(0.433,\,0.511)\) & \(0.5165\) & \(0.4099\) \\
  16 & \textemdash & \textemdash & \textemdash & \(0.5049\) & \(0.3952\) \\
  17 & \textemdash & \textemdash & \textemdash & \(0.4943\) & \(0.3816\) \\
  18 & \textemdash & \textemdash & \textemdash & \(0.4845\) & \(0.3690\) \\
  19 & \textemdash & \textemdash & \textemdash & \(0.4754\) & \(0.3573\) \\
  20 & \textemdash & \textemdash & \(0.396\;(0.359,\,0.436)\) & \(0.4669\) & \(0.3464\) \\
  21 & \textemdash & \textemdash & \textemdash & \(0.4589\) & \(0.3361\) \\
  22 & \textemdash & \textemdash & \textemdash & \(0.4515\) & \(0.3266\) \\
  23 & \textemdash & \textemdash & \textemdash & \(0.4445\) & \(0.3175\) \\
  24 & \textemdash & \textemdash & \textemdash & \(0.4379\) & \(0.3090\) \\
  25 & \textemdash & \textemdash & \textemdash & \(0.4317\) & \(0.3010\) \\
  26 & \textemdash & \textemdash & \textemdash & \(0.4258\) & \(0.2934\) \\
  27 & \textemdash & \textemdash & \textemdash & \(0.4202\) & \(0.2862\) \\
  28 & \textemdash & \textemdash & \textemdash & \(0.4149\) & \(0.2793\) \\
  29 & \textemdash & \textemdash & \textemdash & \(0.4098\) & \(0.2728\) \\
  30 & \textemdash & \textemdash & \textemdash & \(0.4050\) & \(0.2666\) \\
  \bottomrule
\end{tabular}

%% file: figures/effective_transmission/caption.tex
\caption{Effective transmission coefficients near coexistence (upper)
and their logarithmic ratios relative to KGH at the same \(N\) (lower).
The states and values are those of Table~\ref{si:tab:effective-transmission};
each rate is divided by its same-state TST rate, without aligning different
methods' reporting extensions. Symbols show ACF, core-residence and nonlinear
RF results with their saved 95\% intervals scaled by the fixed TST denominator;
the interval types are given in Table~\ref{tab:activated-rates-paper}.
Solid black and dashed violet curves show KGH and harmonic RF through \(N=30\).
Shading ends at the operational divider \(N=7.5\), as in main Fig.~8.
Effective rate ratios need not be bounded by one; the KGH ratio is distinct
from its local dynamical transmission factor. No ACF or core estimate is available at \(N=20\); beyond that point only
theory is shown.}

%% file: figures/off_coexistence_rates/caption.tex
\caption{Pointwise KGH predictions and tension-ACF rates at five extensions
  for each \(N=7,\ldots,12\) and \(15\). Bars show 95\% trajectory-bootstrap
  intervals; the dashed line denotes equality. Open symbols identify five
  states with larger uncertainty or fitting-window sensitivity. All 35 states
  enter the displayed logarithmic RMSE.}

%% file: si_sections/committor_diagnostics.tex
\section{Committor sampling and reaction-coordinate diagnostics}
\label{si:sec:committor}

The committor measures the probability that a configuration will fracture
before returning to an intact state. Repeated trajectories from the same
particle positions can have different outcomes because their initial
velocities and subsequent thermal forces differ. These repeated
launches answer two questions: where does the fracture probability,
averaged over configurations, reach one half, and how much do individual
configuration probabilities differ at that location? The calculations cover
\(N=6,8,10\) at their sampled coexistence states.

\subsection{What probability is being measured?}
\label{si:sec:committor-definition}

The periodic chain contains \(N\) particles and \(N\) bonds. Write
\(l_{k}^{*}=l_{k}/l_{0}\) for dimensionless bond length and \(l^{*}=l_{j}^{*}\) for the
length of a fixed tagged bond \(j\). The launch configurations satisfy
\(\sum_k l_{k}^{*}=N\lambda\), with box length \(L=N\lambda l_{0}\).
A spectator is any bond other than the tagged bond. The tagged
length is fixed when preparing a launch; all bonds evolve during shooting, and a
spectator may fracture. The tagged bond retains its identity even when another bond becomes the
longest.

The intact and fractured target cores are regions within the two basins.
Their boundaries are defined using the equal-sharing adiabatic branch
\begin{equation}
 U_{\mathrm{ad,abs}}^{(N-1)}(l^{*})
 =U(l_{0}l^{*})+(N-1)U\!\left(l_{0}\frac{N\lambda-l^{*}}{N-1}\right).
 \label{si:eq:committor-adiabatic-branch}
\end{equation}
Here one bond has reduced length \(l^{*}\) and the remaining bonds share the rest of
the extension equally. Let \(l_{\mathrm i}^{\mathrm{ad},*}=\lambda\), \(d^{*}\), and
\(l_{\mathrm f}^{\mathrm{ad},*}\) be this branch's intact minimum, saddle and fractured
minimum. This absolute energy and the reduced relative profile of
main-text Sec.~II~B obey
\(E_{\mathrm{ad}}^{(N-1),*}=U_{\mathrm{ad,abs}}^{(N-1)}/\De-Nu^*(\lambda)\)
and therefore have the same stationary lengths. The core boundaries lie
halfway from the saddle to each minimum:
\begin{equation}
 \begin{aligned}
 c_{\mathrm i}^{*}&=\tfrac12(l_{\mathrm i}^{\mathrm{ad},*}+d^{*}),&
 c_{\mathrm f}^{*}&=\tfrac12(d^{*}+l_{\mathrm f}^{\mathrm{ad},*}),\\
 \mathcal C_{\mathrm i}&=\{\max_k l_{k}^{*}\le c_{\mathrm i}^{*}\},&
 \mathcal C_{\mathrm f}&=\{\max_k l_{k}^{*}\ge c_{\mathrm f}^{*}\}.
 \end{aligned}
 \label{si:eq:committor-cores}
\end{equation}
Thus the intact core requires every bond to be short, while entry into the
fractured core requires any bond to reach the longer threshold. This core choice
specifies the event whose probability is measured. The PMF maximum instead
provides a thermodynamic comparison and guides the initial shooting lengths.

For \(a\in\{\mathrm i,\mathrm f\}\), define the first-entry time
\(\tau_a=\inf\{t\ge0:\bm r(t)\in\mathcal C_a\}\), where
\(\bm r\) denotes all particle positions. The binary outcome is
\begin{equation}
 Y=\mathbf1(\tau_{\mathrm f}<\tau_{\mathrm i}).
 \label{si:eq:committor-outcome}
\end{equation}
It equals one for fracture first and zero for return to the intact core
first. Fracture at either the tagged bond or another bond contributes to
\(Y=1\). Crossing a shooting surface or the PMF maximum alone is not the
stopping event.

Three averages distinguish a complete initial state from a projected
coordinate. With both positions \(\bm r\) and velocities \(\bm v\)
fixed, average over subsequent thermostat randomness \(\mathcal W\):
\begin{equation}
 Q_{\mathrm{comm}}(\bm r,\bm v)
 =\Pr_{\mathcal W}(\tau_{\mathrm f}<\tau_{\mathrm i}\mid\bm r,\bm v).
 \label{si:eq:committor-phase-space}
\end{equation}
Next, let \(M_0\) be the Maxwell--Boltzmann probability distribution of
initial velocity vectors conditioned on zero total momentum. Averaging over
these velocities gives the configuration committor \citep{PetersTrout2006},
\begin{equation}
 q_{\mathrm{comm}}(\bm r)=\int Q_{\mathrm{comm}}(\bm r,\bm v)\,M_0(d\bm v).
 \label{si:eq:committor-configuration}
\end{equation}
Finally, let \(\pi_{l^*}\)
be the thermal probability distribution of configurations with tagged-bond
length \(l^{*}\) and the spectator restriction specified below. The projected
committor and its half-probability crossing are
\begin{equation}
 \overline q(l^{*})=\int q_{\mathrm{comm}}(\bm r)\,\pi_{l^*}(d\bm r),
 \qquad \overline q(l_{1/2}^{*})=\tfrac12.
 \label{si:eq:committor-projection}
\end{equation}
The notation \(\pi_{l^*}(d\bm r)\) describes probability weights over
configurations, and \(M_0(d\bm v)\) does the same for velocity vectors.
At the projected crossing, the fracture probability averaged over
configurations is one half. Individual configurations with the same tagged
length can nevertheless have different fracture probabilities because the
remaining bonds are arranged differently.

\subsection{Initial conditions and shooting samples}
\label{si:sec:committor-sampling}

The temperature is \(450\,\mathrm K\), friction
\(\xi=0.022\,\mathrm{fs}^{-1}\), and integration step
\(0.25\,\mathrm{fs}\). The masses, Morse parameters and periodic bonded
communication settings are those in Sec.~\ref{si:langevin}.
Table~\ref{si:tab:committor-actual-states} lists the sampled states and
core boundaries.

\begin{table}[htbp]
 \centering\small
 \caption{Sampled coexistence states and fixed boundaries in dimensionless
 bond length. Here \(d^{*}\) is both the adiabatic saddle and the spectator
 cutoff used when sampling launch configurations.}
 \label{si:tab:committor-actual-states}
 \begin{tabular}{rcccc}
 \toprule
 \(N\) & \(\lambda\) & \(d^{*}\) & \(c_{\mathrm i}^{*}\) & \(c_{\mathrm f}^{*}\) \\
 \midrule
 \RegisteredCommittorStateRows
 \bottomrule
 \end{tabular}
\end{table}

\paragraph{Conditional position ensemble.}
At a chosen tagged length \(l_{j}^{*}=l^{*}\), the spectators share the remaining
length \(N\lambda-l^{*}\). They are sampled jointly, since extending one
leaves less length for the others. With \(\beta=(k_{\mathrm B}T)^{-1}\),
the continuum target measure in bond coordinates is
\begin{equation}
 \begin{aligned}
 \pi_{l^*}(d\bm l_{-j}^{*})={}&
 \frac{\exp[-\beta\sum_{k\ne j}U(l_{0}l_{k}^{*})]}
      {Z_{N-1}^{\mathrm i,*}(N\lambda-l^{*};d^{*})}\\
 &\times\delta\!\left(\sum_{k\ne j}l_{k}^{*}-(N\lambda-l^{*})\right)
 \prod_{k\ne j}\bigl[\mathbf1(0<l_{k}^{*}<d^{*})\,dl_{k}^{*}\bigr].
 \end{aligned}
 \label{si:eq:committor-ensemble}
\end{equation}
Here \(\bm l_{-j}^{*}\) contains the reduced spectator lengths, the delta function
enforces their sum, and \(Z_{N-1}^{\mathrm i,*}\) normalizes the distribution. The tagged
energy is constant at fixed \(l^{*}\) and cancels from the conditional
probability. The fixed cutoff \(d^{*}\) applies only to spectators; it neither
restricts the tagged length nor changes with it.

The implemented sampler approximates this measure on a dimensionless grid
of spacing \(0.0005\) over \([0.4,3]\); the allowed spectator range here
is therefore \([0.4,d^{*})\). Restricted Boltzmann weights and partition
functions generate successive spectator lengths at a constrained discrete
total. The two grid totals bracketing \(N\lambda-l^{*}\) are selected with
weights proportional to their partition densities times their interpolation
fractions. After sampling, a uniform correction enforces the exact total;
the sampler checks support and randomly permutes the spectators. This
construction approximates the continuum distribution in
Eq.~\eqref{si:eq:committor-ensemble}; an exact final sum alone does not
establish continuum sampling accuracy.

To construct a launch, convert \(l_{k}^{*}\) to \(l_{k}=l_{0}l_{k}^{*}\), then
place particles by cumulative addition of these lengths in the periodic
box \(L=N\lambda l_{0}\). The final bond closes across the boundary.
The positions and bond connectivity are written to the LAMMPS data file.
Fixing an arbitrary overall translation maps the bond ensemble to the
position ensemble in Eq.~\eqref{si:eq:committor-projection}. These
thermally sampled bond lengths are generally unequal. The same positions
are reused for a configuration's repeated launches; no bond is held fixed
or confined by the spectator cutoff during the ensuing trajectory.

\paragraph{Velocities and trajectories.}
For equal particle masses \(m\), draw each longitudinal velocity independently
from the same Gaussian distribution, then subtract their mean:
\begin{equation}
 \begin{aligned}
 \widetilde v_i&\sim
     \mathcal N(0,k_{\mathrm B}T/m),\qquad i=1,\ldots,N,\\
 v_i&=\widetilde v_i-\frac1N\sum_{k=1}^{N}\widetilde v_k,
 \qquad \bm v\sim M_0,\qquad \sum_{i=1}^N m v_i=0.
 \end{aligned}
 \label{si:eq:committor-velocities}
\end{equation}
The resulting velocities are jointly Gaussian, with zero total momentum
and fluctuating kinetic energy. No rescaling to a fixed kinetic energy or
positive-velocity selection is applied. Subsequent motion uses
\texttt{fix nve} with ordinary \texttt{fix langevin}, damping time
\(1/\xi\), and an independent thermostat seed for each shot. Its centered
uniform random-force draws have the Langevin variance; the Gaussian
initial-velocity law is a separate sampling step. Zero total momentum is
imposed at launch, rather than continuously on the random forces.

Core entry is checked every integration step. A trajectory stops at the
first entry or at \(6000\,\mathrm{fs}\); a trajectory reaching neither
core would be censored, rather than counted as an intact outcome. All
270336 production shots reached a core, with no failed, censored or initially
absorbed launches. The largest observed first-hit time was
\(438\,\mathrm{fs}\). These observations describe the finite samples.

\paragraph{Production sampling.}
The crossing sample estimates the projected committor at five fixed tagged
lengths spanning probability one half. These outcomes determine the reported
crossing. The dispersion sample fixes the tagged bond at the
estimated crossing \(\widehat l_{1/2}^{*}\) and draws fresh configurations
from \(\pi_{\widehat l_{1/2}^{*}}\), and also samples two flanking surfaces at
\(\widehat l_{1/2}^{*}\pm0.035\). These samples measure variation between
configurations at each fixed length. Table~\ref{si:tab:committor-allocation}
gives the allocation per chain.

\begin{table}[htbp]
 \centering\small
 \caption{Production shooting allocation per chain. At each surface, \(C\) counts
 independent configurations, \(V\) the velocity vectors per configuration,
 and \(R\) the independent noise realizations per velocity vector.
 The final column includes all surfaces in the sample.}
 \label{si:tab:committor-allocation}
 \begin{tabular}{lrrrrr}
 \toprule
 Sample & Surfaces & \(C\) & \(V\) & \(R\) & Shots \\
 \midrule
 Crossing & 5 & 512 & 16 & 1 & 40960 \\
 Dispersion & 3 & 512 & 16 & 2 & 49152 \\
 \bottomrule
 \end{tabular}
\end{table}

Using fresh configurations separates the crossing estimate from the estimate of
configuration dispersion there. The dispersion-sample mean is therefore not
forced to equal one half, and its profile does not replace the crossing
estimate. The reported uncertainties condition on the surfaces actually
chosen; they do not include uncertainty in selecting those surfaces.

\paragraph{From probability integrals to measured averages.}
At one surface, let \(c\), \(v\) and \(r\) index configuration,
velocity draw and noise realization. The scalar index \(r\) is distinct
from the position vector \(\bm r\). For each configuration
\(\bm r_c\) and velocity vector \(\bm v_{cv}\), record
\(Y_{cvr}=\mathbf1(\tau_{\mathrm f}^{(cvr)}<\tau_{\mathrm i}^{(cvr)})\).
The three probability averages above become
\begin{equation}
 \begin{aligned}
 \widehat Q_{\mathrm{comm},cv}&=\frac1R\sum_{r=1}^R Y_{cvr},\\
 \widehat q_c&=\frac1V\sum_{v=1}^V\widehat Q_{\mathrm{comm},cv}
             =\frac1{VR}\sum_{v=1}^V\sum_{r=1}^R Y_{cvr},\\
 \widehat{\overline q}(l^{*})&=\frac1C\sum_{c=1}^C\widehat q_c
             =\frac1{CVR}\sum_{c=1}^C\sum_{v=1}^V\sum_{r=1}^R Y_{cvr}.
 \end{aligned}
 \label{si:eq:committor-sampling-averages}
\end{equation}
These estimate \(Q_{\mathrm{comm}}(\bm r_c,\bm v_{cv})\), \(q_{\mathrm{comm}}(\bm r_c)\), and
\(\overline q(l^{*})\), respectively. Sampling from \(M_0\) and \(\pi_{l^*}\)
supplies the probability weighting. The equal, fully resolved allocations used
here justify these simple averages. A finite-shot \(\widehat q_c\)
remains noisy, even at fixed true \(q_{\mathrm{comm}}(\bm r_c)\).

\subsection{Estimating the projected crossing}
\label{si:sec:committor-crossing}

At shooting length \(l_{s}^{*}\), let \(n_{\mathrm f,s}\) count fracture-first outcomes
and \(n_s\) count resolved trajectories. Equation~\eqref{si:eq:committor-sampling-averages}
gives the empirical projected committor
\(\widehat{\overline q}_s\equiv\widehat{\overline q}(l_{s}^{*})=n_{\mathrm f,s}/n_s\),
with \(n_s=CVR\). Find two adjacent probabilities surrounding one half
and interpolate between their coordinates. Writing
\(\widetilde{\overline q}_s\) for the fitted nondecreasing profile,
\begin{equation}
 \widehat l_{1/2}^{*}=l_{s-1}^{*}+
 \frac{1/2-\widetilde{\overline q}_{s-1}}
      {\widetilde{\overline q}_s-\widetilde{\overline q}_{s-1}}
 (l_{s}^{*}-l_{s-1}^{*}).
 \label{si:eq:committor-interpolation}
\end{equation}
The ratio specifies the fraction of the interval from \(l_{s-1}^{*}\) to
\(l_{s}^{*}\) at which the connecting line reaches one half.

If measured fractions decrease locally, isotonic regression \citep{Ayer1955} pools
neighboring success and shot counts until the fitted profile is
nondecreasing. No pooling was needed for the observed profiles, so their
empirical and fitted probabilities coincide. Local monotonicity is a
modelling assumption. A flat segment exactly at one half identifies a
range rather than a unique crossing; without a bracket, no crossing is
reported.

Uncertainty is estimated by resampling whole configurations with
replacement separately at each crossing-sample surface, retaining all linked
shots and re-estimating the crossing. The 2.5th and 97.5th percentiles of
2000 bootstrap roots give the marginal 95\% intervals in
Table~\ref{si:tab:committor-locations}. Every bootstrap replicate had
a unique bracketed crossing. These intervals describe sampling uncertainty
conditional on the realized crossing-sample surfaces; they exclude sensitivity
to interpolation choice and numerical discretization.

\begin{table}[htbp]
 \centering\small
 \setlength{\tabcolsep}{5pt}
 \caption{Same-state transition locations in \(l^{*}=l/l_{0}\). The primary
 crossing uses the crossing sample and monotone linear interpolation.
 Intervals are conditional sampling intervals and exclude surface-selection,
 interpolation-method and numerical discretization uncertainty.}
 \label{si:tab:committor-locations}
 \begin{tabular}{rcccc}
 \toprule
 \(N\) & \(l_{\mathrm{ad}}^{\ddagger,*}\) & \(l_{\mathrm s}^{\mathrm{site},*}\) &
 \(\widehat l_{1/2}^{*}\) & 95\% interval \\
 \midrule
 \RegisteredCommittorLocationRows
 \bottomrule
 \end{tabular}
\end{table}

\begin{figure}[!htbp]
 \centering
 \includegraphics[width=16cm]{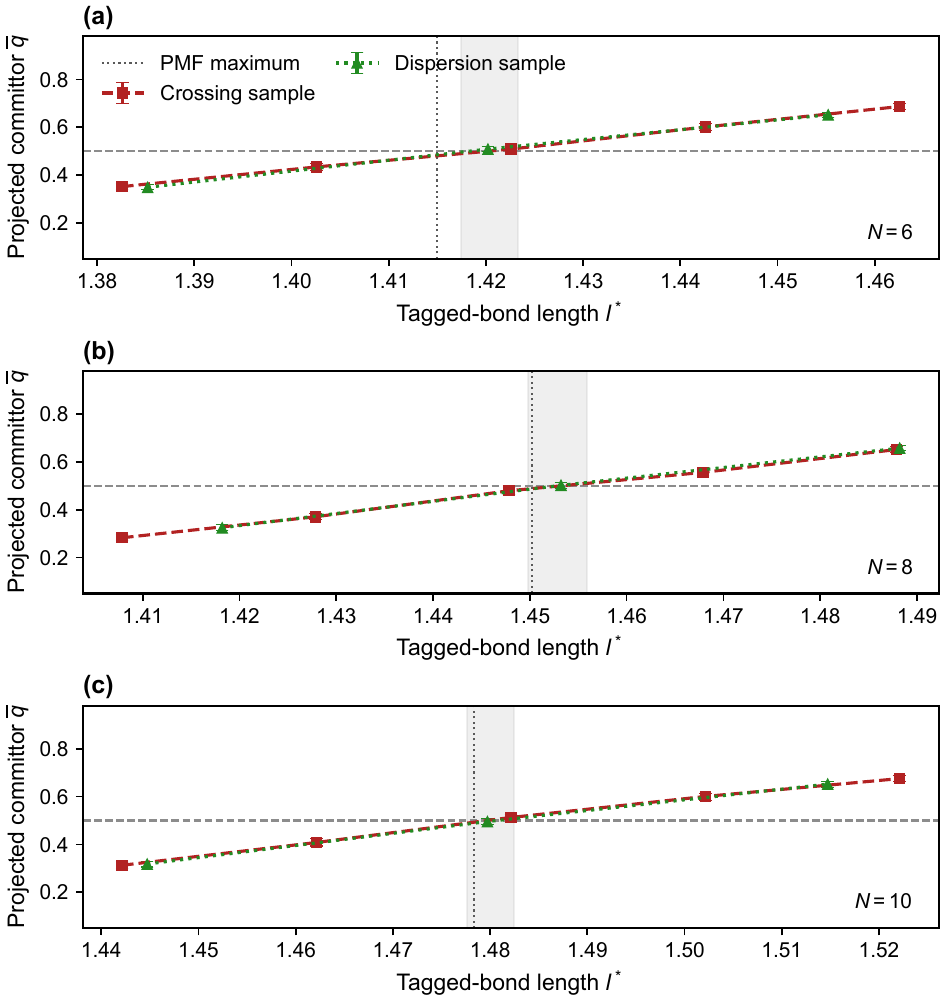}
 \input{figures/committor/caption_profiles}
 \label{si:fig:committor-profiles}
\end{figure}

Each crossing lies much closer to the site-PMF maximum than to the
adiabatic saddle. The \(N=6\) nominal interval excludes its PMF marker,
while the \(N=8,10\) intervals include theirs. The PMF describes equilibrium
free energy; the committor also depends on the cores and dynamics, so
coincidence is not required. The small residual offsets remain sensitive
to interpolation choice, and the sampling intervals do not bound numerical
discretization error. They therefore do not resolve a physical displacement
from the PMF maximum.
Figure~\ref{si:fig:committor-profiles} shows
the measured profiles from the two production samples.

\FloatBarrier

\subsection{Estimating configuration dispersion}
\label{si:sec:committor-dispersion}

The desired quantity is the standard deviation of configuration
committors at fixed tagged length,
\(\sigma_q(l^{*})=\{\operatorname{Var}_{\bm r\sim\pi_{l^*}}[q_{\mathrm{comm}}(\bm r)]\}^{1/2}\).
Observed success fractions \(\widehat q_c\) contain both this variation
and finite-shooting variability. A configuration with true \(q=1/2\),
for example, need not yield exactly 16 successes in 32 shots. Its raw
fraction is an estimate of its committor, not a directly observed probability.

Let \(q_c=q_{\mathrm{comm}}(\bm r_c)\) and
\(p_{cv}=Q_{\mathrm{comm}}(\bm r_c,\bm v_{cv})\). Applying conditional variance to the
configuration, velocity and noise hierarchy gives, for equal allocations,
\begin{equation}
 \begin{aligned}
 \operatorname{Var}(\widehat q_c)={}&
 \underbrace{\operatorname{Var}_c(q_c)}_{\text{configuration spread}}
 +\underbrace{\frac1V\operatorname E_c[\operatorname{Var}_v(p_{cv}\mid c)]}
             _{\text{finite velocity sampling}}\\
 &+\underbrace{\frac1{VR}\operatorname E_{c,v}[p_{cv}(1-p_{cv})]}
              _{\text{finite noise sampling}}.
 \end{aligned}
 \label{si:eq:committor-variance-decomposition}
\end{equation}
The first term is the target variance. Drawing more independent velocities
reduces the second term; repeating noise with the same velocity only
reduces the third. This distinction motivates retaining the velocity groups
when estimating configuration dispersion.

\paragraph{Observations and model.}
At each dispersion-sample surface, \(C=512\) configurations have \(V=16\)
velocity draws and \(R=2\) noise realizations per draw. Define
\begin{equation}
 n_{\mathrm f,cv}=\sum_{r=1}^R Y_{cvr}\in\{0,1,2\},\qquad R_{cv}=2.
 \label{si:eq:committor-count-matrix}
\end{equation}
Here \(R_{cv}\) counts resolved noise shots in group \((c,v)\); all
shots reached a core. The observations form a \(512\times16\) matrix of
counts with a matching matrix of totals. A row contains all 32 shots
from one configuration. The two shots in each column of that row share
both positions and a complete initial velocity vector, while their
thermostat-noise sequences differ.

Define the logistic function \(\operatorname{logit}^{-1}(z)=(1+e^{-z})^{-1}\), which maps a real
number to a probability. At a fixed surface, the adopted model is
\begin{equation}
 \begin{aligned}
 n_{\mathrm f,cv}\mid u_c,w_{cv}&\sim
   \operatorname{Binomial}(R_{cv},p_{cv}),\qquad
 p_{cv}=\operatorname{logit}^{-1}(\mu+u_c+w_{cv}),\\
 u_c&\sim\mathcal N(0,s_{\mathrm c}^2),\qquad
 w_{cv}\sim\mathcal N(0,s_{\mathrm v}^2).
 \end{aligned}
 \label{si:eq:committor-nested-model}
\end{equation}
The intercept \(\mu\) sets the central logit,
\(\operatorname{logit}p=\log[p/(1-p)]\). The configuration effect
\(u_c\) is shared by all velocity groups in row \(c\); the velocity
effect \(w_{cv}\) is shared by the two noise shots in that group.
Different configuration effects are independent, and the velocity effects
are mutually independent and independent of them before observing outcomes.
Their standard deviations \(s_{\mathrm c}\) and \(s_{\mathrm v}\)
are on the logit scale. Conditional on both effects, independent noise
realizations give the binomial count.

The counts are observed, while \(u_c\), \(w_{cv}\) and \(p_{cv}\)
are latent quantities. In particular, \(w_{cv}\) is a statistical effect
on fracture probability, not the physical velocity vector or a Maxwell
velocity variance. Gaussian effects, additive logits and a common
velocity-effect variance within each surface are assumptions of this model.

\paragraph{Likelihood and fitting.}
Write \(\phi_s\) for the zero-mean Gaussian density with standard
deviation \(s\). At specified effects \(u,w\), the probability assigned
to one observed count is
\begin{equation}
 \begin{aligned}
 B_{cv}(u,w)={}&{R_{cv}\choose n_{\mathrm f,cv}}
 [\operatorname{logit}^{-1}(\mu+u+w)]^{n_{\mathrm f,cv}}\\
 &\times[1-\operatorname{logit}^{-1}(\mu+u+w)]^{R_{cv}-n_{\mathrm f,cv}}.
 \end{aligned}
 \label{si:eq:committor-binomial-factor}
\end{equation}
For \(R_{cv}=2\) and \(p=\operatorname{logit}^{-1}(\mu+u+w)\), zero, one and two successes
have probabilities \((1-p)^2\), \(2p(1-p)\), and \(p^2\).
The factor of two counts the two possible orders of a mixed outcome. At
\(p=1/2\) these probabilities are \(1/4,1/2,1/4\); this example does
not assume that every group has probability one half.

Averaging over the unobserved effects gives the full marginal likelihood,
\begin{equation}
 \mathcal L(\mu,s_{\mathrm c},s_{\mathrm v})
 =\prod_{c=1}^C\int\phi_{s_{\mathrm c}}(u)
   \prod_{v=1}^V\left[
     \int\phi_{s_{\mathrm v}}(w)B_{cv}(u,w)\,dw
   \right]du.
 \label{si:eq:committor-likelihood}
\end{equation}
Read this expression from the inside out. The inner integral averages the
probability of one count over its velocity effect. For a fixed
configuration effect, the product combines that row's velocity groups.
The outer integral averages the entire row probability over its shared
configuration effect, and the final product combines independent rows.
The velocity product must remain inside the configuration integral:
integrating a separate \(u\) for each velocity group would discard their
shared configuration.

For candidate values of \(\mu,s_{\mathrm c},s_{\mathrm v}\), 24-node
Gauss--Hermite quadrature evaluates each Gaussian integral. The logarithms
of the row probabilities are summed to form the log likelihood. Quadrature
nodes are weighted integration points, not additional simulated
configurations or velocities. Maximizing this likelihood fits three
population parameters separately at each surface; the individual latent
effects are integrated out rather than freely fitted one by one.

The calculation uses bounded L-BFGS-B \citep{Byrd1995} from multiple starts, with
\(\mu\in[-12,12]\) and positive standard deviations represented by
\(\log s_{\mathrm c},\log s_{\mathrm v}\in[-8,3]\). Null fits start
at velocity SDs 0.2 and 0.8; positive configuration-SD fits start at 0.1,
0.5 and 1.0, using the fitted null values for the other parameters. The
best converged finite candidate is retained, including the exact
\(s_{\mathrm c}=0\) solution if a positive-SD fit does not improve the
likelihood. These numerical choices specify the calculation; convergence
alone does not establish model adequacy.

\paragraph{From fitted parameters to probability dispersion.}
For a configuration represented by effect \(u\), average over velocity
effects to obtain
\begin{equation}
 q(u)=\int \operatorname{logit}^{-1}(\mu+u+w)\phi_{s_{\mathrm v}}(w)\,dw.
 \label{si:eq:committor-probability-dispersion}
\end{equation}
This is the model counterpart of the velocity average in
Eq.~\eqref{si:eq:committor-configuration}. The resulting configuration
probabilities have population mean and variance
\begin{equation}
 \begin{aligned}
 \overline q_{\mathrm{fit}}&=\int q(u)\phi_{s_{\mathrm c}}(u)\,du,\\
 \sigma_q^2&=\int[q(u)-\overline q_{\mathrm{fit}}]^2
                     \phi_{s_{\mathrm c}}(u)\,du.
 \end{aligned}
 \label{si:eq:committor-model-moments}
\end{equation}
With fitted parameters substituted, the calculation evaluates \(q(u)\)
at configuration quadrature nodes, uses their weights to obtain its mean
and variance, and takes the square root to report \(\widehat\sigma_q\).
The latent SD \(s_{\mathrm c}\) measures variation in logit effects;
\(\sigma_q\) measures variation in configuration fracture probabilities
after averaging velocities. They are different quantities. The integral
in Eq.~\eqref{si:eq:committor-probability-dispersion} also means that the
final distribution of \(q\) is generally not an ordinary logit-normal
distribution.

\paragraph{Uncertainty and measured results.}
The bootstrap samples 512 configuration rows with replacement at each
surface, retaining all 16 velocity groups and their linked noise outcomes.
For each of 1000 resampled matrices, the model is refitted and its mean
and SD recomputed. Their 2.5th and 97.5th percentiles give marginal 95\%
intervals. All replicates succeeded. This resampling preserves the
within-configuration dependence; its intervals are conditional on the selected
surface and the statistical model and exclude uncertainty in choosing
that surface and in numerical discretization.

\begin{table}[htbp]
 \centering\small
 \setlength{\tabcolsep}{4pt}
 \caption{Configuration probabilities at the selected crossings. The empirical
 mean is the average of the recorded outcomes; the fitted mean and SD use
 the hierarchical model. Brackets give marginal conditional 95\% intervals
 for the fitted population summaries.}
 \label{si:tab:root-committor-width}
 \begin{tabular}{rccc}
 \toprule
 \(N\) & Empirical mean & Fitted mean [95\% interval] &
 \(\widehat\sigma_q\) [95\% interval] \\
 \midrule
 \RegisteredCommittorDispersionRows
 \bottomrule
 \end{tabular}
\end{table}

Table~\ref{si:tab:root-committor-width} distinguishes the empirical and
fitted means: likelihood fitting need not reproduce the aggregate fraction
exactly. Neither mean is constrained to one half at the independently
estimated crossing. The measured configuration SDs are approximately
\(\RegisteredCommittorSdSix\), \(\RegisteredCommittorSdEight\), and
\(\RegisteredCommittorSdTen\) for \(N=6,8,10\).
An SD describes variation between configuration probabilities; its
confidence interval describes uncertainty in estimating that population
spread. The interval is not a range of probabilities containing individual
configurations.

The fitted dispersion depends on the assumed statistical model. Repeated
velocities separate configuration differences from within-configuration
variation, and the two noise shots per velocity provide the additional
information from repetition needed to distinguish velocity and noise effects.
This modest repetition does not determine full latent distribution shapes.

\FloatBarrier

%% file: figures/committor/caption_profiles.tex
\caption{Projected committor profiles at the sampled coexistence states for
(a) \(N=6\), (b) \(N=8\), and (c) \(N=10\).
The crossing sample locates the projected half-probability crossing;
the dispersion sample uses fresh configurations at that estimated crossing
and two flanking lengths. Only the crossing sample enters the reported
primary crossing estimate. Both curves show production data.
Vertical dotted lines mark the same-state site-PMF maxima; gray bands show
the conditional 95\% crossing intervals. The horizontal dashed
line marks probability one half. Point bars are marginal 95\%
configuration-bootstrap intervals; connecting lines guide the eye.
All coordinates are dimensionless, \(l^*=l/l_0\).}

%% file: si_sections/kinetic_crossover.tex
\section{Two-state accuracy and excursion timescales}
\label{si:sec:kinetic-crossover}

A chain can cross a dividing surface several times before settling in the
opposite basin. A two-state description replaces this detailed motion by
transitions between intact and fractured states with constant rates. The analysis asks
over which observation times this approximation describes the simulated
chain, and how the time spent in unsuccessful excursions compares with the
time over which the populations relax. The first question is tested using
independent trajectory groups; the second is answered by identifying and
timing individual excursions. The calculations cover \(N=6\)--10 at
\(T=450\,\mathrm K\).

\subsection{Preparing trajectories and identifying the two states}
\label{si:sec:crossover-trajectories}

The analysis uses dimensionless bond lengths \(l_{k}^{*}=l_{k}/l_{0}\) and the maximum
\(l_{\max}^{*}(t)=\max_k l_{k}^{*}(t)\). As in Sec.~\ref{si:sec:committor}, \(l_{\mathrm i}^{\mathrm{ad},*}=\lambda\),
\(d^{*}\), and \(l_{\mathrm f}^{\mathrm{ad},*}\) denote the intact minimum, saddle,
and fractured minimum of \(U_{\mathrm{ad,abs}}^{(N-1)}\)
[Eq.~\eqref{si:eq:committor-adiabatic-branch}]. These landmarks define
regions lying inside the two basins, called cores. For a core fraction
\(\eta\),
\begin{equation}
 \begin{aligned}
 c_{\mathrm i}^{*}(\eta)&=l_{\mathrm i}^{\mathrm{ad},*}
       +\eta(d^{*}-l_{\mathrm i}^{\mathrm{ad},*}),\\
 c_{\mathrm f}^{*}(\eta)&=l_{\mathrm f}^{\mathrm{ad},*}
       +\eta(d^{*}-l_{\mathrm f}^{\mathrm{ad},*}),\\
 \mathcal C_{\mathrm i}(\eta)&=\{l_{\max}^{*}\le c_{\mathrm i}^{*}(\eta)\},&
 \mathcal C_{\mathrm f}(\eta)&=\{l_{\max}^{*}\ge c_{\mathrm f}^{*}(\eta)\}.
 \end{aligned}
 \label{si:eq:crossover-cores}
\end{equation}
The primary choice, \(\eta=0.5\), is exactly the core construction in
Eq.~\eqref{si:eq:committor-cores}; the boundaries are then written \(c_{\mathrm i}^{*}\) and
\(c_{\mathrm f}^{*}\) without the argument. At the shared \(N=6,8,10\) states,
the primary boundaries are also numerically identical to those used for
the committor. Values \(\eta=0.25\) and 0.75 place the boundaries deeper
inside the basins or closer to the saddle, respectively, and test how much
the measured rates depend on this choice.

Each simulation has the fixed extension listed in
Table~\ref{si:tab:crossover-coordinates}. Half of the trajectories start
in each basin. An initial bond configuration is sampled from the
fixed-extension canonical distribution restricted to that basin:
\begin{equation}
 \rho_a(d\bm l^{*})=\frac{1}{Z_{a}^{*}}
  e^{-\beta\sum_k U(l_{0}l_{k}^{*})}
  \delta\!\left(\sum_kl_{k}^{*}-N\lambda\right)
  \mathbf1_{\mathcal B_a}(\bm l^{*})
  \prod_k\bigl[\mathbf1(l_{k}^{*}>0)\,dl_{k}^{*}\bigr],
 \quad a\in\{\mathrm i,\mathrm f\},
 \label{si:eq:crossover-initial-ensemble}
\end{equation}
where \(\beta=(k_{\mathrm B}T)^{-1}\), \(Z_{a}^{*}\) is the basin-restricted
normalizer in reduced bond coordinates,
and \(\mathcal B_{\mathrm i}=\{l_{\max}^{*}\le d^{*}\}\),
\(\mathcal B_{\mathrm f}=\{l_{\max}^{*}>d^{*}\}\). Preparation uses these entire basins,
not just their smaller cores. The sum constraint means that the bond
lengths must be sampled together.

To sample Eq.~\eqref{si:eq:crossover-initial-ensemble}, a
random amount of length is exchanged between two randomly selected bonds,
\(l_{j}^{\prime,*}=l_{j}^{*}+\delta l^{*}\), \(l_{k}^{\prime,*}=l_{k}^{*}-\delta l^{*}\),
where a prime denotes a proposed bond length. This leaves their sum
unchanged. A proposal that gives a nonpositive bond length or falls outside the selected
basin is rejected; otherwise its acceptance probability is \citep{Metropolis1953}
\begin{equation}
 A=\min\{1,\exp(-\beta\Delta U)\},
 \label{si:eq:crossover-metropolis}
\end{equation}
where \(\Delta U\) is the change in the two bonds' Morse energy. The
proposal is uniform and symmetric about zero. Its width is adjusted
during 20000 warmup proposals and then held fixed. Each independent
sampler supplies 4000 further recorded configurations separated by 20
proposals; the last is used for one trajectory. Sampling diagnostics
compare these independent samplers before launching dynamics. A cyclic
relabeling places the longest bond first, without changing the
maximum-bond observables considered here.

Particle positions are formed by cumulatively adding \(l_{0}l_{k}^{*}\) in the
periodic box of length \(N\lambda l_{0}\), as described in Sec.~\ref{si:sec:committor}.
Initial velocities follow the zero-total-momentum Maxwell distribution
in Eq.~\eqref{si:eq:committor-velocities}. The ensuing Langevin trajectories
are unrestricted, with friction \(\xi=0.022\,\mathrm{fs}^{-1}\),
integration step \(0.25\,\mathrm{fs}\), and an independent noise realization
for every trajectory. The 5632 trajectories have independent initial
configurations, velocities and thermostat seeds.

The maximum bond length \(l_{\max}^{*}\) is recorded every \(1\,\mathrm{fs}\). The last-core label
\(z_{\mathrm{LC}}(t)\in\{\mathrm i,\mathrm f\}\) changes only when the opposite core is
reached: between the cores it retains the identity of the last core
visited \citep{Schutte2011}. Thus a brief outward crossing followed by a return does not
itself change the state label. The initial segment before the first
observed core visit is excluded. Sensitivity checks also retain
every second or fourth sample, always starting from the same initial
time, and reconstruct the labels on each coarser trace. Together with the
three core fractions, the observation intervals
\(\Delta t_{\mathrm{obs}}=1,2,4\,\mathrm{fs}\) give nine conditions
\(c_{\mathrm{an}}=(\eta,\Delta t_{\mathrm{obs}})\). Thus \(c_{\mathrm{an}}\) denotes an analysis
condition: \(\eta\) determines the core boundaries and
\(\Delta t_{\mathrm{obs}}\) the observation interval. These are nine
analyses of the same trajectories.

\begin{table}[htbp]
  \centering\small
  \setlength{\tabcolsep}{3.5pt}
  \caption{Simulated states and coordinates used in the crossover analysis.
  Coordinate columns are in \(\text{\AA}\): \(l_{0}c_{\mathrm i}^{*}\) and
  \(l_{0}c_{\mathrm f}^{*}\) are the primary core boundaries,
  \(s_{\mathrm{ad}}=l_{0}d^{*}\) is the adiabatic surface,
  \(s_{\mathrm{PMF}}=l_{0}l_{\mathrm s}^{\mathrm{site},*}\) is the
  intact-conditioned tagged-site PMF maximum, and
  \(s_{\mathrm{out}}=2s_{\mathrm{PMF}}-s_{\mathrm{ad}}\) is the outer surface.
  The last two surfaces are used for the excursion comparison below.}
  \label{si:tab:crossover-coordinates}
  \begin{tabular}{rrrrrrr}
  \toprule
  \(N\)&\(\lambda\)&\(l_{0}c_{\mathrm i}^{*}\)&\(s_{\mathrm{ad}}\)&\(s_{\mathrm{PMF}}\)&\(s_{\mathrm{out}}\)&\(l_{0}c_{\mathrm f}^{*}\)\\
  \midrule
  \RegisteredCrossoverCoordinateRows
  \bottomrule
  \end{tabular}
\end{table}

\subsection{From recorded state labels to transition rates}
\label{si:sec:crossover-rate-estimation}

The observation lag \(\tau_{\mathrm{lag}}\) separates the two state labels being
compared. It is distinct from the interval \(\Delta t_{\mathrm{obs}}\)
between stored samples: a lag may span many samples. For each trajectory,
the analysis examines all pairs \(z_{\mathrm{LC}}(t),z_{\mathrm{LC}}(t+\tau_{\mathrm{lag}})\) for which both times are recorded
and the last-core label is known. Pairs can overlap in time but never join
different trajectories.

Before running the dynamics, trajectories are divided equally into
calibration and validation groups within each preparation batch and
initial basin. Alternating sampler indices determine the assignment.
Let \(\mathcal D\) denote either set of trajectories. Calibration uses
only the primary condition
\(c_{\mathrm{an},0}=(\eta,\Delta t_{\mathrm{obs}})=(0.5,1\,\mathrm{fs})\) at a fixed
calibration lag \(\tau_{\mathrm{lag},0}\) to determine the reference
rates \(k_{\nu}^{\mathrm{core},0}\). The index \(\nu\) denotes the transition direction:
\(\mathrm{if}\) for intact to fractured and \(\mathrm{fi}\) for fractured
to intact. Validation uses independent trajectories at all nine conditions,
including \(c_{\mathrm{an},0}\), over the tested lags. Every validation estimate is
compared with the reference rate for the same direction. Calibration
and validation therefore differ in their trajectories, while the primary
analysis condition is common to both.

For trajectory \(r\) under condition \(c_{\mathrm{an}}\), write the label as
\(z_{\mathrm{LC},r,c_{\mathrm{an}}}(t)\). Let \(\mathcal T_{r,c_{\mathrm{an}}}(\tau_{\mathrm{lag}})\) contain
all starting times on the selected observation grid for which both
\(t\) and \(t+\tau_{\mathrm{lag}}\) are recorded and have known labels.
For starting state \(a\) and endpoint state \(b\), the transition counts
and probability estimates are
\begin{equation}
 \begin{aligned}
 C_{ab}^{\mathcal D,c_{\mathrm{an}}}(\tau_{\mathrm{lag}})
   &=\sum_{r\in\mathcal D}\ \sum_{t\in\mathcal T_{r,c_{\mathrm{an}}}(\tau_{\mathrm{lag}})}
     \mathbf1[z_{\mathrm{LC},r,c_{\mathrm{an}}}(t)=a]\,\mathbf1[z_{\mathrm{LC},r,c_{\mathrm{an}}}(t+\tau_{\mathrm{lag}})=b],\\
 \widehat P_{ab}^{\mathcal D,c_{\mathrm{an}}}(\tau_{\mathrm{lag}})
   &=\frac{C_{ab}^{\mathcal D,c_{\mathrm{an}}}(\tau_{\mathrm{lag}})}
     {C_{a\mathrm i}^{\mathcal D,c_{\mathrm{an}}}(\tau_{\mathrm{lag}})+
      C_{a\mathrm f}^{\mathcal D,c_{\mathrm{an}}}(\tau_{\mathrm{lag}})},
 \qquad a,b\in\{\mathrm i,\mathrm f\}.
 \end{aligned}
 \label{si:eq:crossover-transition-counts}
\end{equation}
The counts are pooled before division. For an illustrative example, suppose
1000 eligible pairs start intact at a lag of 20 fs: 120 end fractured
and 880 end intact. Then
\[
 \widehat P_{\mathrm{if}}(20\,\mathrm{fs})
   =\frac{120}{880+120}=0.12.
\]
Thus 12\% of these intact-start observations have a fractured label
20 fs later. These illustrative counts do not imply exactly one transition
in each intact--fractured pair: several switches may occur between the
endpoints. Likewise, an intact--intact pair may include fracture followed
by return. An endpoint probability therefore differs from a first-passage
probability or a count of individual surface crossings.

To understand how these probabilities determine rates, consider a
continuous-time two-state model with intact-to-fractured rate
\(k_{\mathrm{if}}^{\mathrm{core}}\) and reverse rate \(k_{\mathrm{fi}}^{\mathrm{core}}\). If
\(p_{\mathrm f}(t)\) is the fraction carrying the fractured last-core
label in an ensemble of trajectories, then
\begin{equation}
 \frac{dp_{\mathrm f}}{dt}
   =k_{\mathrm{if}}^{\mathrm{core}}(1-p_{\mathrm f})-k_{\mathrm{fi}}^{\mathrm{core}}p_{\mathrm f}.
 \label{si:eq:crossover-population-equation}
\end{equation}
The first term adds fractured population; the second removes it. Writing
\(k_{\mathrm{eff}}^{\mathrm{core}}=k_{\mathrm{if}}^{\mathrm{core}}+k_{\mathrm{fi}}^{\mathrm{core}}\) and
\(\Pi_{\mathrm f}^{\mathrm{LC}}=k_{\mathrm{if}}^{\mathrm{core}}/k_{\mathrm{eff}}^{\mathrm{core}}\), the solution is
\begin{equation}
 p_{\mathrm f}(t)-\Pi_{\mathrm f}^{\mathrm{LC}}
  =[p_{\mathrm f}(0)-\Pi_{\mathrm f}^{\mathrm{LC}}]e^{-k_{\mathrm{eff}}^{\mathrm{core}}t}.
 \label{si:eq:crossover-population-relaxation}
\end{equation}
The population-relaxation time is
\(\tau_{\mathrm{pop}}=1/k_{\mathrm{eff}}^{\mathrm{core}}\), where \(k_{\mathrm{eff}}^{\mathrm{core}}=k_{\mathrm{if}}^{\mathrm{core}}+k_{\mathrm{fi}}^{\mathrm{core}}\).
Population relaxation means the loss of an initial excess in one state.
Its time constant \(\tau_{\mathrm{pop}}\) is the time for that excess to fall by a factor
\(e\), rather than the mean waiting time \(1/k_{\mathrm{if}}^{\mathrm{core}}\) for a
single forward transition. Here \(\Pi_{\mathrm f}^{\mathrm{LC}}\) is the
long-time fractured-label fraction in the two-state description. It is
estimated from the transition probabilities below. It need not equal
one half because the last-core label depends on the trajectory's history,
whereas thermodynamic occupancy uses an instantaneous basin classification.

Starting the model entirely in either state gives
\begin{equation}
 P_{\mathrm{if}}(\tau_{\mathrm{lag}})=\Pi_{\mathrm f}^{\mathrm{LC}}(1-e^{-k_{\mathrm{eff}}^{\mathrm{core}}\tau_{\mathrm{lag}}}),
 \qquad
 P_{\mathrm{fi}}(\tau_{\mathrm{lag}})=(1-\Pi_{\mathrm f}^{\mathrm{LC}})(1-e^{-k_{\mathrm{eff}}^{\mathrm{core}}\tau_{\mathrm{lag}}}).
 \label{si:eq:crossover-model-probabilities}
\end{equation}
Inverting these expressions for each measured pair of probabilities gives
\begin{equation}
 \begin{aligned}
 \widehat k_{\mathrm{eff}}^{\mathrm{core},\mathcal D,c_{\mathrm{an}}}(\tau_{\mathrm{lag}})
  &=-\frac{1}{\tau_{\mathrm{lag}}}\log\!\left[
    1-\widehat P_{\mathrm{if}}^{\mathcal D,c_{\mathrm{an}}}(\tau_{\mathrm{lag}})
     -\widehat P_{\mathrm{fi}}^{\mathcal D,c_{\mathrm{an}}}(\tau_{\mathrm{lag}})\right],\\
 \widehat \Pi_{\mathrm f}^{\mathrm{LC},\mathcal D,c_{\mathrm{an}}}(\tau_{\mathrm{lag}})
  &=\frac{\widehat P_{\mathrm{if}}^{\mathcal D,c_{\mathrm{an}}}(\tau_{\mathrm{lag}})}
    {\widehat P_{\mathrm{if}}^{\mathcal D,c_{\mathrm{an}}}(\tau_{\mathrm{lag}})
     +\widehat P_{\mathrm{fi}}^{\mathcal D,c_{\mathrm{an}}}(\tau_{\mathrm{lag}})},\\
 \widehat k_{\mathrm{if}}^{\mathrm{core},\mathcal D,c_{\mathrm{an}}}(\tau_{\mathrm{lag}})
  &=\widehat k_{\mathrm{eff}}^{\mathrm{core},\mathcal D,c_{\mathrm{an}}}(\tau_{\mathrm{lag}})\,
    \widehat \Pi_{\mathrm f}^{\mathrm{LC},\mathcal D,c_{\mathrm{an}}}(\tau_{\mathrm{lag}}),\\
 \widehat k_{\mathrm{fi}}^{\mathrm{core},\mathcal D,c_{\mathrm{an}}}(\tau_{\mathrm{lag}})
  &=\widehat k_{\mathrm{eff}}^{\mathrm{core},\mathcal D,c_{\mathrm{an}}}(\tau_{\mathrm{lag}})\,
    [1-\widehat \Pi_{\mathrm f}^{\mathrm{LC},\mathcal D,c_{\mathrm{an}}}(\tau_{\mathrm{lag}})].
 \end{aligned}
 \label{si:eq:crossover-implied-rates}
\end{equation}
Thus a validation rate is obtained by taking \(\mathcal D\) to be the
validation trajectory set in Eqs.~\eqref{si:eq:crossover-transition-counts} and
\eqref{si:eq:crossover-implied-rates}, at the chosen lag and condition.
The inversion requires
\(0<\widehat P_{\mathrm{if}}+\widehat P_{\mathrm{fi}}<1\).
Here times and lags are physical, and the inferred rates have units of
inverse time. If the two-state approximation
is accurate, these rates should be approximately independent of lag;
the test examines their remaining lag dependence.

Both estimators use last-core labels. Residence rates
[Eq.~\eqref{si:eq:core-residence-rate}] count completed switches per unit
time carrying the origin-state label; crossover rates are inferred from
finite-lag transition probabilities. With sufficient sampling and temporal
resolution, they should agree when a two-state Markov description is adequate.

\subsection{Calibrating the population clock and testing its accuracy}
\label{si:sec:crossover-validation}

At the primary condition
\(c_{\mathrm{an},0}=(\eta,\Delta t_{\mathrm{obs}})=(0.5,1\,\mathrm{fs})\), the
calibrated directional references define the population clock:
\begin{equation}
 \begin{aligned}
 k_{\nu}^{\mathrm{core},0}&=\widehat k_{\nu}^{\mathrm{core},\mathrm{calibration},c_{\mathrm{an},0}}(\tau_{\mathrm{lag},0}),
 \qquad \nu\in\{\mathrm{if},\mathrm{fi}\},\\
 k_{\mathrm{eff}}^{\mathrm{core},0}&=k_{\mathrm{if}}^{\mathrm{core},0}+k_{\mathrm{fi}}^{\mathrm{core},0},\qquad
 \tau_{\mathrm{pop}}=\frac{1}{k_{\mathrm{eff}}^{\mathrm{core},0}}.
 \end{aligned}
 \label{si:eq:crossover-clock}
\end{equation}
Here \(k_{\mathrm{eff}}^{\mathrm{core},0}\) is the calibrated value of the model's total rate \(k_{\mathrm{eff}}^{\mathrm{core}}\).
This is an operational clock because it is defined by the specified
measurement procedure. It provides a common time unit for comparing
chains: \(\tau_{\mathrm{lag}}/\tau_{\mathrm{pop}}\) is the observation lag as a fraction
of a population-relaxation time. Its interpretation as a relaxation time
of the simulated chain is assessed by the independent validation below.

Let \(k_{\mathrm{eff}}^{\mathrm{ACF}}=\xi k_{\mathrm{eff}}^{\mathrm{ACF},*}\) be the
dimensional counterpart of the reduced tension-ACF decay rate in
Sec.~\ref{si:simulation}, Eq.~\eqref{si:eq:acf}.
The planning time \(\tau_{\mathrm{plan}}=1/k_{\mathrm{eff}}^{\mathrm{ACF}}\)
sets each trajectory's duration to \(500\tau_{\mathrm{plan}}\), rounded
upward to a whole 1-fs sample, and the calibration lag
\(\tau_{\mathrm{lag},0}\) to \(\tau_{\mathrm{plan}}/2\), rounded upward
to a multiple of 4 fs. Table~\ref{si:tab:crossover-clocks} lists this
planning input and \(\tau_{\mathrm{pop}}\), calibrated independently
from the crossover trajectories through Eq.~\eqref{si:eq:crossover-clock}.

The lag grid combines \(4,8,16,32,64,128,256,512,1024\,\mathrm{fs}\)
up to \(\tau_{\mathrm{plan}}\), with the fractions
\(0.005,\allowbreak 0.01,\allowbreak 0.02,\allowbreak 0.03,\allowbreak 0.05,\allowbreak 0.07,\allowbreak 0.1,\allowbreak 0.15,\allowbreak 0.2,\allowbreak 0.3,\allowbreak 0.4,\allowbreak 0.5,\allowbreak 0.7,\allowbreak 1\)
of \(\tau_{\mathrm{plan}}\). All targets are rounded upward to multiples
of 4 fs and duplicate lags are removed. This makes every lag available
on all three sampling grids.

\begin{table}[htbp]
  \centering\small
  \setlength{\tabcolsep}{4pt}
  \caption{Trajectory allocation, planning time and calibrated population
  time. Half the trajectories at each \(N\) belong to calibration and half
  to validation, with equal intact/fractured preparation within each group.
  All times are in fs; intervals are 95\% whole-trajectory bootstrap
  intervals.}
  \label{si:tab:crossover-clocks}
  \begin{tabular}{rrrrl}
  \toprule
  \(N\)&Trajectories&\(\tau_{\mathrm{plan}}\)&\(\tau_{\mathrm{lag},0}\)&\(\tau_{\mathrm{pop}}\) [95\% interval]\\
  \midrule
  \RegisteredCrossoverPopulationRows
  \bottomrule
  \end{tabular}
\end{table}

The same count-to-rate calculation is applied to the validation group
at every tested lag and all nine conditions, including \(c_{\mathrm{an},0}\).
Each result is compared with the
reference for the same direction \(k_{\nu}^{\mathrm{core},0}\):
\begin{equation}
 \delta_{\nu c_{\mathrm{an}}}(\tau_{\mathrm{lag}})
   =\frac{\widehat k_{\nu}^{\mathrm{core},\mathrm{validation},c_{\mathrm{an}}}(\tau_{\mathrm{lag}})}{k_{\nu}^{\mathrm{core},0}}-1,
 \qquad
 E_N(\tau_{\mathrm{lag}})=\max_{\nu,c_{\mathrm{an}}}|\delta_{\nu c_{\mathrm{an}}}(\tau_{\mathrm{lag}})|.
 \label{si:eq:crossover-error}
\end{equation}
The direction index \(\nu\) selects either forward or reverse switching;
\(c_{\mathrm{an}}\) selects the core fraction and sampling interval. Thus \(E_N=0.05\)
means that the largest measured absolute relative discrepancy among the
18 rate estimates is 5\%. Taking the maximum makes an accuracy statement
apply to every tested direction and condition; an average could hide
an individual disagreement. \(E_N\) describes measured disagreement,
before accounting for sampling uncertainty. The main figure plots
\(E_N(\tau_{\mathrm{lag}})\) against
\(\tau_{\mathrm{lag}}/\tau_{\mathrm{pop}}\).
The core fractions \(\eta=0.25,0.5,0.75\), each at observation intervals
of 1, 2 and 4 fs, enter this maximum and the uncertainty bound \(U_N\)
defined below; they are not shown as separate curves.

\subsection{Sampling uncertainty and supported accuracy intervals}
\label{si:sec:crossover-uncertainty}

Sampling uncertainty is quantified with \(B=4000\) bootstrap samples,
indexed by \(b=1,\ldots,B\), by resampling whole trajectories with
replacement within each batch,
prepared basin and calibration/validation group. Each group retains its
original size, and trajectories keep their calibration or validation
role. All linked observations, time pairs and excursions from a selected
trajectory remain together across lags and conditions, retaining their
correlations. In each resample, the calibration trajectories determine
new reference rates \(k_{\nu}^{\mathrm{core},0}\) and a new clock \(\tau_{\mathrm{pop}}\);
the validation trajectories determine new rate estimates at every lag
and condition. Comparing these resampled estimates gives a new value of
\(\delta_{\nu c_{\mathrm{an}}}(\tau_{\mathrm{lag}})\), so uncertainty in both the
reference and validation rates is retained.
Let \(s_{\nu c_{\mathrm{an}}}(\tau_{\mathrm{lag}})\) be the bootstrap standard deviation
of this relative discrepancy.
A single threshold accounts jointly for the tested comparisons. It is drawn
from the bootstrap distribution of the maximum standardized error
\citep{RomanoWolf2005}. Each sample \(b\) gives
\begin{equation}
 Z_{\mathrm{boot}}^{(b)}=\max_{\tau_{\mathrm{lag}},\nu,c_{\mathrm{an}}}
   \frac{|\delta_{\nu c_{\mathrm{an}}}^{(b)}(\tau_{\mathrm{lag}})-\delta_{\nu c_{\mathrm{an}}}(\tau_{\mathrm{lag}})|}
        {s_{\nu c_{\mathrm{an}}}(\tau_{\mathrm{lag}})}.
 \label{si:eq:crossover-bootstrap-maximum}
\end{equation}
The threshold is the 95th percentile of all \(B\) maxima:
\[
 Z_{\mathrm{boot},0.95}=\operatorname{Quantile}_{0.95}
          \!\left(\{Z_{\mathrm{boot}}^{(b)}\}_{b=1}^{B}\right).
\]
Here 0.95 specifies the percentile level, not a sample index \(b\).
This threshold gives the upper curve
\begin{equation}
 U_N(\tau_{\mathrm{lag}})=\max_{\nu,c_{\mathrm{an}}}
       \left\{|\delta_{\nu c_{\mathrm{an}}}(\tau_{\mathrm{lag}})|+
               Z_{\mathrm{boot},0.95}s_{\nu c_{\mathrm{an}}}(\tau_{\mathrm{lag}})\right\}.
 \label{si:eq:crossover-upper-bound}
\end{equation}
Whereas \(E_N(\tau_{\mathrm{lag}})\) is the maximum measured discrepancy,
\(U_N(\tau_{\mathrm{lag}})\) is an upper bound that additionally accounts
for sampling uncertainty. The maximum in
Eq.~\eqref{si:eq:crossover-bootstrap-maximum} makes the bound approximately
95\% simultaneous across the tested lags, directions and conditions within
each \(N\). It therefore supports inspecting the whole tested range,
rather than only one lag chosen in advance.

A broad supported interval contains at least three consecutive
tested lags spanning a factor of four, throughout which \(U_N\) lies
below a chosen relative-error tolerance. The primary accuracy criterion
is 5\%.
Table~\ref{si:tab:crossover-windows} gives the 5\% intervals.
For \(N=7\), the first supported endpoint is 64 fs, preceded by 40 fs
on the tested grid. For \(N=8\)--10, the smallest tested lag already
satisfies the criterion, so the start of the accurate regime cannot be
located more precisely. An interval for a selected endpoint divided by
\(\tau_{\mathrm{pop}}\) describes uncertainty in the clock at that
endpoint, not uncertainty in a physical onset time.

\begin{table}[htbp]
  \centering\small
  \caption{Broad windows supported at 5\% accuracy. Endpoints are discrete
  tested lags in fs.}
  \label{si:tab:crossover-windows}
  \begin{tabular}{rrr}
  \toprule
  \(N\)&Start&Stop\\
  \midrule
  6&\multicolumn{2}{c}{Unresolved broad window}\\
  \RegisteredCrossoverFiveWindowRows
  \bottomrule
  \end{tabular}
\end{table}

For \(N=6\), the discrepancy at 88 fs is
\(\RegisteredCrossoverSixError\%\), with upper bound
\(\RegisteredCrossoverSixErrorUpper\%\). The individual lags that satisfy the 5\% criterion
from 128 to 440 fs do not span the required factor of four, but the results support
a broad 10\% interval from
\(\RegisteredCrossoverSixTenStart\) to \(\RegisteredCrossoverSixTenStop\) fs.
These results describe the accuracy of the two-state
approximation over the tested times, without assigning an exact critical
chain length.

\subsection{Measuring failed returns}
\label{si:sec:crossover-excursions}

An excursion starts when a trajectory leaves one core and ends when it
first enters either core. Only excursions that cross a specified
counting surface \(s\) are retained; the primary choice is
\(s=s_{\mathrm{ad}}=l_{0}d^{*}\), compared with the physical maximum \(l_{0}l_{\max}^{*}(t)\).
An episode is a failed return if it ends in its origin core, and a
successful passage if it reaches the opposite core. Several surface
crossings during one episode still constitute a single excursion.

For episode \(e\), let \(t_e^{\mathrm{out}}\) be the last sampled time
inside the origin core before departure and \(t_e^{\mathrm{in}}\) the
first subsequent sampled arrival in either core. Its duration is
\begin{equation}
 D_e=t_e^{\mathrm{in}}-t_e^{\mathrm{out}}.
 \label{si:eq:crossover-excursion-duration}
\end{equation}
This duration includes travel from the origin core to the counting surface
and onward to a core; it does not begin at the surface
crossing. If a qualifying episode is still unfinished when its trajectory
ends, its final outcome and full duration are unknown. Such right-censored
episodes are counted separately and excluded from completed-event
probabilities and duration summaries
(Table~\ref{si:tab:crossover-counts}).

For origin \(a\) and surface \(s\), denote the numbers of completed
failed and successful excursions by \(n_{\mathrm{return},a}(s)\)
and \(n_{\mathrm{pass},a}(s)\). The estimated return probability is
\begin{equation}
 \widehat P_{\mathrm{return},a}(s)=
 \frac{n_{\mathrm{return},a}(s)}
      {n_{\mathrm{return},a}(s)+n_{\mathrm{pass},a}(s)}.
 \label{si:eq:crossover-return-probability}
\end{equation}
These counts pool qualifying episodes from all trajectories, including
both the calibration and validation groups, and distinguish the actual
origin core of each episode. This origin need not be the basin in which
that trajectory was initially prepared.

For failed returns from each origin, sort the completed durations
\(D_{(1)}\le\cdots\le D_{(n)}\). The median and 90th percentile are empirical
quantiles with linear interpolation between adjacent ordered values:
\begin{equation}
 h=1+(n-1)p,\quad j=\lfloor h\rfloor,\quad
 \widehat D_p=(1-h+j)D_{(j)}+(h-j)D_{(j+1)},
 \qquad p=0.5,\ 0.9.
 \label{si:eq:crossover-duration-quantile}
\end{equation}
At an integer rank the corresponding ordered value is used. Dividing
\(\widehat D_p\) by \(\tau_{\mathrm{pop}}\) gives the fraction of a
population-relaxation time occupied by such an excursion. The main
figure's second panel uses failed returns at the adiabatic surface;
Figure~\ref{si:fig:return-duration-robustness} also shows physical
return durations and their dependence on the counting surface.

The same whole-trajectory bootstrap supplies uncertainty for these
statistics. For each sample \(b\), episodes are recounted and the
duration statistic \(\widehat D_p^{(b)}\) and calibration clock
\(\tau_{\mathrm{pop}}^{(b)}\) are recomputed from that same sample, then divided:
\begin{equation}
 \left(\frac{\widehat D_p}{\tau_{\mathrm{pop}}}\right)^{(b)}
   =\frac{\widehat D_p^{(b)}}{\tau_{\mathrm{pop}}^{(b)}}.
 \label{si:eq:crossover-paired-duration}
\end{equation}
The superscript \((b)\) labels a bootstrap sample, not an exponent.
Using the same trajectory draws retains the dependence between numerator
and denominator. The 2.5th and 97.5th percentiles of the resulting ratio
distribution give the reported pointwise 95\% bootstrap intervals.

\begin{table}[htbp]
 \centering\small
 \caption{Primary adiabatic-surface excursion counts at 1-fs sampling.
 Failed returns and successful passages are completed episodes;
 the last column counts final right-censored qualifying episodes.
 Uncertainty resamples trajectories, not individual episodes.}
 \label{si:tab:crossover-counts}
 \begin{tabular}{rrrr}
 \toprule
 \(N\)&Returns&Successful passages&Right-censored\\
 \midrule
 \multicolumn{4}{c}{Intact origin}\\
 \RegisteredCrossoverIntactCountRows
 \midrule
 \multicolumn{4}{c}{Fractured origin}\\
 \RegisteredCrossoverFracturedCountRows
 \bottomrule
 \end{tabular}
\end{table}

For the adiabatic surface, the median failed-return durations divided by
\(\tau_{\mathrm{pop}}\) decrease from
\(\RegisteredCrossoverRatioSixI/\RegisteredCrossoverRatioSixF\)
(intact/fractured origin) at \(N=6\) to
\(\RegisteredCrossoverRatioTenI/\RegisteredCrossoverRatioTenF\) at
\(N=10\). Physical return durations increase modestly, but the population
time increases much faster. Individual failed excursions therefore occupy
a progressively smaller part of the population-relaxation time. This
comparison measures separation between the specified excursion and
population timescales.

\subsection{How the counting surface selects excursions}
\label{si:sec:crossover-surfaces}

The excursion calculation is repeated for the adiabatic surface
\(s_{\mathrm{ad}}\), the intact-conditioned tagged-site PMF maximum
\(s_{\mathrm{PMF}}\), and the outer surface
\(s_{\mathrm{out}}=2s_{\mathrm{PMF}}-s_{\mathrm{ad}}\)
(Table~\ref{si:tab:crossover-coordinates}). In these excursion figures,
the primary cores at \(\eta=0.5\), trajectories, last-core labels and
calibrated population clock stay fixed. Only the
surface that an excursion must cross to be counted changes. An
intact-origin excursion must travel farther outward to qualify at
\(s_{\mathrm{out}}\); a fractured-origin excursion reaches it earlier on
its inward journey.

\begin{figure}[htbp]
 \centering
 \includegraphics[width=16cm]{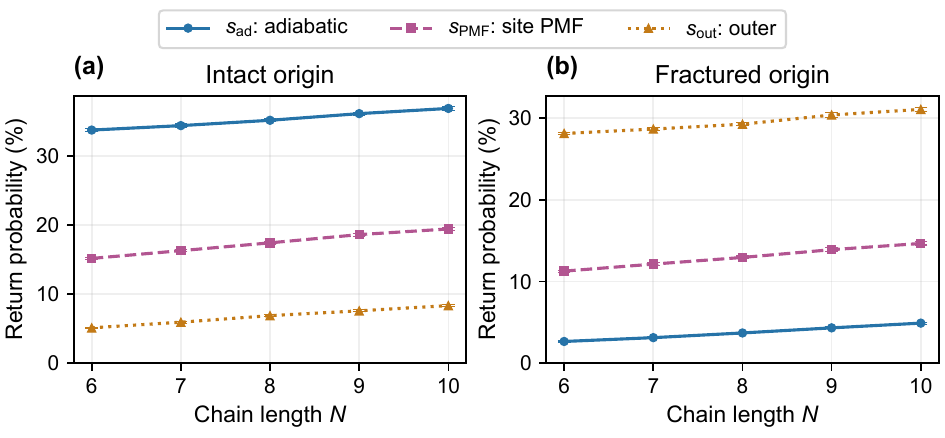}
 \caption{Completed-event return probabilities for intact (a) and fractured
 (b) origins, using the three qualifying surfaces and fixed primary cores.
 Bars are pointwise 95\% whole-trajectory bootstrap intervals; lines guide
 the eye. Moving the surface changes the selected excursion ensemble,
 not the underlying trajectories or population clock.}
 \label{si:fig:surface-return-probabilities}
\end{figure}

At every sampled \(N\), moving the surface outward reduces the
intact-origin return probability and increases the fractured-origin one
(Fig.~\ref{si:fig:surface-return-probabilities}). At \(N=6\), the
intact-minus-fractured gap at the PMF surface is
\(\RegisteredCrossoverPmfGap\) percentage points, with 95\% interval
\([\RegisteredCrossoverPmfGapLower,\RegisteredCrossoverPmfGapUpper]\).
The PMF surface thus reduces the directional imbalance relative to the
adiabatic choice. Equal probabilities are not required there: the two
excursion ensembles are selected by their different core histories,
whereas the PMF describes a tagged-site equilibrium distribution.
Likewise, Eq.~\eqref{si:eq:crossover-return-probability} is a completed-event
fraction, not a reactive-flux transmission amplitude.

\begin{figure}[H]
 \centering
 \includegraphics[width=16cm]{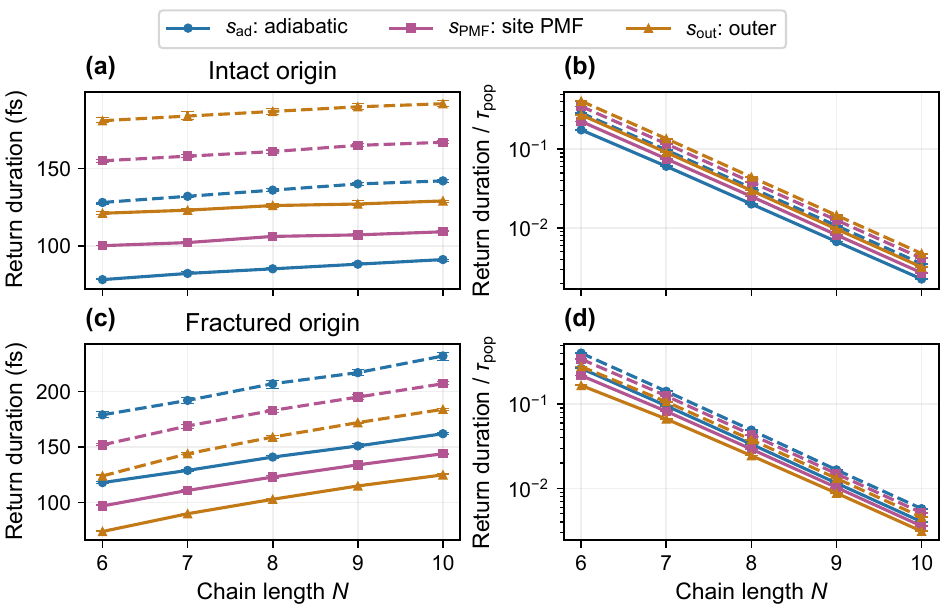}
 \caption{Median (solid) and 90th-percentile (dashed) completed failed-return
 duration in fs (left panels) and divided by the population time (right),
 for intact (upper panels) and fractured (lower) origins. The three qualifying surfaces use the same
 primary cores and clock. Bars are pointwise 95\% whole-trajectory
 bootstrap intervals; duration/clock covariance is retained. Lines guide
 the eye.}
 \label{si:fig:return-duration-robustness}
\end{figure}

Figure~\ref{si:fig:return-duration-robustness} shows how this selection
changes the median and 90th-percentile failed-return durations, in fs
and relative to the fixed population clock. All three surfaces lie
between the primary core boundaries. At \(N=6\), the outer surface lies
beyond the alternative \(\eta=0.75\) fractured-core boundary, so the
three-surface comparison uses only the primary \(\eta=0.5\) cores.

\FloatBarrier

%% file: references.bib
@article{RomanoWolf2005,
  author  = {Romano, Joseph P. and Wolf, Michael},
  title   = {Exact and Approximate Stepdown Methods for Multiple Hypothesis Testing},
  journal = {Journal of the American Statistical Association},
  volume  = {100},
  number  = {469},
  pages   = {94--108},
  year    = {2005},
  doi     = {10.1198/016214504000000539}
}

@article{Crist1984,
  author  = {Crist, Jr., B. and Oddershede, J. and Sabin, J. R. and
             Perram, J. W. and Ratner, M. A.},
  title   = {Polymer fracture---A simple model for chain scission},
  journal = {Journal of Polymer Science: Polymer Physics Edition},
  volume  = {22},
  pages   = {881--897},
  year    = {1984},
  doi     = {10.1002/pol.1984.180220510}
}

@article{Oliveira1998,
  author  = {Oliveira, F. A.},
  title   = {Transition-state analysis for fracture nucleation in polymers:
             The {Lennard-Jones} chain},
  journal = {Physical Review B},
  volume  = {57},
  pages   = {10576--10582},
  year    = {1998},
  doi     = {10.1103/PhysRevB.57.10576}
}

@article{PuthurSebastian2002,
  author  = {Puthur, Rosabella and Sebastian, K. L.},
  title   = {Theory of polymer breaking under tension},
  journal = {Physical Review B},
  volume  = {66},
  pages   = {024304},
  year    = {2002},
  doi     = {10.1103/PhysRevB.66.024304}
}

@article{PetersTrout2006,
  author  = {Peters, Baron and Trout, Bernhardt L.},
  title   = {Obtaining reaction coordinates by likelihood maximization},
  journal = {The Journal of Chemical Physics},
  volume  = {125},
  pages   = {054108},
  year    = {2006},
  doi     = {10.1063/1.2234477}
}

@article{Sain2006,
  author  = {Sain, A. and Dias, C. L. and Grant, M.},
  title   = {Rupture of an extended object: A many-body {Kramers}
             calculation},
  journal = {Physical Review E},
  volume  = {74},
  pages   = {046111},
  year    = {2006},
  doi     = {10.1103/PhysRevE.74.046111}
}

@article{Lee2009,
  author  = {Lee, C. F.},
  title   = {Thermal breakage of a discrete one-dimensional string},
  journal = {Physical Review E},
  volume  = {80},
  pages   = {031134},
  year    = {2009},
  doi     = {10.1103/PhysRevE.80.031134}
}

@article{Ghosh2010,
  author  = {Ghosh, A. and Dimitrov, D. I. and Rostiashvili, V. G. and
             Milchev, A. and Vilgis, T. A.},
  title   = {Thermal breakage and self-healing of a polymer chain under
             tensile stress},
  journal = {The Journal of Chemical Physics},
  volume  = {132},
  pages   = {204902},
  year    = {2010},
  doi     = {10.1063/1.3427245}
}

@article{Theodorou2024,
  author  = {Theodorou, Doros N.},
  title   = {Simple model for the fracture of a polymer chain:
             Single-bond potential of mean force and tension-based rate
             constants for chain rupture},
  journal = {The Journal of Chemical Physics},
  volume  = {161},
  pages   = {184905},
  year    = {2024},
  doi     = {10.1063/5.0239831}
}

@article{Kramers1940,
  author  = {Kramers, H. A.},
  title   = {Brownian motion in a field of force and the diffusion model
             of chemical reactions},
  journal = {Physica},
  volume  = {7},
  pages   = {284--304},
  year    = {1940},
  doi     = {10.1016/S0031-8914(40)90098-2}
}

@article{GroteHynes1980,
  author  = {Grote, Richard F. and Hynes, James T.},
  title   = {The stable states picture of chemical reactions. {II}.
             {Rate} constants for condensed and gas phase reaction models},
  journal = {The Journal of Chemical Physics},
  volume  = {73},
  pages   = {2715--2732},
  year    = {1980},
  doi     = {10.1063/1.440485}
}

@article{Zhurkov1965,
  author  = {Zhurkov, S. N.},
  title   = {Kinetic concept of the strength of solids},
  journal = {International Journal of Fracture Mechanics},
  volume  = {1},
  pages   = {311--323},
  year    = {1965},
  doi     = {10.1007/BF03545562}
}

@incollection{Bennett1977,
  author    = {Bennett, Charles H.},
  title     = {Molecular Dynamics and Transition State Theory: The Simulation
               of Infrequent Events},
  booktitle = {Algorithms for Chemical Computations},
  series    = {ACS Symposium Series},
  volume    = {46},
  pages     = {63--97},
  publisher = {American Chemical Society},
  year      = {1977},
  doi       = {10.1021/bk-1977-0046.ch004}
}

@article{Chandler1978,
  author  = {Chandler, David},
  title   = {Statistical mechanics of isomerization dynamics in liquids and
             the transition state approximation},
  journal = {The Journal of Chemical Physics},
  volume  = {68},
  number  = {6},
  pages   = {2959--2970},
  year    = {1978},
  doi     = {10.1063/1.436049}
}

@article{Vernerey2018,
  author  = {Vernerey, F. J. and Brighenti, R. and Long, R. and Shen, T.},
  title   = {Statistical Damage Mechanics of Polymer Networks},
  journal = {Macromolecules},
  volume  = {51},
  pages   = {6609--6622},
  year    = {2018},
  doi     = {10.1021/acs.macromol.8b01052}
}

@inproceedings{Anogiannakis2015,
  author    = {Anogiannakis, S. D. and Tzoumanekas, C. and Georgilas, V. and
               Theodorou, D. N.},
  title     = {Molecular modeling of deformation and fracture of polymer networks},
  booktitle = {Book of Abstracts of the 16th International Conference on
               Deformation, Yield, and Fracture of Polymers ({DYFP2015})},
  editor    = {Gong, J. P. and Govaert, L.},
  publisher = {Eindhoven University of Technology},
  address   = {Eindhoven},
  pages     = {21--24},
  year      = {2015}
}

@article{Sgouros2025,
  author  = {Sgouros, A. P. and Knippenberg, S. and Bocahut, A. and
             Rauscher, P. M. and Sikora, B. and Caputo, S. and Choi, H.-S. and
             Finsy, V. and Guillaume, M. and Theodorou, D. N.},
  title   = {Mesoscale modelling of polymer-mediated adhesion: application
             to tack tests},
  journal = {Molecular Systems Design \& Engineering},
  volume  = {10},
  pages   = {394--412},
  year    = {2025},
  doi     = {10.1039/D4ME00199K}
}

@article{Terzis2002,
  author  = {Terzis, A. F. and Theodorou, D. N. and Stroeks, A.},
  title   = {Entanglement Network of the Polypropylene/Polyamide Interface.
             3. Deformation to Fracture},
  journal = {Macromolecules},
  volume  = {35},
  pages   = {508--521},
  year    = {2002},
  doi     = {10.1021/ma010691z}
}

@article{Suhir2013,
  author  = {Suhir, E. and Kang, S.-M.},
  title   = {{Boltzmann Arrhenius Zhurkov (BAZ)} Model in
             Physics-of-Materials Problems},
  journal = {Modern Physics Letters B},
  volume  = {27},
  number  = {13},
  pages   = {1330009},
  year    = {2013},
  doi     = {10.1142/S0217984913300093}
}

@article{Morse1929,
  author  = {Morse, Philip M.},
  title   = {Diatomic Molecules According to the Wave Mechanics.
             {II}. Vibrational Levels},
  journal = {Physical Review},
  volume  = {34},
  pages   = {57--64},
  year    = {1929},
  doi     = {10.1103/PhysRev.34.57}
}

@article{Eyring1935,
  author  = {Eyring, Henry},
  title   = {The Activated Complex in Chemical Reactions},
  journal = {The Journal of Chemical Physics},
  volume  = {3},
  pages   = {107--115},
  year    = {1935},
  doi     = {10.1063/1.1749604}
}

@article{FritschButland1984,
  author  = {Fritsch, F. N. and Butland, J.},
  title   = {A Method for Constructing Local Monotone Piecewise Cubic Interpolants},
  journal = {SIAM Journal on Scientific and Statistical Computing},
  volume  = {5},
  number  = {2},
  pages   = {300--304},
  year    = {1984},
  doi     = {10.1137/0905021}
}

@article{Virtanen2020,
  author  = {Virtanen, Pauli and Gommers, Ralf and Oliphant, Travis E. and
             Haberland, Matt and Reddy, Tyler and Cournapeau, David and
             Burovski, Evgeni and Peterson, Pearu and Weckesser, Warren and
             Bright, Jonathan and {van der Walt}, St{\'e}fan J. and
             Brett, Matthew and Wilson, Joshua and Millman, K. Jarrod and
             Mayorov, Nikolay and Nelson, Andrew R. J. and Jones, Eric and
             Kern, Robert and Larson, Eric and Carey, C J and
             Polat, {\.I}lhan and Feng, Yu and Moore, Eric W. and
             {VanderPlas}, Jake and Laxalde, Denis and Perktold, Josef and
             Cimrman, Robert and Henriksen, Ian and Quintero, E. A. and
             Harris, Charles R. and Archibald, Anne M. and
             Ribeiro, Ant{\^o}nio H. and Pedregosa, Fabian and
             {van Mulbregt}, Paul and {SciPy 1.0 Contributors}},
  title   = {{SciPy} 1.0: Fundamental Algorithms for Scientific Computing in {Python}},
  journal = {Nature Methods},
  volume  = {17},
  pages   = {261--272},
  year    = {2020},
  doi     = {10.1038/s41592-019-0686-2}
}

@article{Thompson2022,
  author  = {Thompson, Aidan P. and others},
  title   = {{LAMMPS} -- a flexible simulation tool for particle-based materials
             modeling at the atomic, meso, and continuum scales},
  journal = {Computer Physics Communications},
  volume  = {271},
  pages   = {108171},
  year    = {2022},
  doi     = {10.1016/j.cpc.2021.108171}
}

@article{Langer1969,
  author = {Langer, J. S.},
  title = {Statistical theory of the decay of metastable states},
  journal = {Annals of Physics},
  volume = {54},
  pages = {258--275},
  year = {1969},
  doi = {10.1016/0003-4916(69)90153-5}
}

@article{Bartsch2005,
  author  = {Bartsch, Thomas and Uzer, T. and Hernandez, Rigoberto},
  title   = {Stochastic transition states: Reaction geometry amidst noise},
  journal = {The Journal of Chemical Physics},
  volume  = {123},
  number  = {20},
  pages   = {204102},
  year    = {2005},
  doi     = {10.1063/1.2109827}
}

@article{Zhou2025,
  author  = {Zhou, Qingyuan and Bakaev, Artur and Lavacchi, Laura and
             Netz, Roland R. and Dalton, Benjamin A.},
  title   = {Rapid state-recrossing kinetics and slow escape kinetics in
             non-{Markovian} systems},
  journal = {Physical Review E},
  volume  = {111},
  pages   = {064110},
  year    = {2025},
  doi     = {10.1103/PhysRevE.111.064110}
}

@article{Buche2021,
  author = {Buche, Michael R. and Silberstein, Meredith N.},
  title = {Chain breaking in the statistical mechanical constitutive theory of polymer networks},
  journal = {Journal of the Mechanics and Physics of Solids},
  volume = {156},
  pages = {104593},
  year = {2021},
  doi = {10.1016/j.jmps.2021.104593}
}

@article{White2025,
  author = {White, Zachary T. and Smith, Aidan M. and Vernerey, Franck J.},
  title = {Mechanical cooperation between time-dependent and covalent bonds in molecular damage of polymer networks},
  journal = {Communications Physics},
  volume = {8},
  pages = {265},
  year = {2025},
  doi = {10.1038/s42005-025-02192-0}
}

@article{Lazzeri2023,
  author = {Lazzeri, Gianmarco and Jung, Hendrik and Bolhuis, Peter G. and Covino, Roberto},
  title = {Molecular Free Energies, Rates, and Mechanisms from Data-Efficient Path Sampling Simulations},
  journal = {Journal of Chemical Theory and Computation},
  volume = {19},
  pages = {9060--9076},
  year = {2023},
  doi = {10.1021/acs.jctc.3c00821}
}

@article{Slootman2022,
  author = {Slootman, J. and Yeh, C. J. and Millereau, P. and Comtet, J. and Creton, C.},
  title = {A molecular interpretation of the toughness of multiple network elastomers at high temperature},
  journal = {Proceedings of the National Academy of Sciences},
  volume = {119},
  pages = {e2116127119},
  year = {2022},
  doi = {10.1073/pnas.2116127119}
}

@article{Zhao2025,
  author = {Zhao, D. and Cartier, A. and Narita, T. and Lechenault, F. and Creton, C. and Ciccotti, M.},
  title = {Why cutting is easier than tearing elastomers},
  journal = {Nature Communications},
  volume = {16},
  pages = {3203},
  year = {2025},
  doi = {10.1038/s41467-025-58483-1}
}

@article{Lamont2021,
  author = {Lamont, S. C. and Mulderrig, J. and Bouklas, N. and Vernerey, F. J.},
  title = {Rate-Dependent Damage Mechanics of Polymer Networks with Reversible Bonds},
  journal = {Macromolecules},
  volume = {54},
  pages = {10801--10813},
  year = {2021},
  doi = {10.1021/acs.macromol.1c01943}
}

@article{Lamont2025,
  author = {Lamont, S. C. and Bouklas, N. and Vernerey, F. J.},
  title = {Cohesive instability in elastomers: insights from a crosslinked {Van der Waals} fluid model},
  journal = {International Journal of Fracture},
  volume = {249},
  pages = {20},
  year = {2025},
  doi = {10.1007/s10704-025-00840-8}
}

@article{Barney2022,
  author = {Barney, C. W. and Ye, Z. and Sacligil, I. and McLeod, K. R. and
            Zhang, H. and Tew, G. N. and Riggleman, R. A. and Crosby, A. J.},
  title = {Fracture of model end-linked networks},
  journal = {Proceedings of the National Academy of Sciences},
  volume = {119},
  pages = {e2112389119},
  year = {2022},
  doi = {10.1073/pnas.2112389119}
}

@article{Mao2017,
  author = {Mao, Yunwei and Talamini, Brandon and Anand, Lallit},
  title = {Rupture of polymers by chain scission},
  journal = {Extreme Mechanics Letters},
  volume = {13},
  pages = {17--24},
  year = {2017},
  doi = {10.1016/j.eml.2017.01.003}
}

@article{Sliozberg2014,
  author = {Sliozberg, Y. R. and Hoy, R. S. and Mrozek, R. A. and
            Lenhart, J. L. and Andzelm, J. W.},
  title = {Role of entanglements and bond scission in high strain-rate deformation of polymer gels},
  journal = {Polymer},
  volume = {55},
  pages = {2543--2551},
  year = {2014},
  doi = {10.1016/j.polymer.2014.03.051}
}

@incollection{Paturej2013,
  author = {Paturej, Jaroslaw and Milchev, Andrey and Rostiashvili, Vakhtang G. and Vilgis, Thomas A.},
  title = {Rupture Dynamics of Macromolecules},
  booktitle = {Fracture Mechanics and Statistical Mechanics of Reinforced Elastomeric Blends},
  series = {Lecture Notes in Applied and Computational Mechanics},
  volume = {70},
  pages = {1--42},
  publisher = {Springer},
  year = {2013},
  doi = {10.1007/978-3-642-37910-9_1}
}

@article{Ansari2027,
  author = {Ansari, Mohammad A. and Liechti, Kenneth M. and Makarov, Dmitrii E. and Huang, Rui},
  title = {Energy barriers for reversible chain scission and healing under tension with displacement control},
  journal = {Journal of the Mechanics and Physics of Solids},
  volume = {218},
  pages = {106858},
  year = {2027},
  doi = {10.1016/j.jmps.2026.106858}
}

@article{Schutte2011,
  author = {Sch{\"u}tte, Christof and No{\'e}, Frank and Lu, Jianfeng and
            Sarich, Marco and Vanden-Eijnden, Eric},
  title = {{Markov} state models based on milestoning},
  journal = {The Journal of Chemical Physics},
  volume = {134},
  pages = {204105},
  year = {2011},
  doi = {10.1063/1.3590108}
}

@article{Metropolis1953,
  author = {Metropolis, Nicholas and Rosenbluth, Arianna W. and
            Rosenbluth, Marshall N. and Teller, Augusta H. and Teller, Edward},
  title = {Equation of State Calculations by Fast Computing Machines},
  journal = {The Journal of Chemical Physics},
  volume = {21},
  pages = {1087--1092},
  year = {1953},
  doi = {10.1063/1.1699114}
}

@article{KaplanMeier1958,
  author = {Kaplan, E. L. and Meier, Paul},
  title = {Nonparametric Estimation from Incomplete Observations},
  journal = {Journal of the American Statistical Association},
  volume = {53},
  number = {282},
  pages = {457--481},
  year = {1958},
  doi = {10.1080/01621459.1958.10501452}
}

@article{Ayer1955,
  author = {Ayer, Miriam and Brunk, H. D. and Ewing, G. M. and
            Reid, W. T. and Silverman, Edward},
  title = {An Empirical Distribution Function for Sampling with Incomplete Information},
  journal = {The Annals of Mathematical Statistics},
  volume = {26},
  number = {4},
  pages = {641--647},
  year = {1955},
  doi = {10.1214/aoms/1177728423}
}

@article{Byrd1995,
  author = {Byrd, Richard H. and Lu, Peihuang and Nocedal, Jorge and Zhu, Ciyou},
  title = {A Limited Memory Algorithm for Bound Constrained Optimization},
  journal = {SIAM Journal on Scientific Computing},
  volume = {16},
  number = {5},
  pages = {1190--1208},
  year = {1995},
  doi = {10.1137/0916069}
}

@article{FieldWelsh2007,
  author = {Field, C. A. and Welsh, A. H.},
  title = {Bootstrapping clustered data},
  journal = {Journal of the Royal Statistical Society: Series B (Statistical Methodology)},
  volume = {69},
  number = {3},
  pages = {369--390},
  year = {2007},
  doi = {10.1111/j.1467-9868.2007.00593.x}
}

@misc{PetersenPedersen2012,
  author = {Petersen, Kaare Brandt and Pedersen, Michael Syskind},
  title = {The {Matrix Cookbook}},
  year = {2012},
  month = nov,
  publisher = {Technical University of Denmark},
  note = {Version November 15, 2012},
  url = {https://www2.imm.dtu.dk/pubdb/pubs/3274-full.html}
}

@article{Beech2023,
  author = {Beech, Haley K. and Wang, Shu and Sen, Devosmita and Rota, Dechen and
            Kouznetsova, Tatiana B. and Arora, Akash and Rubinstein, Michael and
            Craig, Stephen L. and Olsen, Bradley D.},
  title = {Reactivity-Guided Depercolation Processes Determine Fracture Behavior
           in End-Linked Polymer Networks},
  journal = {ACS Macro Letters},
  volume = {12},
  pages = {1685--1691},
  year = {2023},
  doi = {10.1021/acsmacrolett.3c00559}
}

@article{Arora2022,
  author = {Arora, Akash and Lin, Tzyy-Shyang and Olsen, Bradley D.},
  title = {Coarse-Grained Simulations for Fracture of Polymer Networks:
           Stress Versus Topological Inhomogeneities},
  journal = {Macromolecules},
  volume = {55},
  pages = {4--14},
  year = {2022},
  doi = {10.1021/acs.macromol.1c01689}
}

@article{AraujoBrassart2025,
  author = {Araujo, Lucas Mangas and Brassart, Laurence},
  title = {Force-biased chemical degradation in rubbery networks:
           Insights from discrete network simulations},
  journal = {Extreme Mechanics Letters},
  volume = {77},
  pages = {102344},
  year = {2025},
  doi = {10.1016/j.eml.2025.102344}
}

@article{Kothari2018,
  author = {Kothari, Konik and Hu, Yuhang and Gupta, Sahil and Elbanna, Ahmed},
  title = {Mechanical Response of Two-Dimensional Polymer Networks:
           Role of Topology, Rate Dependence, and Damage Accumulation},
  journal = {Journal of Applied Mechanics},
  volume = {85},
  pages = {031008},
  year = {2018},
  doi = {10.1115/1.4038883}
}

@article{BucheRimsza2023,
  author = {Buche, Michael R. and Rimsza, Jessica M.},
  title = {Modeling single-molecule stretching experiments using
           statistical thermodynamics},
  journal = {Physical Review E},
  volume = {108},
  pages = {064503},
  year = {2023},
  doi = {10.1103/PhysRevE.108.064503}
}
